\documentclass[acmtog, nonacm]{acmart}
\acmSubmissionID{836}
\acmJournal{TOG}

\usepackage{booktabs}
\usepackage{amsmath}
\usepackage{graphicx}
\usepackage{tabularx}
\usepackage{subcaption}
\usepackage[ruled]{algorithm2e}
\usepackage[svgnames]{xcolor} 
\usepackage{makecell}

\SetAlFnt{\small}
\SetAlCapFnt{\small}
\SetAlCapNameFnt{\small}
\SetAlCapHSkip{0pt}

\author{Etai Sella}
\affiliation{%
  \institution{Tel Aviv University}
  \country{Israel}
}
\affiliation{%
  \institution{Snap Research}
  \country{Israel}
}
\email{etaisella@gmail.com}
\orcid{0009-0002-2119-9808}

\author{Yoav Baron}
\affiliation{%
  \institution{Tel Aviv University}
  \country{Israel}
}
\email{yvbrn13@gmail.com}
\orcid{0009-0001-0075-0616}

\author{Hadar Averbuch-Elor}
\affiliation{%
  \institution{Cornell University}
  \city{New York}
  \state{New York}
  \country{USA}
}
\email{hadarelor@cornell.edu}
\orcid{0000-0003-3476-0940}

\author{Daniel Cohen-Or}
\affiliation{%
  \institution{Tel Aviv University}
  \country{Israel}
}
\affiliation{%
  \institution{Snap Research}
  \country{Israel}
}
\email{cohenor@gmail.com}
\orcid{0000-0001-6777-7445}

\author{Or Patashnik}
\affiliation{%
  \institution{Tel Aviv University}
  \country{Israel}
}
\affiliation{%
  \institution{Snap Research}
  \country{Israel}
}
\email{orpatashnik@gmail.com}
\orcid{0000-0001-7757-6137}

\newbox\jsavebox

\usepackage{array}

\newcommand{\new}[1]{{#1}}

\newcommand{\op}[1]{#1}

\newcommand{\ignorethis}[1]{}

\begin{document}

\title{LooseRoPE: Content-aware Attention Manipulation \\ for Semantic Harmonization}

\begin{abstract}

Recent diffusion-based image editing methods commonly rely on text or high-level instructions to guide the generation process, offering intuitive but coarse control.
In contrast, we focus on explicit, prompt-free editing, where the user directly specifies the modification by cropping and pasting an object or sub-object into a chosen location within an image. 
This operation affords precise spatial and visual control, yet it introduces a fundamental challenge: preserving the identity of the pasted object while harmonizing it with its new context. 
We observe that attention maps in diffusion-based editing models inherently govern whether image regions are preserved or adapted for coherence. 
Building on this insight, we introduce LooseRoPE, a saliency-guided modulation of rotational positional encoding (RoPE) that loosens the positional constraints to continuously control the attention field of view.
By relaxing RoPE in this manner, our method smoothly steers the model’s focus between faithful preservation of the input image and coherent harmonization of the inserted object, enabling a balanced trade-off between identity retention and contextual blending.
Our approach provides a flexible and intuitive framework for image editing, achieving seamless compositional results without textual descriptions or complex user input.

\end{abstract}

\begin{teaserfigure}
    \centering
\includegraphics[width=\textwidth]{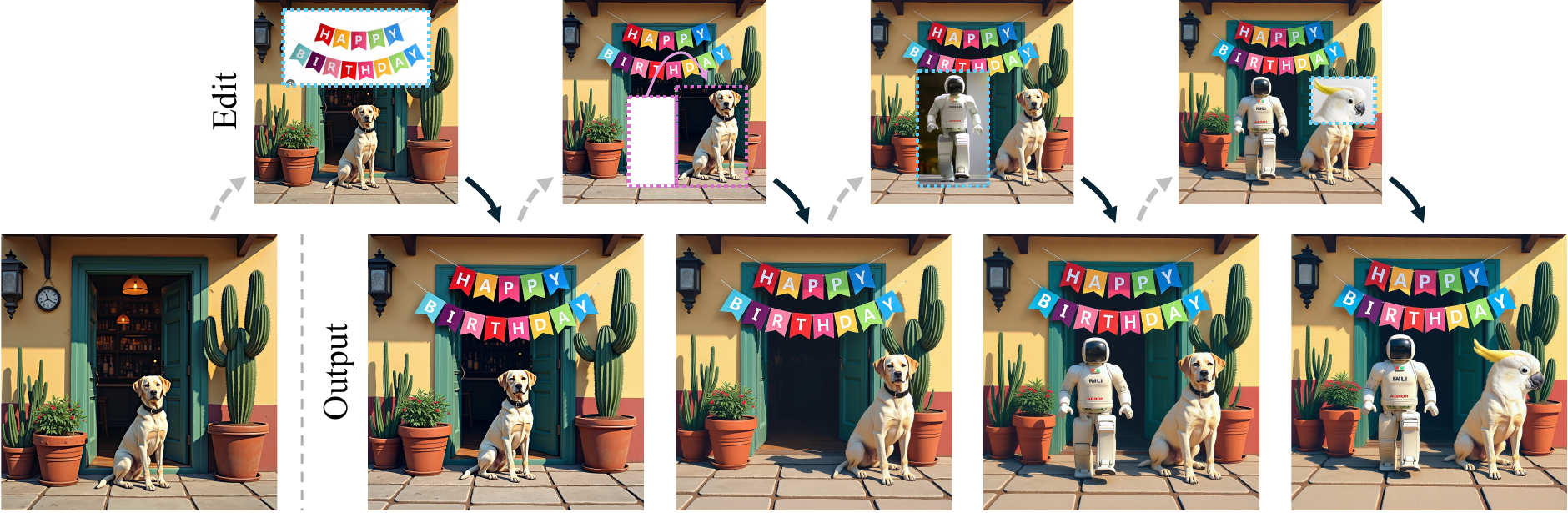}
\caption{We introduce \textbf{LooseRoPE}, a training-free image editing algorithm that turns crudely edited inputs (top row) into coherent, high-quality results (bottom row). In each example, cropped regions are pasted either from other images (\textcolor{cyan}{\textbf{blue frames}}) or moved within the same image (\textcolor{magenta}{\textbf{magenta frames}}), sometimes leaving holes behind. Without any text prompts or additional supervision, LooseRoPE harmonizes the pasted content with its new context, producing seamless and semantically consistent outputs.}
\label{fig:teaser}
\Description{A description of the teaser image for accessibility purposes.}

\end{teaserfigure}

\maketitle

\section{Introduction}
\label{sec:intro}

In recent years, we have witnessed remarkable progress in image editing~\cite{hertz2022prompttoprompt, labs2025flux1kontextflowmatching, brooks2023instructpix2pix, huberman2025image}, largely driven by diffusion models that respond to natural language prompts~\cite{ho2020denoising, song2020score, rombach2022highresolution}. These advances have made image manipulation intuitive and accessible, allowing users to modify content through natural language descriptions. 
Yet, this form of control remains inherently coarse, as many fine-grained aspects of an edit cannot be precisely conveyed through text, such as the exact location, shape, or appearance details of the modification.

To address this challenge, we revisit the compositional editing task and define a setting in which the user directly specifies the modification by cropping and pasting an object or sub-object into a chosen location within a target image (see Figure~\ref{fig:teaser}).
This operation affords precise spatial and visual control, yet it introduces a fundamental challenge: preserving the identity of the pasted object while harmonizing it with its new context. 

Previous approaches to compositional editing often favor one of the two goals at the expense of the other. Classical harmonization methods focus on accurately preserving the pasted object’s appearance, while ensuring local blending and color consistency with the background~\cite{ perez2023poisson, tsai2017deep, cong2020dovenet, jiang2021ssh}.
Yet, these methods typically operate at the pixel or illumination level, and therefore cannot generate substantial semantic or structural adjustments that may be required for a truly coherent composition. In contrast, recent diffusion-based approaches for compositional editing are able to generate globally coherent images~\cite{chen2024anydoor, song2024imprint, wang2024ms}, but often compromise the fidelity of the inserted object, altering its appearance or identity in the process.

\begin{figure}
    \centering
    \setlength{\tabcolsep}{1pt}
    \begin{tabular}{ccccc} 

    & \multicolumn{2}{c}{Neglect} & \multicolumn{2}{c}{Suppression} \\
    
    \raisebox{12pt}{\rotatebox{90}{\centering Input}}
    &
    \includegraphics[width=0.23\linewidth]{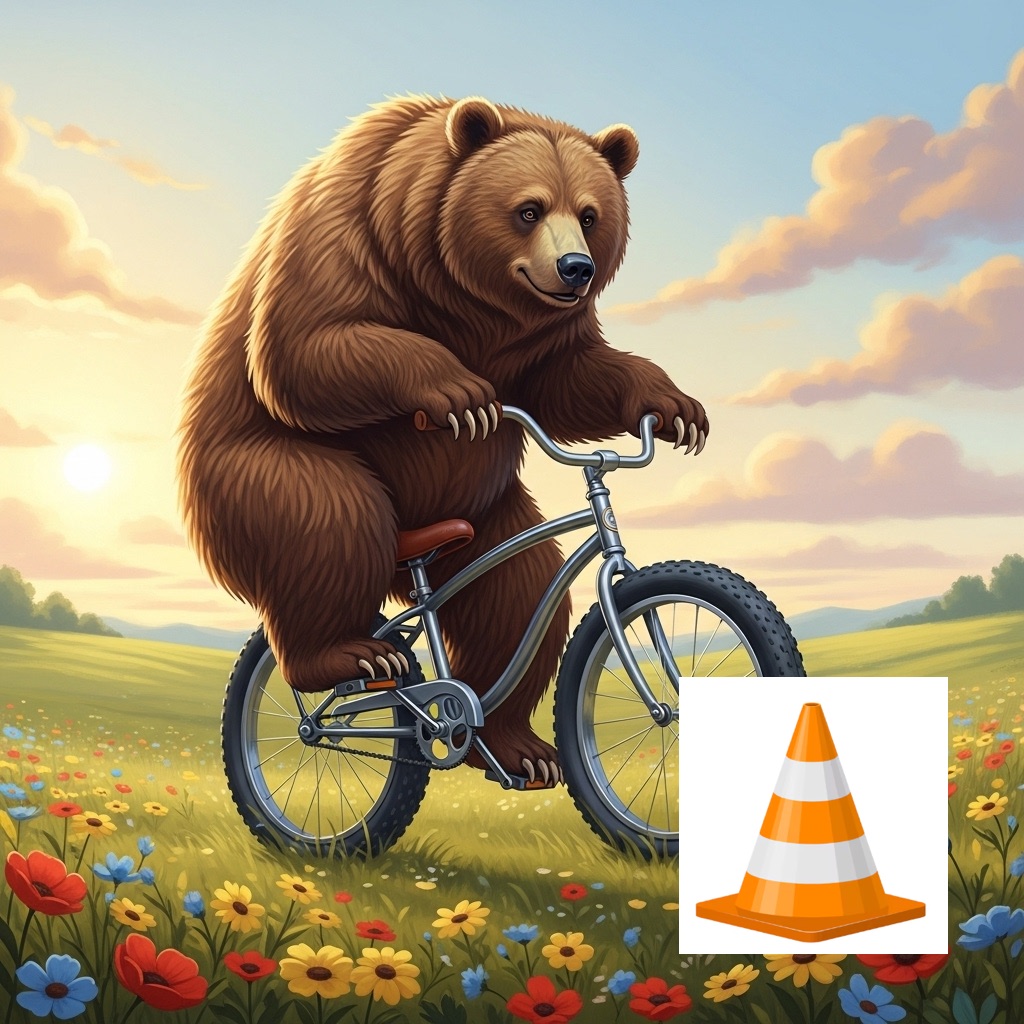}
    &
    \includegraphics[width=0.23\linewidth]{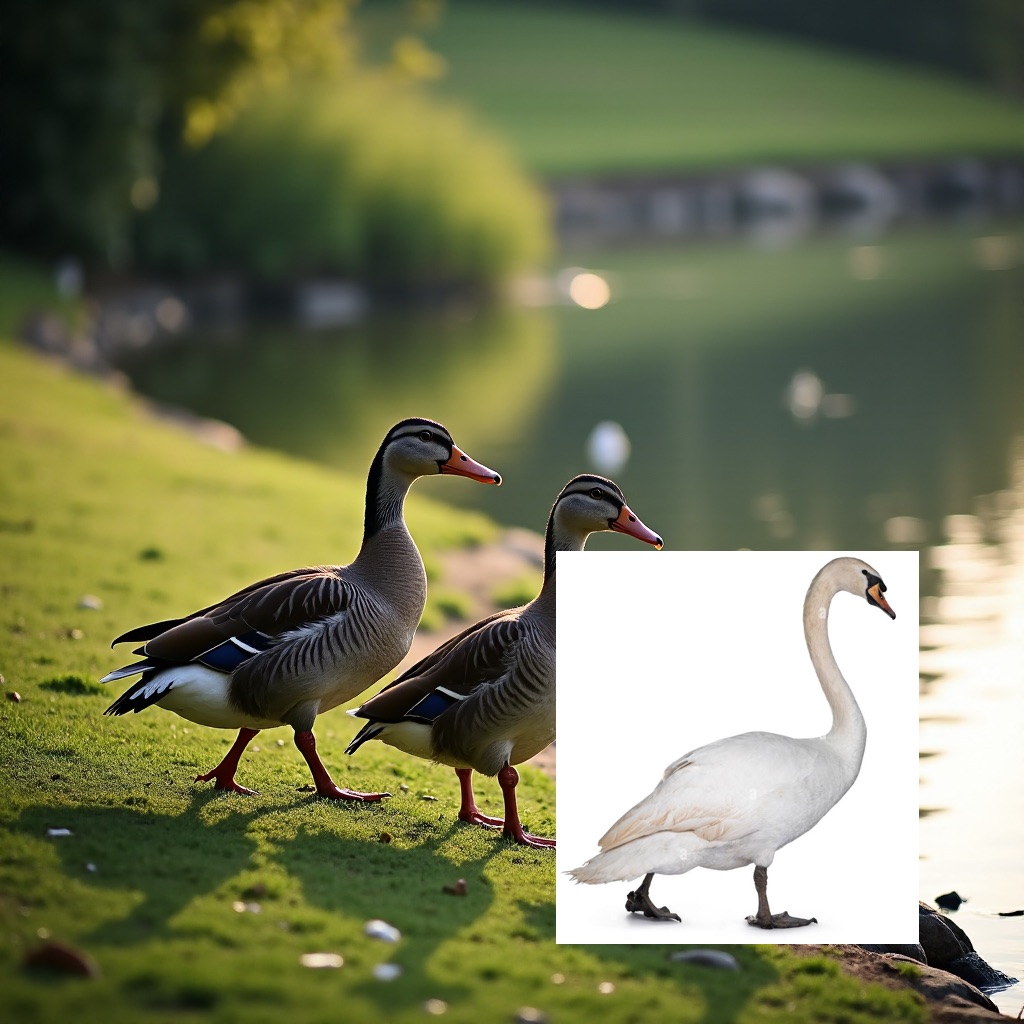}
    & 
    \includegraphics[width=0.23\linewidth]{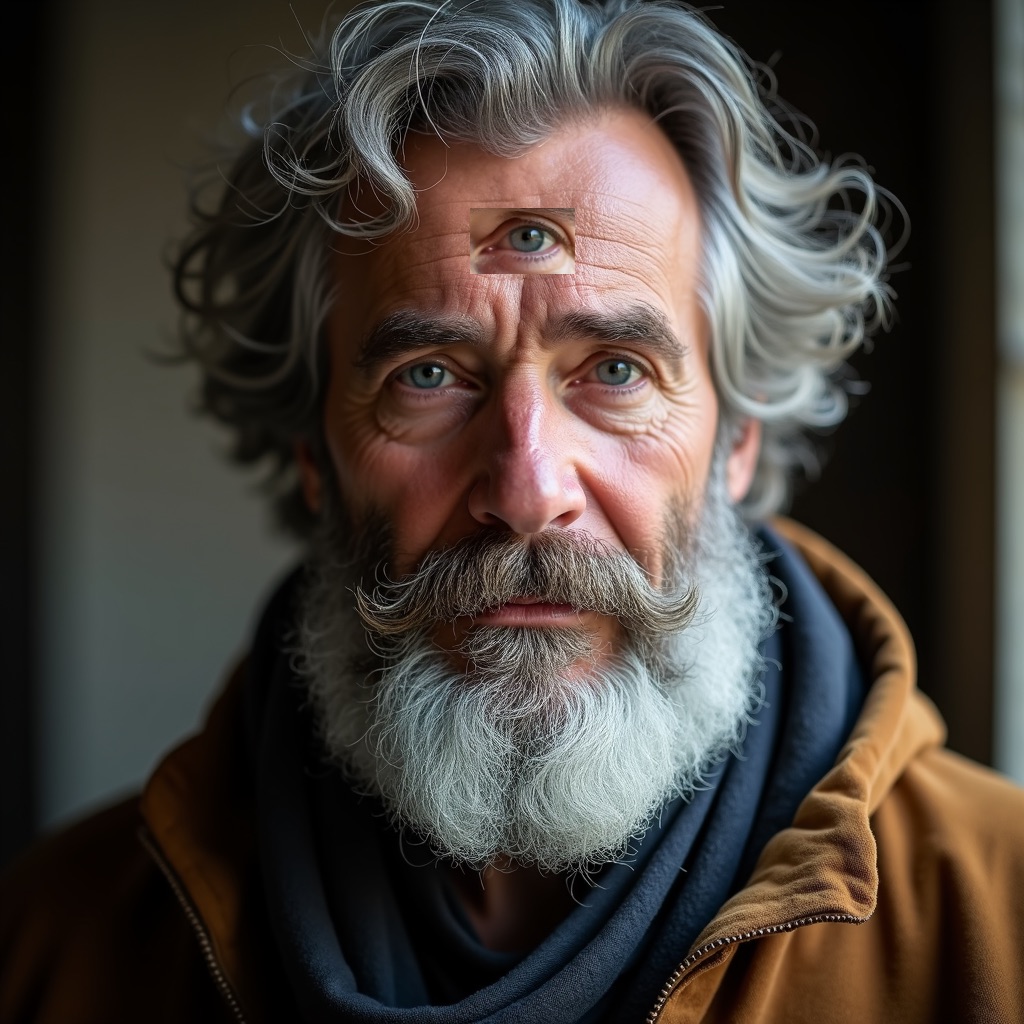}
    &
    \includegraphics[width=0.23\linewidth]{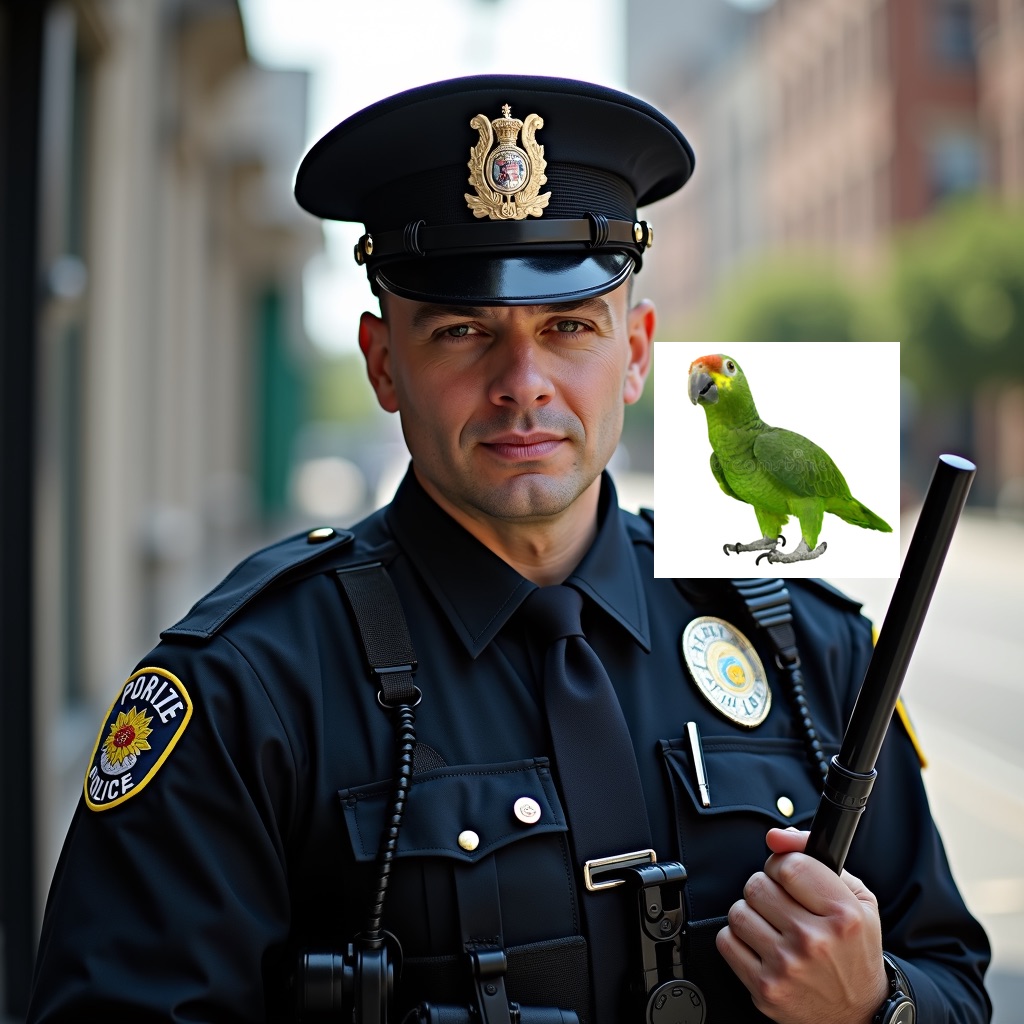}
    \\ %

    \raisebox{8pt}{\rotatebox{90}{\centering Output}}
    &
    \includegraphics[width=0.23\linewidth]{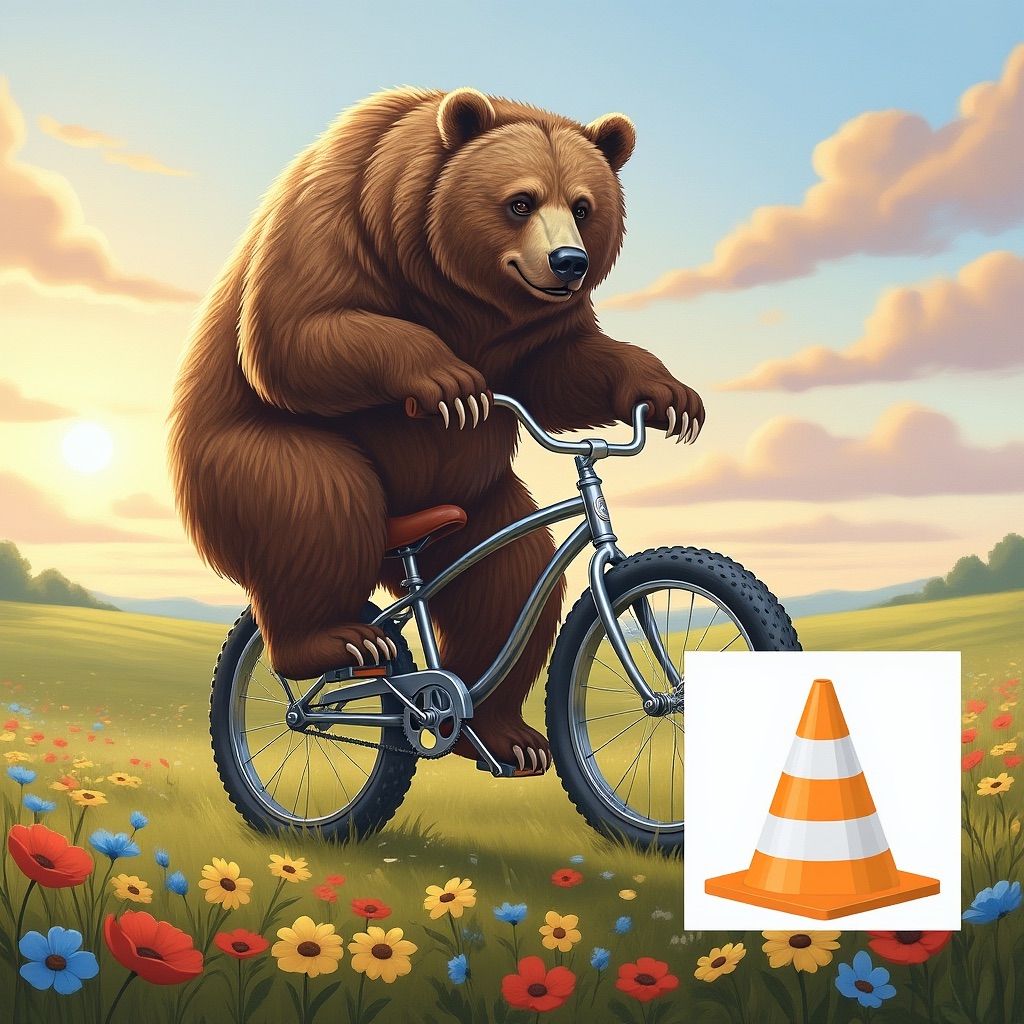}
    &
    \includegraphics[width=0.23\linewidth]{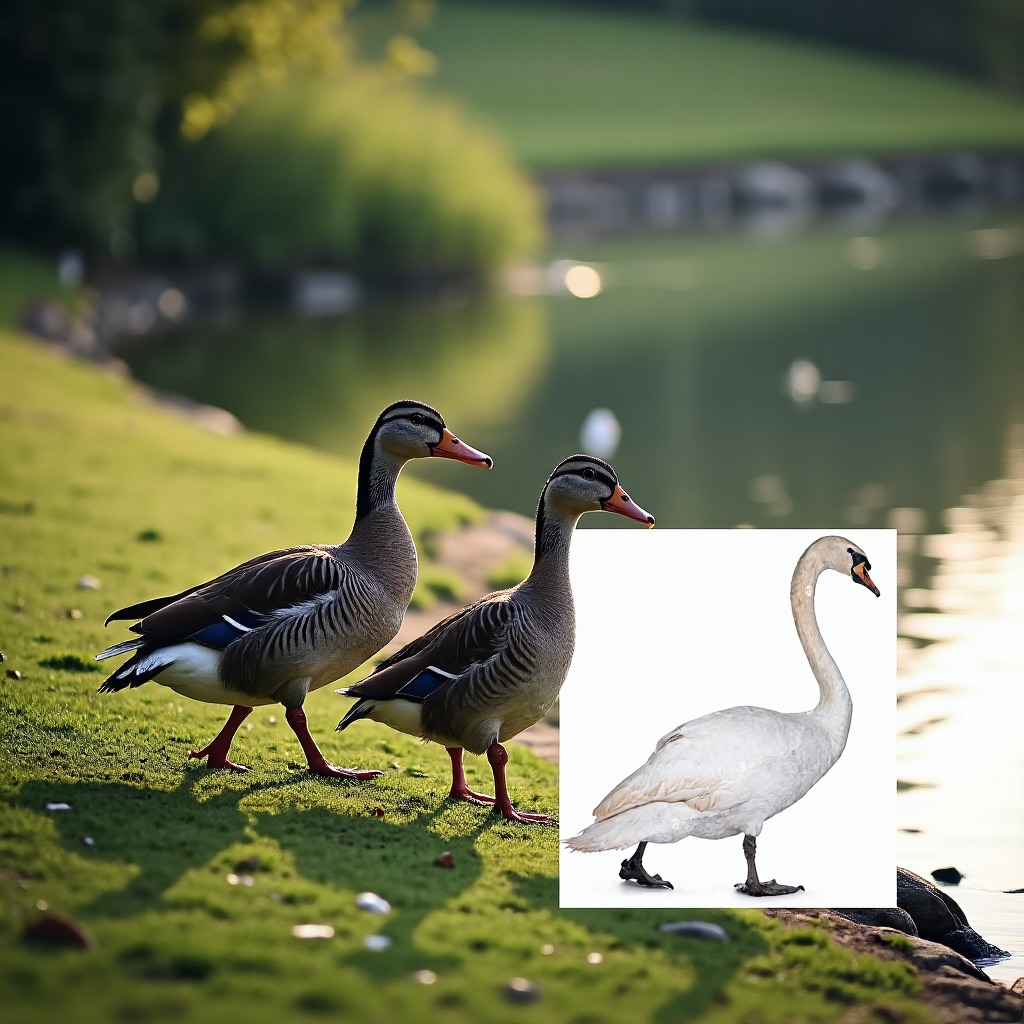}
    &
    \includegraphics[width=0.23\linewidth]{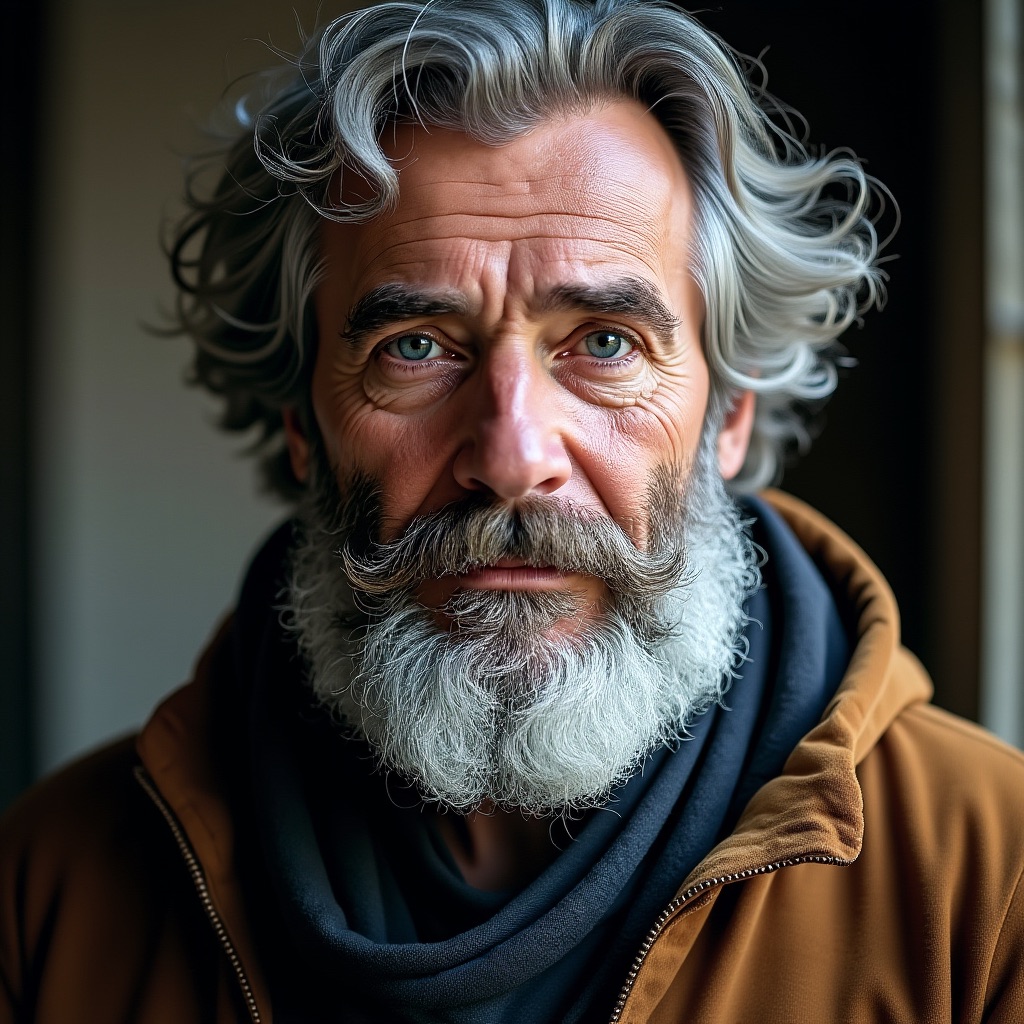}
    &
    \includegraphics[width=0.23\linewidth]{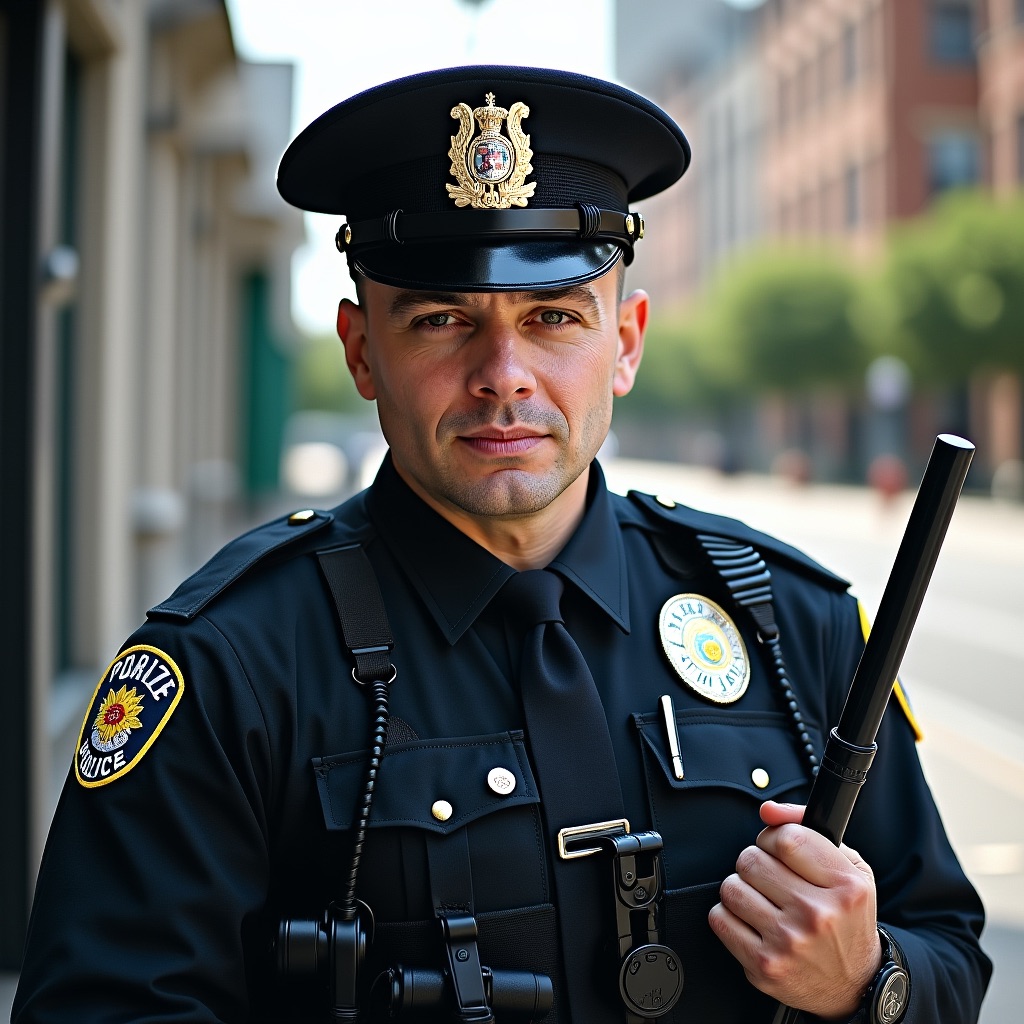}
    \\ %
    
    \end{tabular}
    \vspace{-8pt}
    \caption{Examples of Neglect and Suppression failure modes in vanilla FLUX Kontext. In all the shown examples, we instruct the model with: \emph{``blend the cropped objects into the image in a convincing manner.''}
    }
    \label{fig:failures_kontext}
    \vspace{-12pt}
\end{figure}

Recently, instruction-based editing models have become the leading approach in image editing~\cite{labs2025flux1kontextflowmatching, tan2025ominicontrol, brooks2023instructpix2pix}. These models are effective in maintaining the global layout and preserving the input image content while performing meaningful semantic changes guided by text instructions or image conditions. However, we find that they struggle to balance between these two objectives. When the instruction dominates, the model may \textit{suppress} the inserted object, allowing the generative prior to override its appearance. 
Conversely, when the conditioning on the input image is too strong, the model may \textit{neglect} to blend the inserted object, overemphasizing it at the expense of overall harmonization. 
These two failure modes are demonstrated in Figure~\ref{fig:failures_kontext}.

In this work, we present a method that aims to balance the coherence of the generated image and the preservation of the pasted object, a task we refer to as semantic harmonization. \new{Distinct from traditional photometric harmonization, which focuses on low-level color and illumination matching, our objective is to achieve high-level structural and contextual integration while faithfully maintaining object identity.}
To this end, we analyze the behavior of instruction-based editing models and observe that their attention maps inherently govern whether a given region should be copied from the input image or modified to achieve overall harmonization.
Building on this insight, we introduce LooseRoPE, a saliency-guided modulation of rotational positional encoding (RoPE), which acts as a continuous controller of the attention field of view.
We call our method LooseRoPE as it loosens the positional constraints of RoPE to smoothly steer the model’s focus between faithful preservation of the input image and coherent harmonization of the inserted object, providing control over this tradeoff.

Our approach provides a flexible and intuitive framework for image editing, achieving seamless compositional results without textual descriptions or complex user input.
As illustrated in Figure~\ref{fig:teaser}, our method can even be applied iteratively, performing a series of crop-and-paste operations while maintaining scene coherence.
Across such single or multi-step scenarios, LooseRoPE produces harmonized, coherent results that preserve the original scene and maintain the identity of the pasted object.
\new{While our approach prioritizes semantic and structural consistency over fine-grained lighting adjustments, our evaluations confirm} that controlling attention through positional encoding provides an effective framework for semantically harmonized image editing.

\section{Related Work}
\label{sec:related_work}

Our work lies at the intersection of image harmonization and reference- and layout-guided editing.
Harmonization methods adjust illumination, tone, and color to blend a pasted object with its background while strictly preserving its shape and appearance.
Reference- and layout-guided editing, in contrast, allows users to explicitly control both \textit{where} and \textit{what} to modify by providing spatial cues (e.g., masks, layouts) together with visual references that define the object’s appearance or identity.
In the following, we review related works in both areas and discuss how they relate to our problem.

\paragraph{Image Harmonization.}
Methods for this task aim to adjust the appearance of a composited image so the inserted region naturally fits its new background.
Early approaches focused on low-level adjustments of color, tone, and illumination. Later deep learning-based methods
learned context-aware harmonization from synthetic data~\cite{tsai2017deep, cun2020improving, cong2020dovenet}, or
introduced a self-supervised formulation without annotated masks~\cite{jiang2021ssh}.
More recently, diffusion-based techniques 
extend harmonization toward generative recomposition and lighting-aware adaptation~\cite{lu2023tf, song2023objectstitch, song2024imprint, ren2024relightful}.
While these methods improve the visual realism of composites, they remain limited to low-level appearance adjustment and do not address semantic coherence between the inserted object and its new context.
Our work extends harmonization by enabling appearance and semantic adaptation so the inserted object coherently integrates into its new context while preserving its spatial identity.

A closely related work, Cross-domain Compositing~\cite{hachnochi2023cross}, employs pretrained diffusion models to blend objects across visual domains using localized ILVR-based refinement~\cite{choi2021ilvr}.
While sharing the goal of coherent compositing, it focuses on domain translation and frequency-based blending, whereas we address in-domain semantic harmonization by directly modulating the model’s attention field to balance identity preservation and contextual adaptation.

\paragraph{Reference- and Layout-guided Editing.}
Recent advances in generative models have introduced explicit control mechanisms over both \textit{where} and \textit{what} is synthesized in an image.
Layout-guided synthesis focuses on spatial control, conditioning the generation process on cues such as masks, bounding boxes, depth maps, or keypoints that define object placement or scene structure.
Some methods fine-tune diffusion models to incorporate such layout signals directly~\cite{zhang2023adding, li2023gligen, parmar2024one, mou2023t2i}, achieving strong spatial alignment between conditioning inputs and generated content.
Other approaches enable spatial control in a zero-shot manner, typically by manipulating the internal features of diffusion models along the denoising trajectory~\cite{dahary2025decisive, sella2025instancegen, bar2023multidiffusion, dahary2024yourself, chen2023trainingfree}.

Reference-guided synthesis instead directs generation by conditioning on visual exemplars specifying the desired object’s identity or appearance, allowing models to reproduce precise visual details that are difficult to convey through textual prompts alone.
Such methods can be broadly divided into two categories.
Optimization-based approaches require a per-subject fine-tuning process to embed the reference into the model’s latent space~\cite{gal2022imageworthwordpersonalizing, ruiz2023dreamboothfinetuningtexttoimage, kumari2023multiconceptcustomizationtexttoimagediffusion}.
In contrast, encoder-based methods learn to map reference images directly into conditioning representations, enabling efficient and scalable identity control\new{~\cite{alzayer2025magic, yu2025objectmover, li2023blip2bootstrappinglanguageimagepretraining, tan2025ominicontrol, wei2023elite}}.

Techniques from layout- and reference-guided synthesis have been combined to support reference- and layout-guided editing~\cite{chen2024anydoor, lu2023tf, gu2024swapanythiing}, where both spatial placement and object appearance are explicitly controlled.
Such methods extend the generative capabilities of diffusion models toward compositional and controllable image editing.

\section{Method}
\subsection{Preliminaries}
\label{sec:prelim}

\paragraph{Rotary Positional Embeddings (RoPE).}{
The transformer blocks, which form the core of the diffusion transformer (DiT) architecture~\cite{flux, peebles2023scalable}, are inherently permutation-equivariant and therefore require explicit positional encodings to capture spatial dependencies. 
The \emph{Rotary Positional Embedding} (RoPE)~\citep{su2024roformer} has emerged as an effective method for positional encoding and is employed in most state-of-the-art DiTs.
RoPE represents a position coordinate $m$ as a series of 2D rotations at different frequencies. The number of frequencies is $D = d_{\text{model}} / 2$, where $d_{\text{model}}$ is the hidden model dimension. The angular frequencies usually follow a geometric progression,
\begin{equation}
\theta_d = {\theta_{\text{base}}}^{\tfrac{d}{D-1}}, \qquad d = 0,\ldots,D-1,
\label{eq:rope_theta}
\end{equation}
where $\theta_{\text{base}}$ is a model hyperparameter. 
Each token vector $\mathbf{v}$ is divided into $D$ two-dimensional sub-vectors, $\mathbf{v} = (\mathbf{v_1},\ldots,\mathbf{v_D})$, where $\mathbf{v}_d \in \mathbb{R}^2$ .
Each sub-vector $\mathbf{v}_d$ is then rotated according to its spatial location $m$ as:
\[
\mathbf{v}_d' = e^{i\,(\theta_d m)}\, \mathbf{v}_d,
\]
where the complex exponential denotes a 2D rotation by angle $\theta_d m$. 
For 2D images, RoPE is typically applied \emph{axially}: half of the hidden dimensions encode horizontal positions and the other half vertical ones, enabling independent offsets along each axis~\citep{heo2024rotary}. 

In our work, we augment the RoPE mechanism by introducing an additional \emph{inverse range factor} $r \in [0,1]$ that scales the positional coordinate $m$, yielding:
\[
\mathbf{v}_d' = e^{i\,(\theta_d r m)}\, \mathbf{v}_d.
\]
When $r<1$, the effective spatial distance between tokens is proportionally reduced, bringing them closer in the positional space and thereby broadening the attention field of view. This provides a simple yet effective means of controlling how locally or globally each query attends to surrounding tokens during inference.

}

\begin{figure*}
    \centering
    \includegraphics[width=1.0\textwidth]{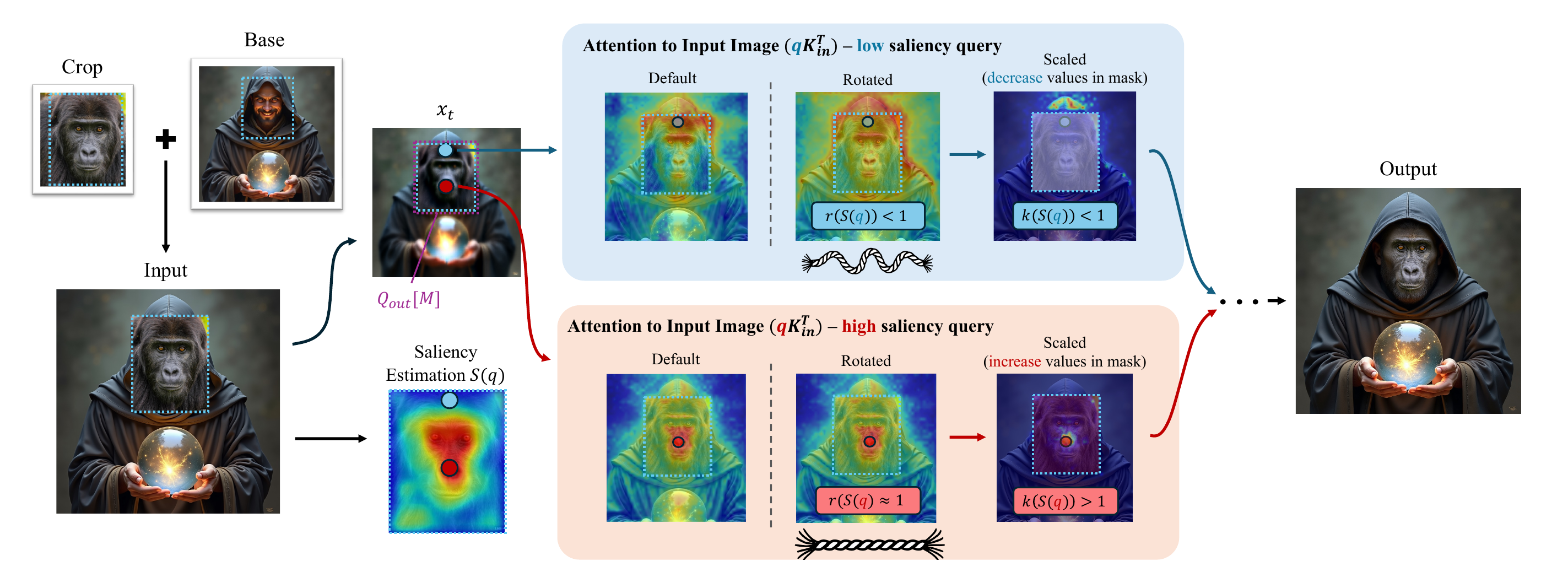}
    \vspace{-20pt}
    \caption{\textbf{Saliency-Guided Attention Manipulation.} 
    Given an image with a crudely pasted crop, we smoothly blend it into the surrounding scene by
    manipulating the attention computation during inference using a saliency map of the cropped region. 
    Output-image queries (within the \textcolor{cyan}{\textbf{dotted blue frame}}) attend to input-image keys using RoPE with a saliency-dependent range factor $r(S(q))$, which scales the positional coordinate and controls the spread of attention (“Rotated”).
The corresponding attention logits in the crop mask are then scaled by $k(S(q))$ (“Scaled”). 
High-saliency queries (\textcolor{red}{\textbf{red}}) have $r(S(q))\!\approx\!1$ and $k(S(q))\!>\!1$, keeping attention localized and preserving identity, evident in the gorilla’s facial expression. 
Low-saliency queries (\textcolor{blue}{\textbf{blue}}) have smaller $r(S(q))$ and $k(S(q))\!<\!1$, broadening attention and reducing crop-internal focus. 
This enables semantic blending with surrounding context, as seen in the forehead query attending to the hood and integrating smoothly in the final result. 
The “Default” attention map is shown for reference only and is not used in our method.
    }
    \vspace{-8pt}
    \label{fig:method}
\end{figure*}

\paragraph{FLUX Kontext}{

This model extends the FLUX~\cite{flux} text-to-image model to support image conditioning, enabling text-guided editing and reference-guided generation.
To achieve this, the input image is encoded into the model’s latent space, tokenized, and the resulting tokens are concatenated with those of the denoised image.
Through the model’s self-attention layers, these conditioning tokens influence the generation process, allowing the model to integrate visual and textual conditions.
In this work, we refer to the tokens of the conditioning image as the input image, and to the tokens of the denoised image as the output image.

\subsection{LooseRoPE}
Our setting assumes an input image $I_\text{in}$ composed of a base image with an additional region crudely pasted on top, along with a binary mask $M$ indicating the pasted area. The pasted region may originate either from another image or from the same image, in which case its removal often leaves a visible hole in the source image. The goal is to produce a harmonized image in which the pasted object or sub-object is seamlessly integrated, without requiring any textual guidance describing the scene or desired edit. 
An overview of our method is depicted in Figure~\ref{fig:method}.

Our method builds on FLUX Kontext~\cite{labs2025flux1kontextflowmatching}; we therefore begin by showing that Kontext alone does not reliably solve the crop-and-paste task and analyzing its failures.
Given input $I_\text{in}$ and the instruction \emph{“blend the cropped objects into the image in a convincing manner”}, Kontext exhibits two failure modes: \emph{neglect}, where the pasted region is barely modified, and \emph{suppression}, where it disappears entirely (Figure~\ref{fig:failures_kontext}).} We investigate these modes by inspecting attention maps during output generation.
Attention maps reveal a clear correlation: neglect involves overly localized attention, while suppression stems from diffuse attention that overlooks the content.
We hypothesize that blending requires an adaptive balance: semantically important regions should attend locally to preserve identity, whereas less salient regions should attend more broadly to the context for coherence.
To this end, we estimate a saliency map to modulate attention during inference, balancing faithful copying with context-aware harmonization.

\paragraph{Saliency Estimation.}A saliency map $S \in [0,1]^{H \times W}$ assigns each pixel a scalar reflecting its relative importance within an image. We seek a map that highlights distinctive features (e.g., facial regions or object details) while suppressing redundant textures or backgrounds. Since instance detection models implicitly capture these significance cues, we extract feature activations $\mathbf{F}_l$ from $L$ early high-resolution layers of a pre-trained network \cite{wu2019detectron2}. For each layer, we compute the feature-norm map $S_l = \|\mathbf{F}_l\|_2$, upsample, and aggregate:
\vspace{-2pt}
\begin{equation}S = \frac{1}{L} \sum_{l=1}^{L} \text{Interp}(S_l).\end{equation}
\vspace{-2pt}
The resulting normalized map serves as a spatial weighting function indicating the relative saliency of each pixel in the cropped region. In cases where a crop originates from the same image, resulting holes are assigned zero saliency.

\begin{algorithm}[t]
\caption{Content-Aware Attention Manipulation}
\label{alg:content_aware_attention}

\KwIn{Saliency map $S$, crop mask $M$, output image queries $Q_{\text{out}}$, input image keys $K_{\text{in}}$, base frequency $\theta_{\text{base}}$, inverse range function $r(\cdot)$, scale factor function $k(\cdot)$}
\KwOut{Updated input image attention weights $W_{\text{in}}$}

\BlankLine %

\For{each query $q$ in $Q_{\text{out}}[M]$}{
    $q_{\text{r}}, K_{\text{in-r}} \gets \text{RoPE}(q, K_{\text{in}}, r(S(q)))$ \tcp*{Rotate}
    
    $W_q = q_{\text{r}} K_{\text{in-r}}^T$ \tcp*{Calculate logits}
    
    $W_q[M] \gets W_q[M] \cdot k(S(q))$ \tcp*{Scale}
    
    $W_{\text{in}}[q] \gets W_q$ \tcp*{Update}
}
\end{algorithm}
\vspace{-8pt}

\paragraph{Content-Aware Attention Manipulation.}

Our mechanism guides the model toward an adaptive balance between copying content from the input image and harmonizing the pasted crop with the surrounding scene.
To achieve this, we modulate the attention weights between the queries within the region of the pasted crop in the \emph{output} image and the corresponding keys from the \emph{input} image, according to the saliency of each query.
\op{This modulation is performed in two stages: first, we apply a RoPE-based manipulation; then, we scale the attention weights.}
We denote the queries in the pasted region as $Q_{\text{out}}[M]$, the keys of the input image as $K_\text{in}$, and the resulting attention weights between them as $W_\text{in} = \operatorname{softmax}({(Q_{\text{out}}[M] K_\text{in}^\top)}/\sqrt{d})$, where $d$ is the feature dimension. Algorithm~\ref{alg:content_aware_attention} summarizes the proposed content-aware attention mechanism. Next, we describe each of the modulation stages in detail.

\begin{figure}
    \centering
    \setlength{\tabcolsep}{1pt}
    \begin{tabular}{ccccc} %
    
    \includegraphics[width=0.19\linewidth]{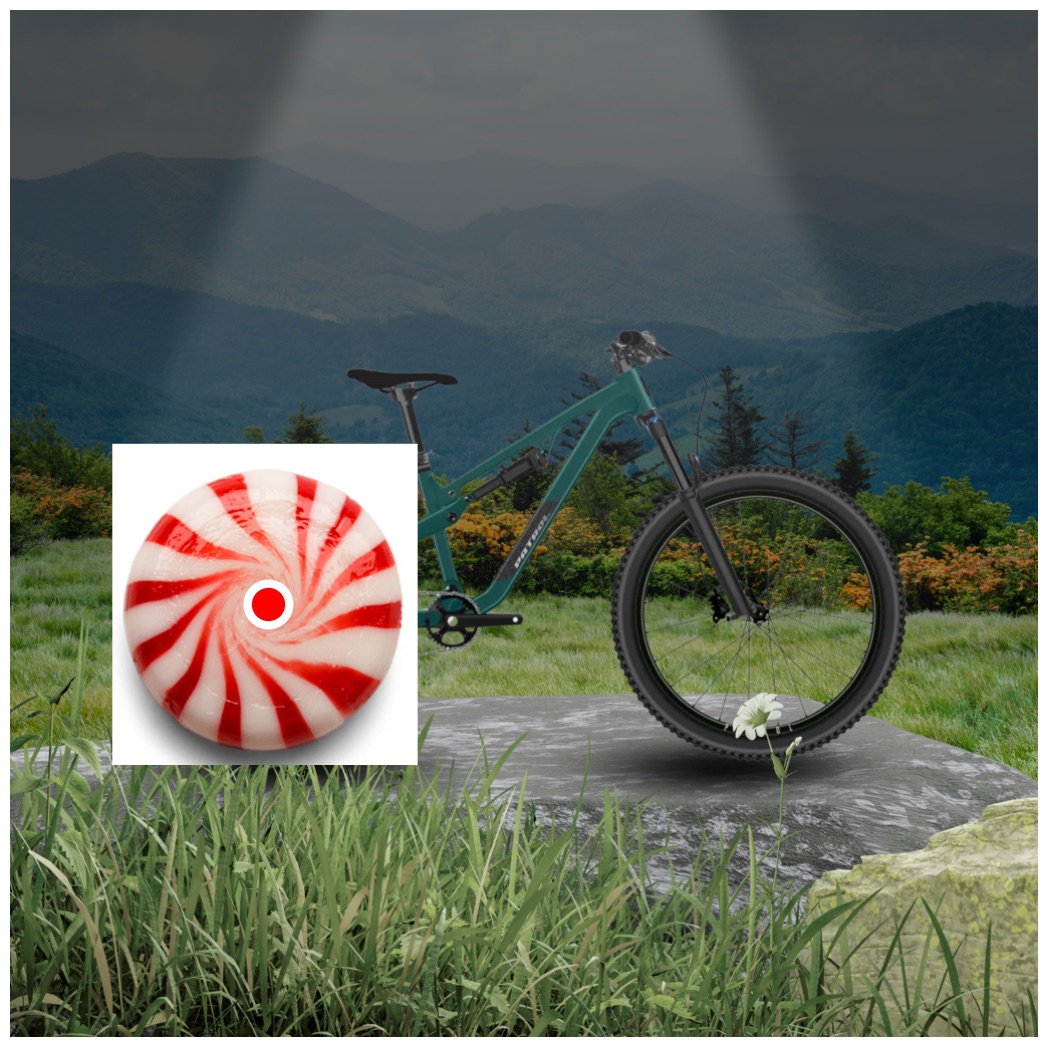}
    &
    \includegraphics[width=0.19\linewidth]{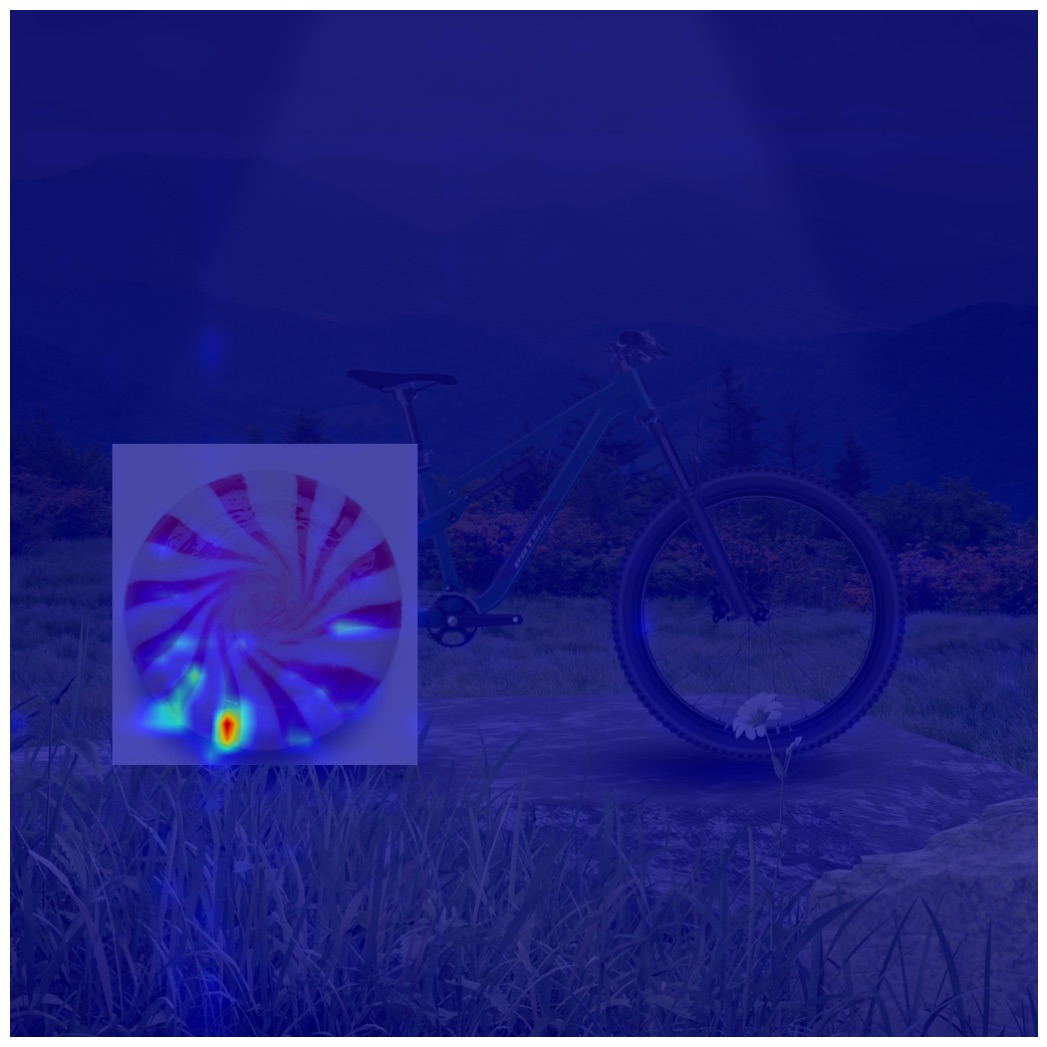}
    &
    \includegraphics[width=0.19\linewidth]{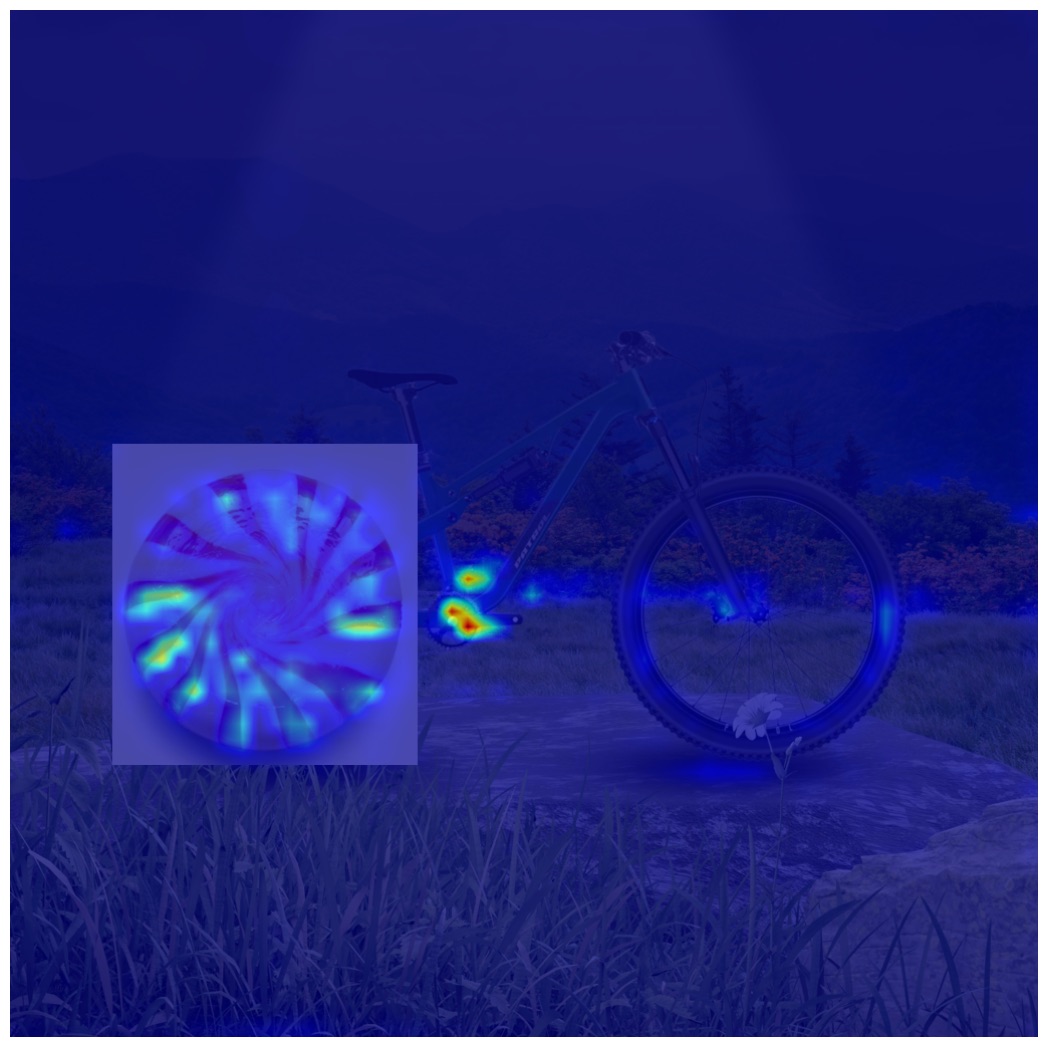}
    &
    \includegraphics[width=0.19\linewidth]{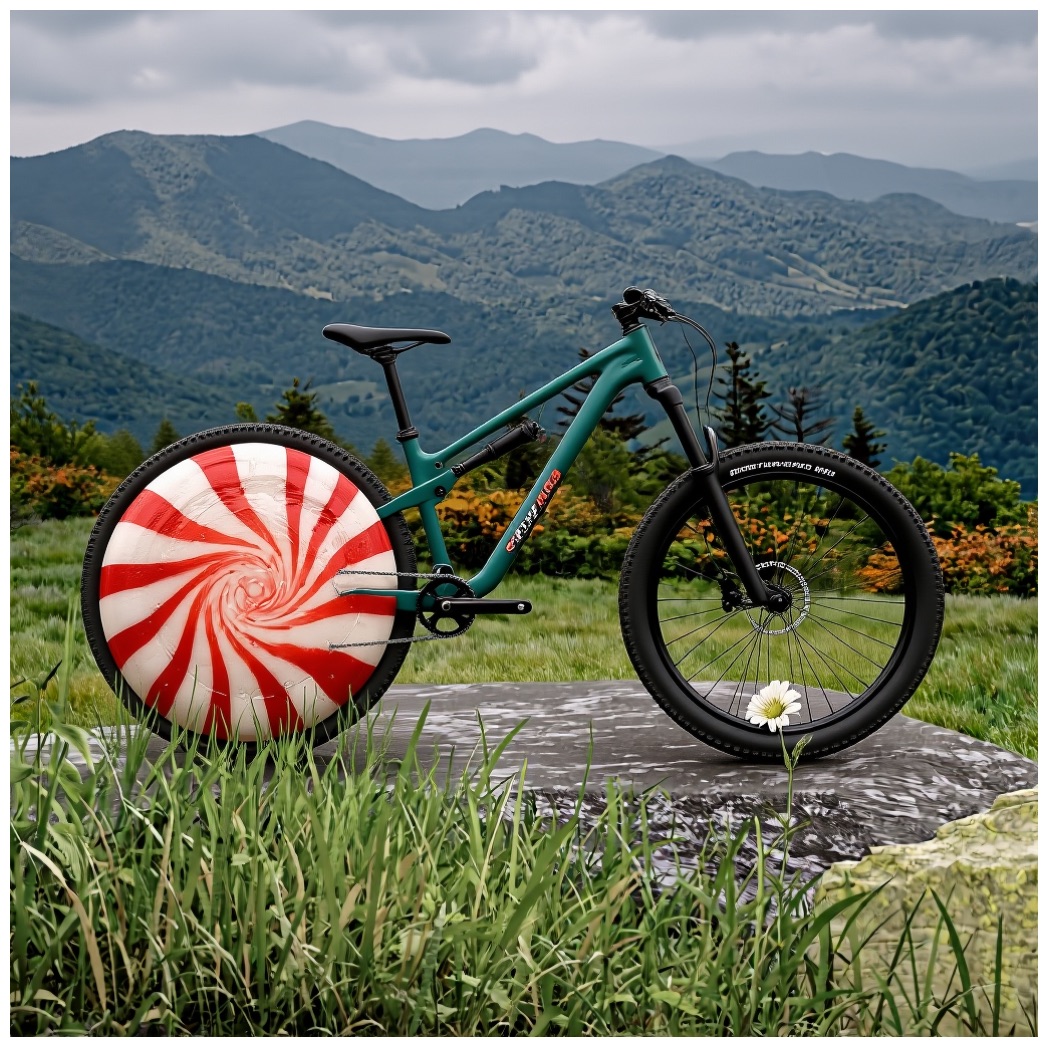}
    &
    \includegraphics[width=0.19\linewidth]{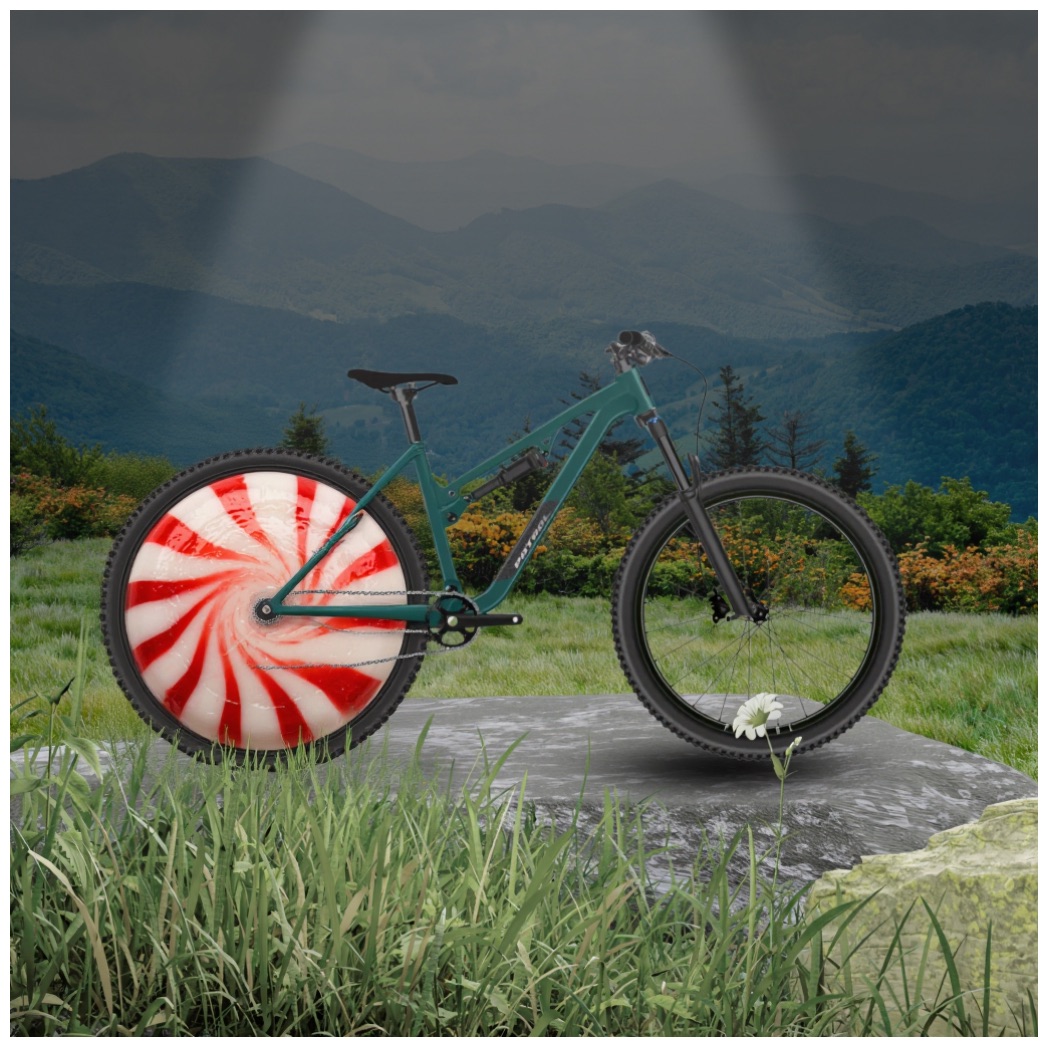}
    \\ %
    
    \includegraphics[width=0.19\linewidth]{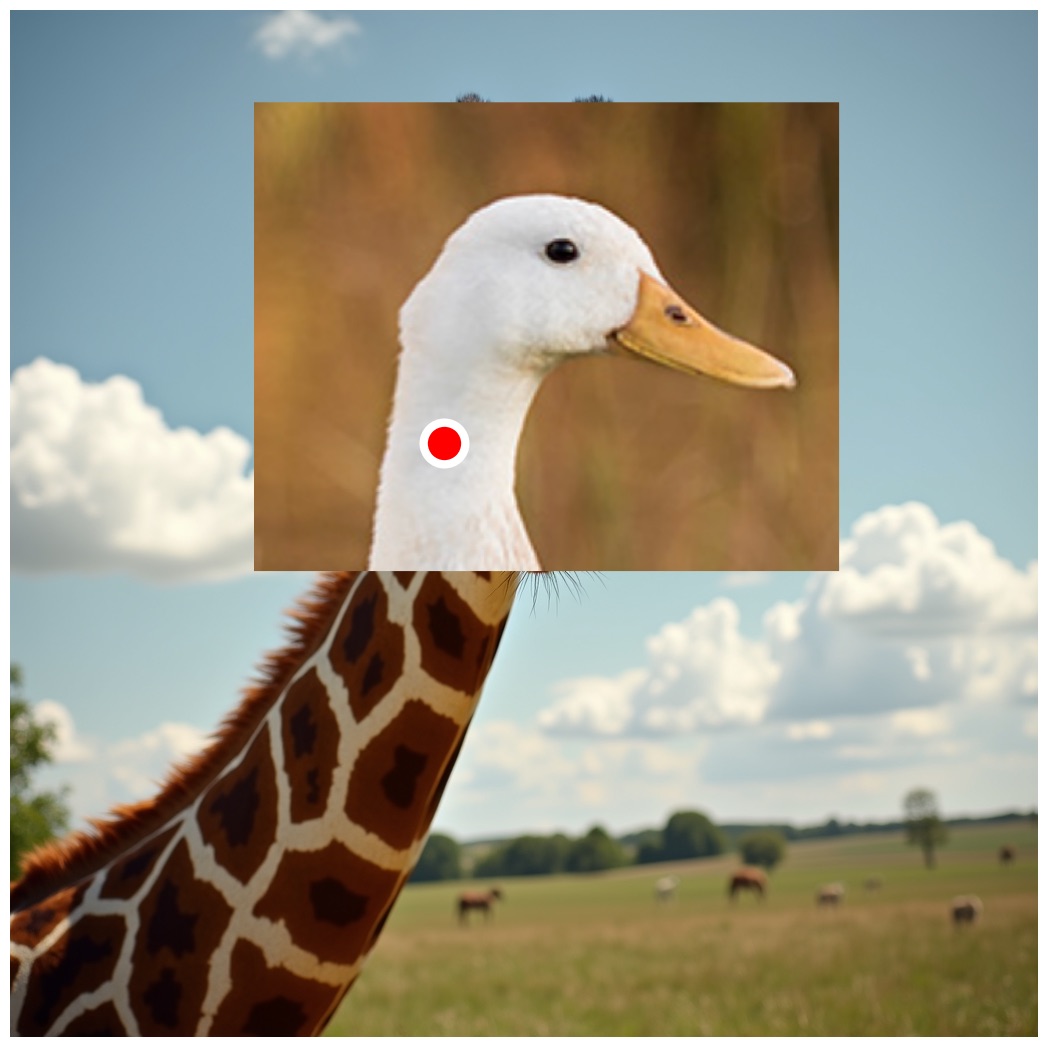}
    &
    \includegraphics[width=0.19\linewidth]{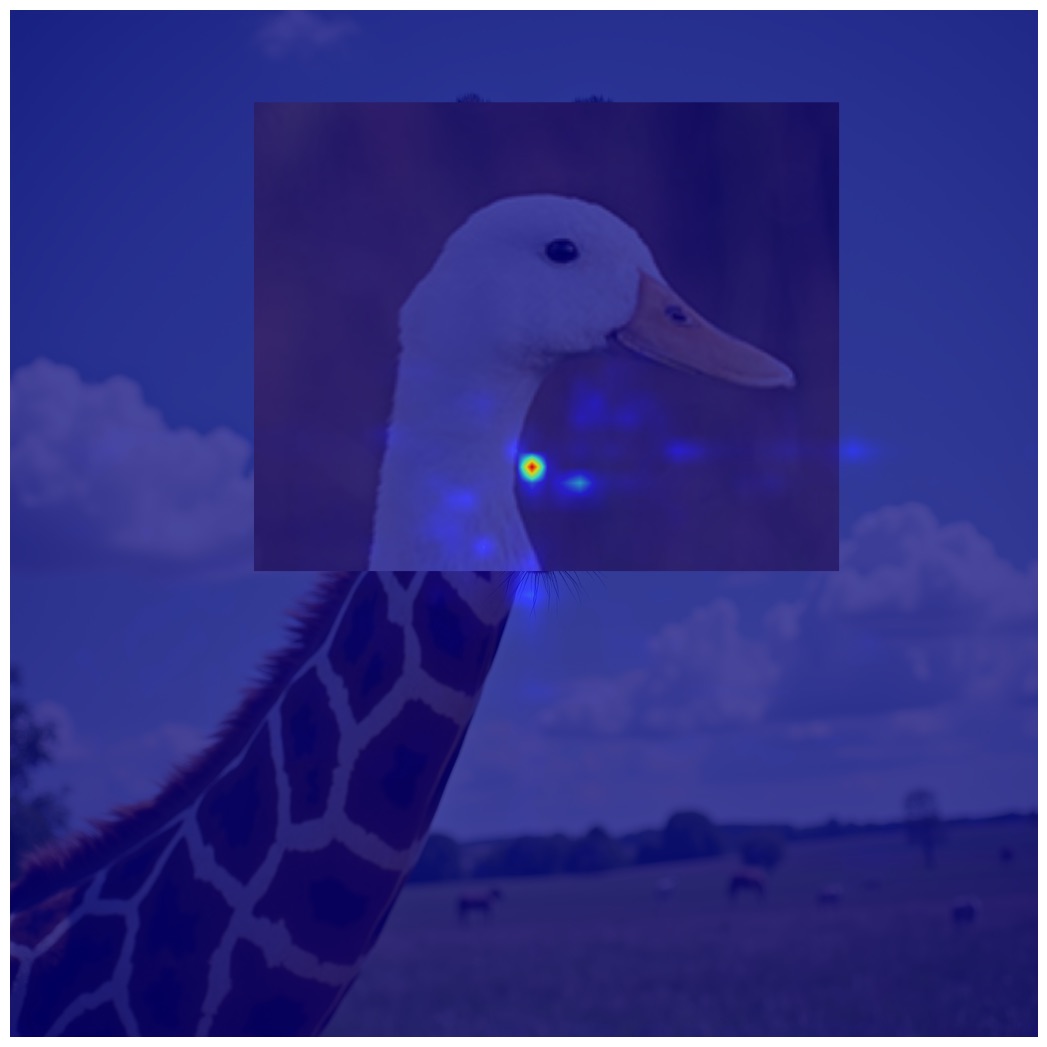}
    &
    \includegraphics[width=0.19\linewidth]{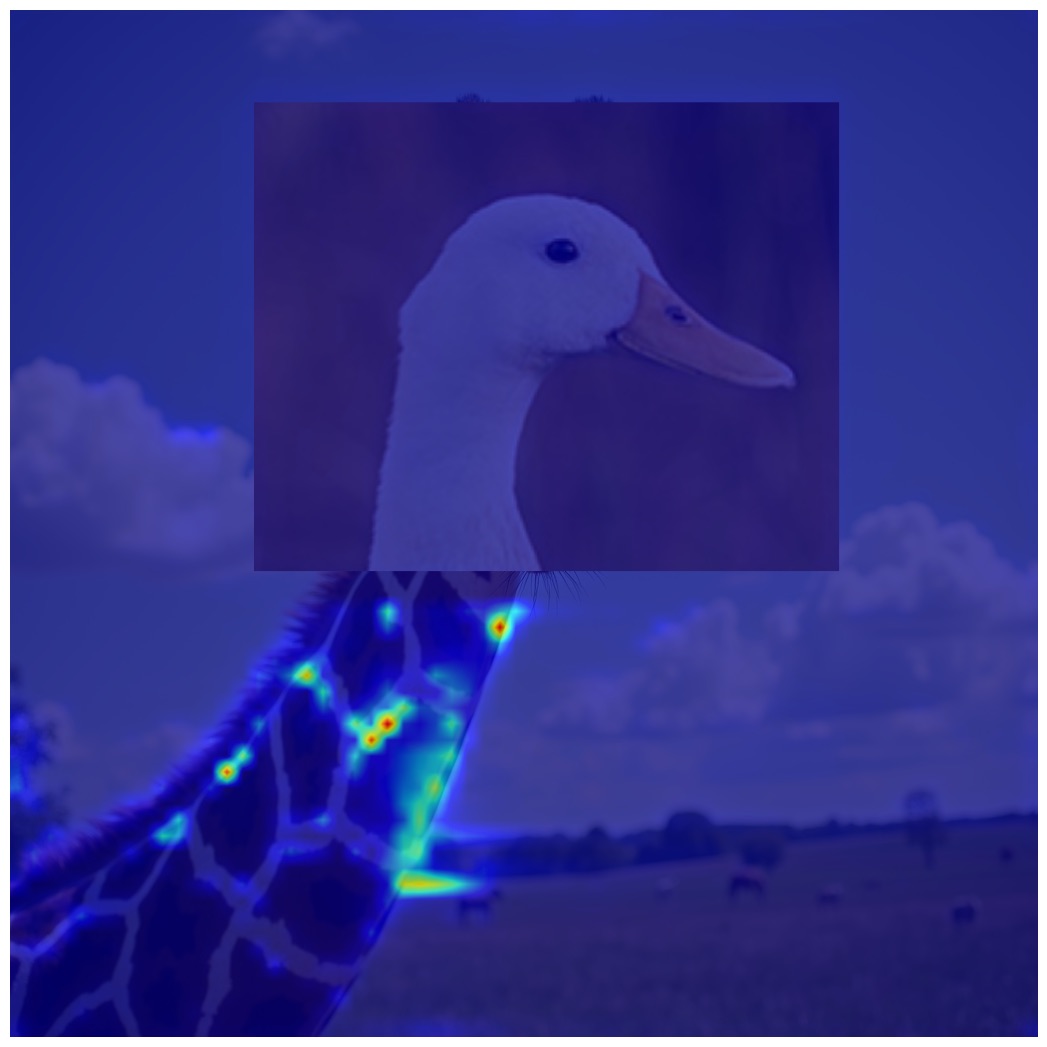}
    &
    \includegraphics[width=0.19\linewidth]{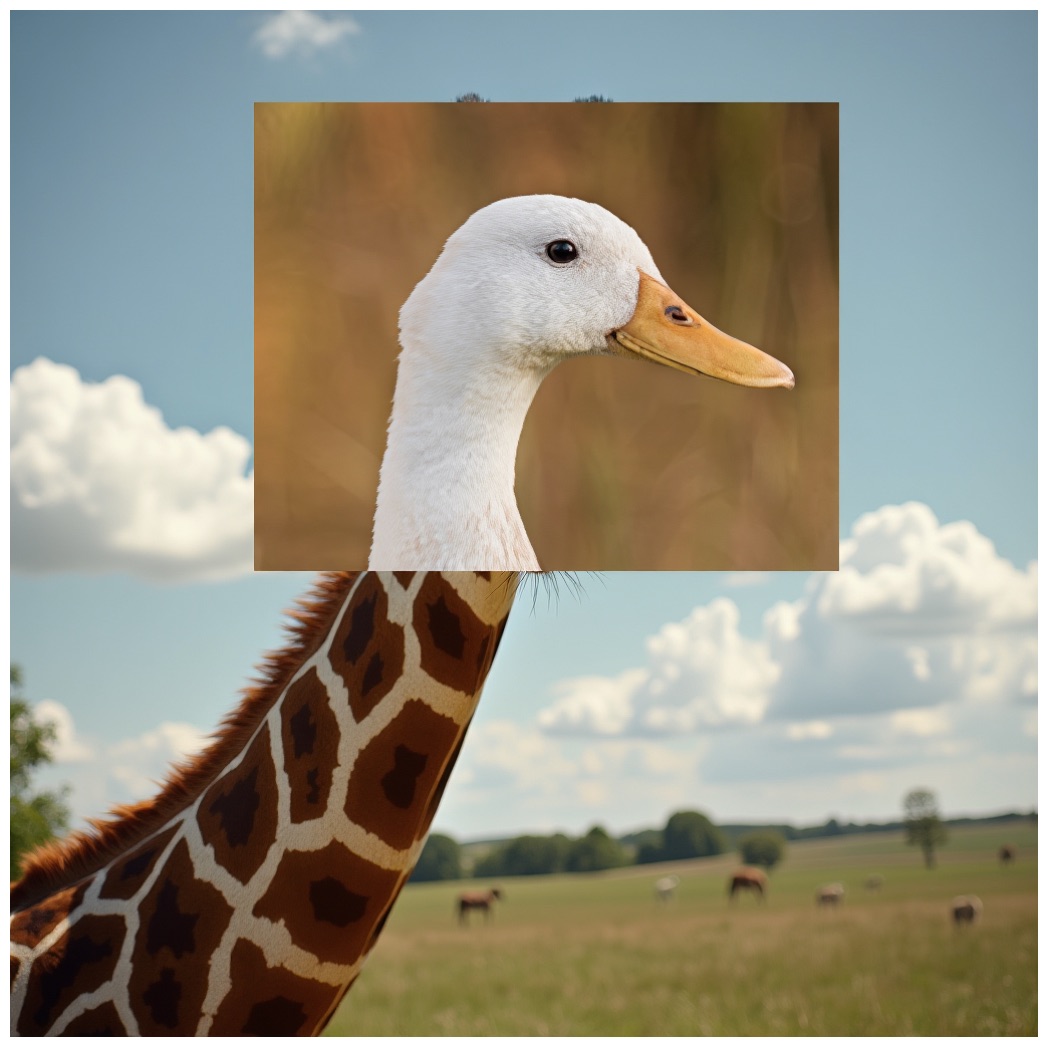}
    &
    \includegraphics[width=0.19\linewidth]{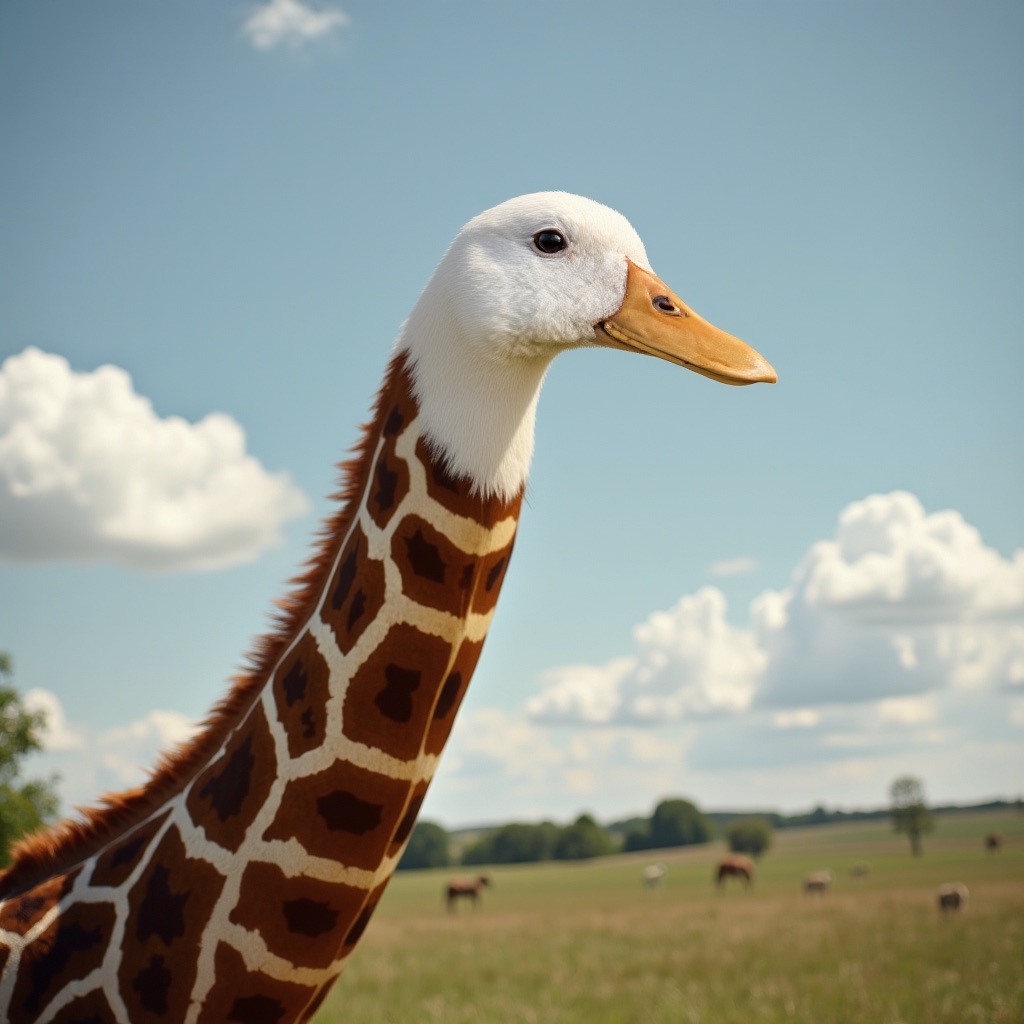}
    \\ %

    \begin{minipage}{0.19\linewidth}
        \centering
        \footnotesize{(a) Input}
    \end{minipage}
    &
    \begin{minipage}{0.19\linewidth}
        \centering
        \footnotesize{(b) Kontext Attn.}
    \end{minipage}
    &
    \begin{minipage}{0.19\linewidth}
        \centering
        \footnotesize{(c) Our Attn.}
    \end{minipage}
    &
    \begin{minipage}{0.19\linewidth}
        \centering
        \footnotesize{(d) Kontext Output}
    \end{minipage}
    &
    \begin{minipage}{0.19\linewidth}
        \centering
        \footnotesize{(e) Our Output}
    \end{minipage}
    
    \end{tabular}
    \vspace{-3pt}
    \caption{
    \textbf{Attention Map Visualization} \ignorethis{(first timestep, selected layers)}.
    \textbf{Top:} For a query on the bike wheel, vanilla Kontext (b) produces highly local attention, whereas our method (c) correctly attends to the gear wheel, enabling coherent blending (e).
    \textbf{Bottom:} For a query on the duck's neck, Kontext (b) again attends locally within the pasted crop. In contrast, our RoPE modification (c) captures the semantic relation to the giraffe's neck, resulting in a seamless blend (e).
}
\vspace{-13pt}
    \label{fig:attn_failure}
\end{figure}

To manipulate the attention weights $W_\text{in}$, we first adjust the RoPE mechanism applied when computing attention between $Q_{\text{out}}[M]$ and $K_\text{in}$. 
As introduced in Section~\ref{sec:prelim}, we augment RoPE with an \emph{inverse range factor} $r \in [0,1]$ that scales down the positional coordinate, thereby controlling how widely a query attends in space.
We leverage this factor by assigning each query $q \in Q_\text{out}[M]$ a \emph{saliency-dependent} inverse range factor $r(S(q))$, where $r(\cdot)$ is a monotonically increasing function of the saliency value $S(q)$ and bounded by 1. 
RoPE is then applied using the modified positional term $r(S(q))\,m$, effectively linking saliency to the attention range: low-saliency queries attend more broadly to encourage contextual blending (see Figure~\ref{fig:attn_failure} and the upper attention map in Figure~\ref{fig:method}), while high-saliency ones remain spatially localized to preserve detail and identity (see lower attention map in Figure~\ref{fig:method} ). 
This saliency-guided modulation enables a smooth transition between semantic adaptation and structural fidelity.

While the RoPE-based manipulation enables queries to capture broader semantic context, it can also introduce undesirable effects: salient regions may lose their identity, as the increased attention range causes them to attend less to their corresponding areas in the original crop, and large background areas within the crop mask may blend insufficiently due to increased attention to other spatially distant background regions.
To mitigate these issues, we introduce a \emph{crop attention factor} $k(S(q)) \in [k_{\text{low}}, k_{\text{high}}]$ that scales the attention weights corresponding to keys within the crop mask.
Let $K_\text{in}[M]$ denote the keys that belong to the pasted crop, and $W_\text{in}[:, M]$ the associated attention weights after RoPE modulation.
For each query $q$, we scale $W_\text{in}[q,M]$, where higher-saliency queries receive stronger scaling (approaching $k_\text{high}$) and less salient ones approach $k_\text{low}$ (see the scaled attention maps in Figure~\ref{fig:method}). 
During early denoising steps, setting $k_\text{high} > 1$ increases attention from salient regions in the output crop to their counterparts in the input image, preventing suppression.
As denoising progresses, $k_\text{high}$ gradually approaches 1, allowing smooth harmonization with the surrounding scene.
Both modulation functions, $r(S(q))$ and $k(S(q))$, are implemented as $\tanh$-based mappings to ensure smooth, high-contrast modulation between salient and non-salient queries, a property we find crucial for stable, high-quality results.

\begin{figure}[t]
    \centering
    \setlength{\tabcolsep}{1pt}
    
    \includegraphics[width=1.0\linewidth]{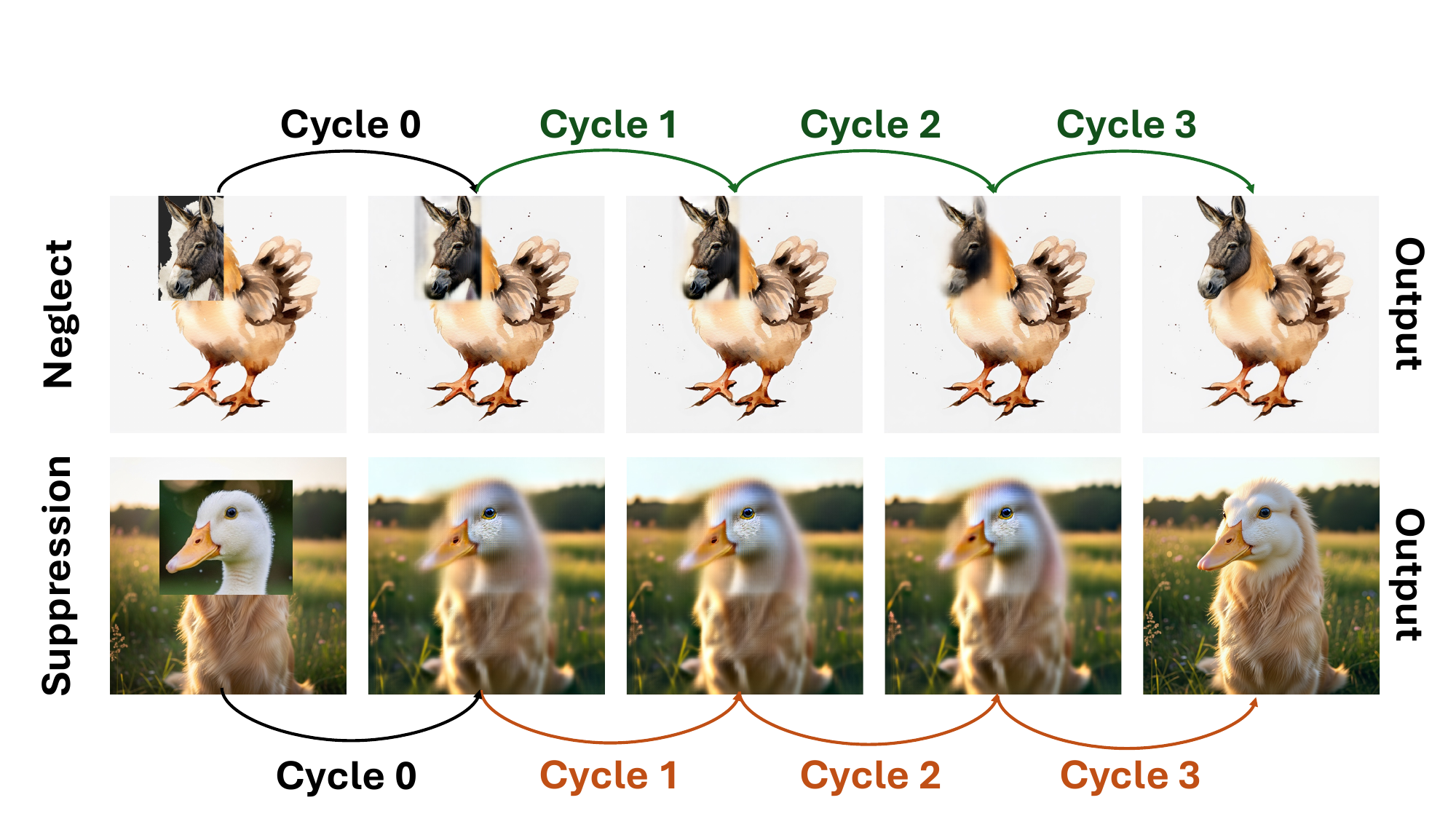}
    \vspace{-22pt}
    \caption{
    \textbf{VLM guided manipulation of attention.} Even inputs that exhibit severe neglect or suppression are eventually edited successfully. \textcolor{DarkGreen}{\textbf{Green}} arrows indicate a downscale in the saliency map (neglect), and \textcolor{orange}{\textbf{Orange}} arrows indicate an upscale (suppression). The figure shows the input, followed by three $\hat{x}_0$ predictions at timestep 2, and our method’s final output.
    \vspace{-14pt}
    }
    \label{fig:vlm}
\end{figure}

\paragraph{VLM Based Parameter Steering.}
While robust, our attention modulation requires adaptive adjustment for optimal blending in certain cases. Specifically, small crops can suffer from \emph{suppression}, while those with distinct backgrounds are prone to \emph{neglect}. Manual hyperparameter tuning can mitigate these effects but often trades performance across different samples. We therefore leverage a Vision-Language Model (VLM) to automate parameter steering during inference. We observe that signs of neglect or suppression appear early in the predicted clean image $\hat{x}_0$, so we query the VLM after initial sampling iterations to classify the blend state as \emph{success}, \emph{neglect}, or \emph{suppression}. If neglect is predicted, we downscale the saliency map; if suppression, we upscale and clip values to $1.0$. Diffusion then restarts with the updated map, looping until the VLM reports success or a maximum number of attempts is reached (Fig.~\ref{fig:vlm}).

\label{sec:method}

\vspace{-2pt}
\section{Experiments}
In this section, we conduct both qualitative and quantitative experiments to assess the effectiveness of our method in semantic harmonization. In the supplementary material, we provide additional implementation details, discuss and present limitations, and show additional results and comparisons.

\begin{figure*}[!htbp]
    \centering
    \setlength{\tabcolsep}{0.5pt} %
    \begin{tabular}{cccccccc}
        \textbf{Input} & TF-ICON & AnyDoor & \new{Qwen-Image-Edit} & FLUX Kontext & Nano Banana & \new{\textbf{Ours-Qwen}} & \textbf{Ours-Kontext} \\

        \includegraphics[width=0.123\linewidth]{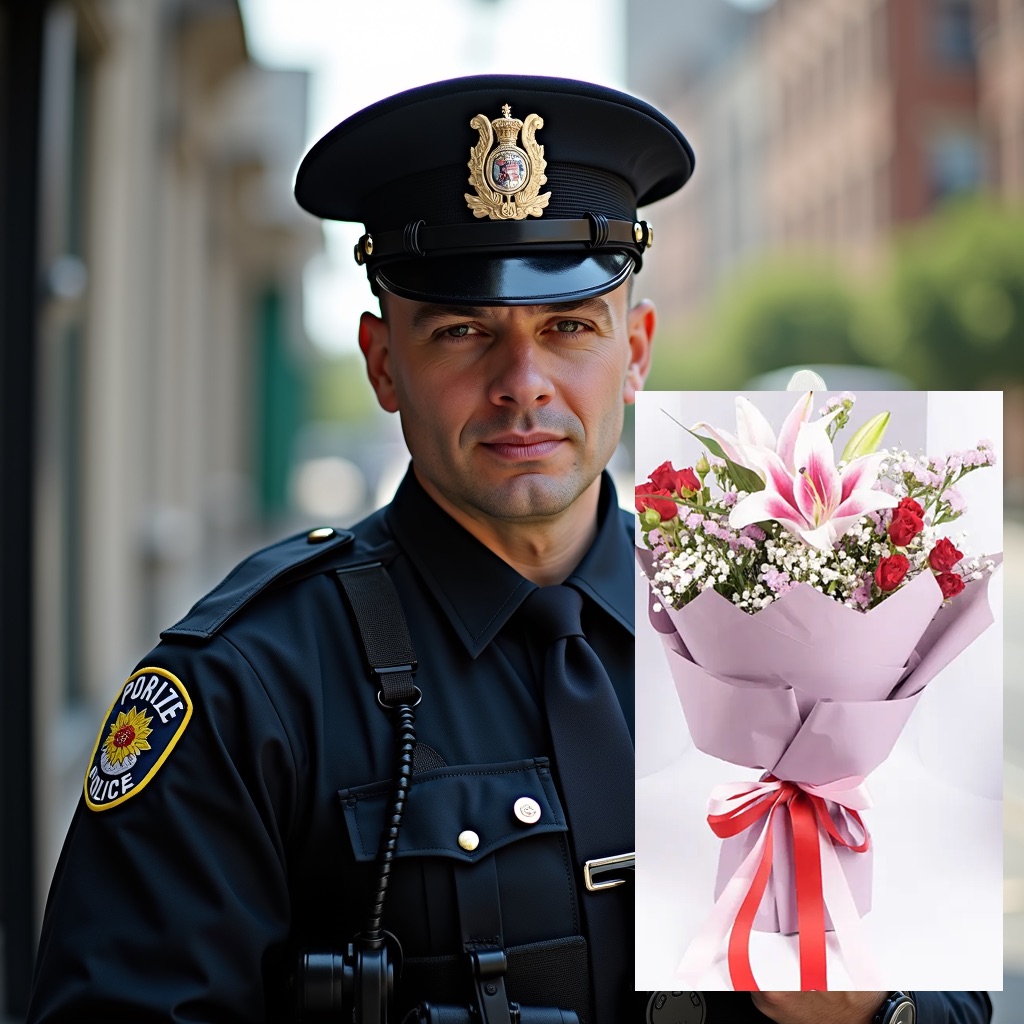} &
        \includegraphics[width=0.123\linewidth]{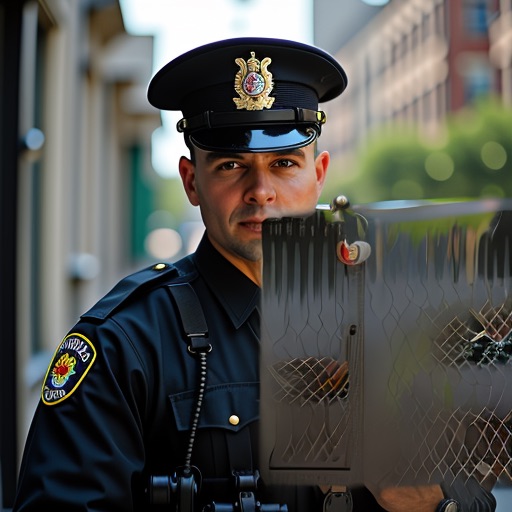} &
        \includegraphics[width=0.123\linewidth]{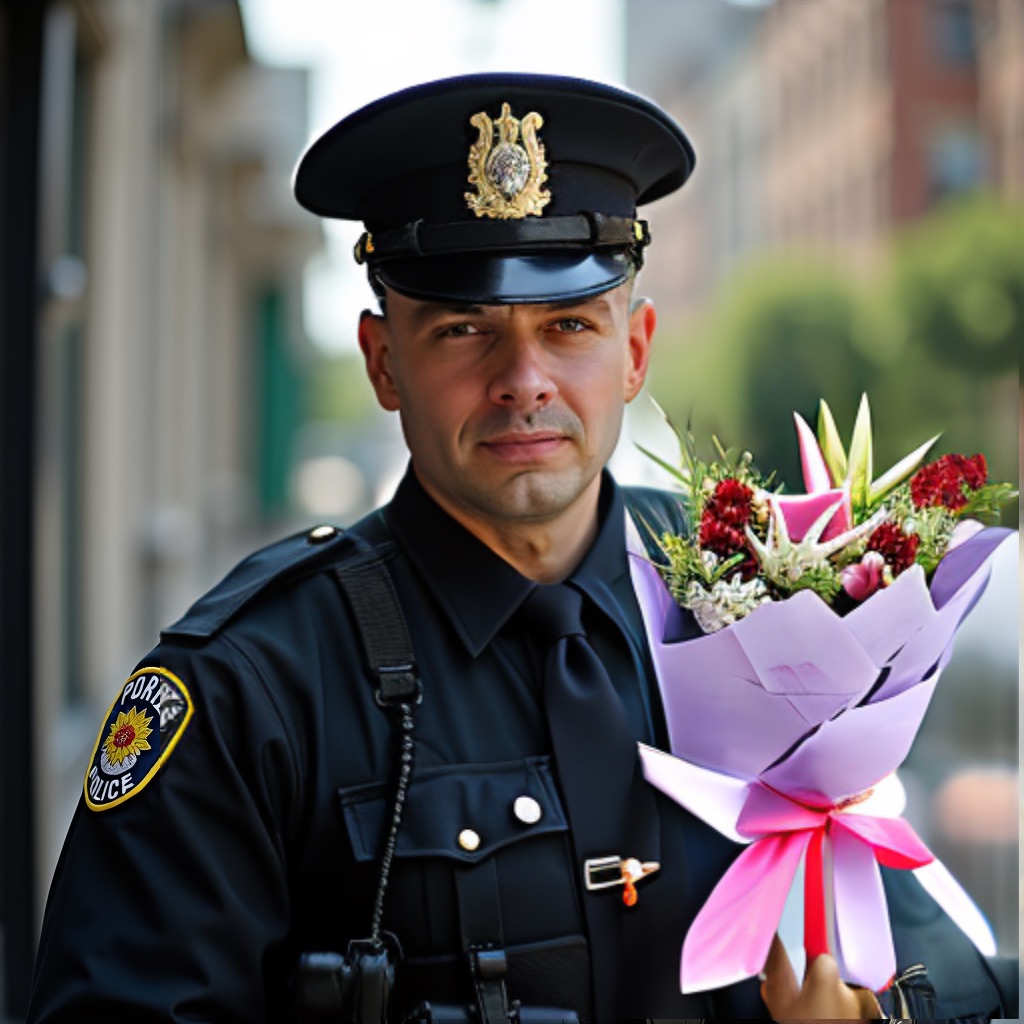} &
        \includegraphics[width=0.123\linewidth]{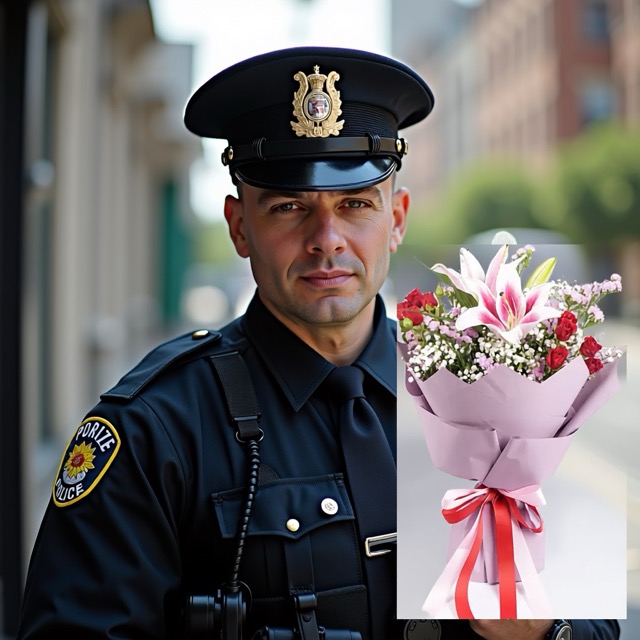} &
        \includegraphics[width=0.123\linewidth]{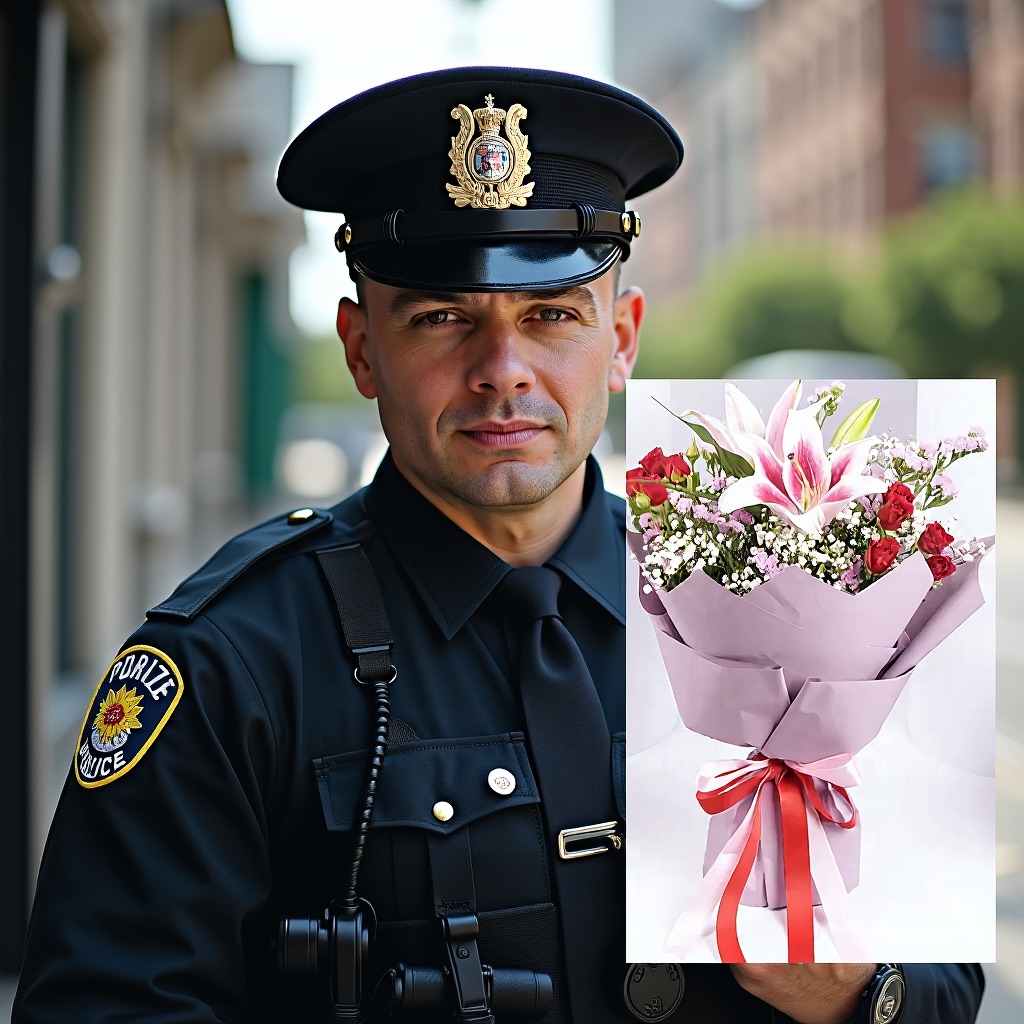} &
        \includegraphics[width=0.123\linewidth]{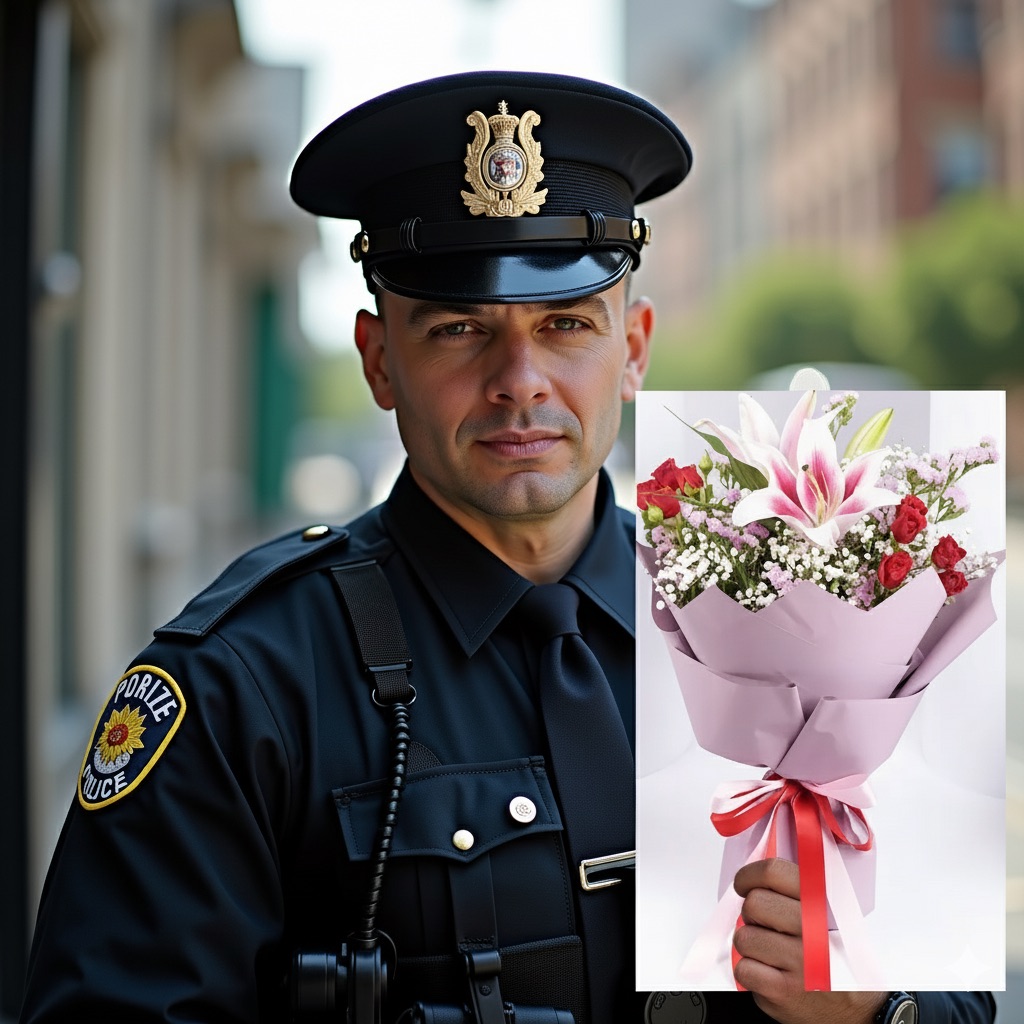} &
        \includegraphics[width=0.123\linewidth]{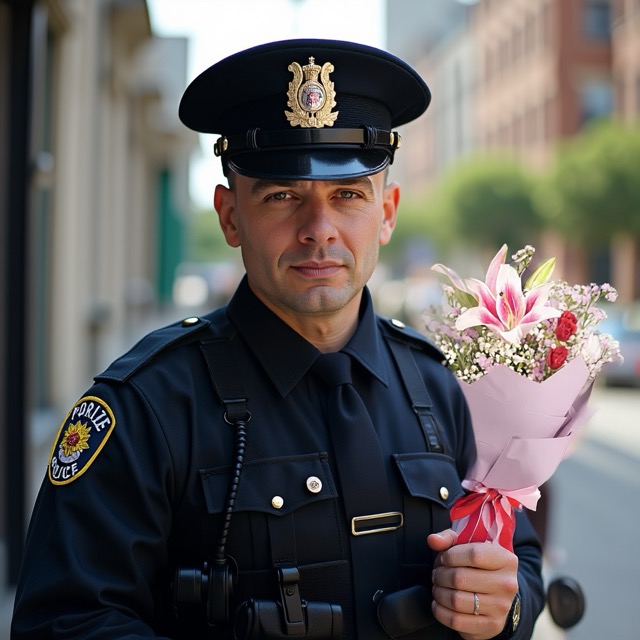} &
        \includegraphics[width=0.123\linewidth]{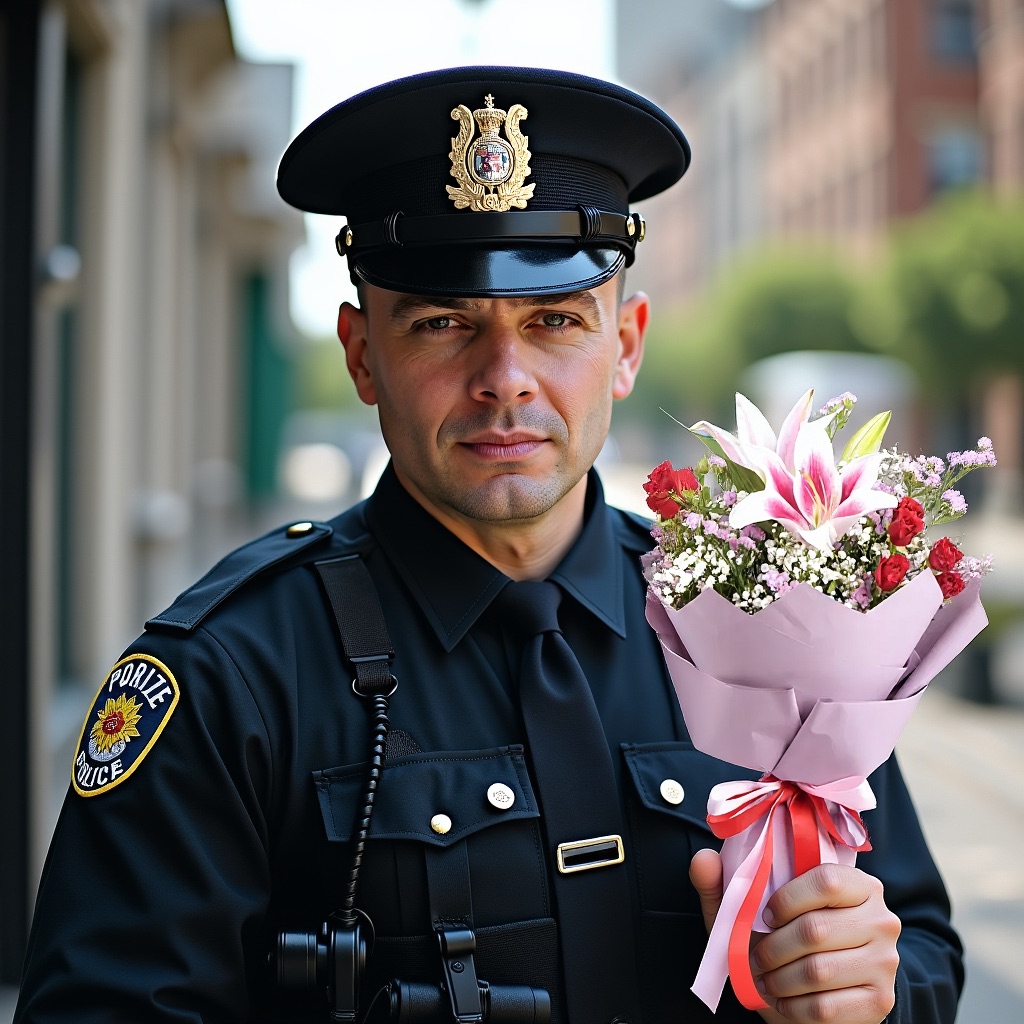} \\

        \includegraphics[width=0.123\linewidth]{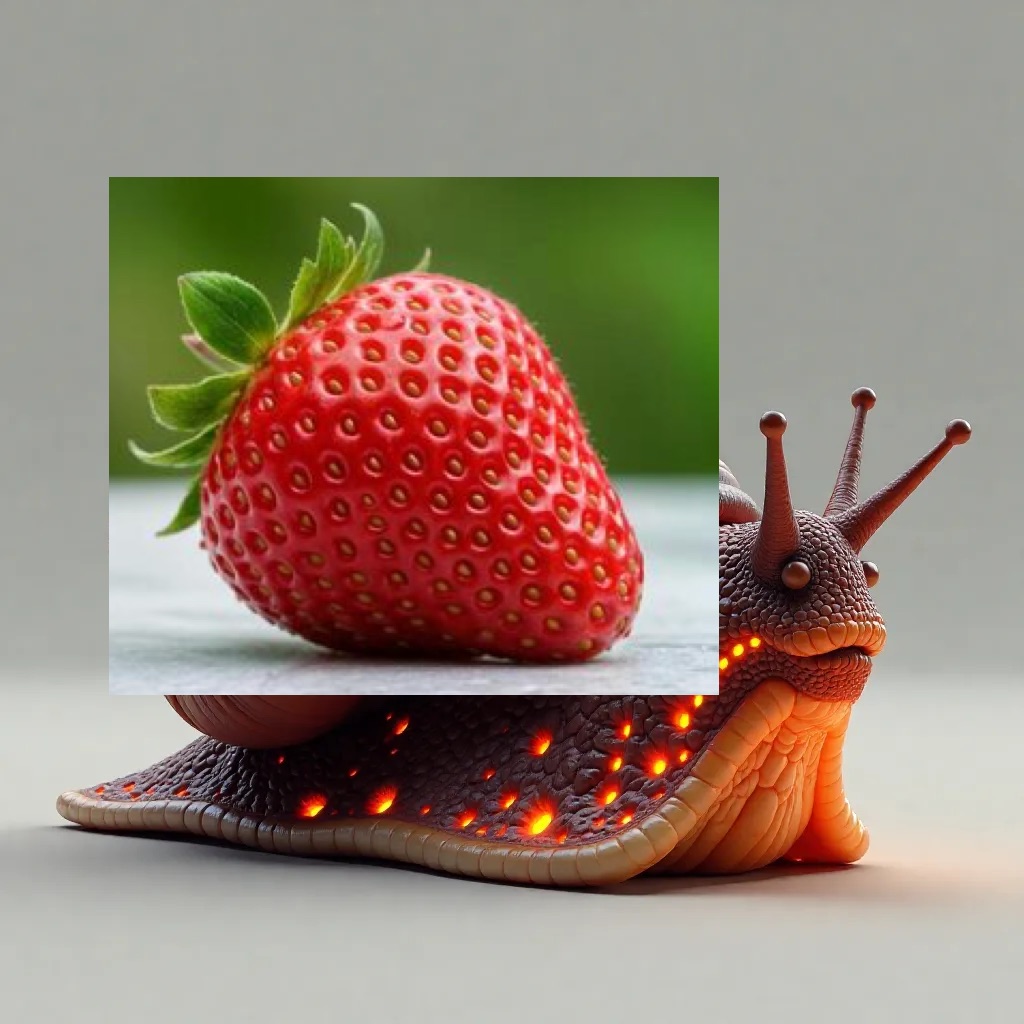} &
        \includegraphics[width=0.123\linewidth]{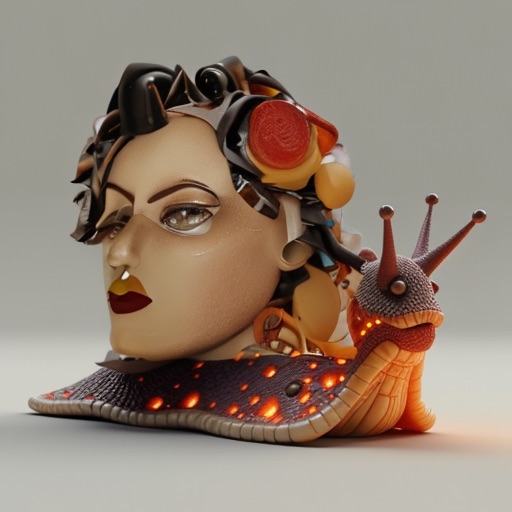} &
        \includegraphics[width=0.123\linewidth]{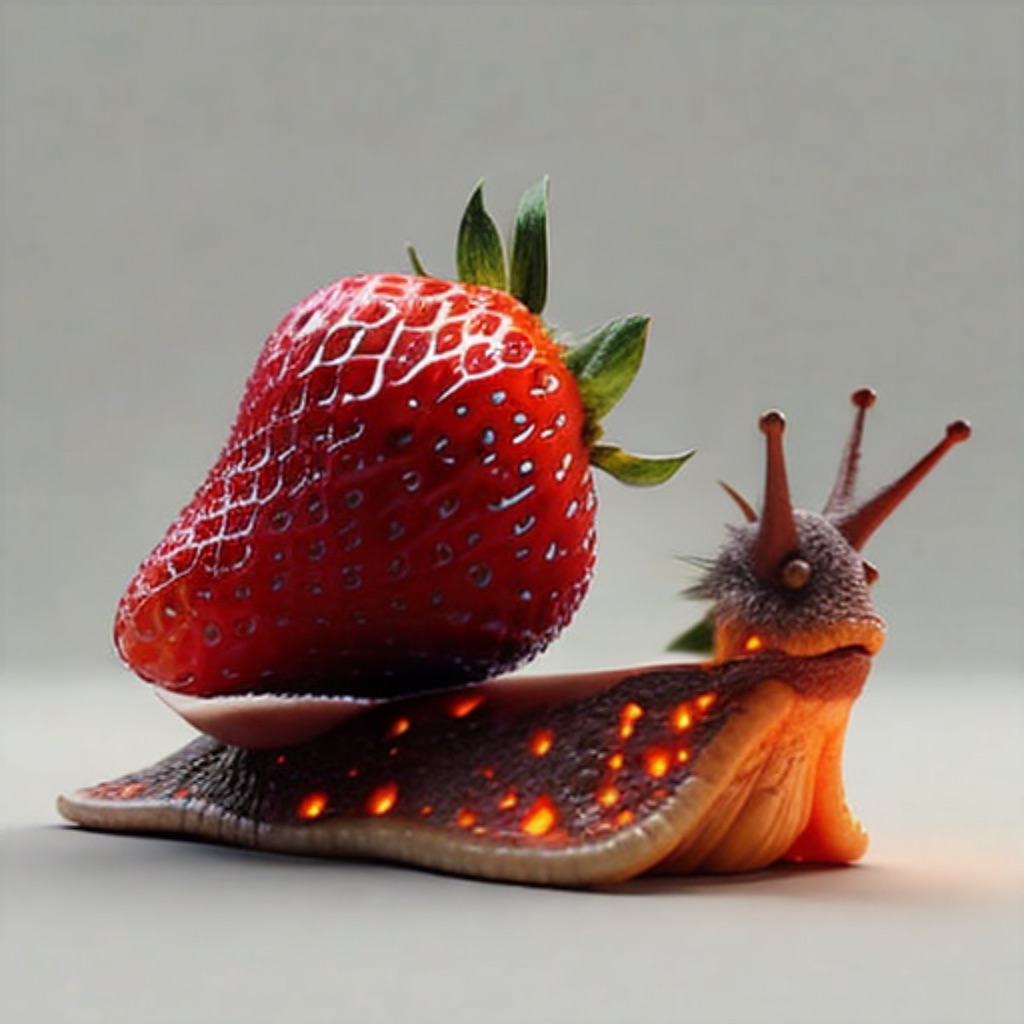} &
        \includegraphics[width=0.123\linewidth]{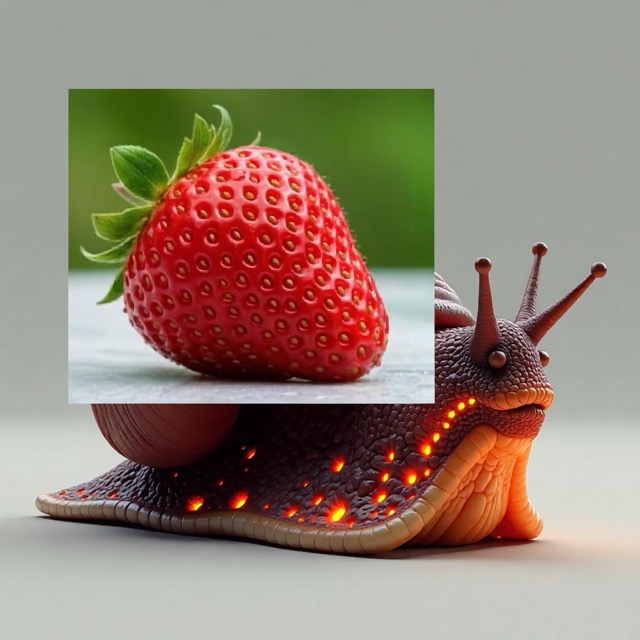} &
        \includegraphics[width=0.123\linewidth]{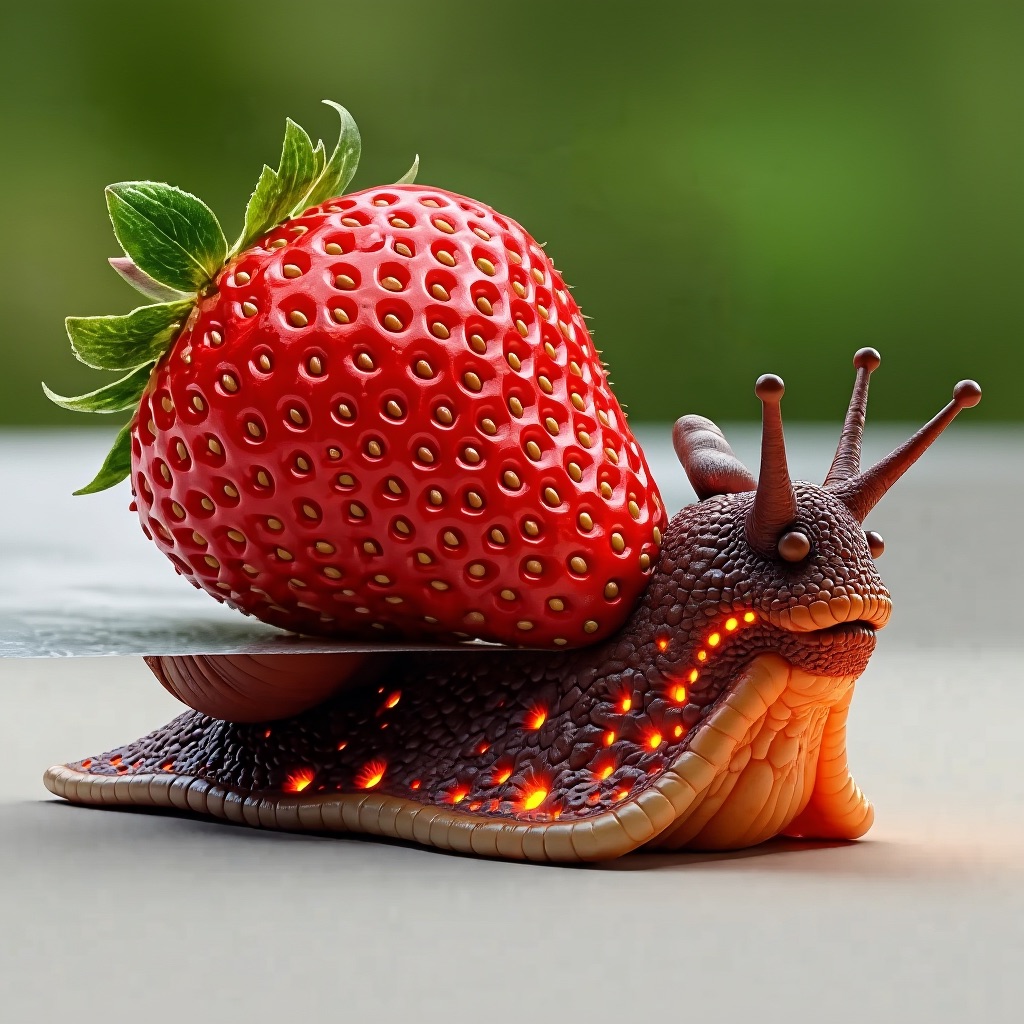} &
        \includegraphics[width=0.123\linewidth]{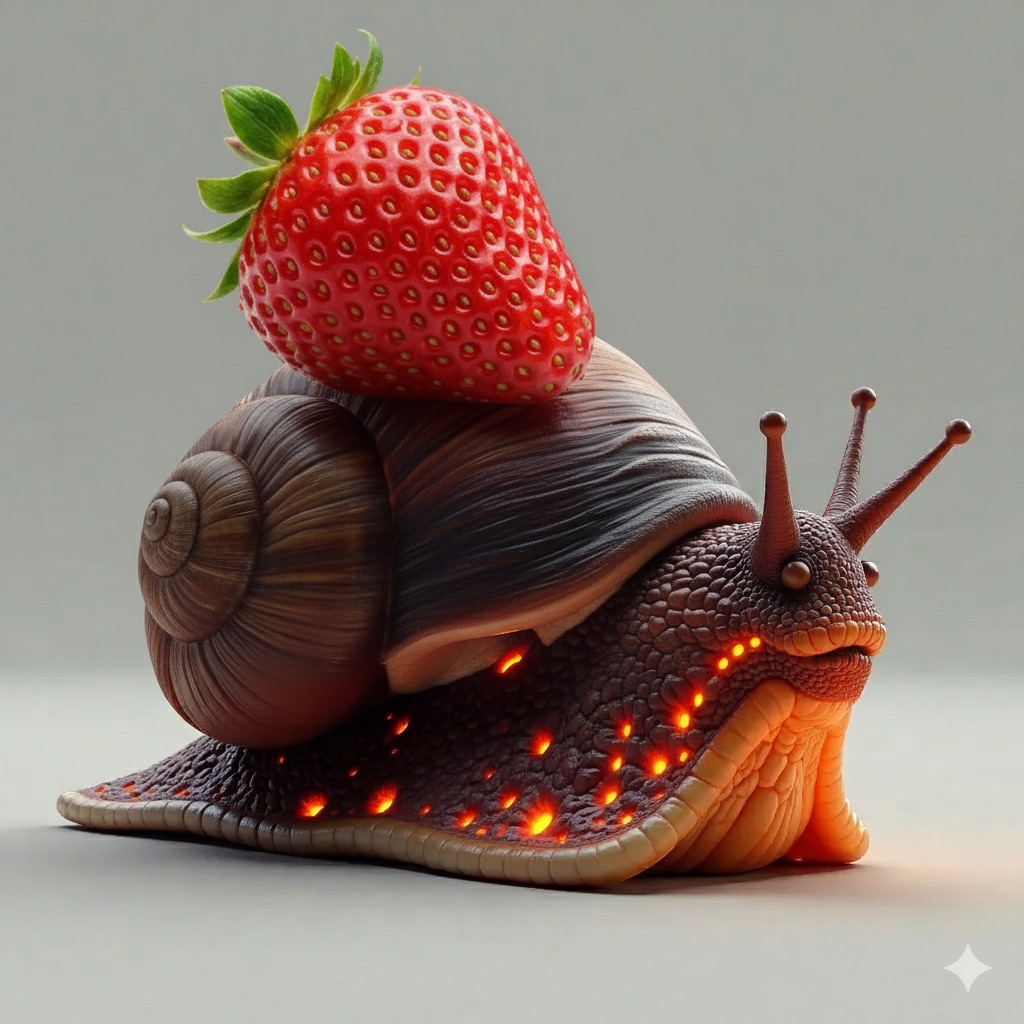} &
        \includegraphics[width=0.123\linewidth]{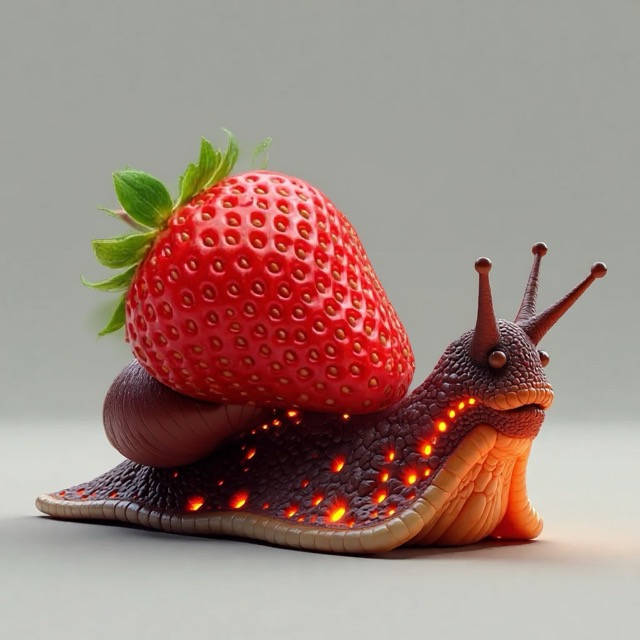} &
        \includegraphics[width=0.123\linewidth]{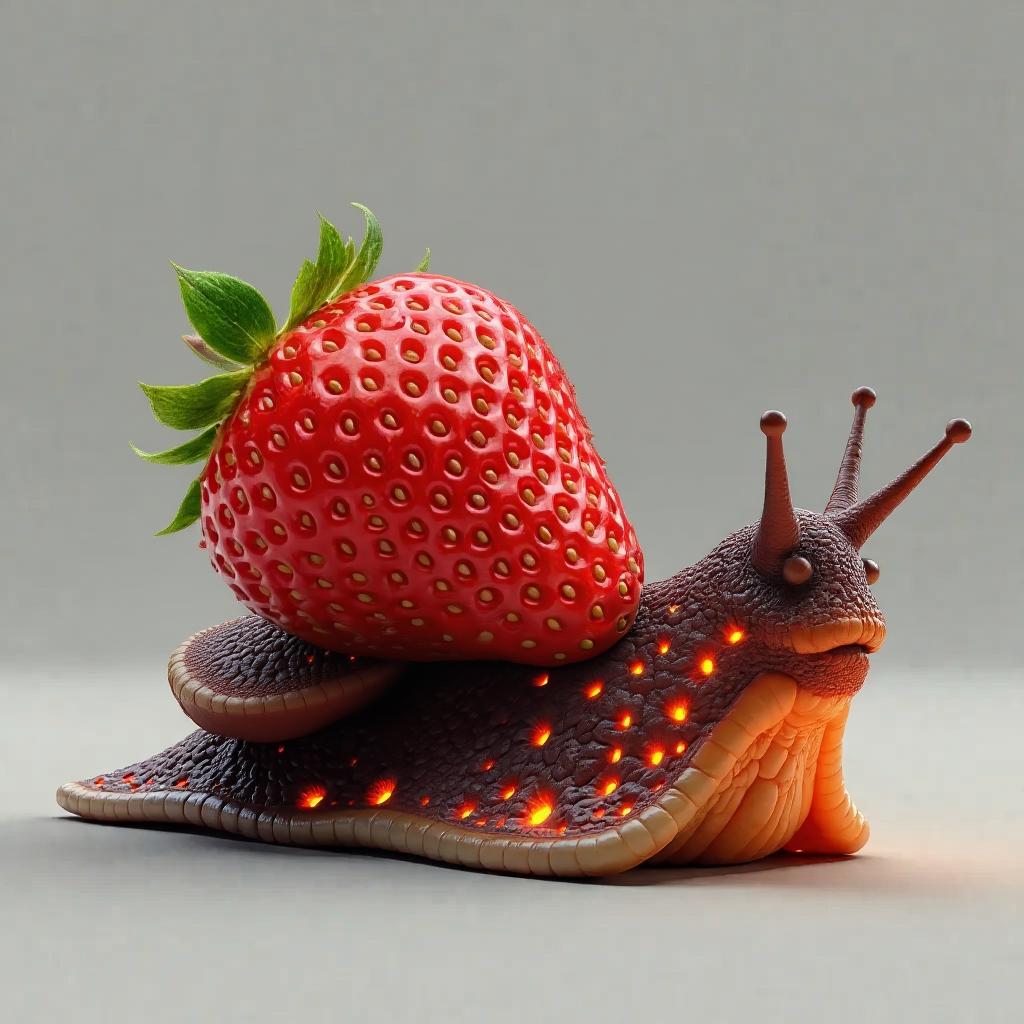} \\

        \includegraphics[width=0.123\linewidth]{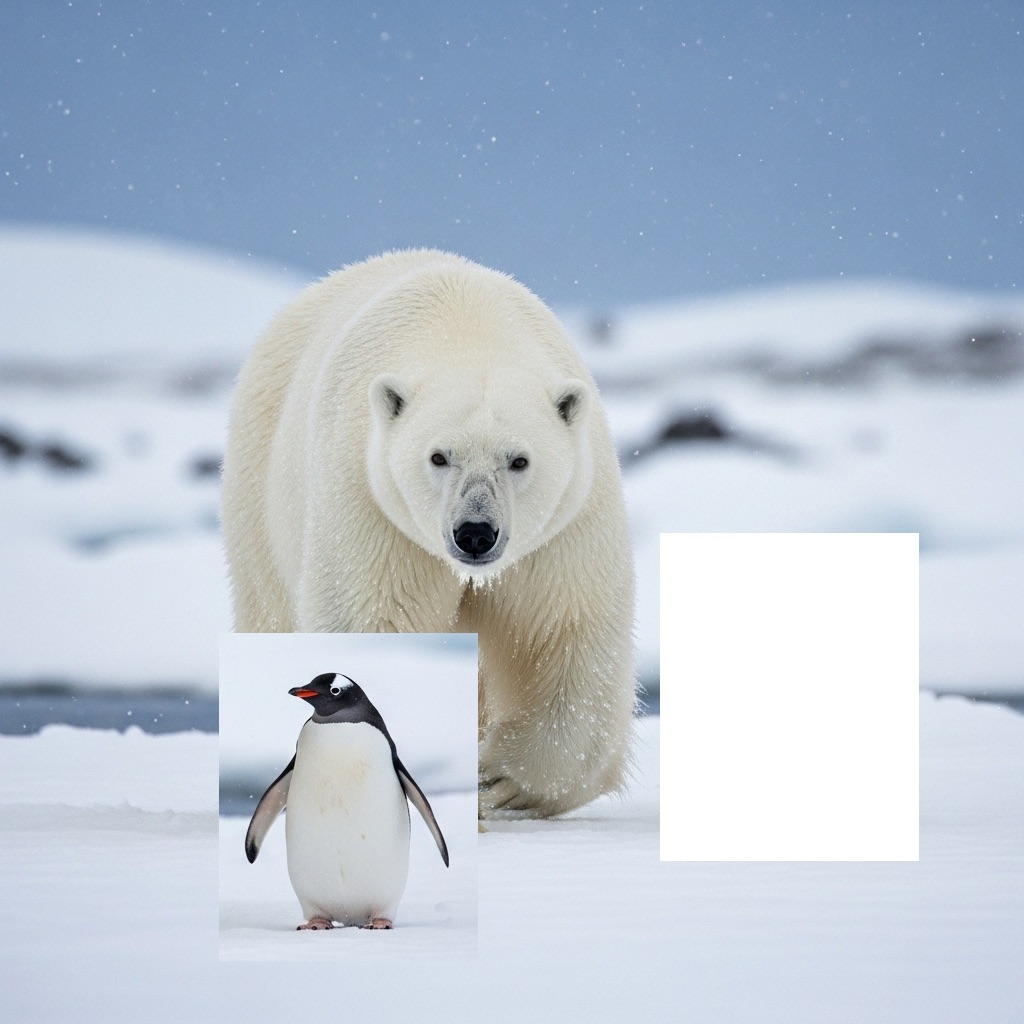} &
        \includegraphics[width=0.123\linewidth]{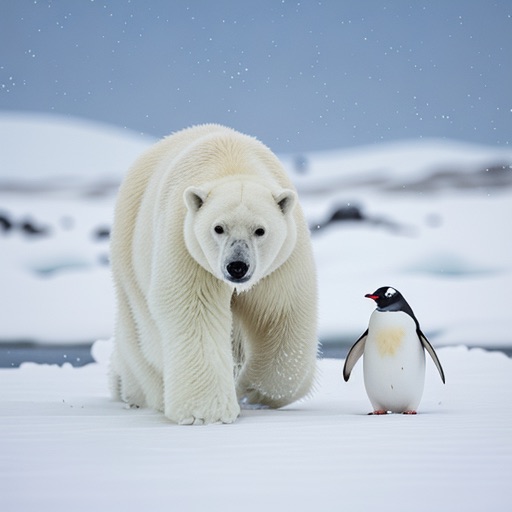} &
        \includegraphics[width=0.123\linewidth]{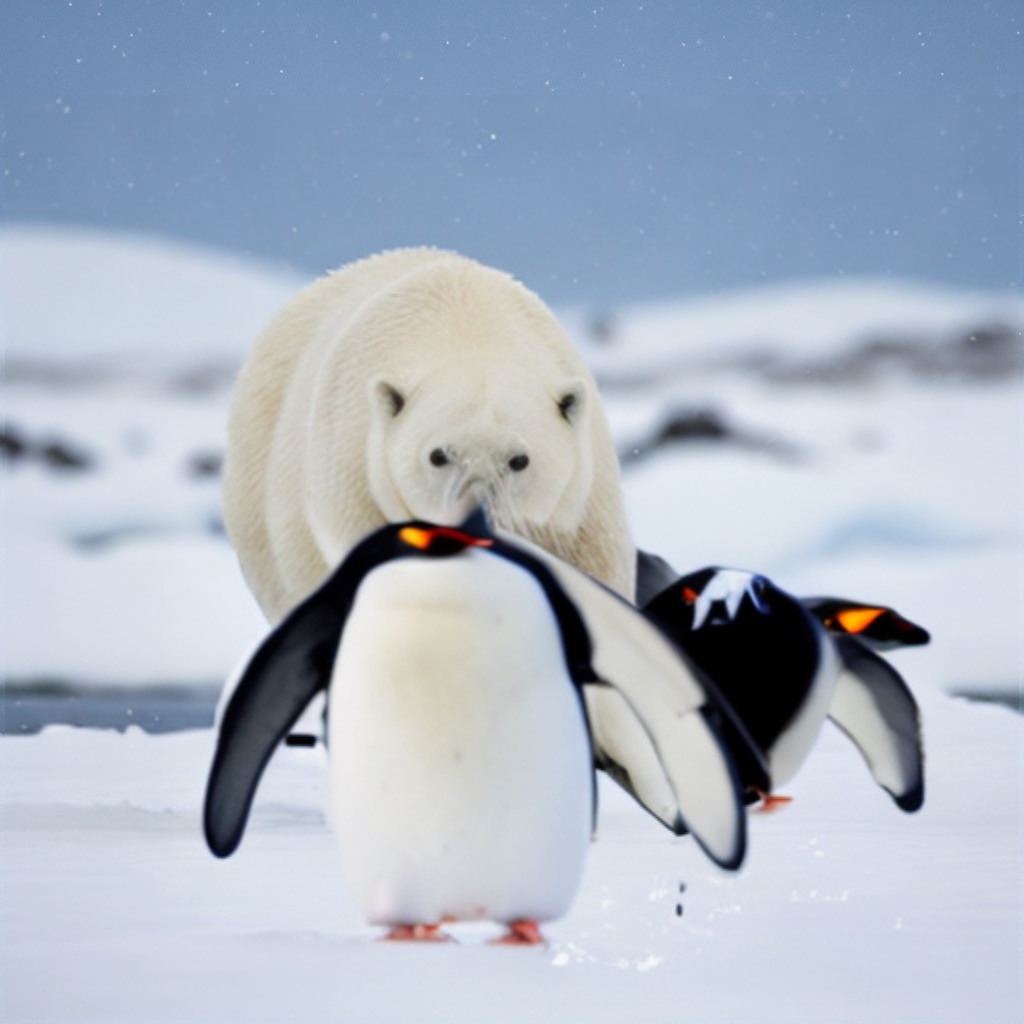} &
        \includegraphics[width=0.123\linewidth]{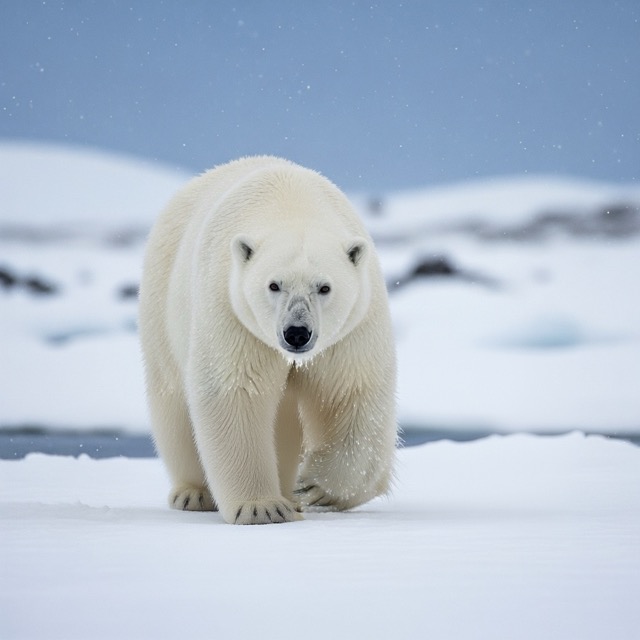} &
        \includegraphics[width=0.123\linewidth]{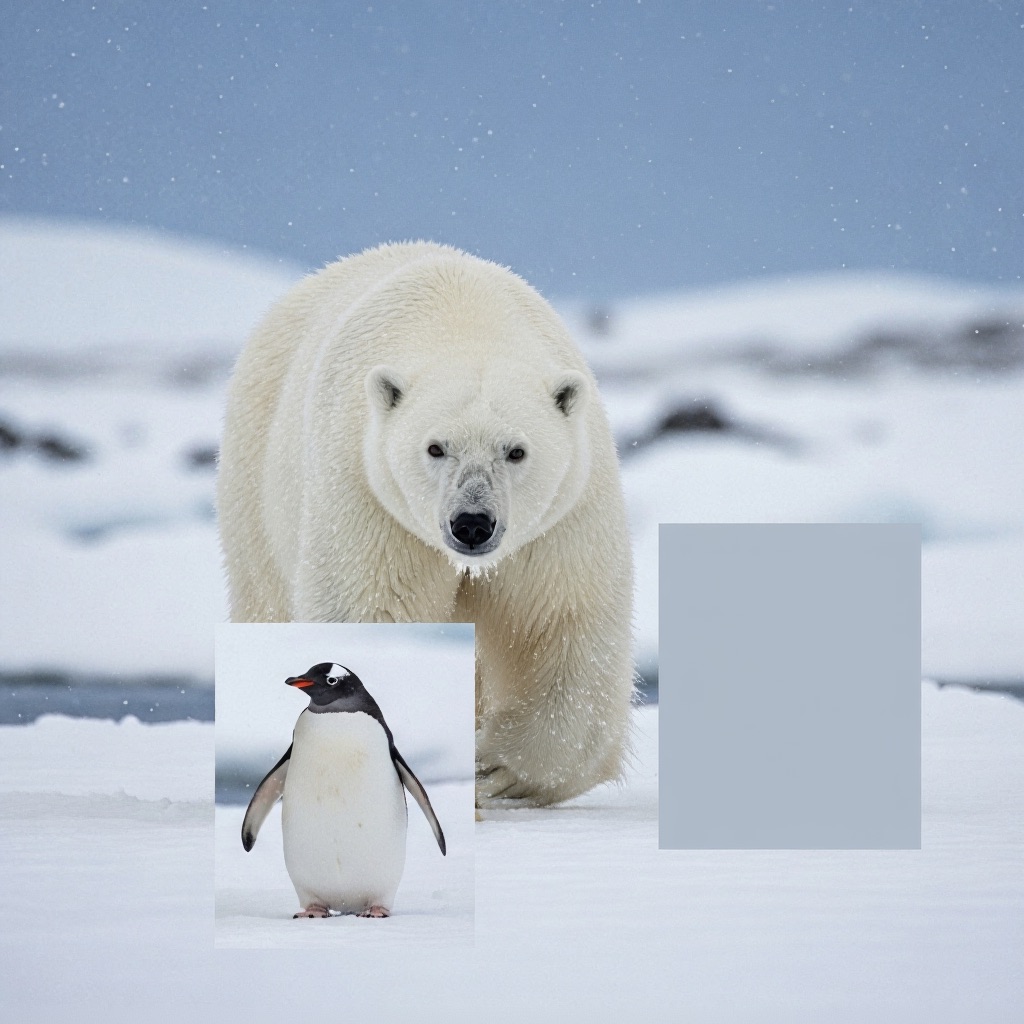} &
        \includegraphics[width=0.123\linewidth]{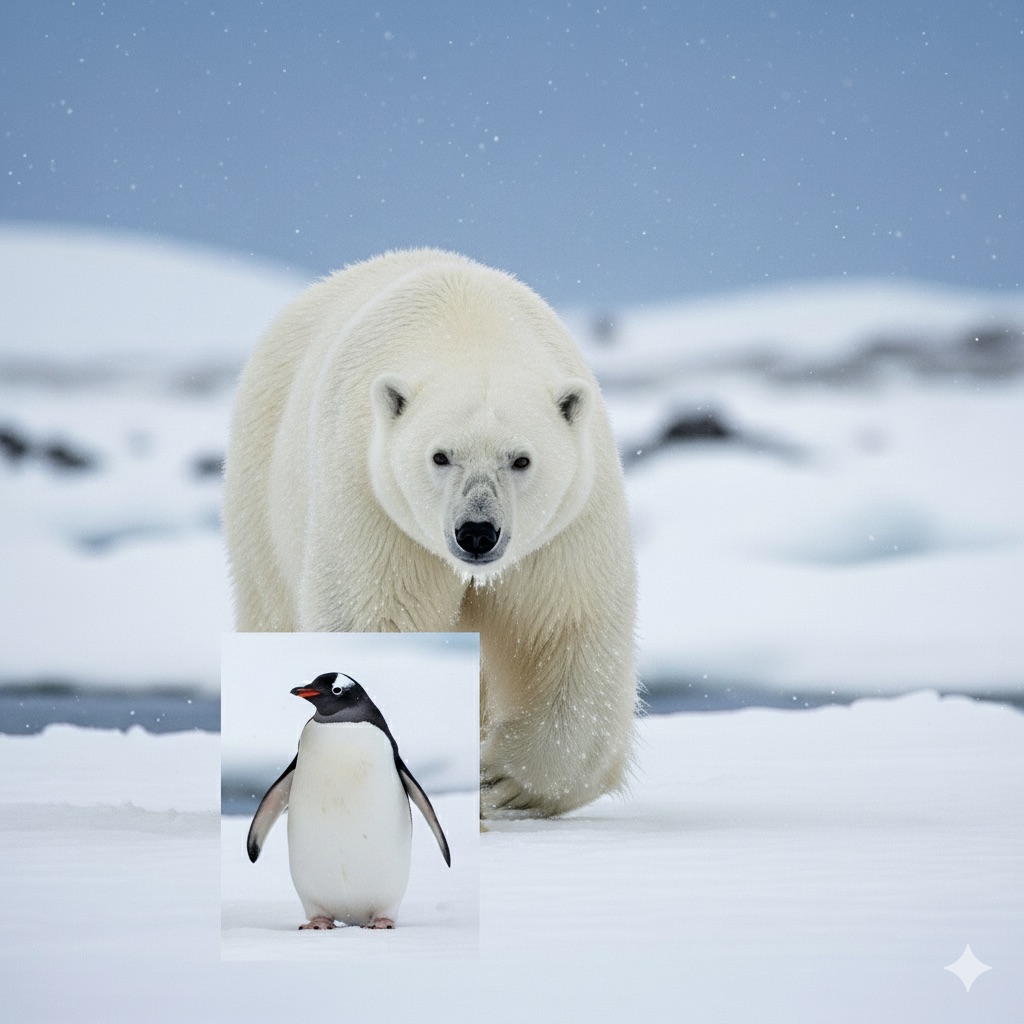} &
        \includegraphics[width=0.123\linewidth]{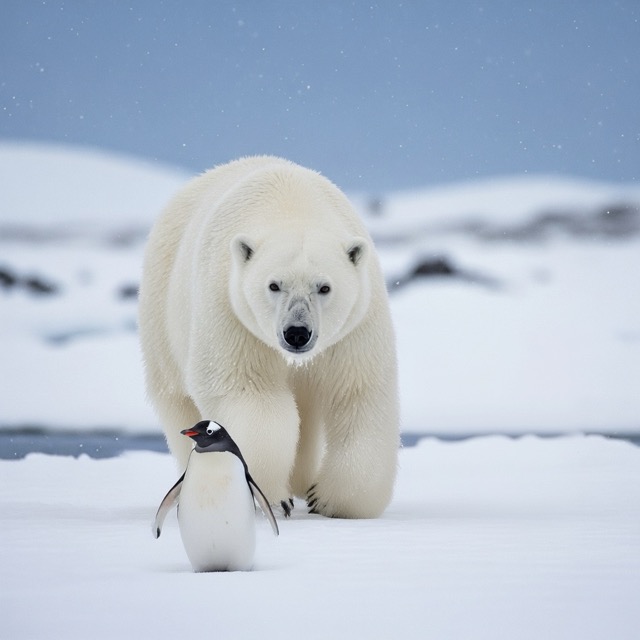} &
        \includegraphics[width=0.123\linewidth]
        {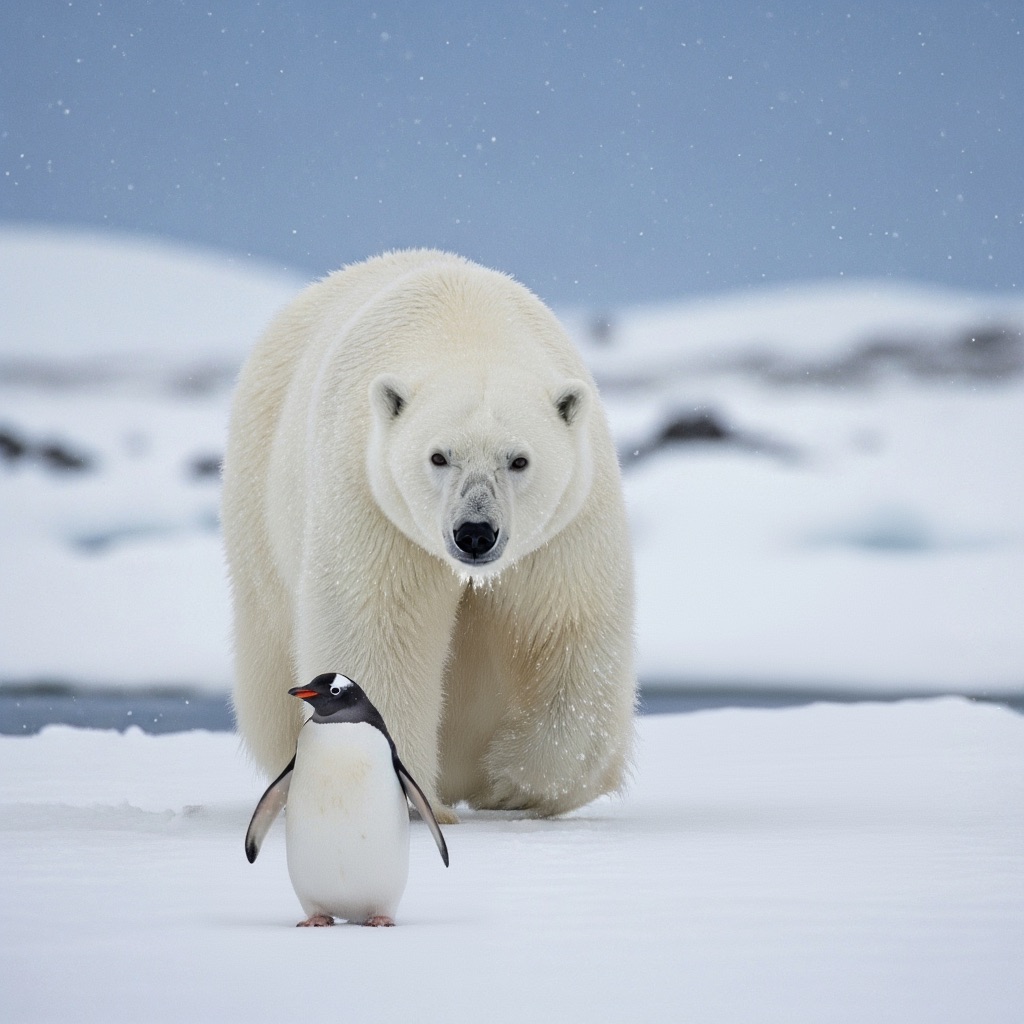} \\

        \includegraphics[width=0.123\linewidth]{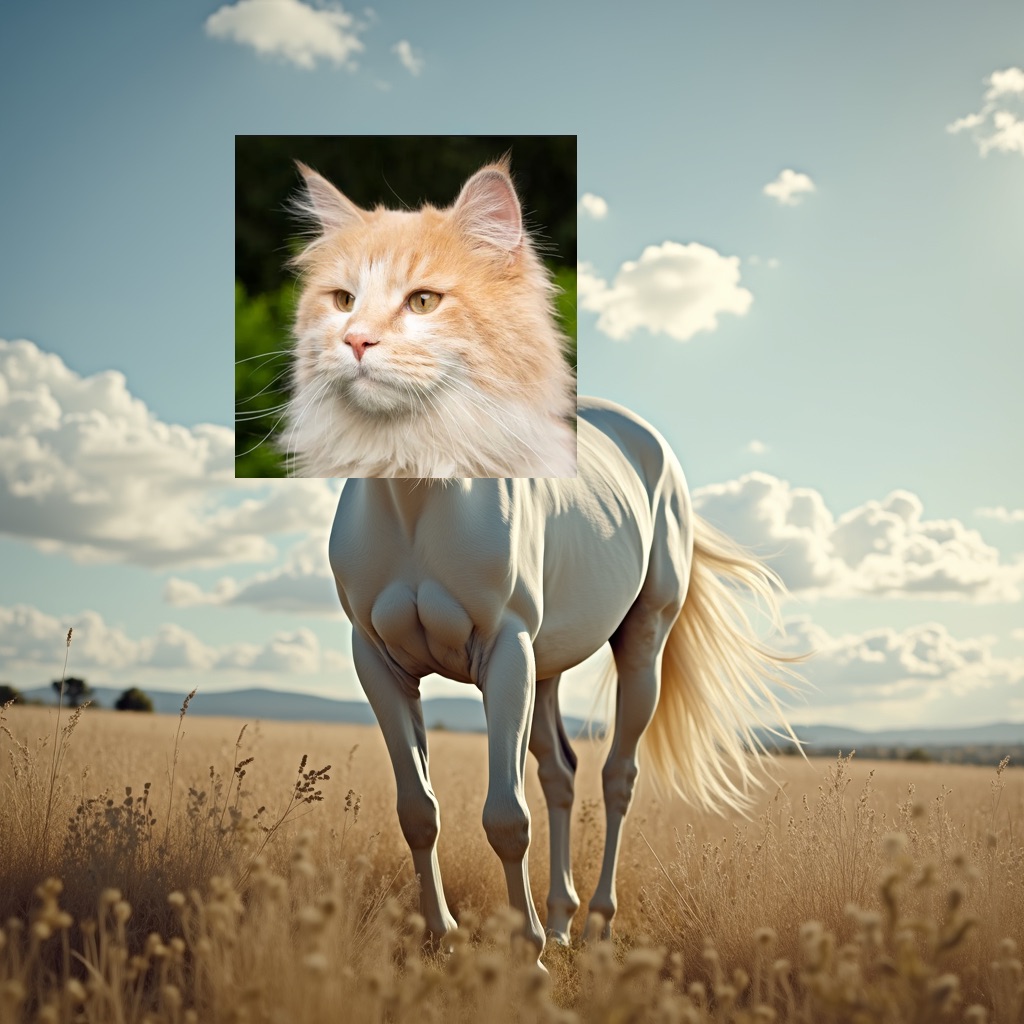} &
        \includegraphics[width=0.123\linewidth]{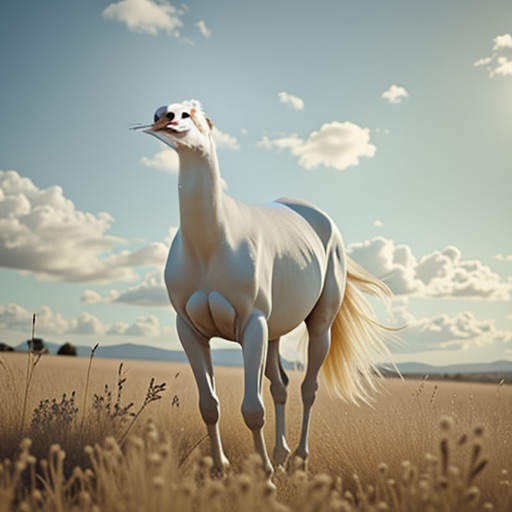} &
        \includegraphics[width=0.123\linewidth]{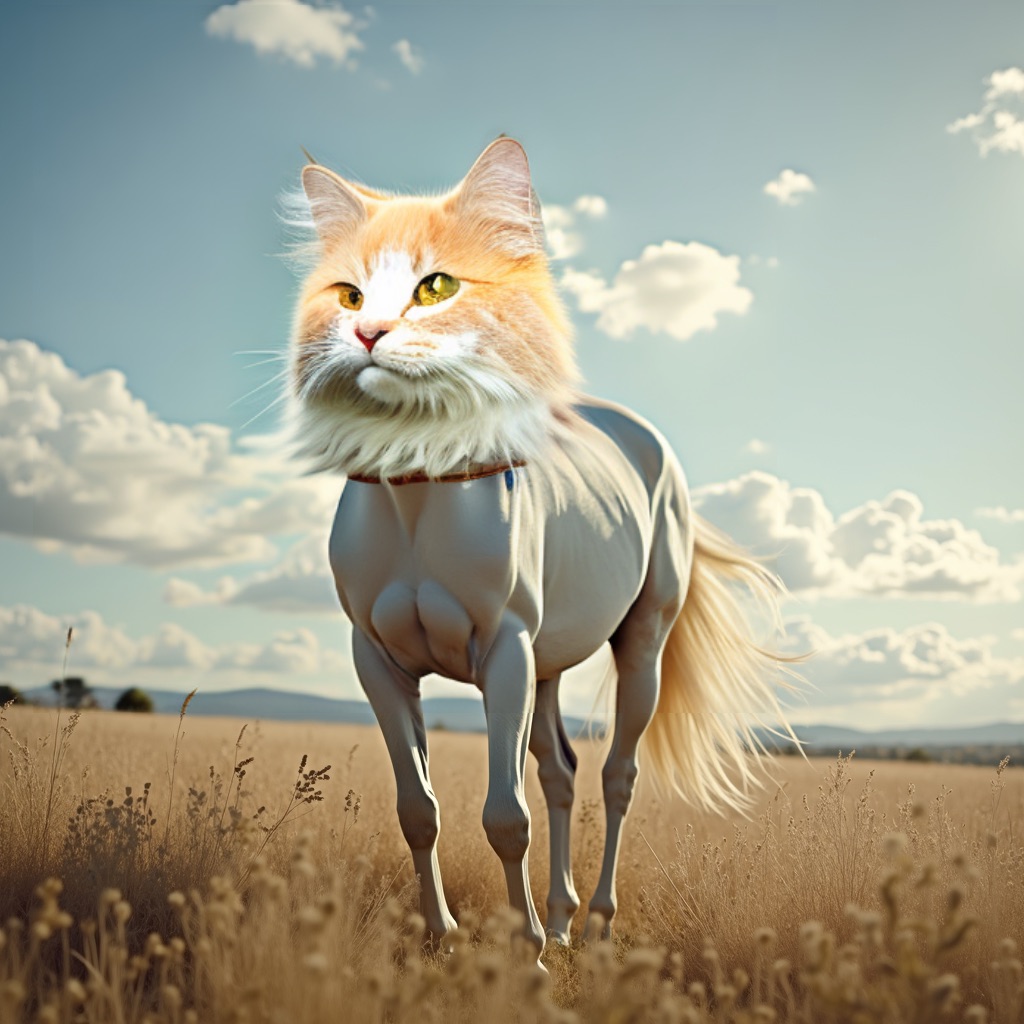} &
        \includegraphics[width=0.123\linewidth]{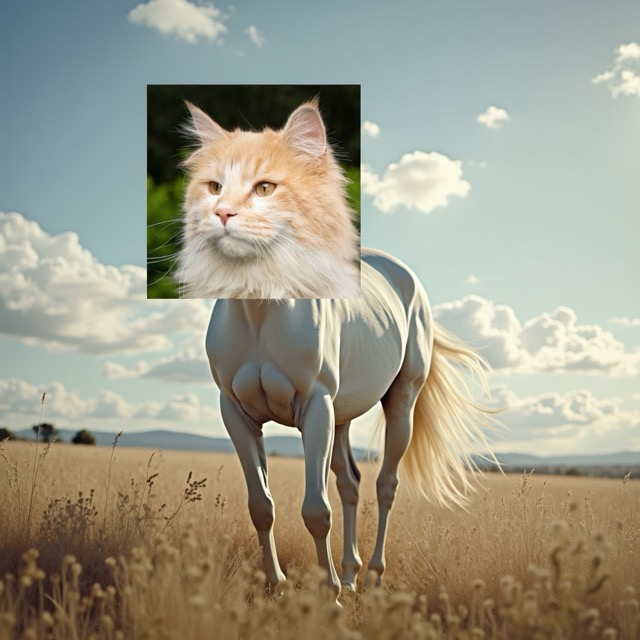} &
        \includegraphics[width=0.123\linewidth]{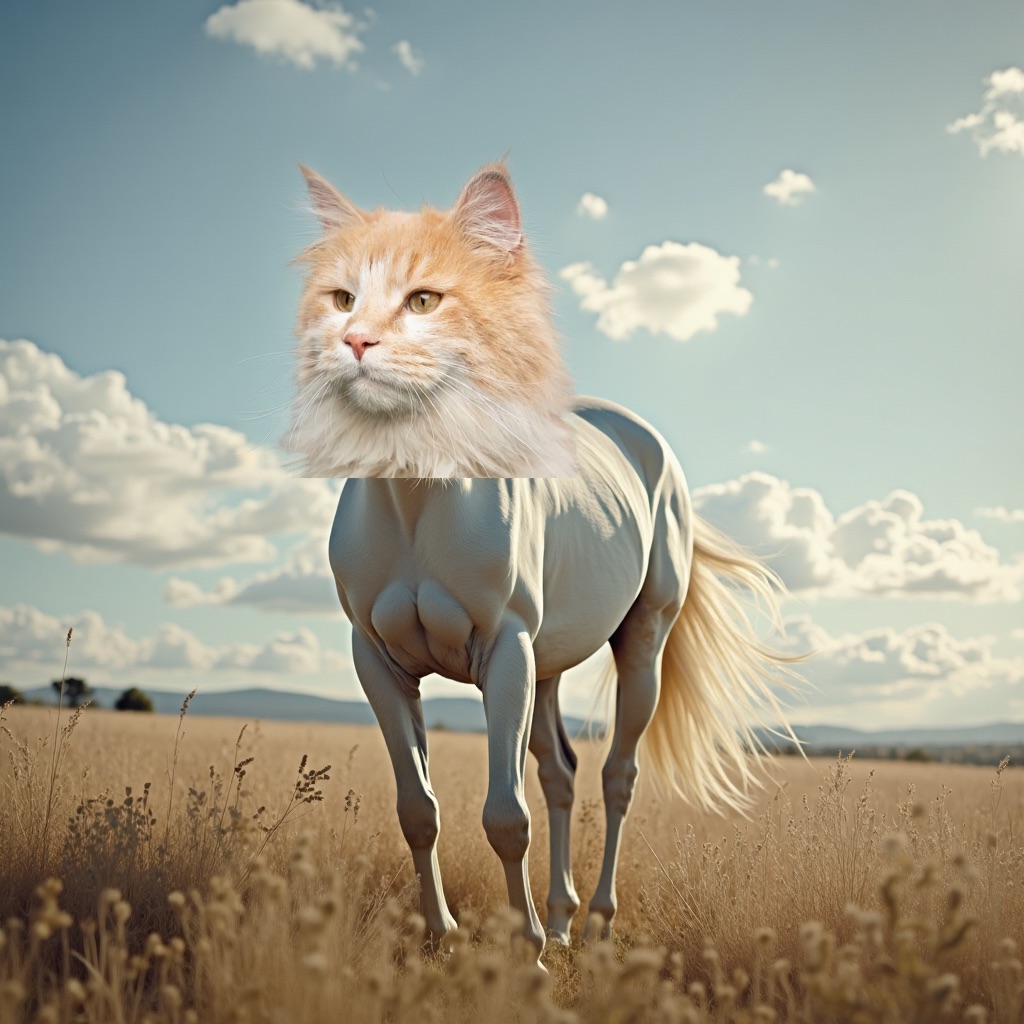} &
        \includegraphics[width=0.123\linewidth]{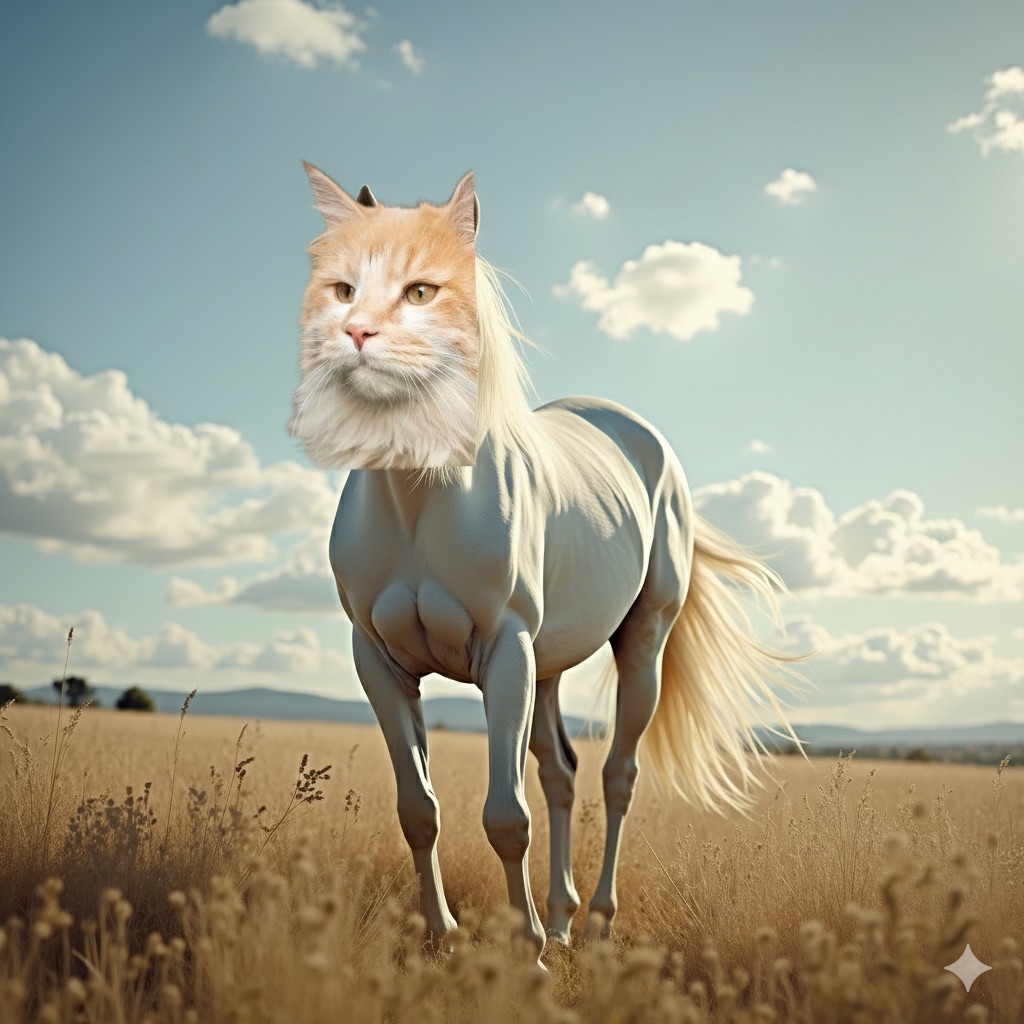} &
        \includegraphics[width=0.123\linewidth]{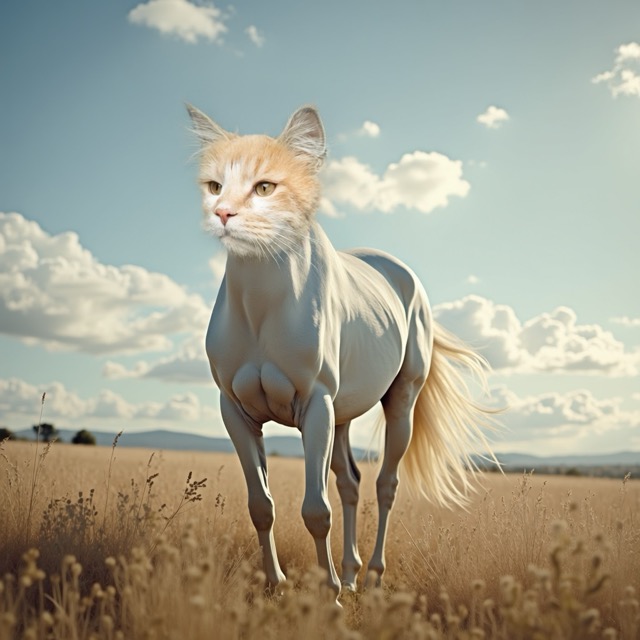} &
        \includegraphics[width=0.123\linewidth]{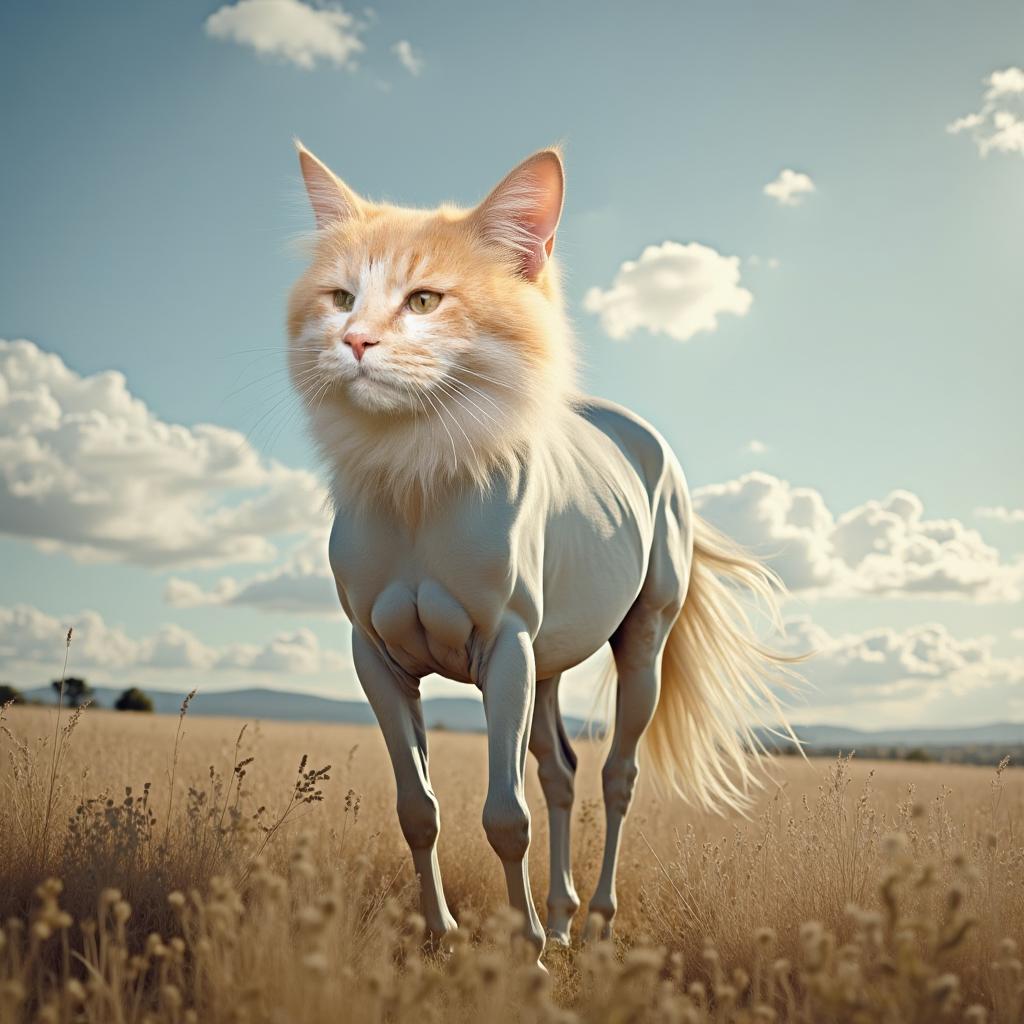} \\

        \includegraphics[width=0.123\linewidth]{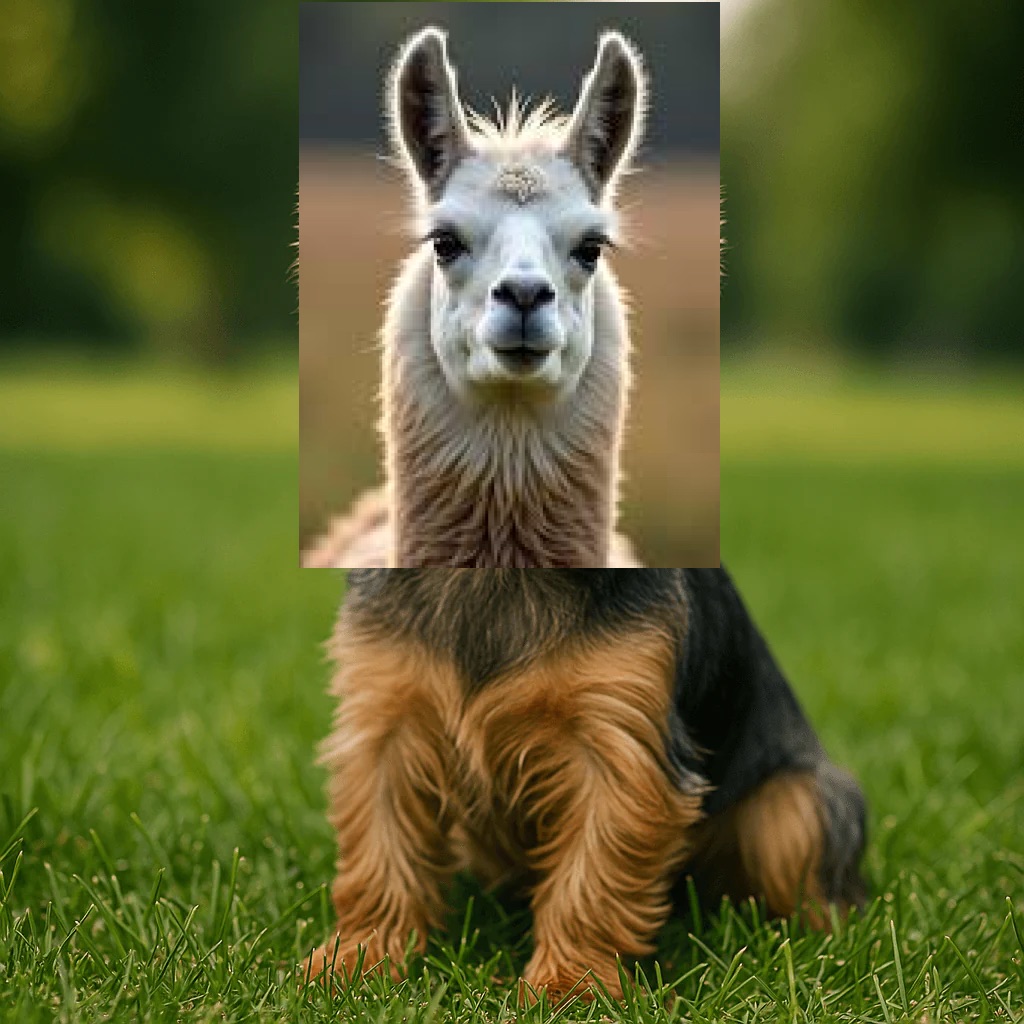} &
        \includegraphics[width=0.123\linewidth]{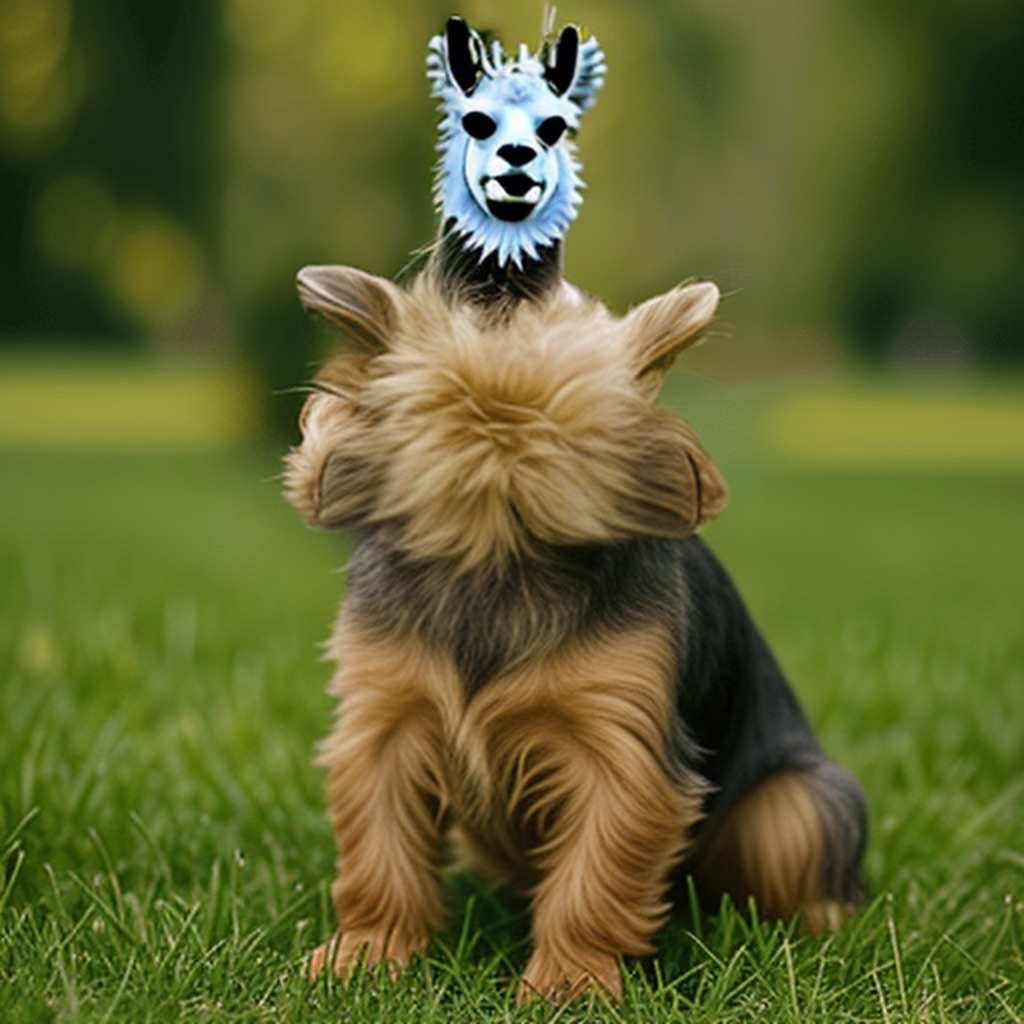} &
        \includegraphics[width=0.123\linewidth]{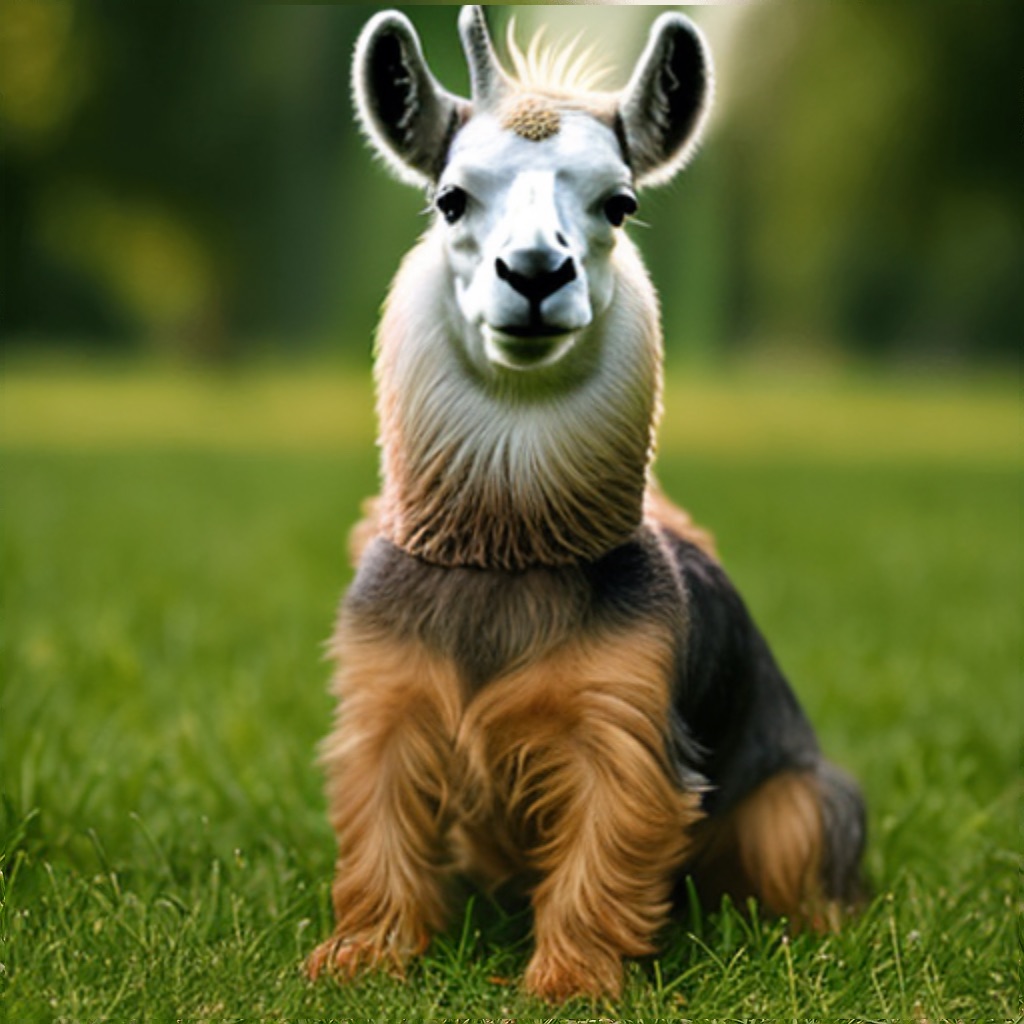} &
        \includegraphics[width=0.123\linewidth]{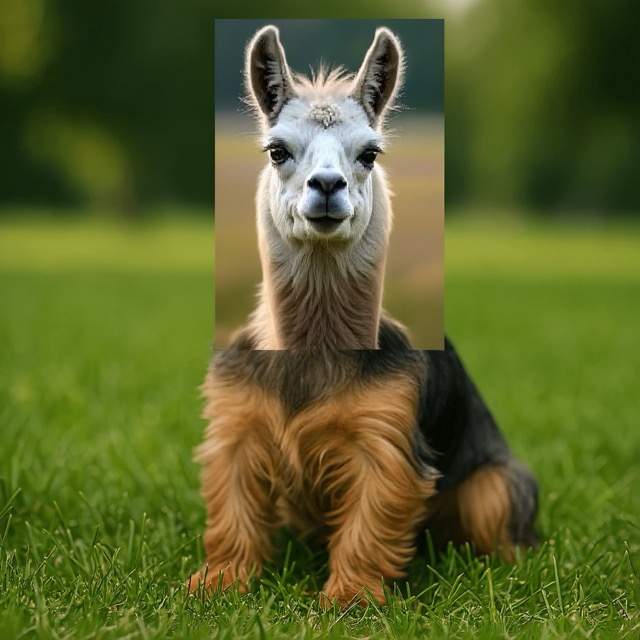} &
        \includegraphics[width=0.123\linewidth]{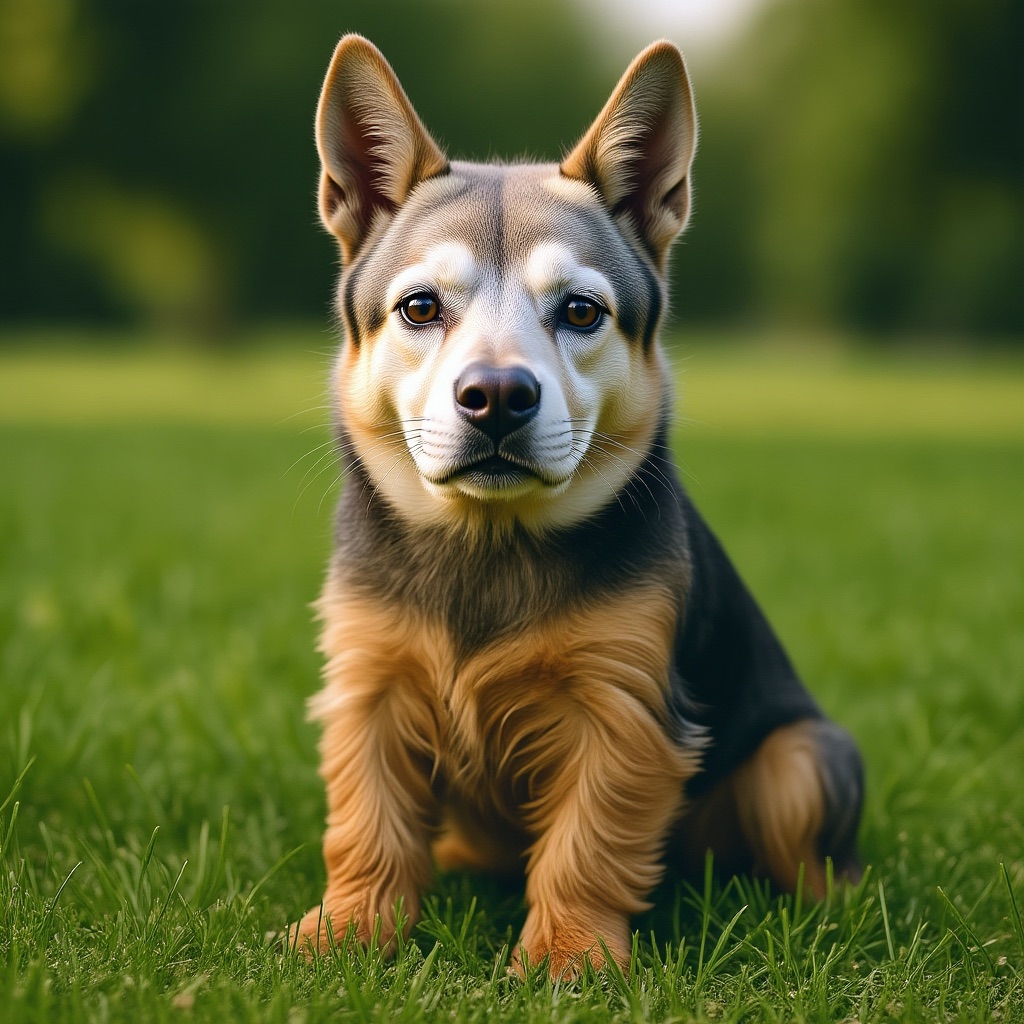} &
        \includegraphics[width=0.123\linewidth]{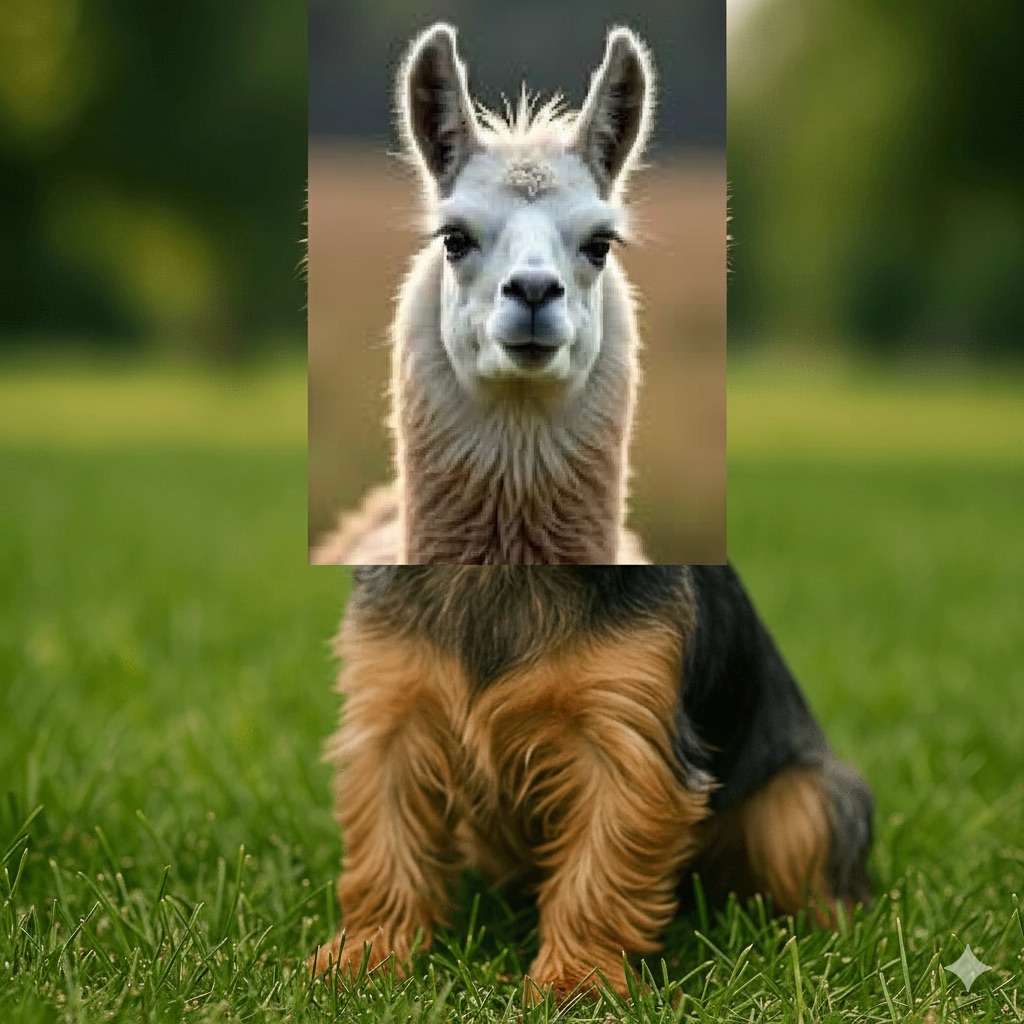} &
        \includegraphics[width=0.123\linewidth]{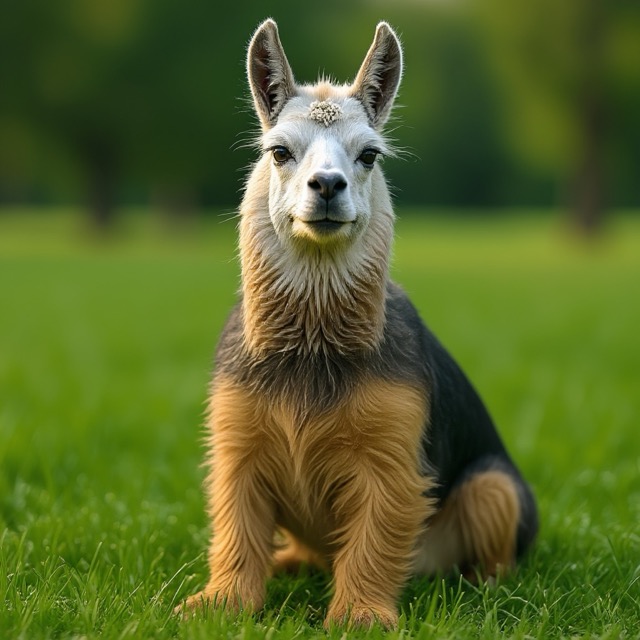} &
        \includegraphics[width=0.123\linewidth]{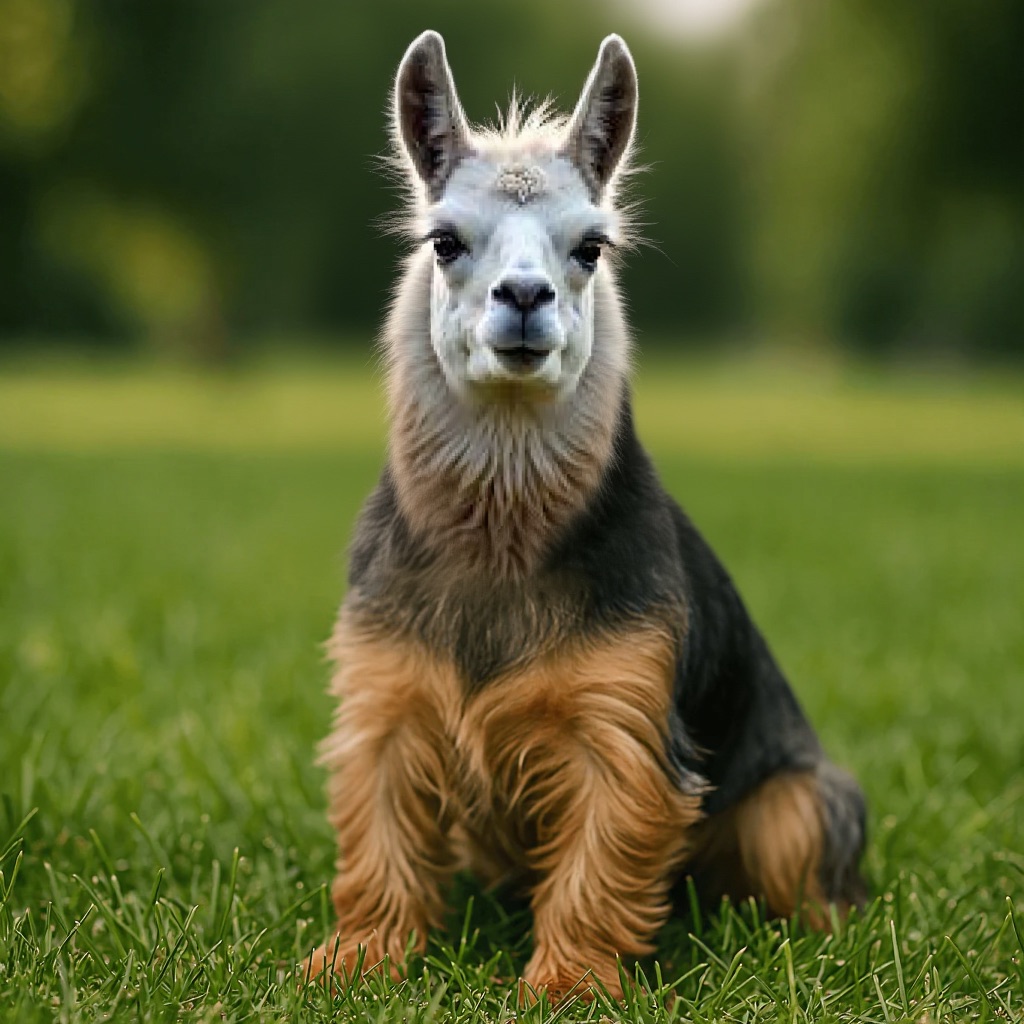} \\

    \end{tabular}
    \caption{\textbf{Qualitative comparison against competing methods.} We compare against the harmonization method TF-ICON~\cite{lu2023tf}, the reference- and layout-guided editing method AnyDoor~\cite{chen2024anydoor}, and high-quality foundation editing models (FLUX Kontext~\cite{labs2025flux1kontextflowmatching}, \new{Qwen-Image-Edit,} Nano Banana~\cite{GoogleDeepMind2025NanoBanana}). \new{We show our method applied to both FLUX Kontext (``Ours-Kontext'') and Qwen-Image-Edit (``Ours-Qwen''), demonstrating its extensibility across backbones.} Our method achieves coherent, semantically consistent blends while preserving object identity.
    }
    \label{fig:qual_eval}
    \vspace{-8pt}
\end{figure*}

\subsection{Benchmark}
While prior benchmarks in image harmonization and compositing have driven impressive progress, they are not directly aligned with our task formulation. The datasets presented in works such as SSH~\cite{jiang2021ssh} and Cross-Domain Compositing~\cite{hachnochi2023cross} primarily evaluate appearance-level consistency, emphasizing adjustments to global color, tone, and illumination. These settings do not require a model to reason about the semantic content of the pasted object, and therefore do not expose the semantic harmonization capabilities central to our approach.
\op{For instance, they do not capture complex compositions such as the “giraffe–duck’’ hybrid in Figure \ref{fig:attn_failure}, where the structure of the duck’s neck must be subtly adjusted to align with the giraffe’s.}

Conversely, benchmarks used in layout- or reference-guided editing, such as AnyDoor~\cite{chen2024anydoor}, consist of concept-location pairs in which the inserted object often differs in pose or structure from the base image. This makes them ill-posed for methods that explicitly preserve original object’s geometry and identity. 

Finally, existing datasets rarely include fine-grained or sub-object edits, such as eyes, animal heads, or accessories like horns or goggles, which our method naturally accommodates. To enable fair evaluation, we construct a new benchmark of 150 diverse compositions spanning both synthetic and natural images, where objects and sub-objects are cropped either from the same image or from distinct sources. Examples are shown in Figures \ref{fig:qual_eval}, \ref{fig:vlm} and \ref{fig:ablation_failure}.

\subsection{Metrics}
The quantitative evaluations of our method reflects the two core objectives of our task: preserving the identity of the pasted content while harmonizing it naturally into the target image. Therefore, we assess performance along two complementary axes: \emph{identity preservation} and \emph{image quality}. 

For image quality, we employ the CLIP-IQA metric~\cite{wang2023exploring}, a no-reference CLIP-based image quality assessment method. 
CLIP-IQA estimates perceptual quality by comparing the image’s CLIP similarity to prompts describing high-quality photographs (e.g., “sharp,” “colorful,” “high contrast”) and low-quality ones (e.g., “noisy,” “blurry,”), providing an interpretable quality score. For identity preservation, we report the Learned Perceptual Image Patch Similarity (LPIPS) score~\cite{zhang2018unreasonable}, computed both over the entire image and specifically over the cropped foreground regions. 

\vspace{-2pt}
\subsection{Comparison against Baselines}
\label{sec:baseline_comp}

Our method bridges traditional image harmonization and more flexible reference- or layout-guided editing approaches. Harmonization methods focus on adjusting color, illumination, and appearance to produce visually coherent composites, but they provide limited control over object semantics or shape. In contrast, layout- or reference-guided methods enable greater semantic flexibility and allow more expressive edits, yet they often compromise identity preservation when integrating a pasted object into a new scene.

Accordingly, we evaluate LooseRoPE against representative methods from both categories. For harmonization-based methods, we include TF-ICON~\cite{lu2023tf}, a diffusion-based harmonization method that jointly inverts the foreground and background latents before blending them into a unified image. For reference- and layout-guided editing, we compare with AnyDoor~\cite{chen2024anydoor}, which employs an identity-preserving encoder for object insertion, and SwapAnything~\cite{gu2024swapanythiing}, which swaps an object in an image with a given concept, while keeping the context unchanged.

In addition, we report results using the base editing backbones, FLUX Kontext~\cite{labs2025flux1kontextflowmatching} \new{and Qwen-Image-Edit \cite{wu2025qwen}} to isolate the contribution of our method, and provide qualitative comparisons against a state-of-the-art proprietary system, NanoBanana~\cite{GoogleDeepMind2025NanoBanana}, to contextualize our method’s visual quality relative to high-end commercial models.

\begin{figure*}[t]
    \centering
    \setlength{\tabcolsep}{1pt}
    \begin{tabular}{cc} 
    
    \includegraphics[width=0.45\linewidth]{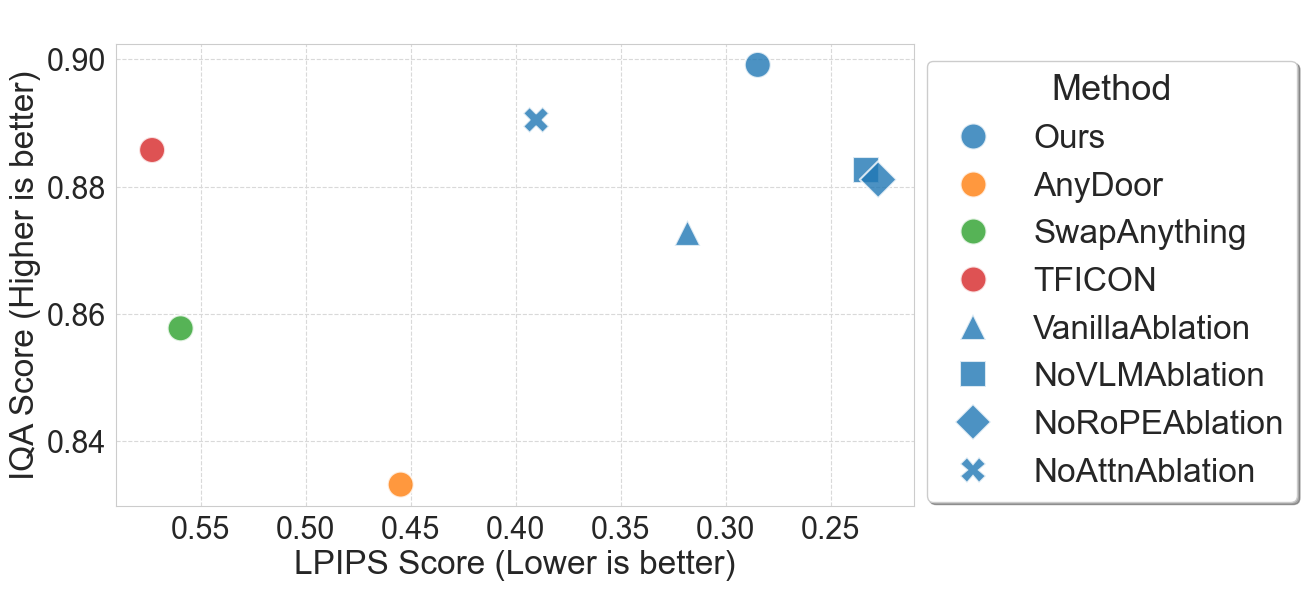}
    &
    \includegraphics[width=0.45\linewidth]{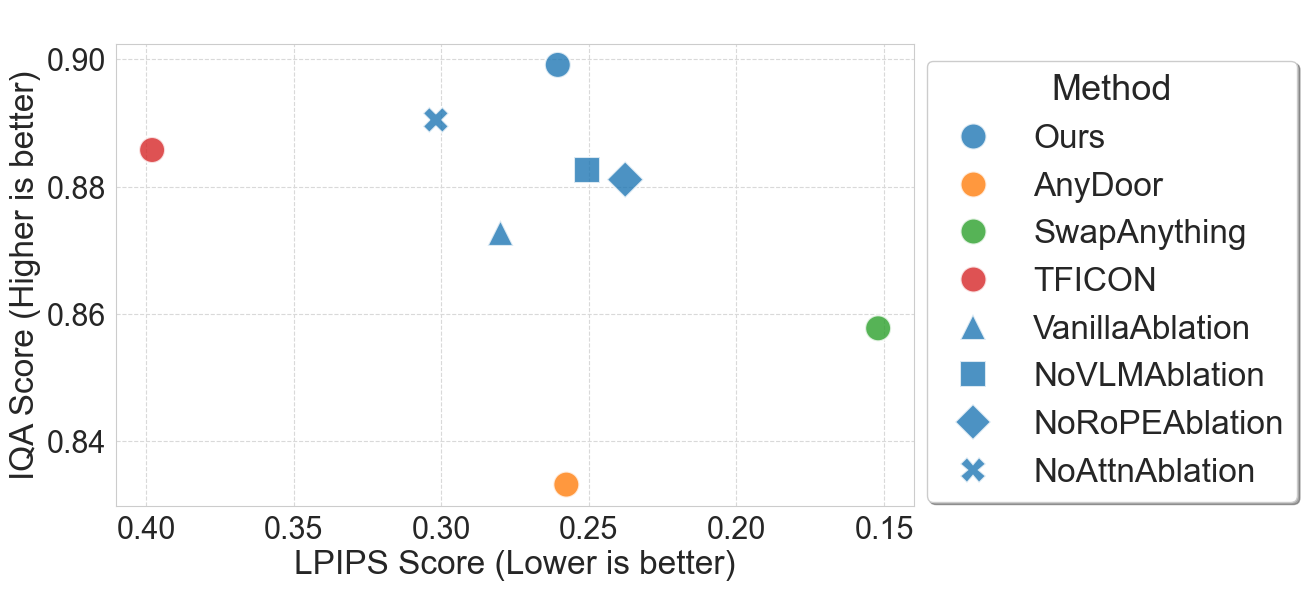}

    \\
    
    \begin{minipage}{0.45\linewidth}
        \centering
        \footnotesize{CLIP IQA vs Crop foreground LPIPS scores}
    \end{minipage}
    &
    \begin{minipage}{0.45\linewidth}
        \centering
        \footnotesize{CLIP IQA vs Full image LPIPS scores}
    \end{minipage}
    
    \end{tabular}
    \vspace{-6pt}
    \caption{\textbf{Quantitative analysis of methods and ablations.} 
    Left: CLIP-IQA score vs. LPIPS computed on the estimated foreground within the cropped region. Right: CLIP-IQA score vs. LPIPS computed over the entire image. Our method preserves the subject’s identity inside the crop while maintaining overall image quality, whereas other methods either preserve the input (low LPIPS) but sacrifice global quality (low CLIP-IQA) by neglecting the blending instruction, or maintain global quality by suppressing the crop. %
    } 
    \vspace{-8pt}

    \label{fig:quant}
\end{figure*}

Figure \ref{fig:qual_eval} presents a qualitative comparison against all competing baselines. \new{In addition to results obtained with FLUX Kontext (``Ours-Kontext''), we also include examples of LooseRoPE applied to Qwen-Image-Edit (``Ours-Qwen'') in this figure, showing that our approach extends beyond a single backbone and transfers to other RoPE-based image editing models. Unless stated otherwise, results of our method and its ablated versions throughout the paper are obtained using FLUX Kontext.} The examples presented in this figure show that while competing methods often fall into either neglect (Nano Banana on top row) or suppression (SwapAnything on bottom row), our method manages to steer between these modes, achieving high quality coherent blends. Furthermore, it is evident that our method excels at preserving identity and placing the cropped objects in their assigned locations. Competing methods, while sometimes producing coherent blends, struggle with identity preservation (see the raised strawberry in NanoBanana on the second row from the top).

Figure~\ref{fig:quant} presents a quantitative comparison against AnyDoor~\cite{chen2024anydoor}, TF-ICON~\cite{lu2023tf}, SwapAnything~\cite{gu2024swapanythiing} and FLUX Kontext~\cite{labs2025flux1kontextflowmatching}. As can be seen, our method achieves high CLIP-IQA scores while maintaining moderate LPIPS values, reflecting a balanced trade-off between visual quality and identity preservation. 
Notably, very low LPIPS scores \op{over the entire image} often indicate neglect, where the model fails to meaningfully integrate the pasted region.

\vspace{-7pt}
\paragraph{User study.} 
Since automatic metrics do not always fully capture perceptual quality or the nuances of identity preservation, we complement our quantitative evaluation with a user study. The study follows a standard two-alternative forced-choice format.
Users were each shown 20 questions, each containing an input image, an output image produced by our method and another produced by one of the competing baselines. Users were instructed to rate the outputs according to: identity preservation, blending coherence, placement location accuracy, and overall quality. We collect results from 27 users, resulting in a total of 540 responses per category. As can be seen in Table \ref{tab:win_rates}, our method outperforms all baselines across all categories.

\begin{table}[t]
    \centering %
    \small %
    \setlength{\tabcolsep}{3.5pt} %
    
    \caption{
    \textbf{Ours vs. Baseline Win Rates.} We report the percentage of user study votes in which our method was preferred over competing baselines. Users evaluated the edits according to four criteria: identity preservation, blending coherence, placement accuracy, and overall quality.
    }
    \label{tab:win_rates} %
    \vspace{-4pt}
    \begin{tabular}{l c c c c}
        \toprule
        \textbf{Baseline} & \makecell{Identity \\ Pres.} & Blending & Placement & \makecell{Overall} \\
        \midrule
        AnyDoor       & 66.07 & 58.93 & 66.96 & 63.39 \\
        Swap Anything & 63.39 & 50.89 & 67.86 & 55.36 \\
        TF-ICON       & 74.23 & 74.23 & 75.26 & 81.44 \\
        Kontext       & 59.82 & 65.18 & 65.18 & 65.18 \\
        \bottomrule
    \end{tabular}
    \vspace{-8pt}
\end{table}

\subsection{Ablations}
\begin{figure}[t]
    \centering
    \setlength{\tabcolsep}{1pt}
    \begin{tabular}{ccccc} %

    \includegraphics[width=0.195\linewidth]{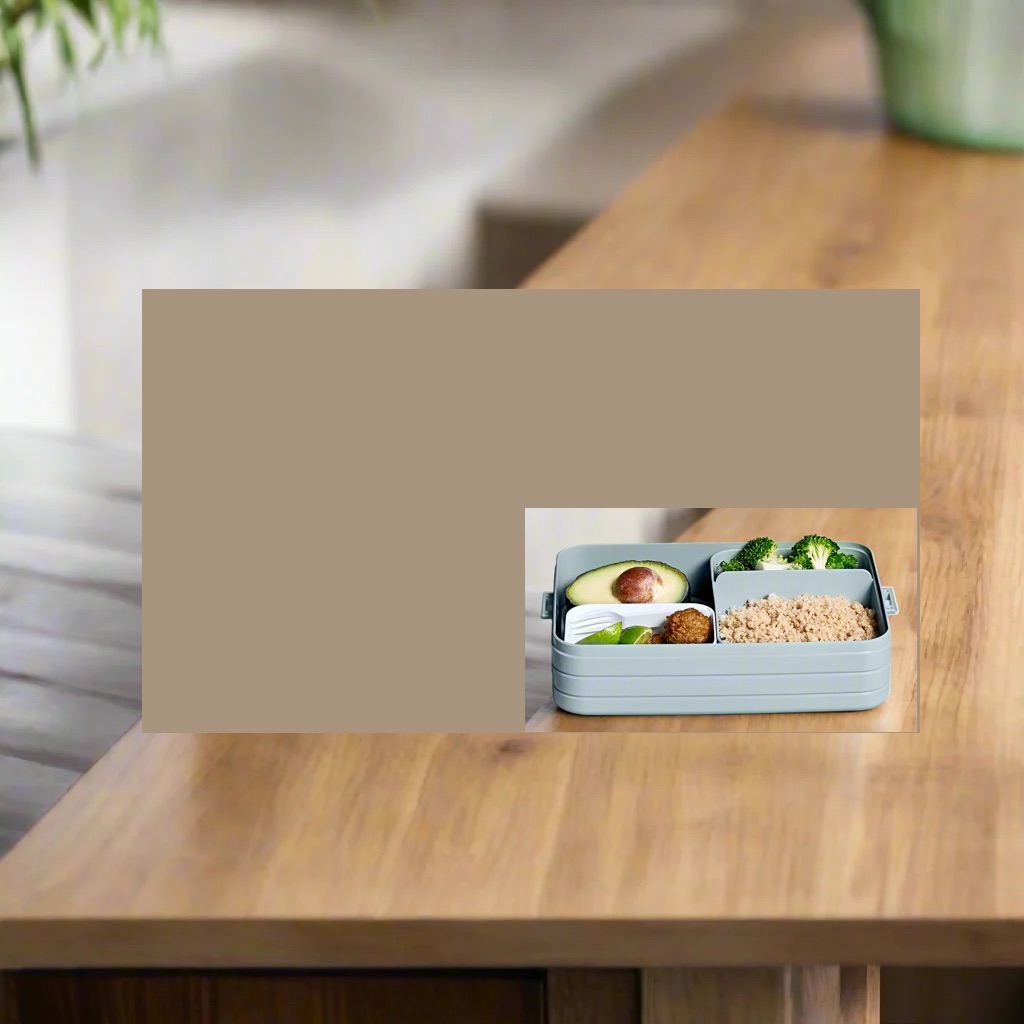}
    &
    \includegraphics[width=0.195\linewidth]{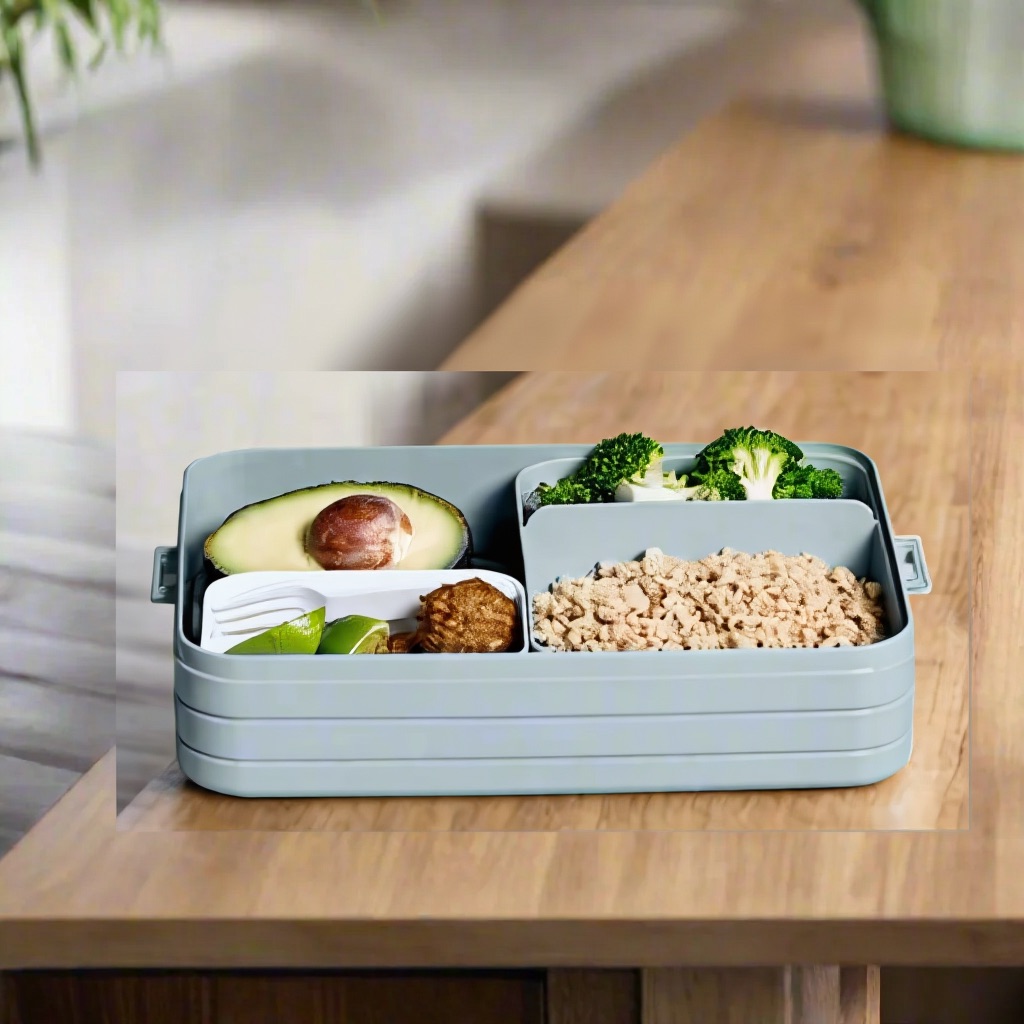}
    &
    \includegraphics[width=0.195\linewidth]{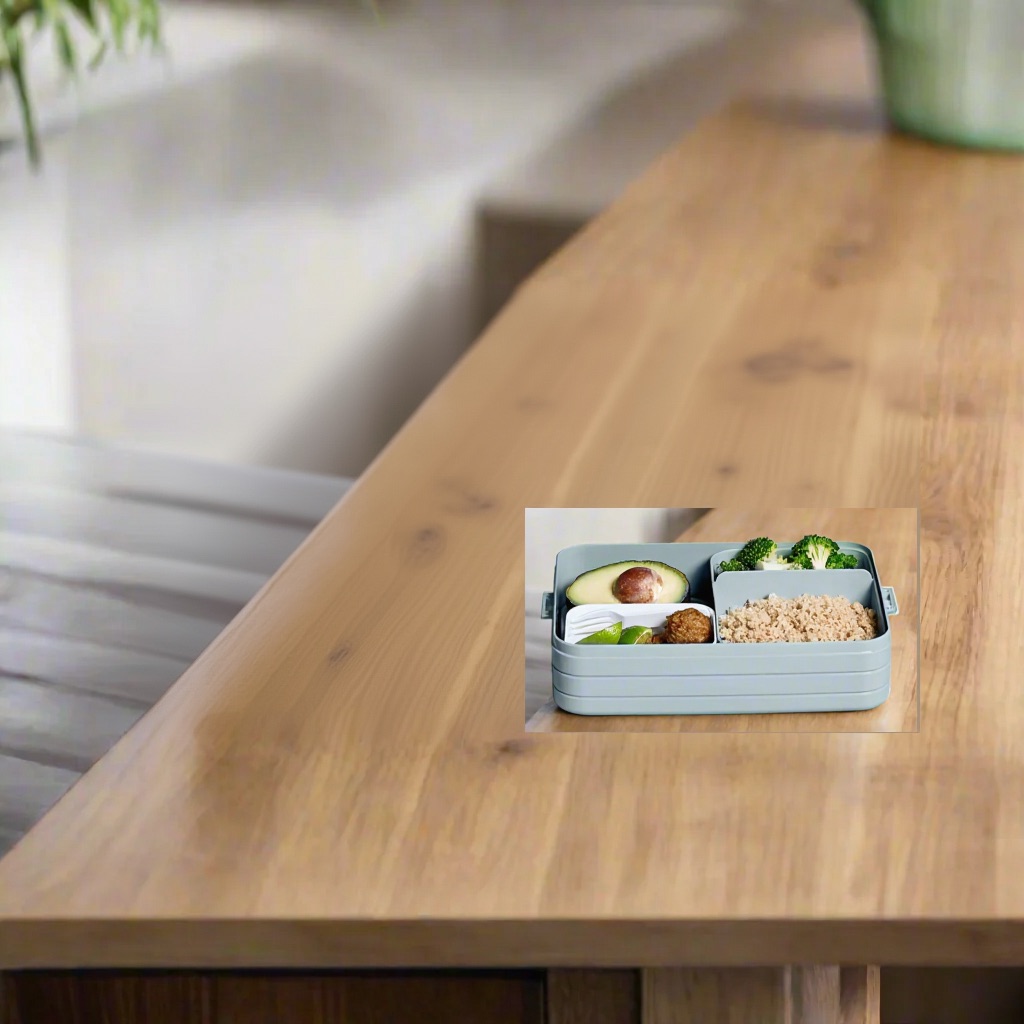}
    &
    \includegraphics[width=0.195\linewidth]{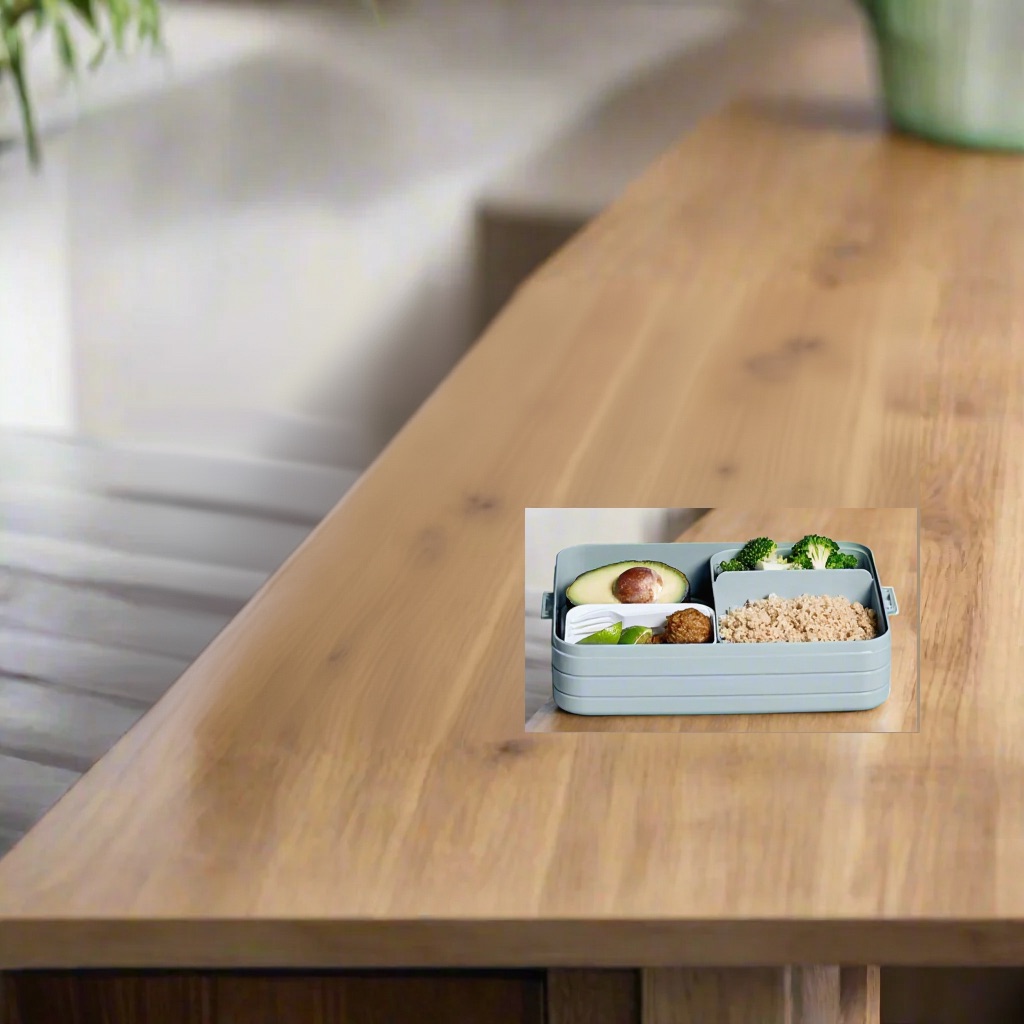}
    &
    \includegraphics[width=0.195\linewidth]{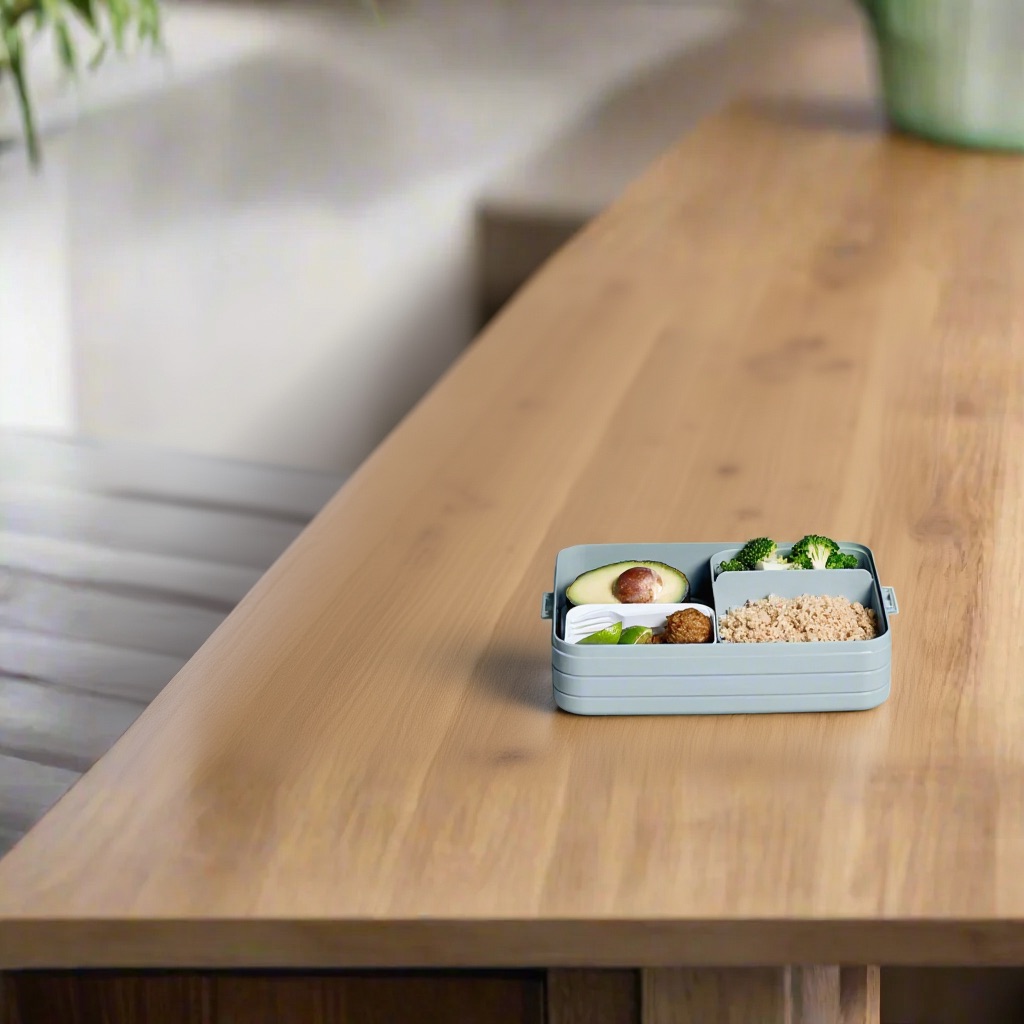}

\\
    
    \includegraphics[width=0.195\linewidth]{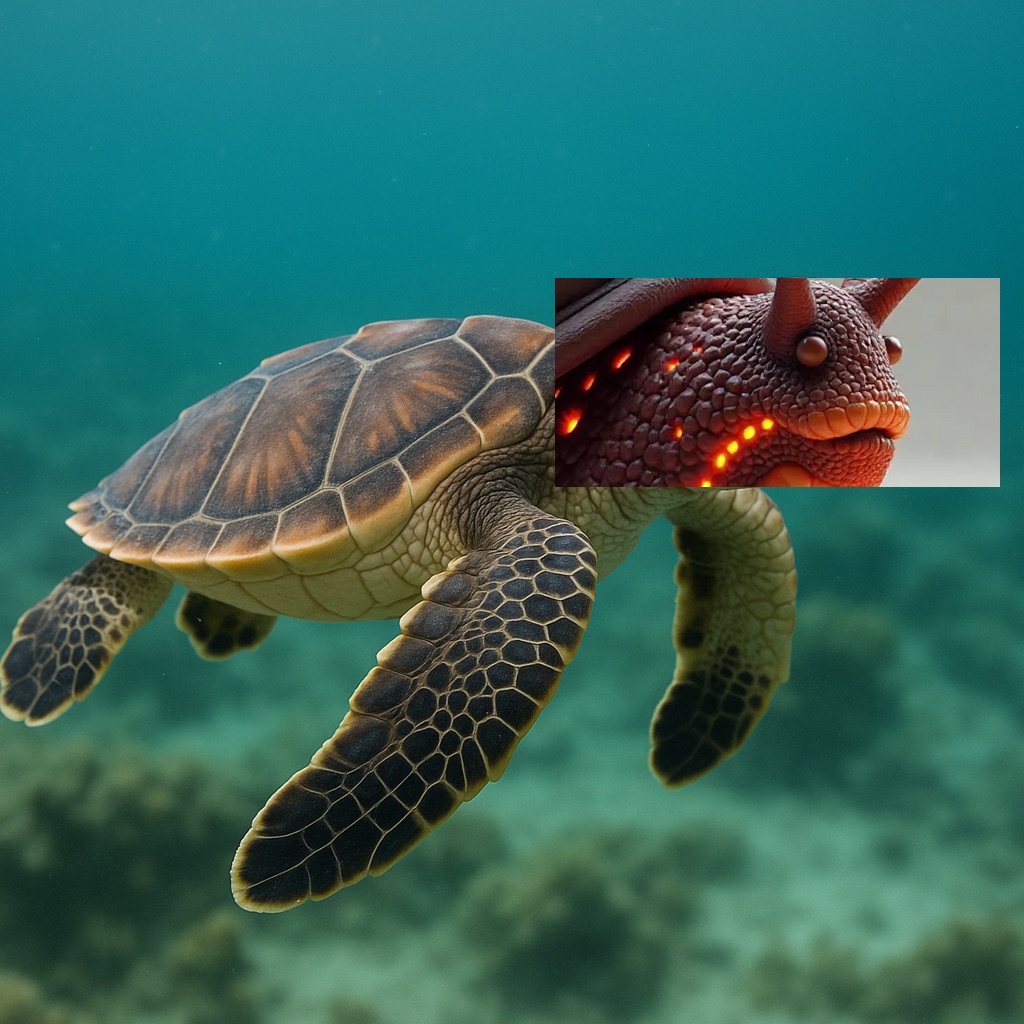}
    &
    \includegraphics[width=0.195\linewidth]{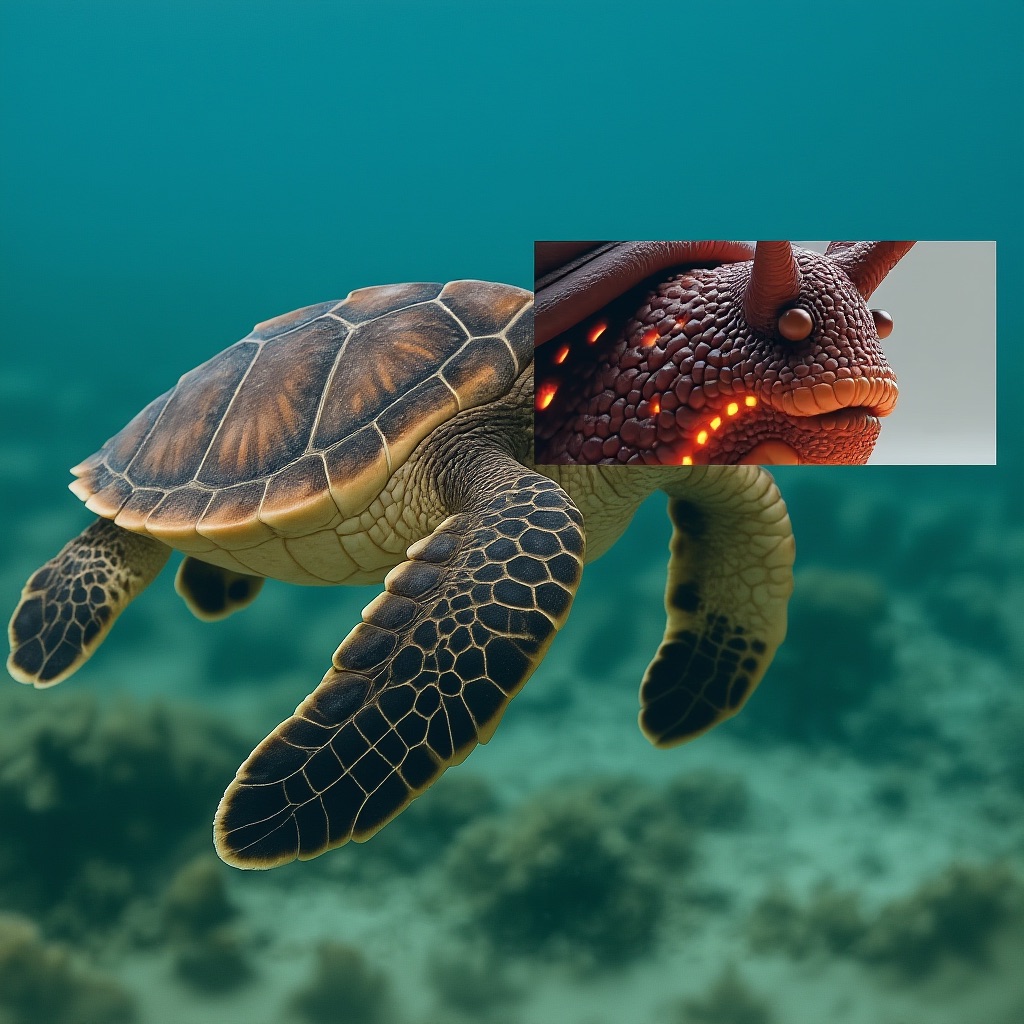}
    &
    \includegraphics[width=0.195\linewidth]{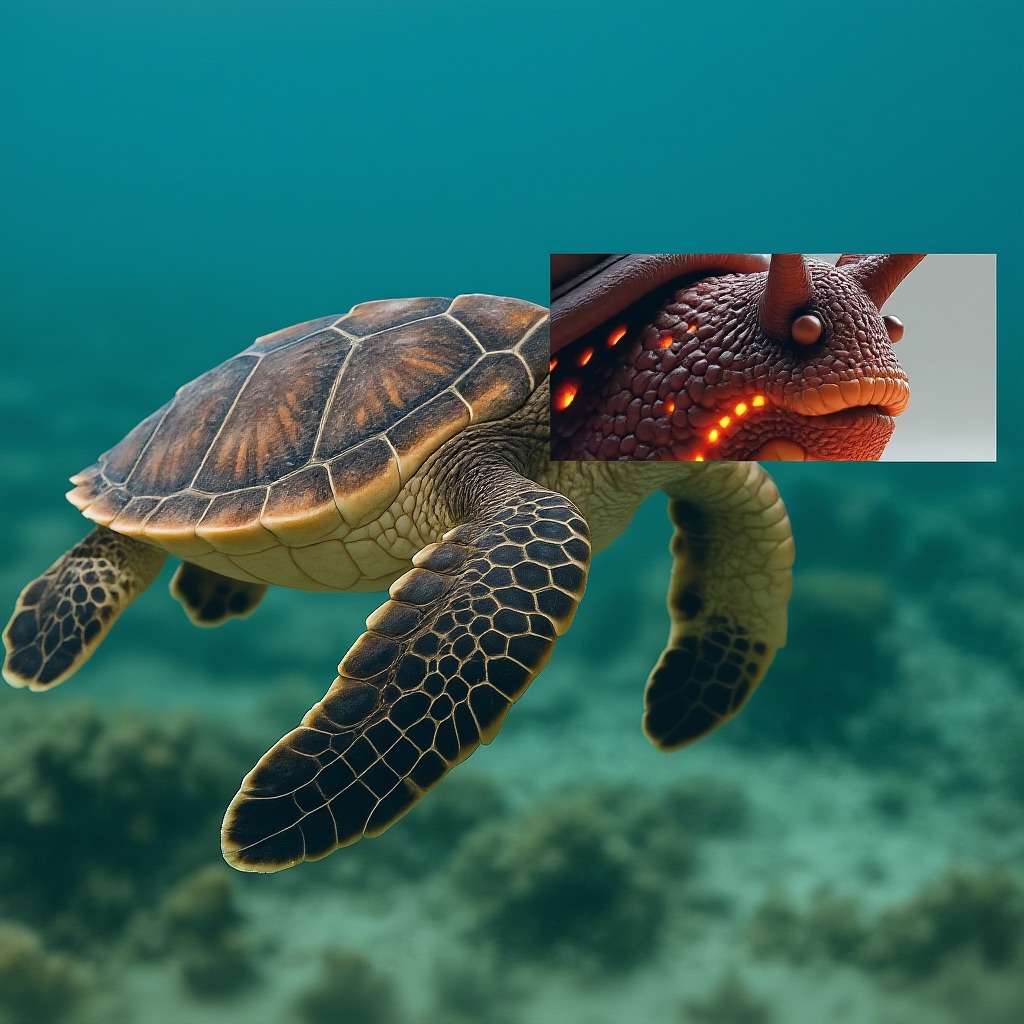}
    &
    \includegraphics[width=0.195\linewidth]{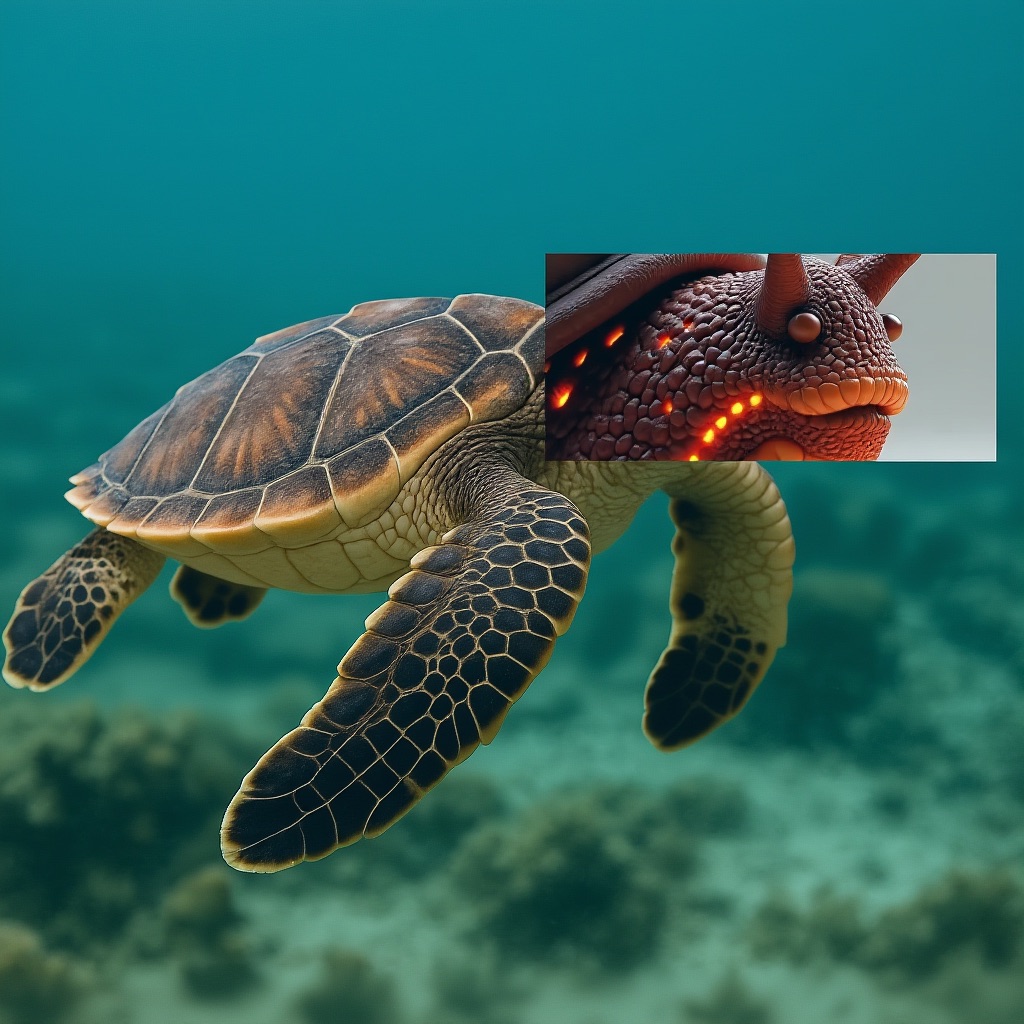}
    &
    \includegraphics[width=0.195\linewidth]{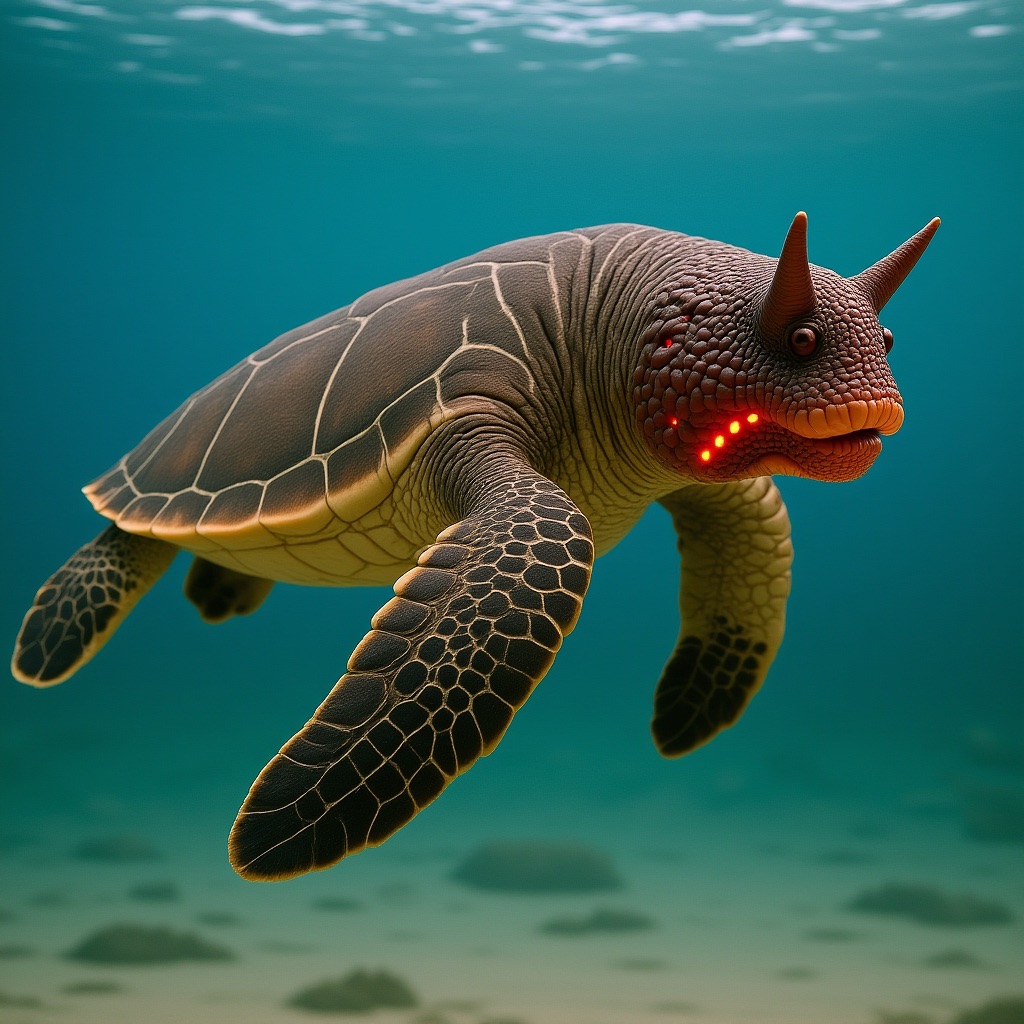}    
    
    \\ %

    \begin{minipage}{0.195\linewidth}
        \centering
        \footnotesize{Input}
    \end{minipage}
    &
    \begin{minipage}{0.195\linewidth}
        \centering
        \footnotesize{w/o attn scaling}
    \end{minipage}
    &
    \begin{minipage}{0.195\linewidth}
        \centering
        \footnotesize{w/o RoPE scaling}
    \end{minipage}
    &
    \begin{minipage}{0.195\linewidth}
        \centering
        \footnotesize{w/o VLM}
    \end{minipage}
    &
    \begin{minipage}{0.195\linewidth}
        \centering
        \footnotesize{LooseRoPE}
    \end{minipage}
    
    \end{tabular}
    \vspace{-6pt}
    \caption{
    \textbf{Ablation effects.} Ablation experiments demonstrate the necessity of each component. 
    In the lunch box translation, removing the attention scaling factor causes the edit to expand beyond the intended region. Ablating RoPE position scaling or VLM guidance prevents the background from being harmonized properly. 
    In the complex edit on the bottom row, all three components are required to overcome neglect. Removing any component causes the edit to fail, whereas our full method achieves a clean blend.}
    \vspace{-6pt}
    \label{fig:ablation_failure}
\end{figure}

To assess the contribution of each component in our framework, we independently remove the saliency-guided attention scaling (“w/o attn scaling”), the saliency-guided RoPE modulation (“w/o RoPE scaling”), and the VLM-based parameter adjustment (“w/o VLM”). 
Their quantitative impact is shown in Figure~\ref{fig:quant}, with corresponding qualitative examples in Figure~\ref{fig:ablation_failure}. \\
The results indicate that all components are necessary to achieve an optimal balance between image quality and identity preservation as high CLIP-IQA scores coupled with moderate LPIPS values signify effective blending. 
While the “w/o VLM” and “w/o RoPE scaling” variants show slightly lower LPIPS scores, this typically reflects neglect rather than genuine improvement in fidelity. 
The qualitative results support this observation: removing attention scaling leads to spatial drift, where the pasted content expands beyond its intended area (see the \emph{lunchbox} example, top row), while removing RoPE scaling or the VLM controller results in partial (top row) or complete (bottom row) neglect \op{in blending} the pasted object.

\label{sec:experiments}

\section{Conclusion}
\label{sec:conclusion}

We presented a prompt-free editing framework, where a user simply crops an object and injects it into a new image without any textual input. This direct operation raises the core challenge of integrating an often unnatural patch so that it blends seamlessly while retaining the source object’s identity. LooseRoPE achieves this balance by modulating positional encoding according to saliency, guiding attention to adaptively shift between preservation and harmonization.

At a broader level, our approach embodies graceful, adaptive control of attention: adjusting its field of view in response to image content rather than external prompts. This perspective points toward more general and interpretable forms of visual control, where attention itself becomes the medium of fine-grained generation.

Future exploration may extend this framework to videos, where maintaining temporal coherence during object insertion remains a central challenge. Another promising direction is to enable multiple, interrelated crops within a single scene, allowing complex compositional interactions. On a more conceptual level, deepening our understanding of the model’s internal attention mechanisms could lead to context-aware modulation, where the model dynamically recognizes and corrects its own inconsistencies.

\section{Acknowledgments}
This work was supported by the Israel Science Foundation (grants no. 2492/20 and 1473/24), Len Blavatnik and the Blavatnik family foundation, and the U.S-Israel Binational Science Foundation (application no. 2022363).
\clearpage

\bibliographystyle{ACM-Reference-Format}
\bibliography{main}
\clearpage

\begin{figure*}[!htp]
    \centering
    \setlength{\tabcolsep}{1pt} %
    \begin{tabular}{ccccccc}
        Input & Kontext & Ours && Input & Kontext & Ours \\

        \includegraphics[width=0.165\linewidth]{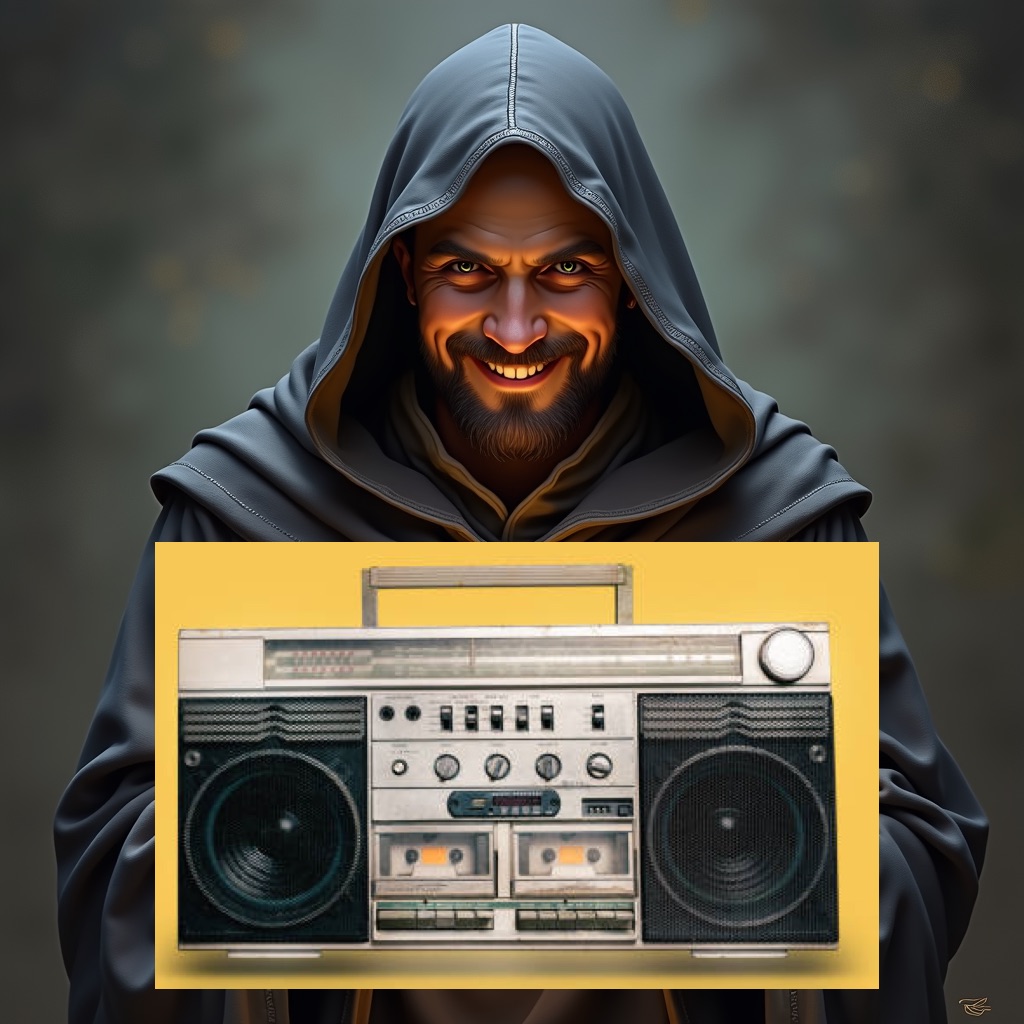} &
        \includegraphics[width=0.165\linewidth]{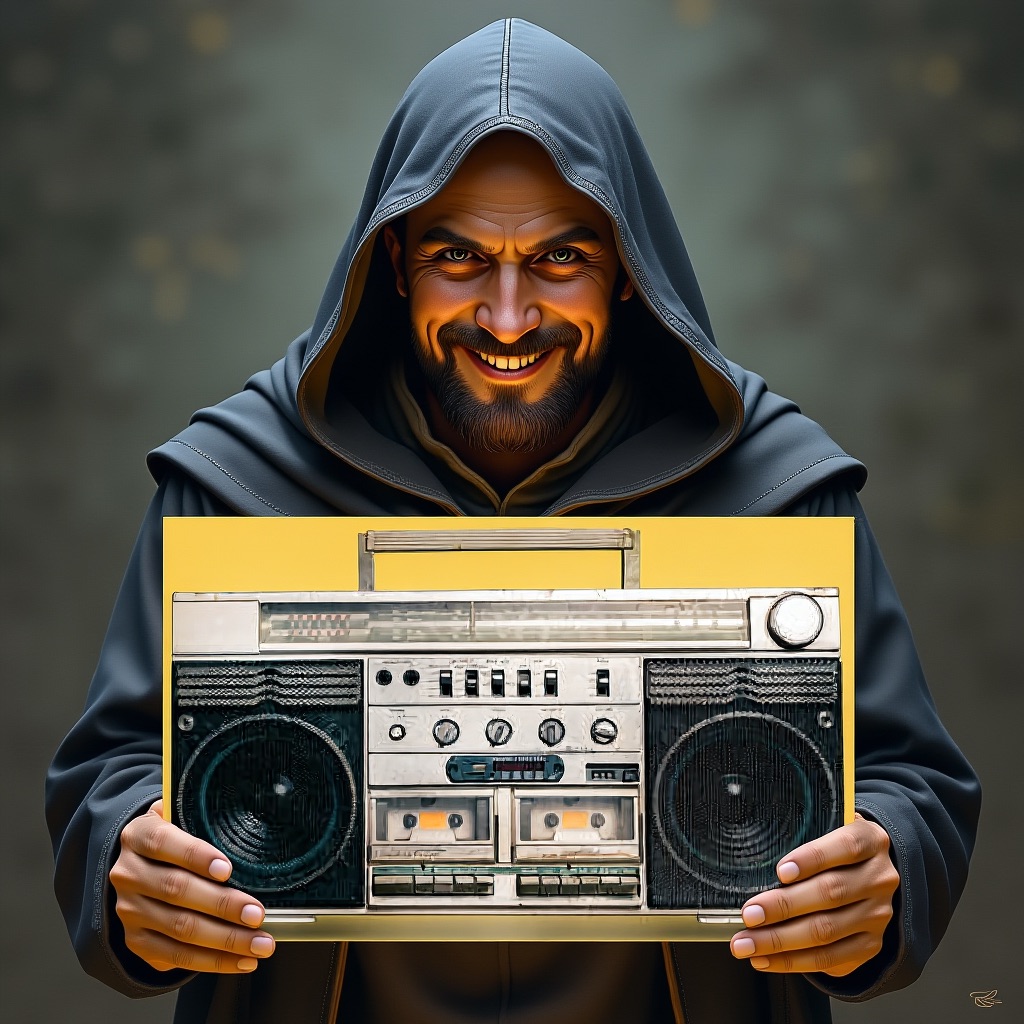} &
        \includegraphics[width=0.165\linewidth]{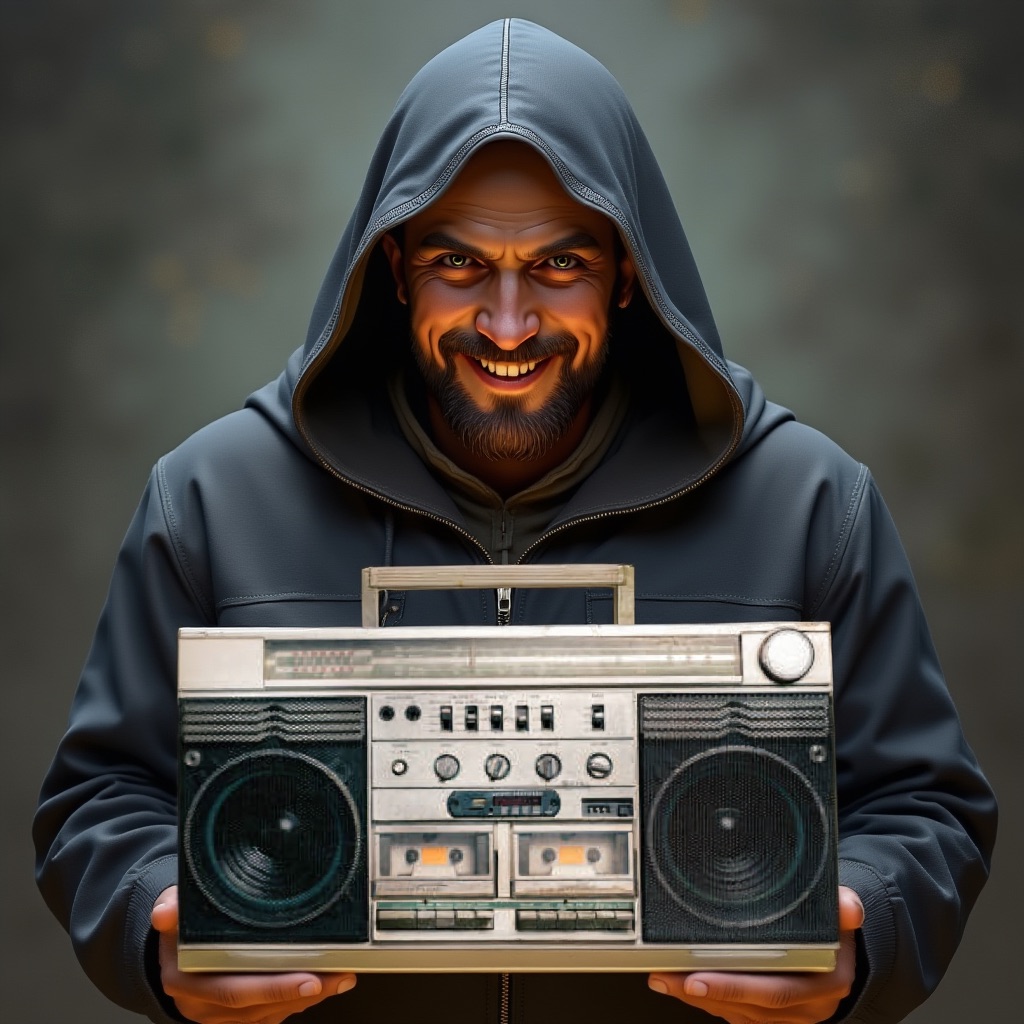} &
        &
        \includegraphics[width=0.165\linewidth]{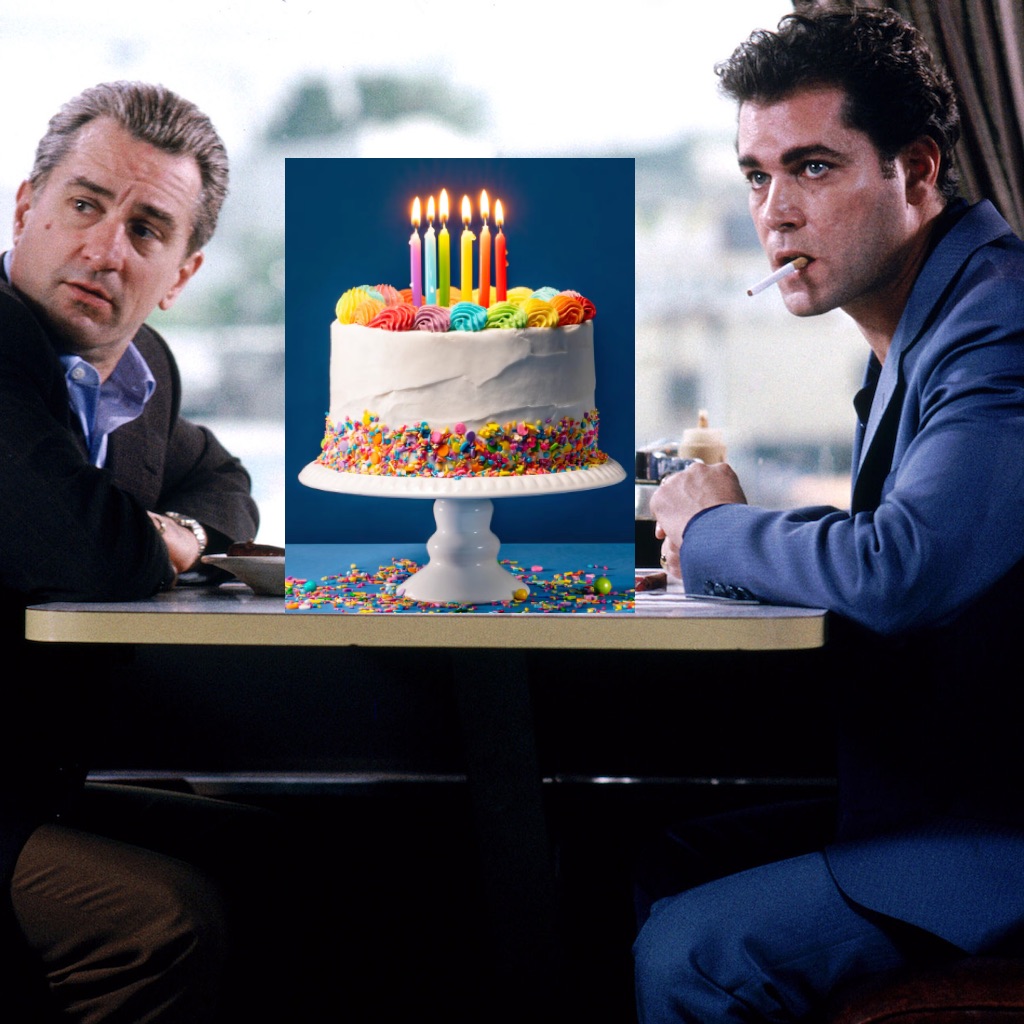} &
        \includegraphics[width=0.165\linewidth]{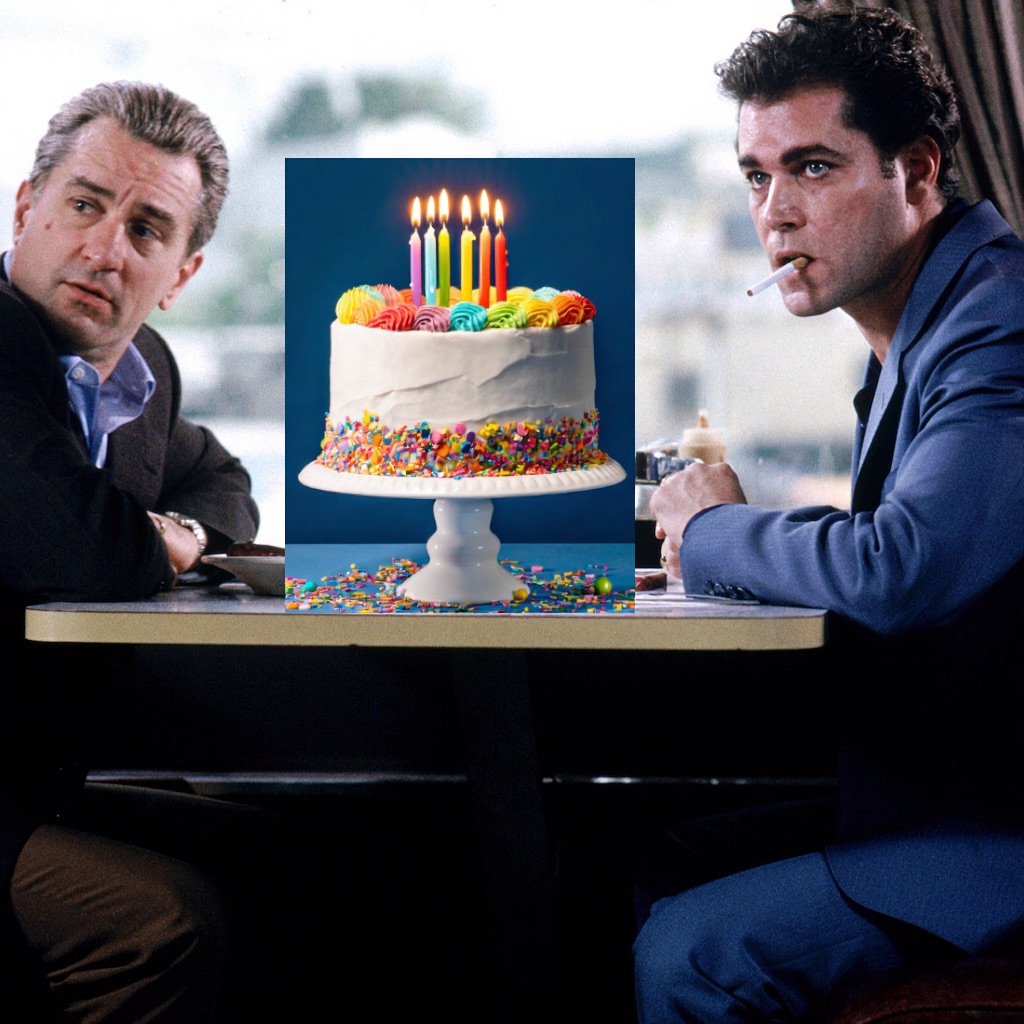} &
        \includegraphics[width=0.165\linewidth]{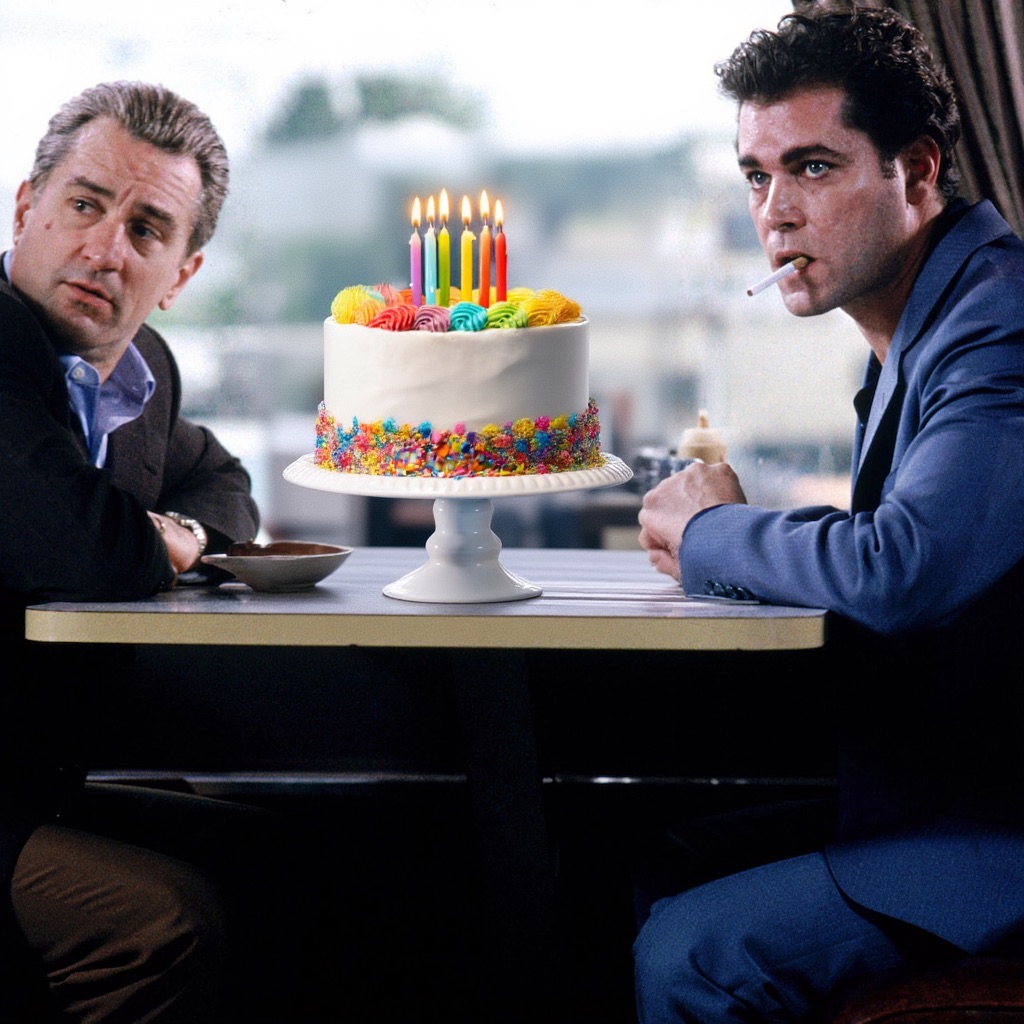} \\

        \includegraphics[width=0.165\linewidth]{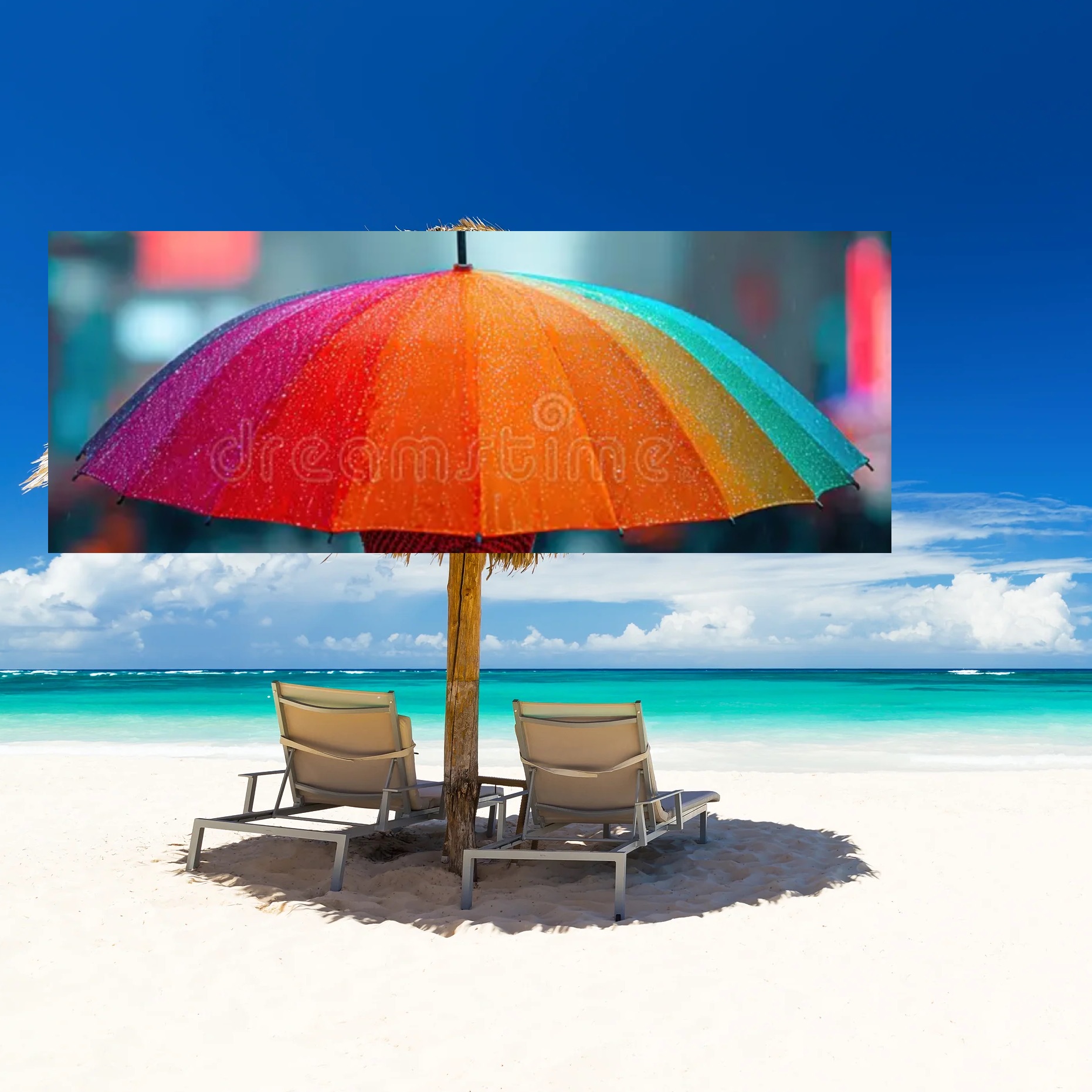} &
        \includegraphics[width=0.165\linewidth]{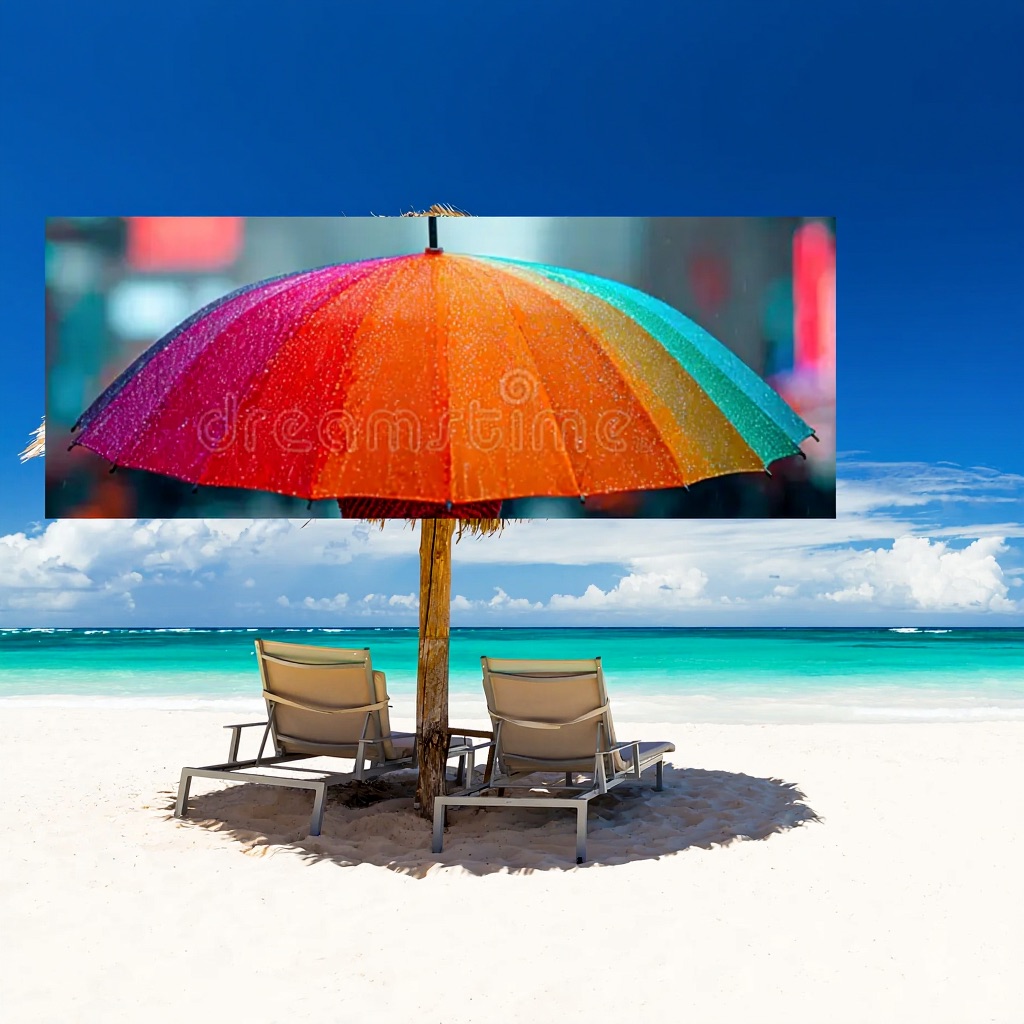} &
        \includegraphics[width=0.165\linewidth]{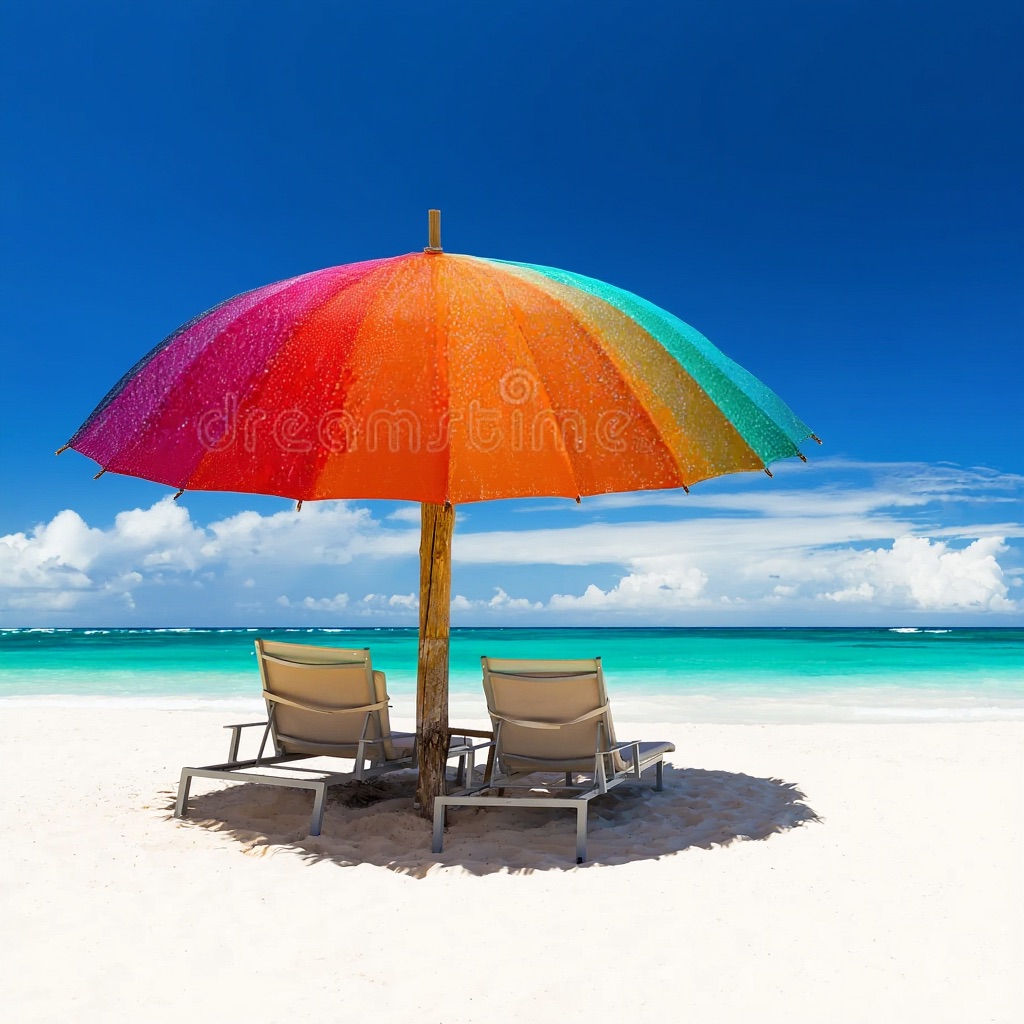} &
        &
        \includegraphics[width=0.165\linewidth]{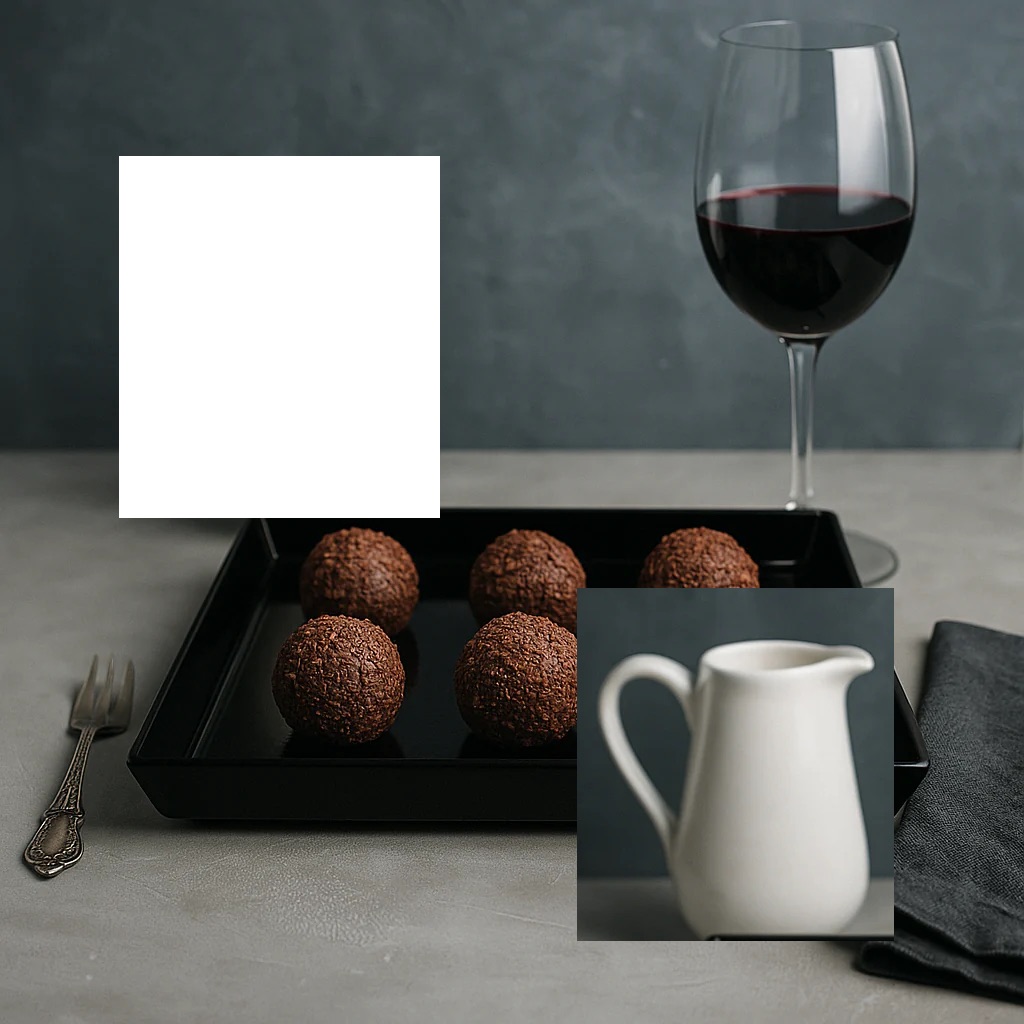} &
        \includegraphics[width=0.165\linewidth]{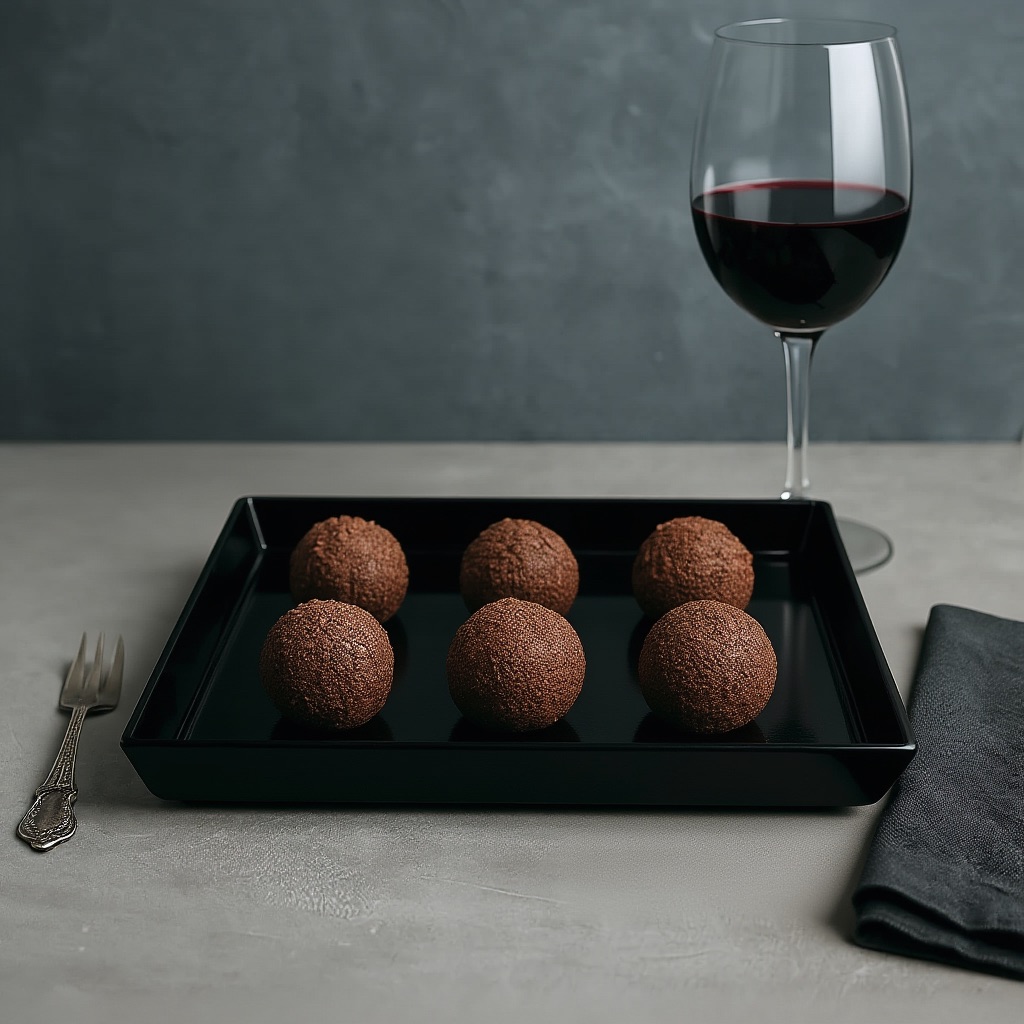} &
        \includegraphics[width=0.165\linewidth]{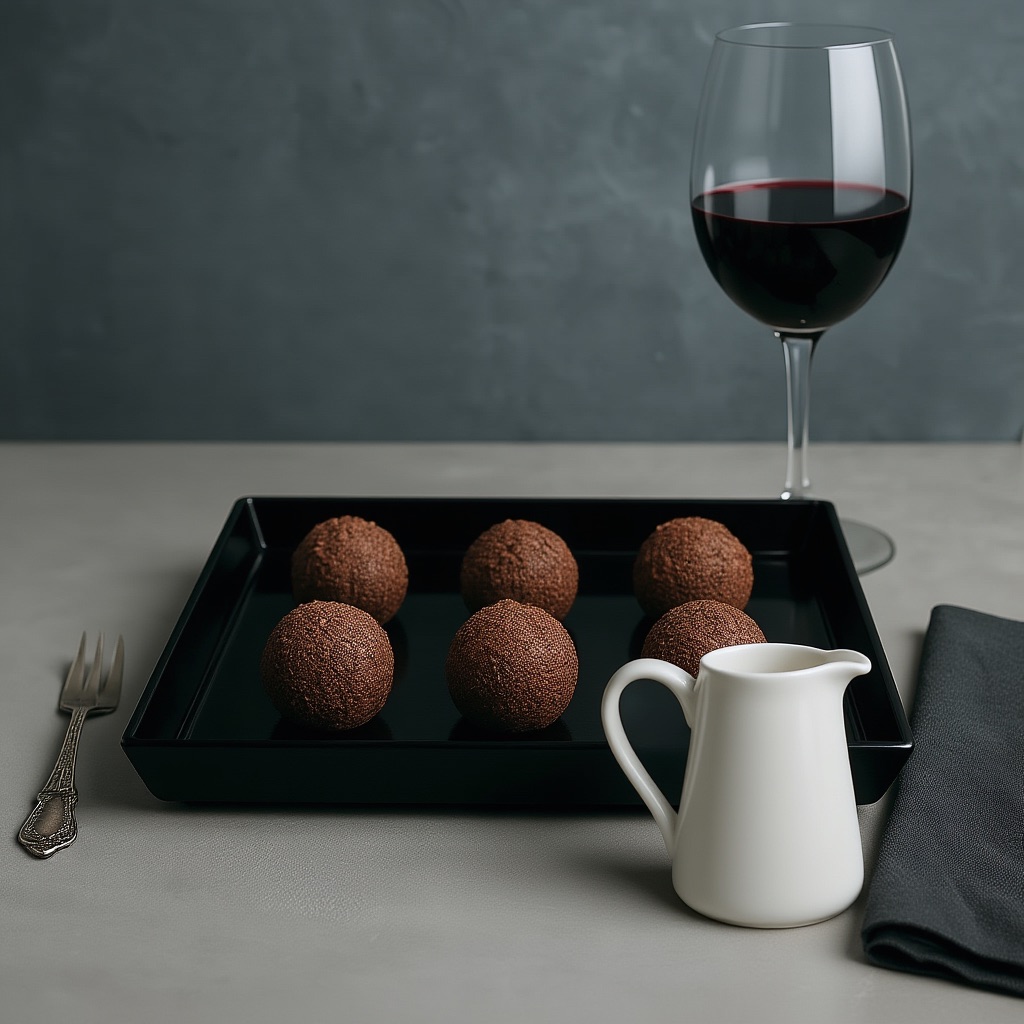} 
        \\

        \includegraphics[width=0.165\linewidth]{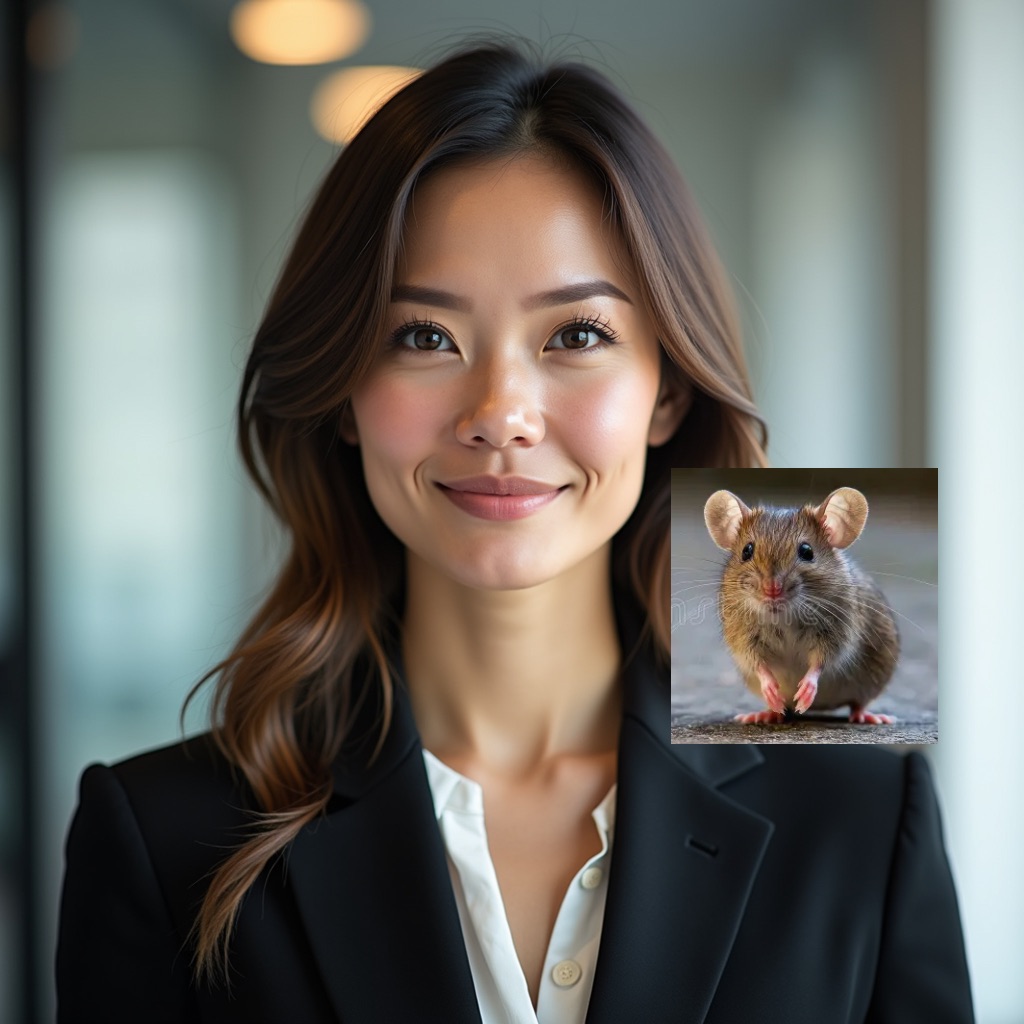} &
        \includegraphics[width=0.165\linewidth]{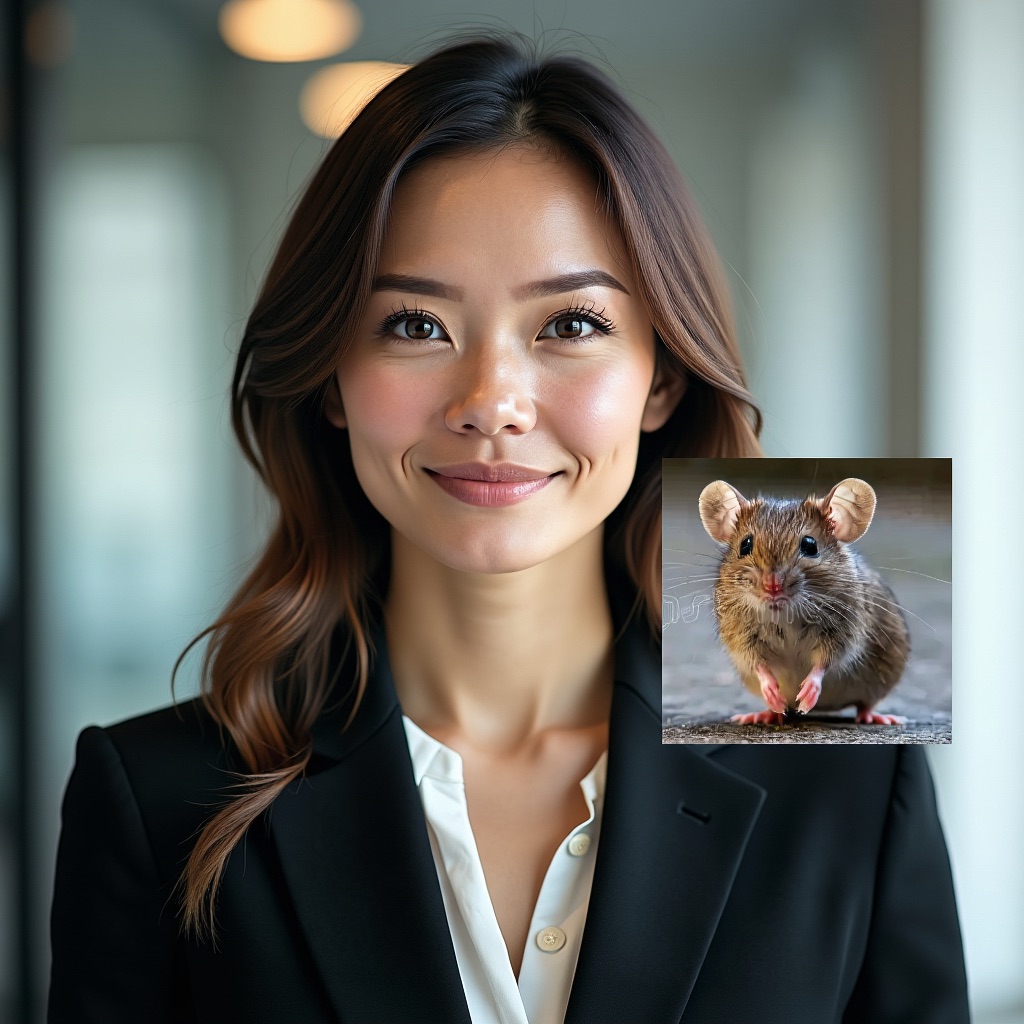} &
        \includegraphics[width=0.165\linewidth]{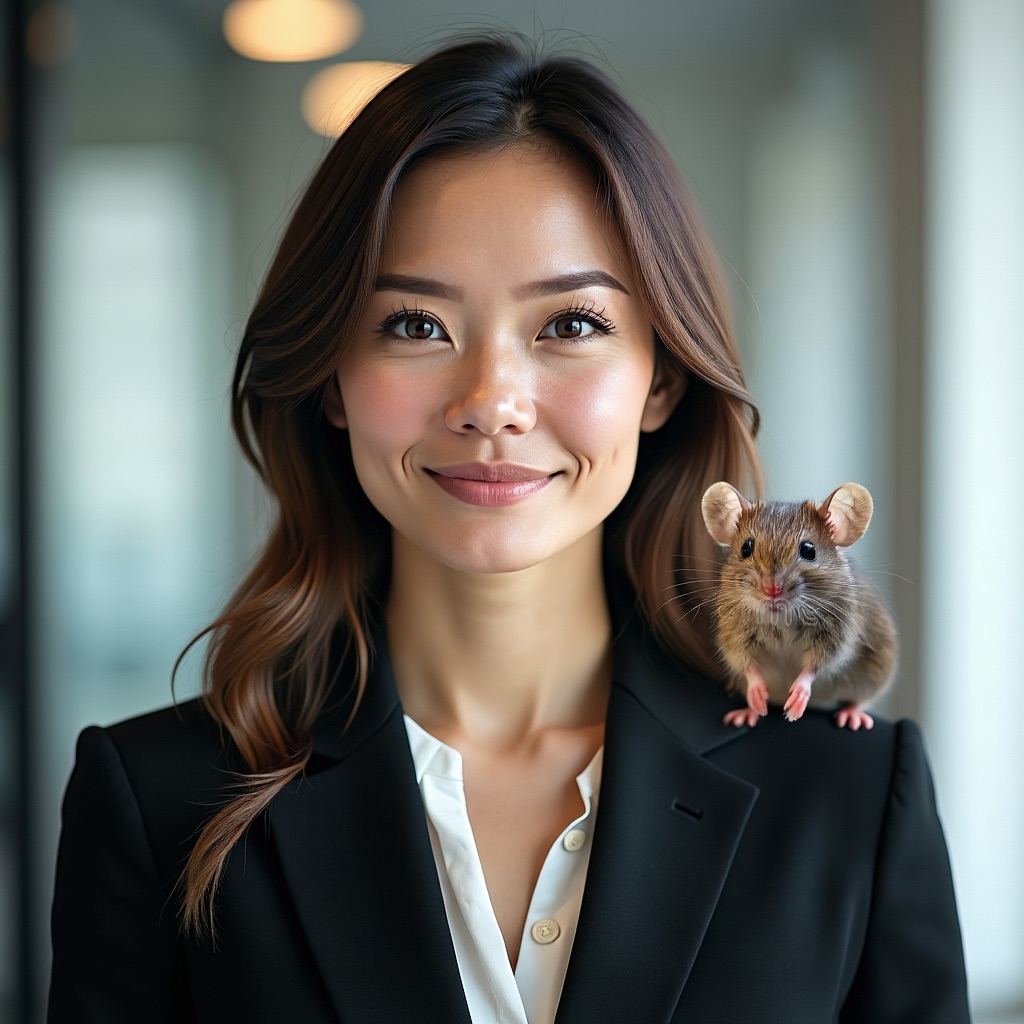} &
        &
        \includegraphics[width=0.165\linewidth]{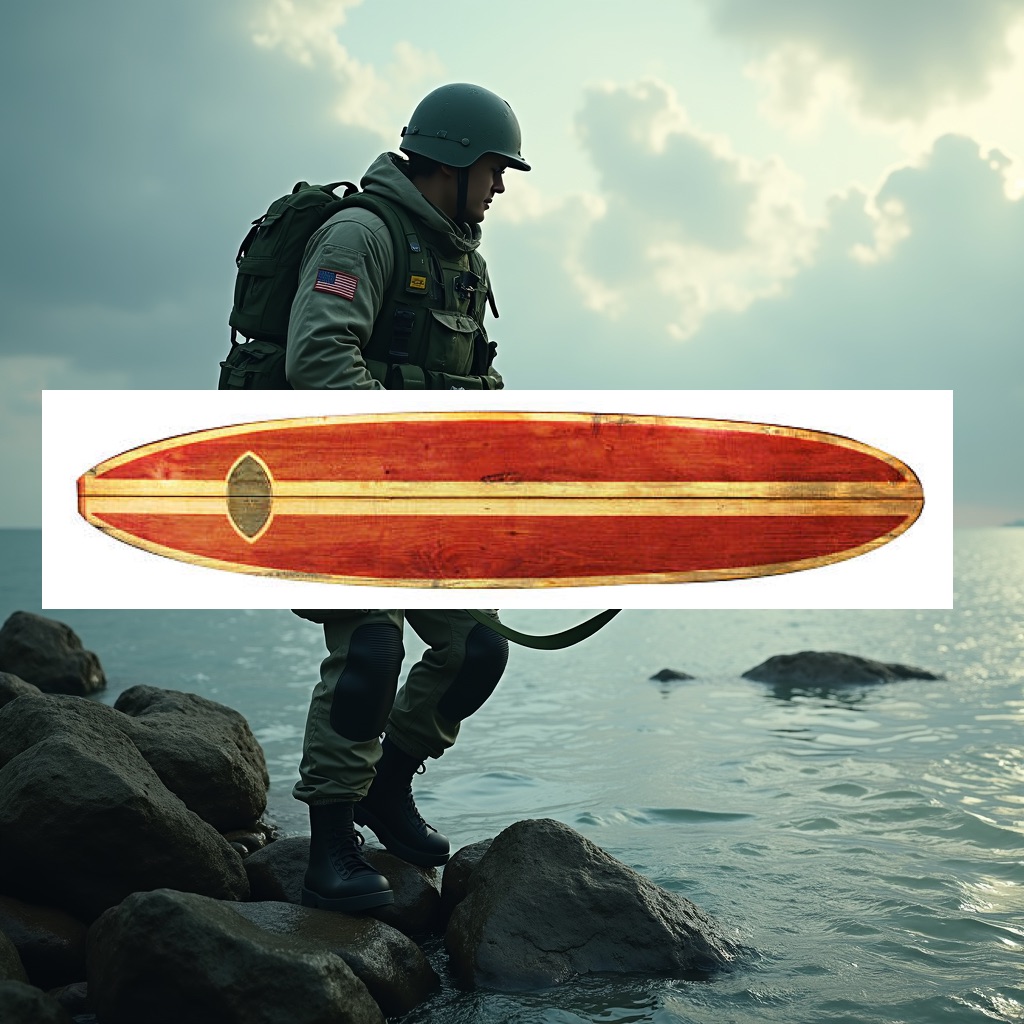} &
        \includegraphics[width=0.165\linewidth]{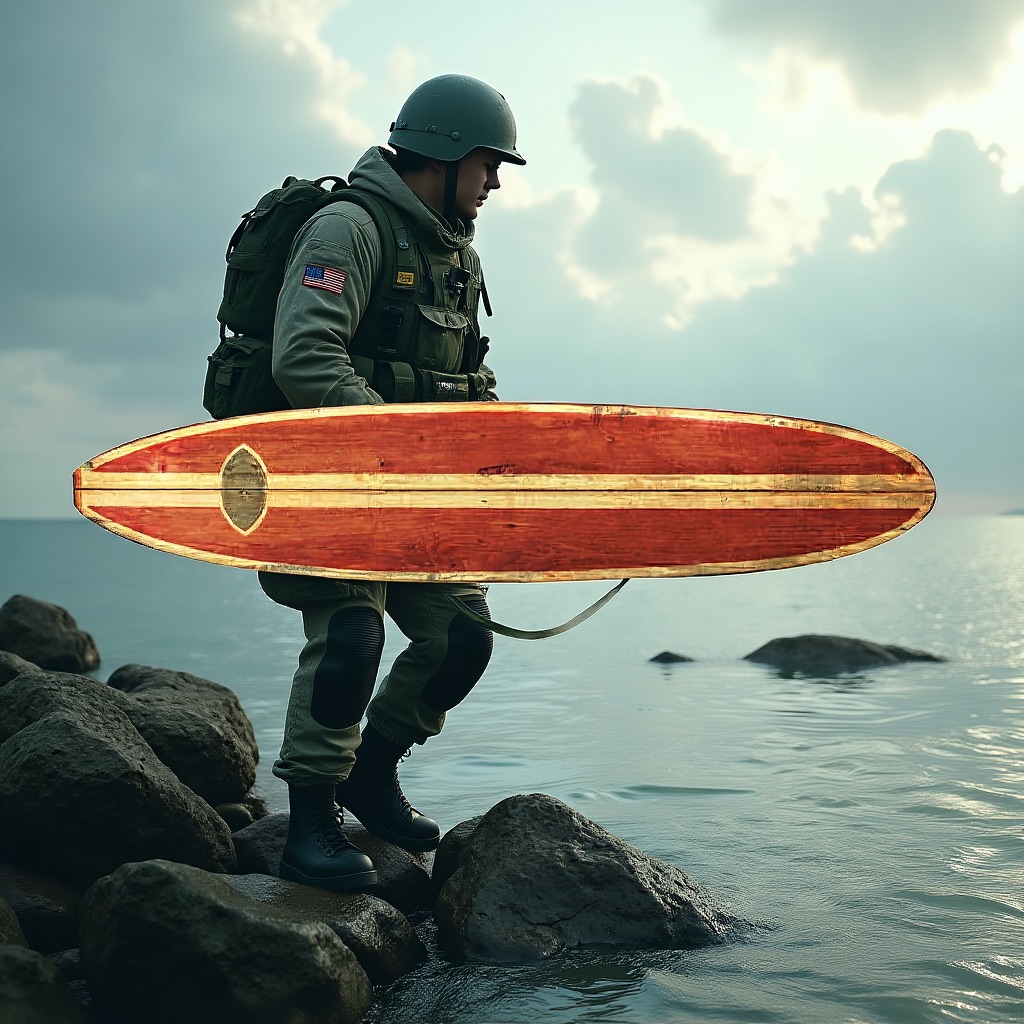} &
        \includegraphics[width=0.165\linewidth]{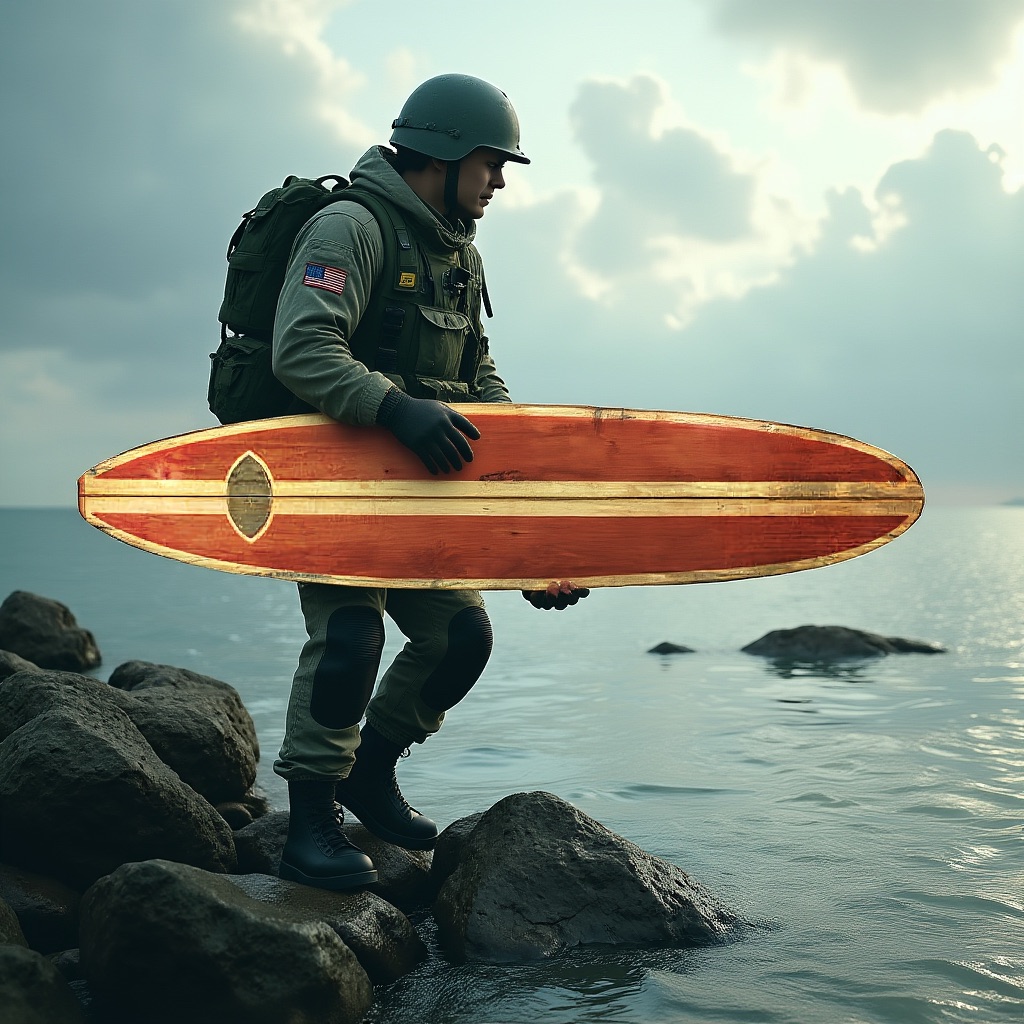} \\

        \includegraphics[width=0.165\linewidth]{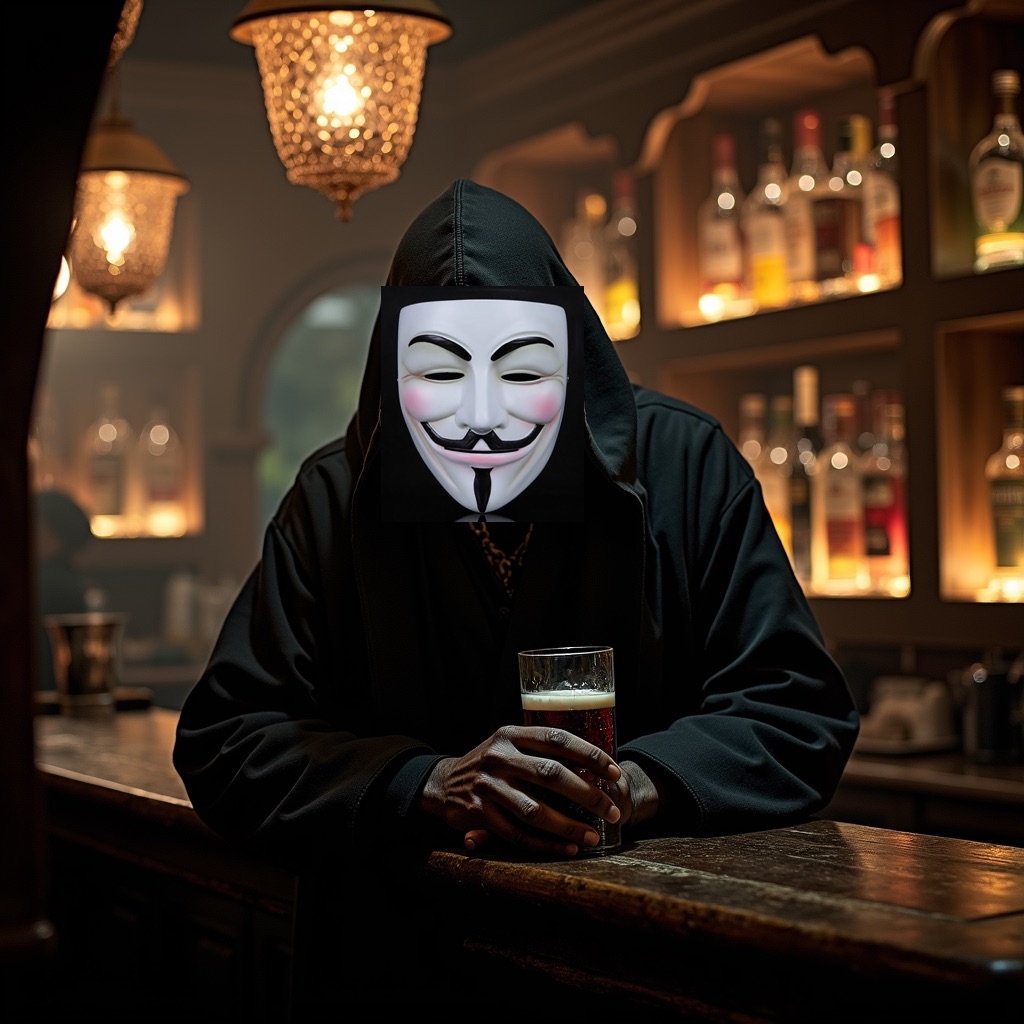} &
        \includegraphics[width=0.165\linewidth]{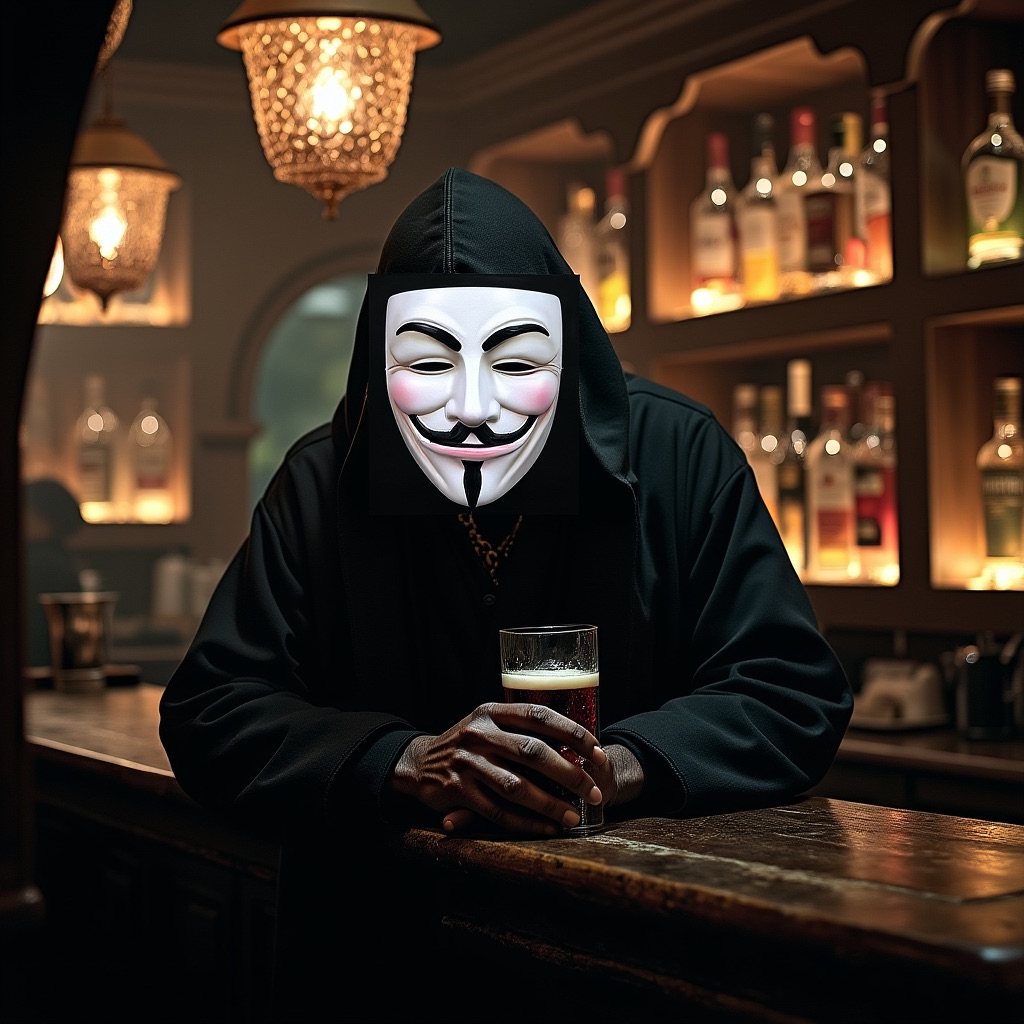} &
        \includegraphics[width=0.165\linewidth]{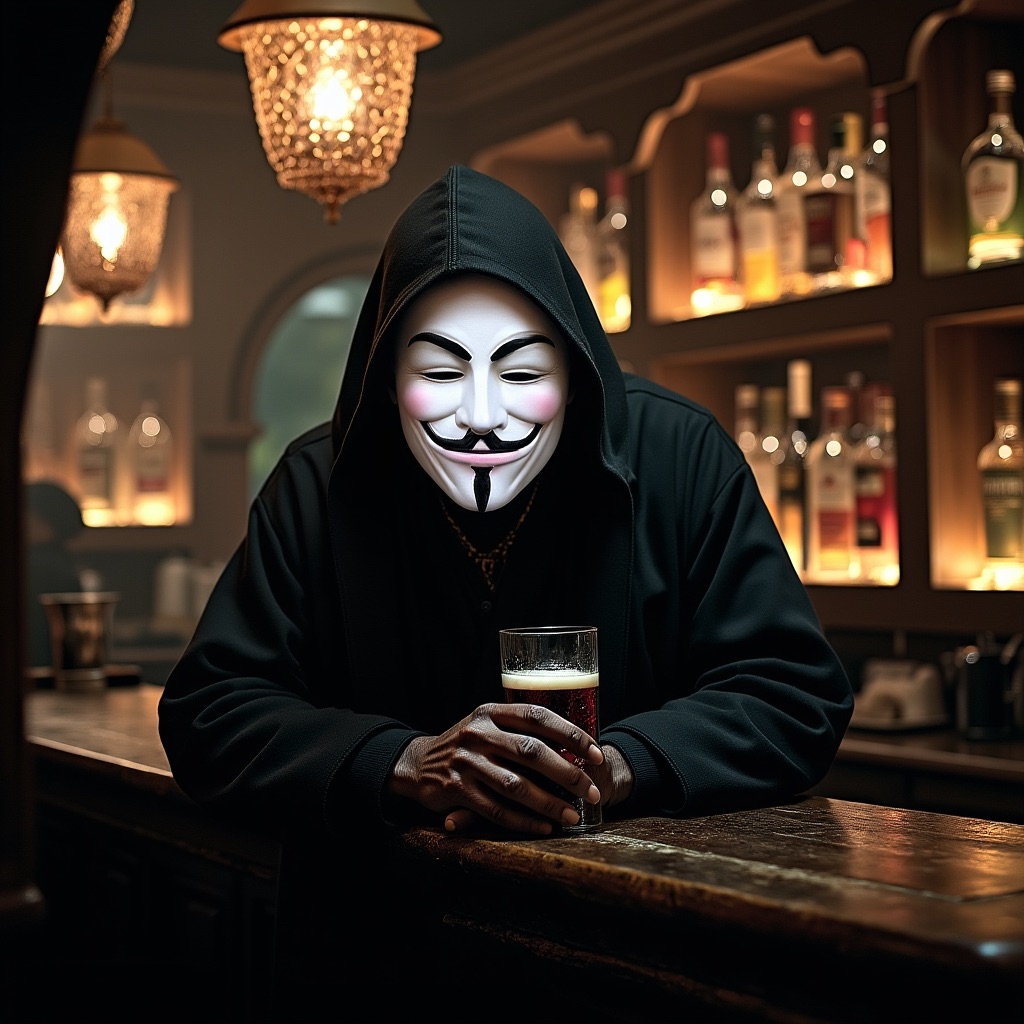} &
        &
        \includegraphics[width=0.165\linewidth, trim=130mm 130mm 0mm 0mm, clip]{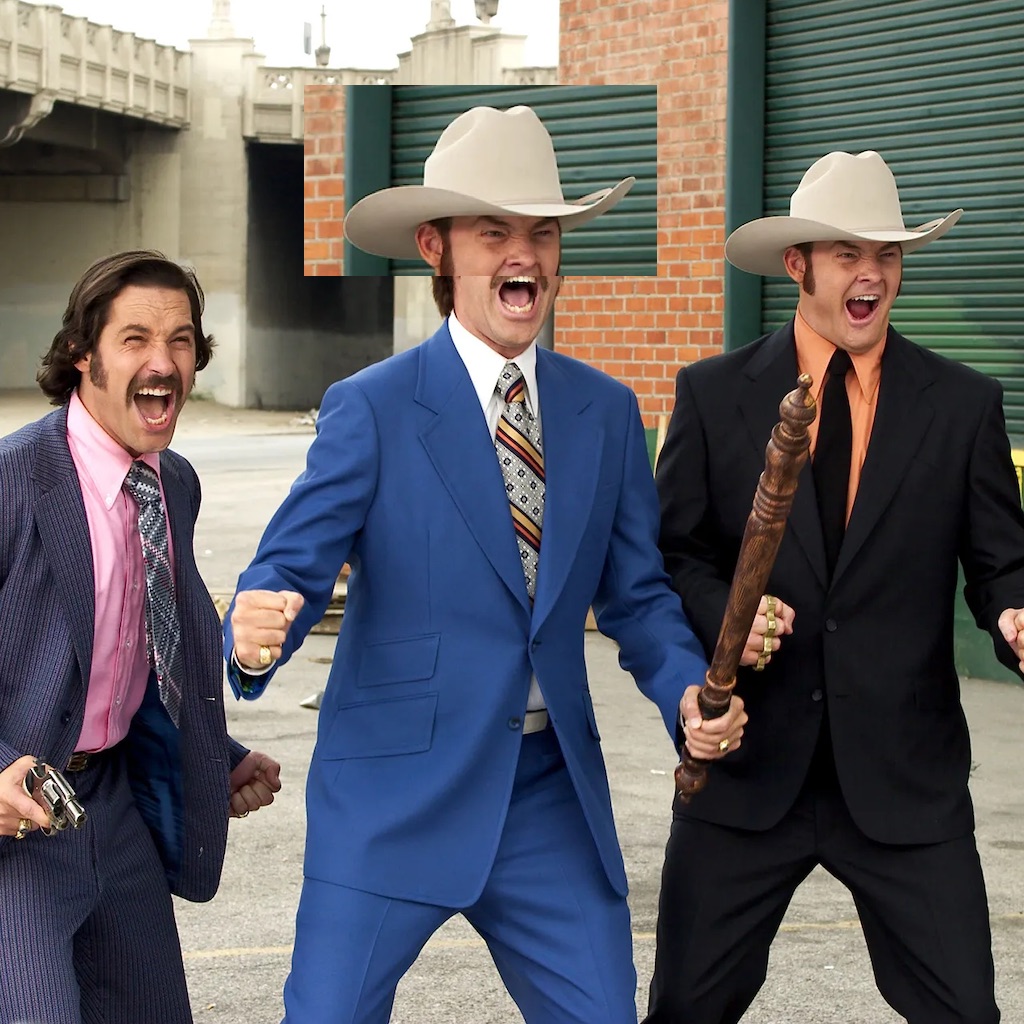} &
        \includegraphics[width=0.165\linewidth, trim=130mm 130mm 0mm 0mm, clip]     {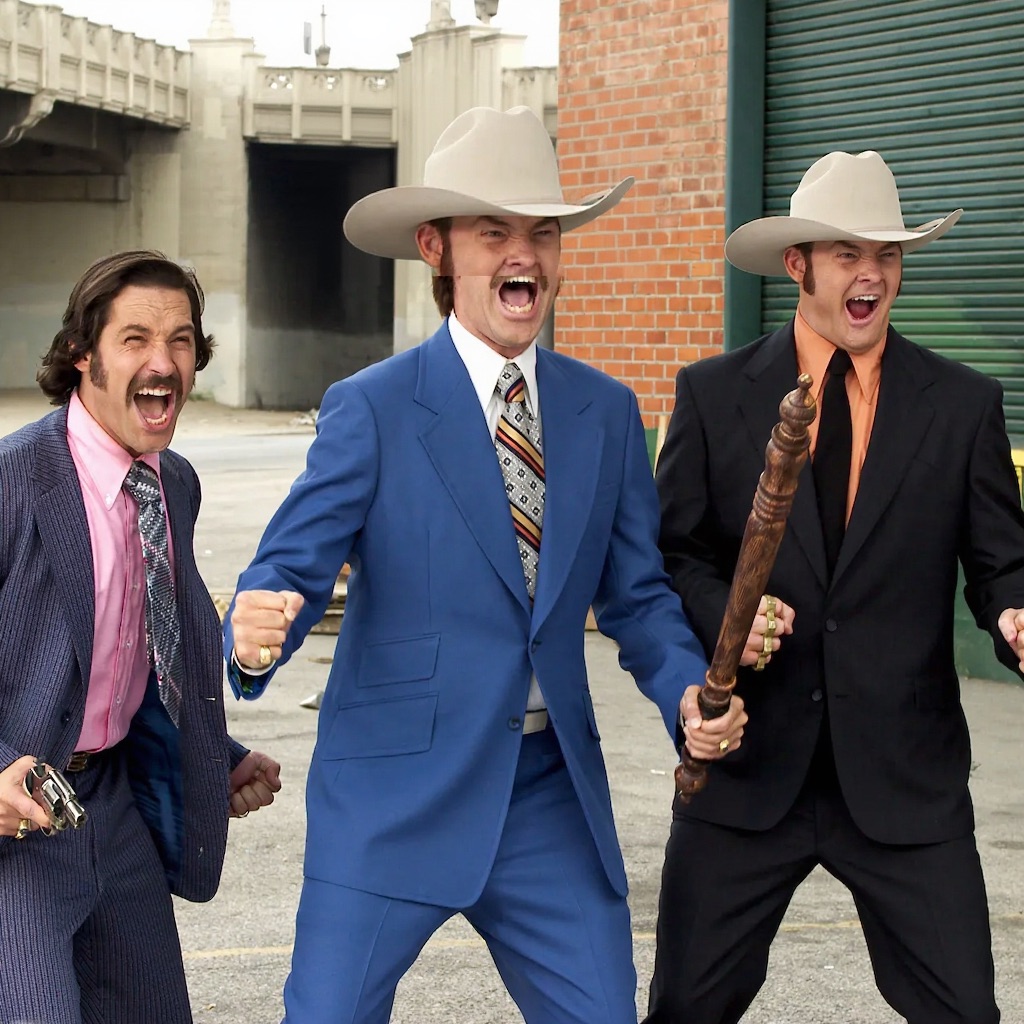} &
        \includegraphics[width=0.165\linewidth, trim=130mm 130mm 0mm 0mm, clip]     {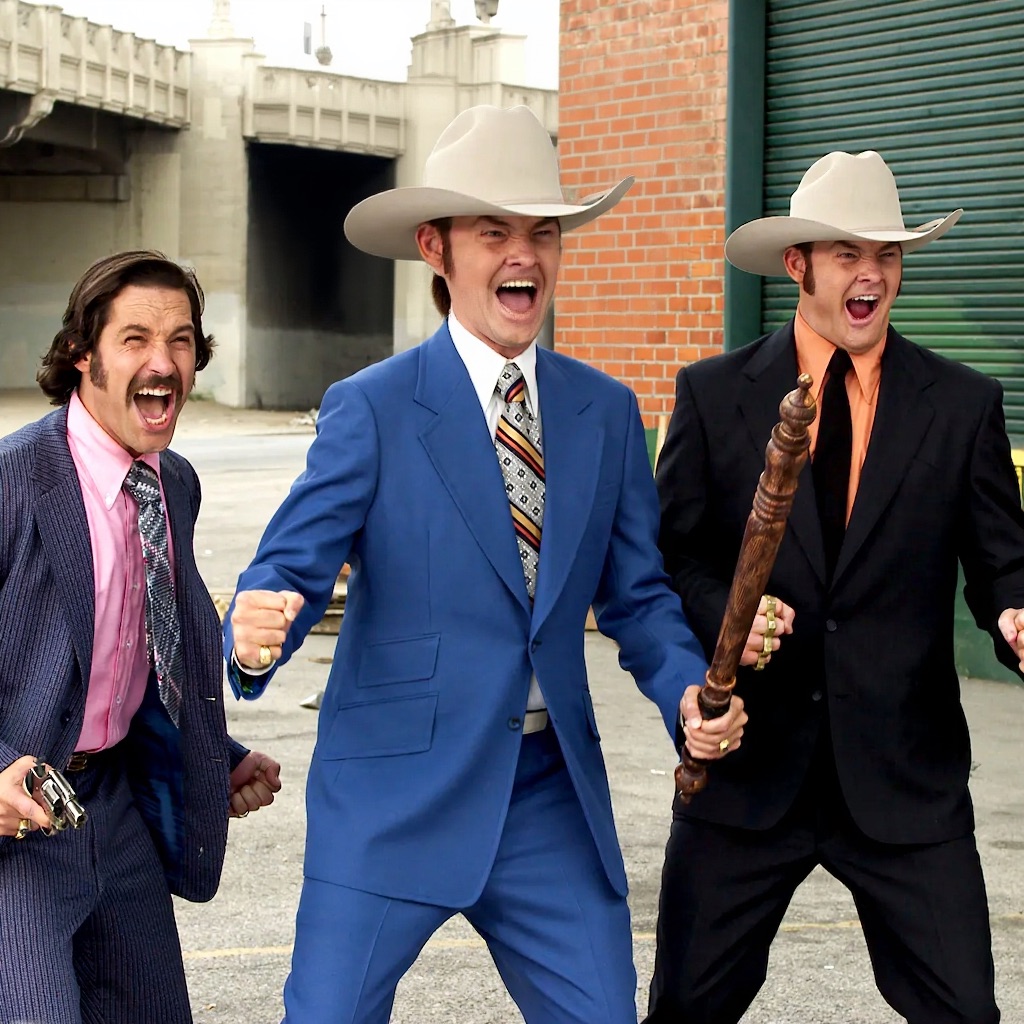} \\

        \includegraphics[width=0.165\linewidth, trim=50mm 50mm 50mm 50mm, clip]{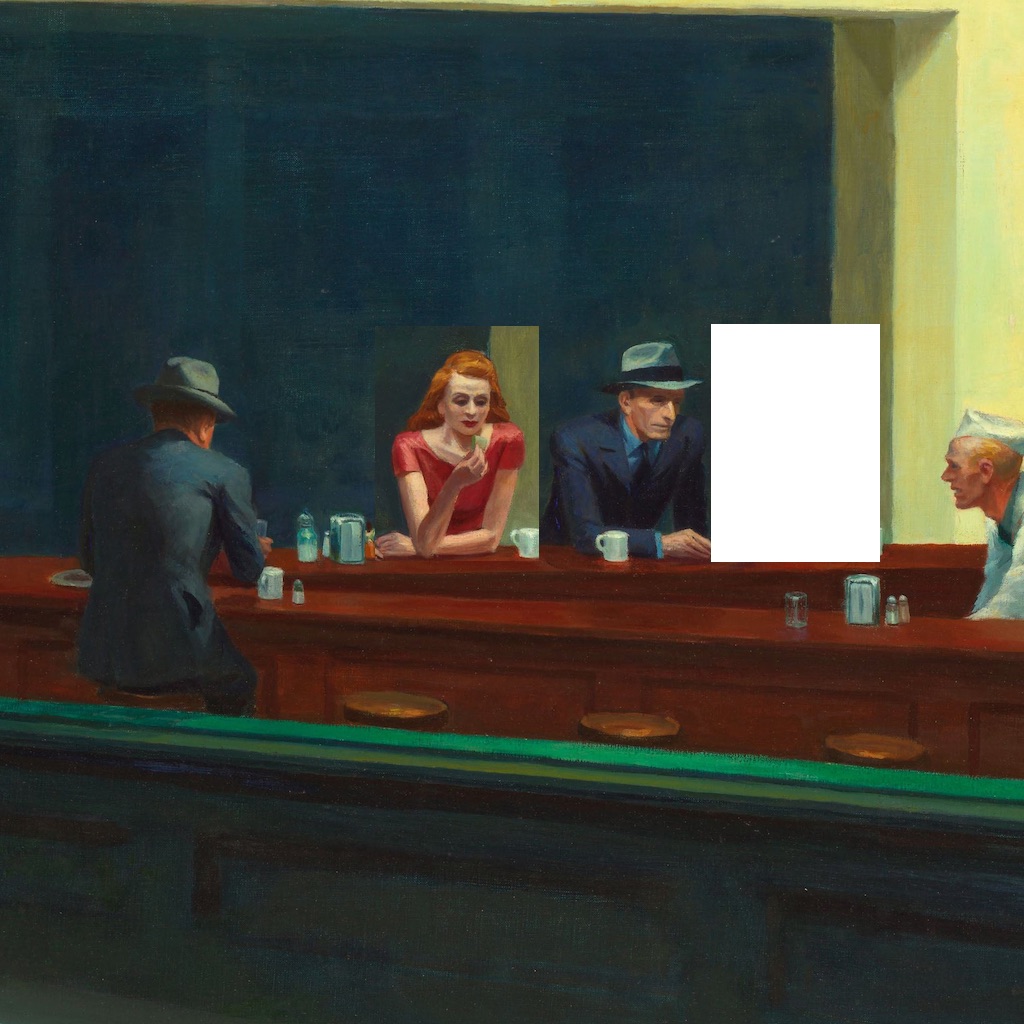} &
        \includegraphics[width=0.165\linewidth, trim=50mm 50mm 50mm 50mm, clip]{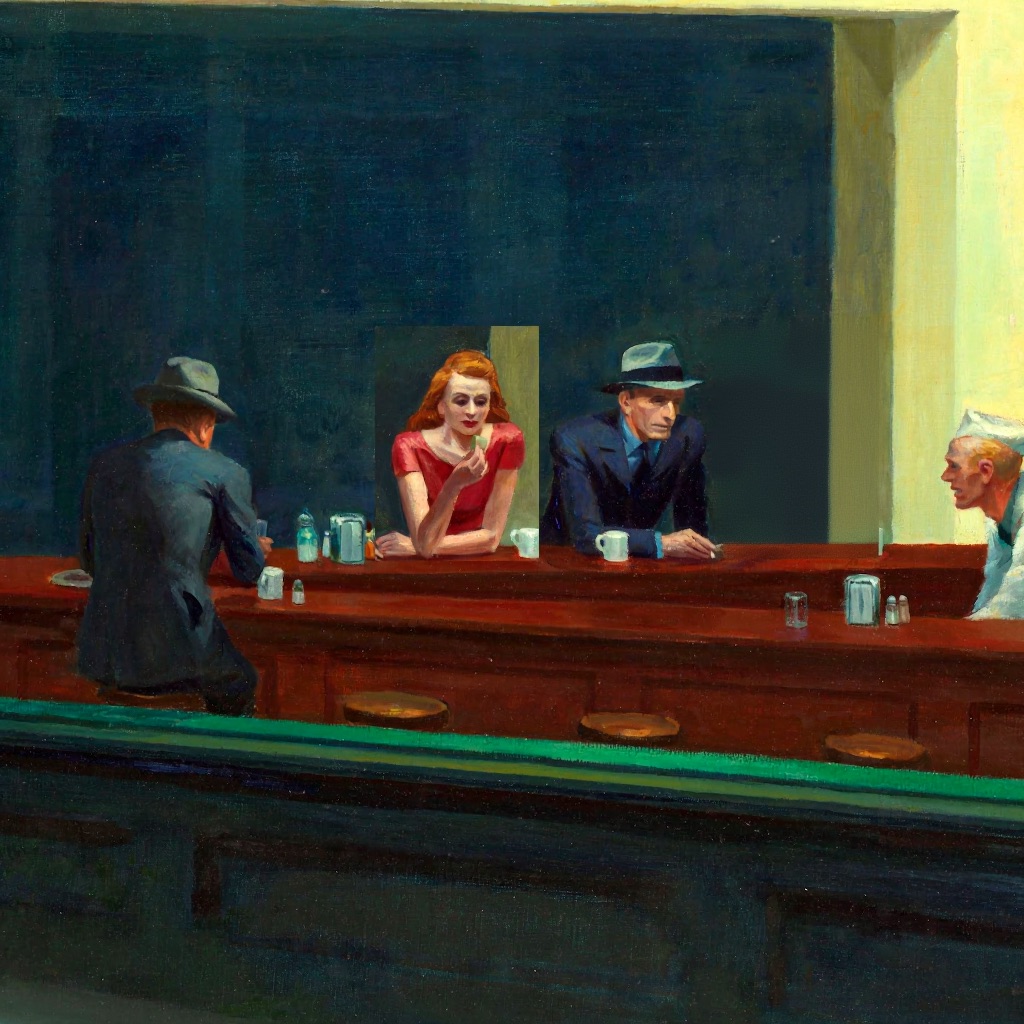} &
        \includegraphics[width=0.165\linewidth, trim=50mm 50mm 50mm 50mm, clip]{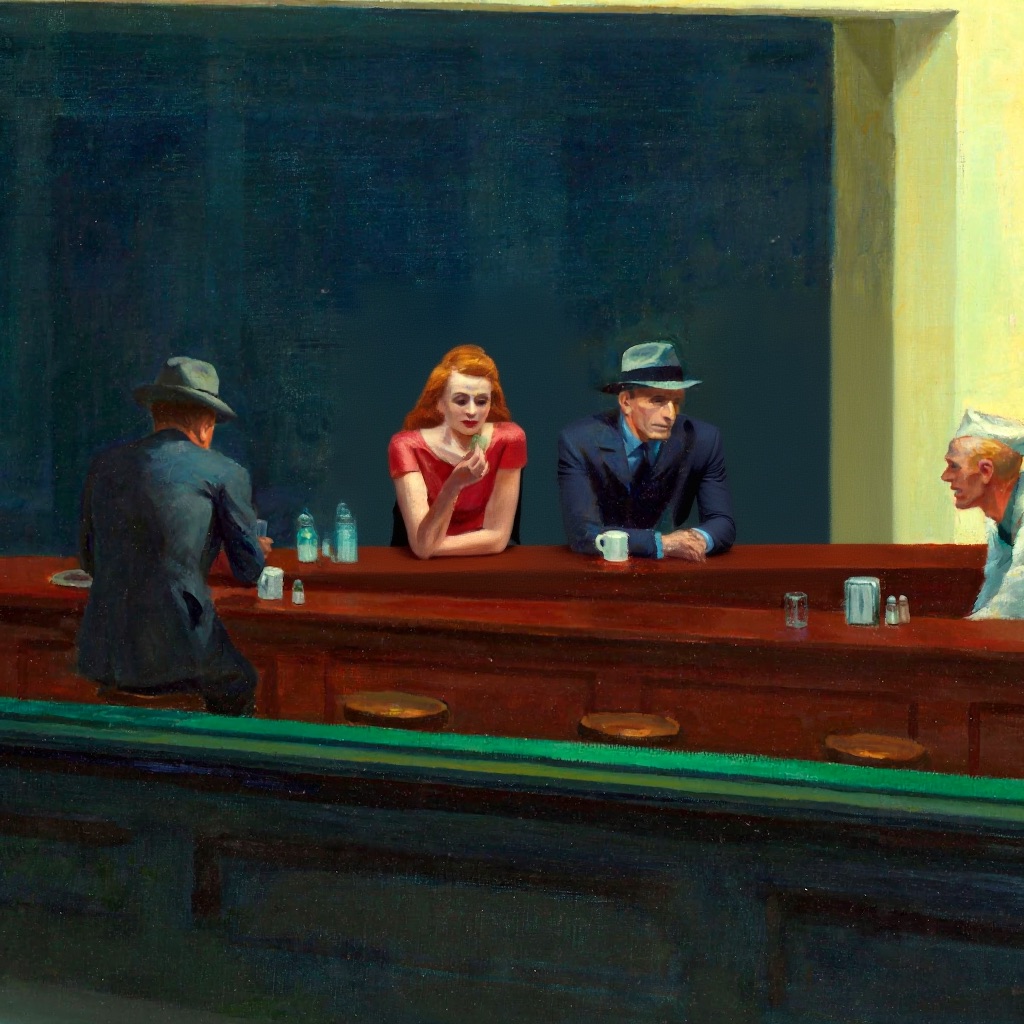} &
        &
        \includegraphics[width=0.165\linewidth]{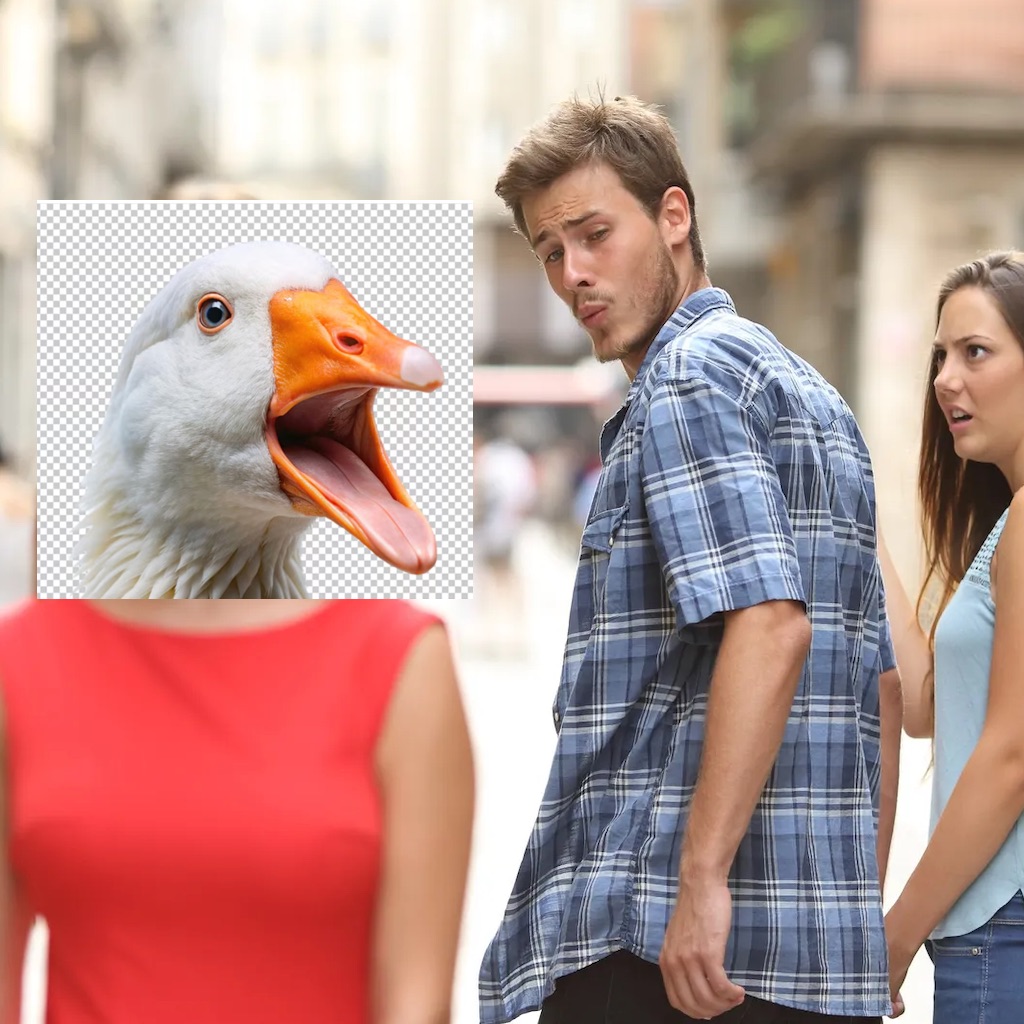} &
        \includegraphics[width=0.165\linewidth]{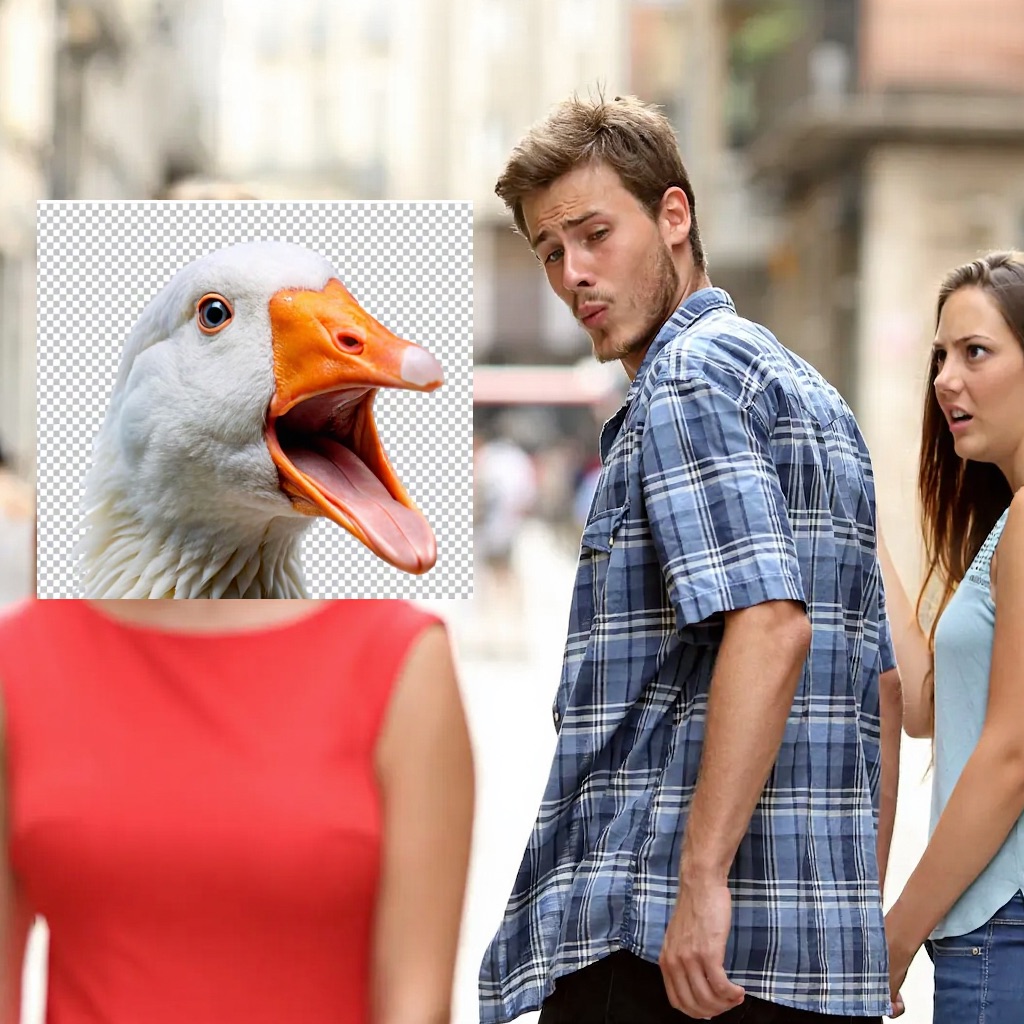} &
        \includegraphics[width=0.165\linewidth]{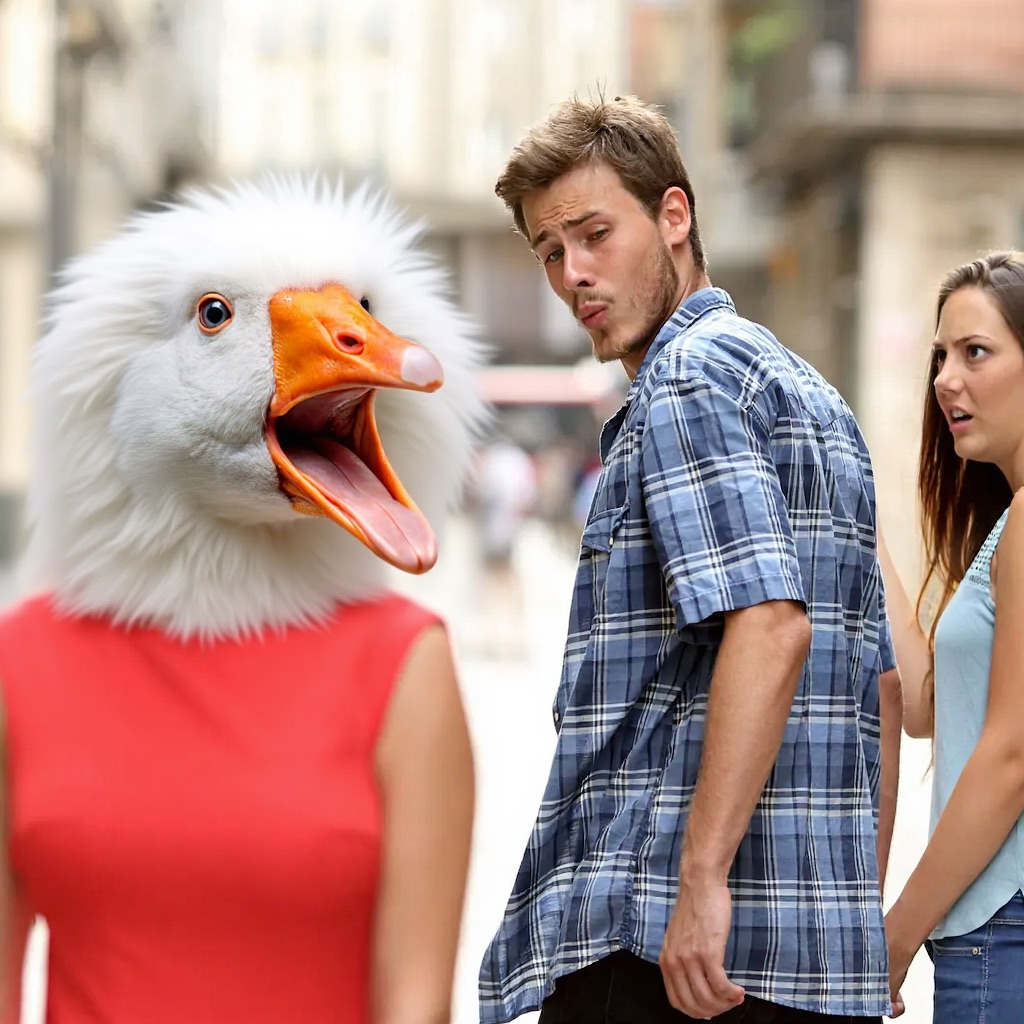} \\

        \includegraphics[width=0.165\linewidth]{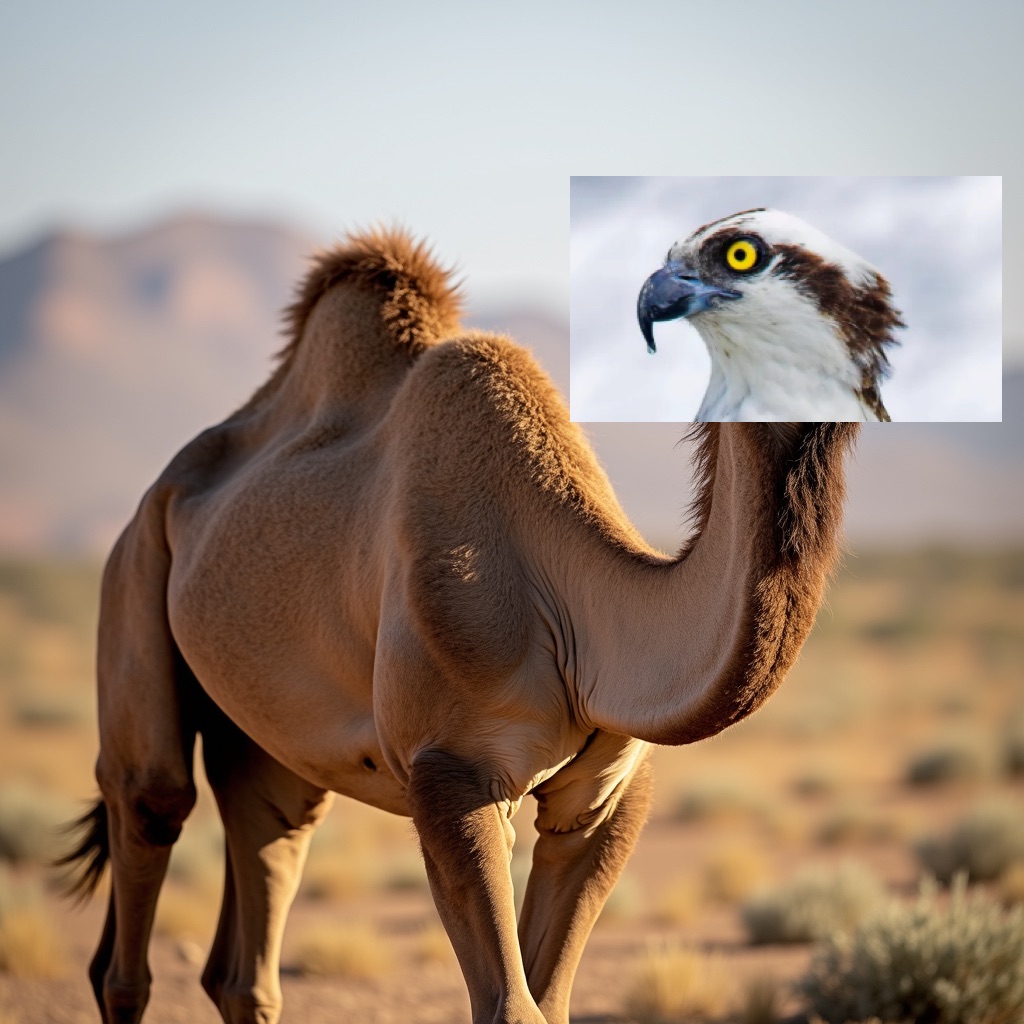} &
        \includegraphics[width=0.165\linewidth]{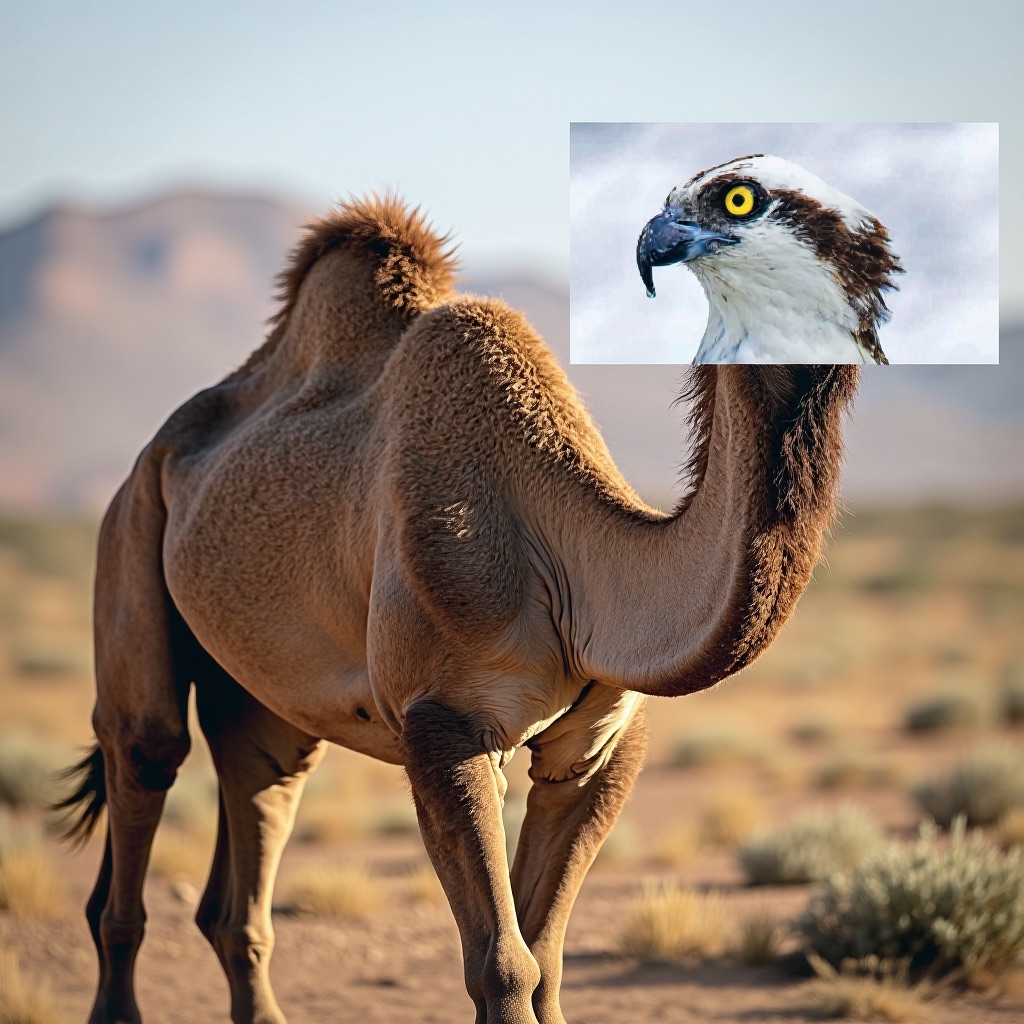} &
        \includegraphics[width=0.165\linewidth]{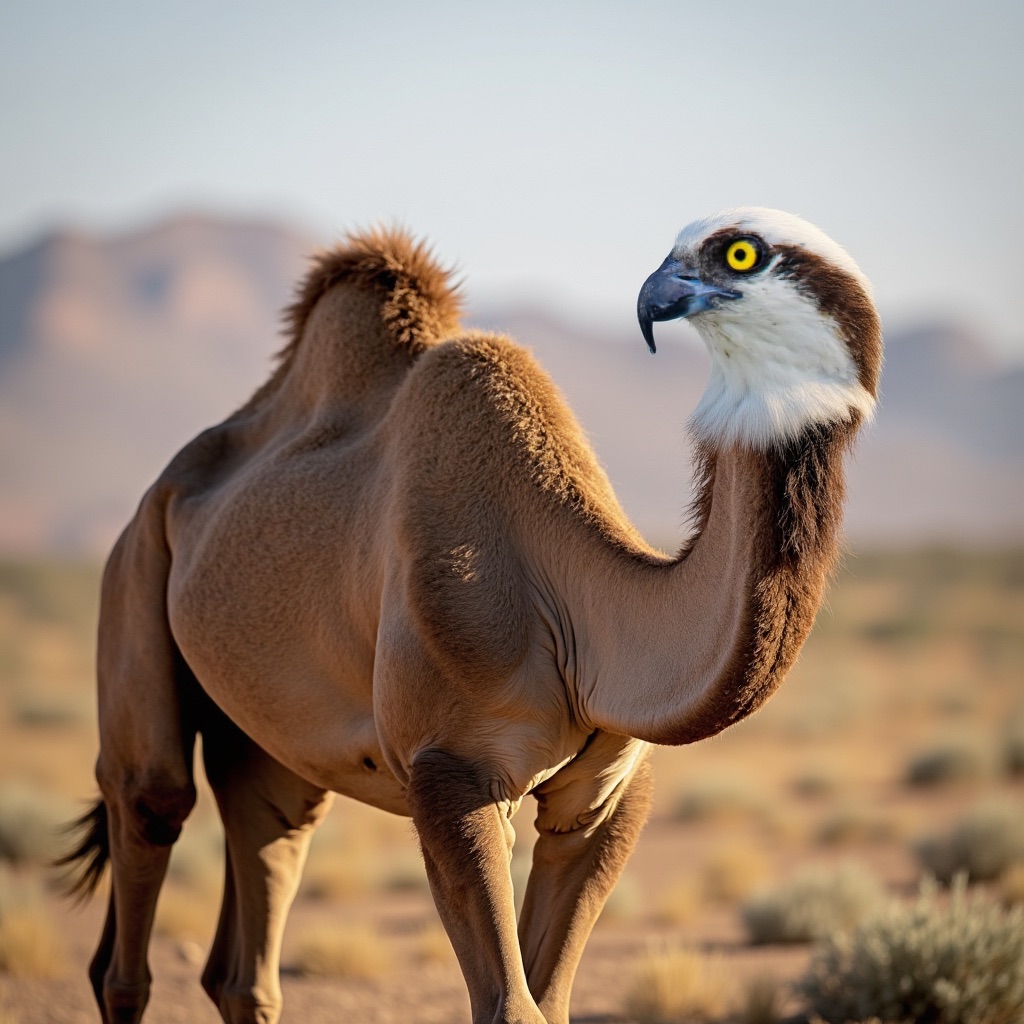} &
        &
        \includegraphics[width=0.165\linewidth, trim=70mm 50mm 30mm 50mm, clip]{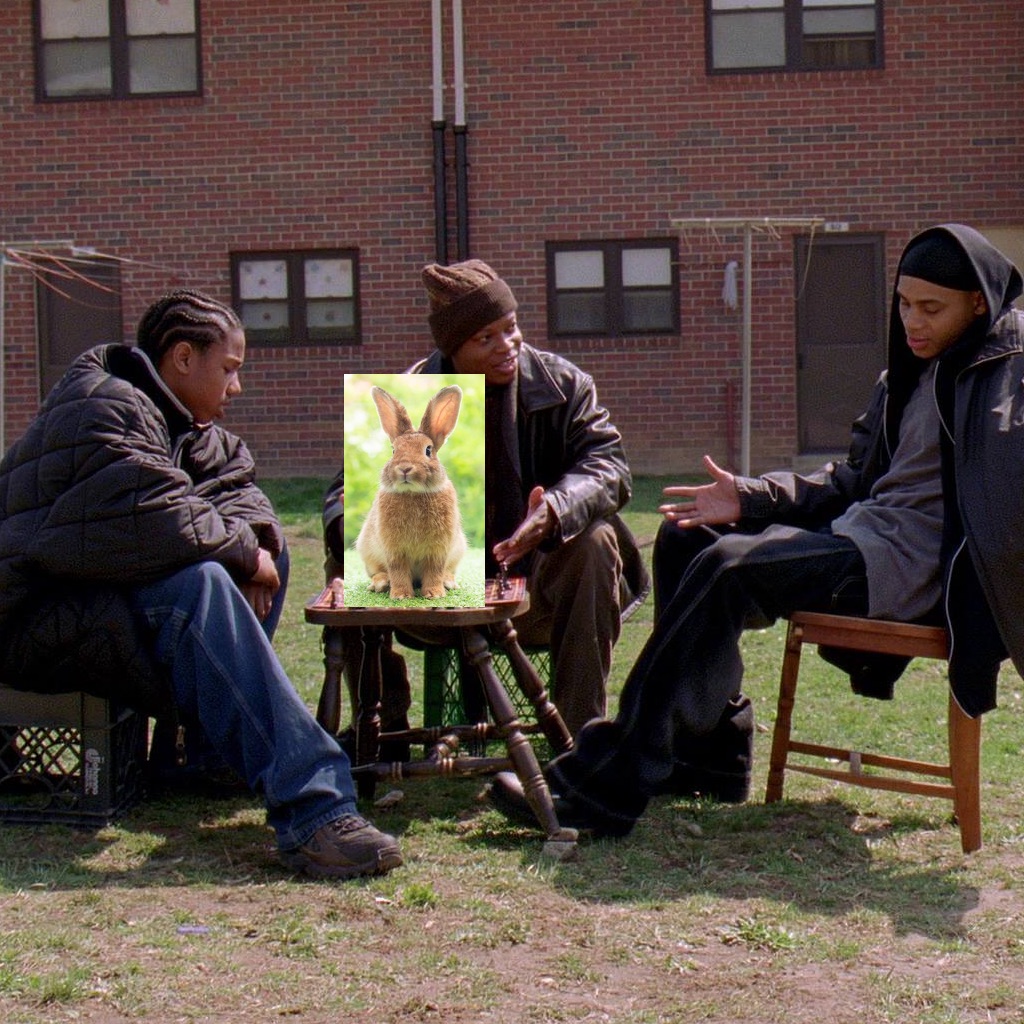} &
        \includegraphics[width=0.165\linewidth, trim=70mm 50mm 30mm 50mm, clip]{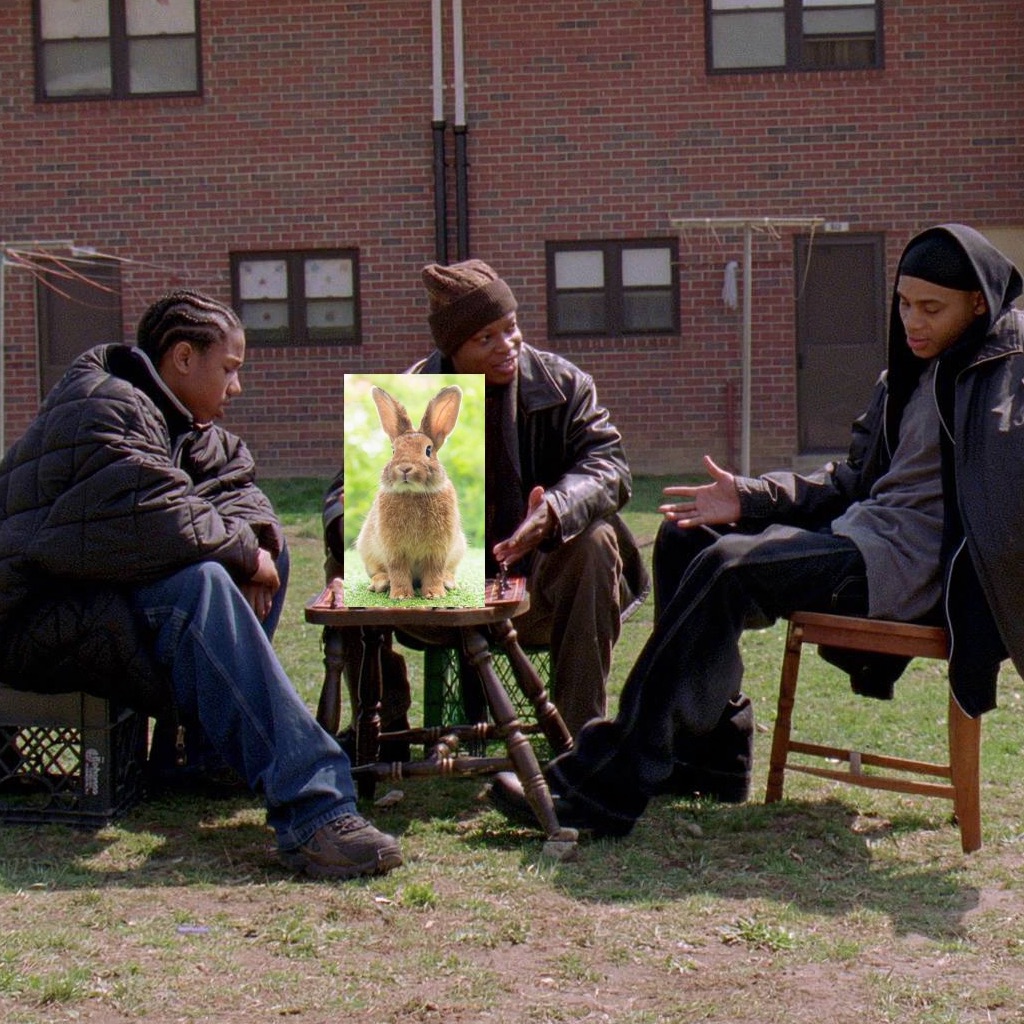} &
        \includegraphics[width=0.165\linewidth, trim=70mm 50mm 30mm 50mm, clip]{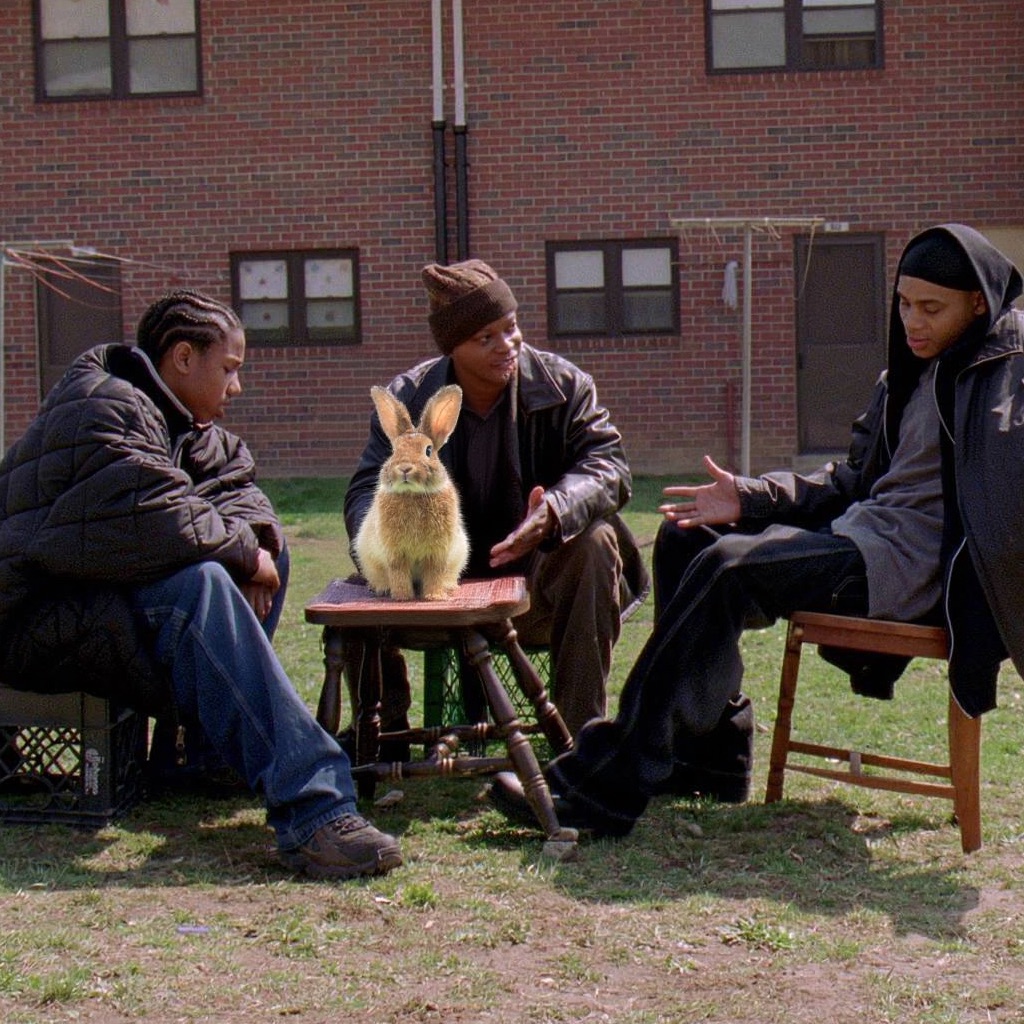} \\

        \includegraphics[width=0.165\linewidth]{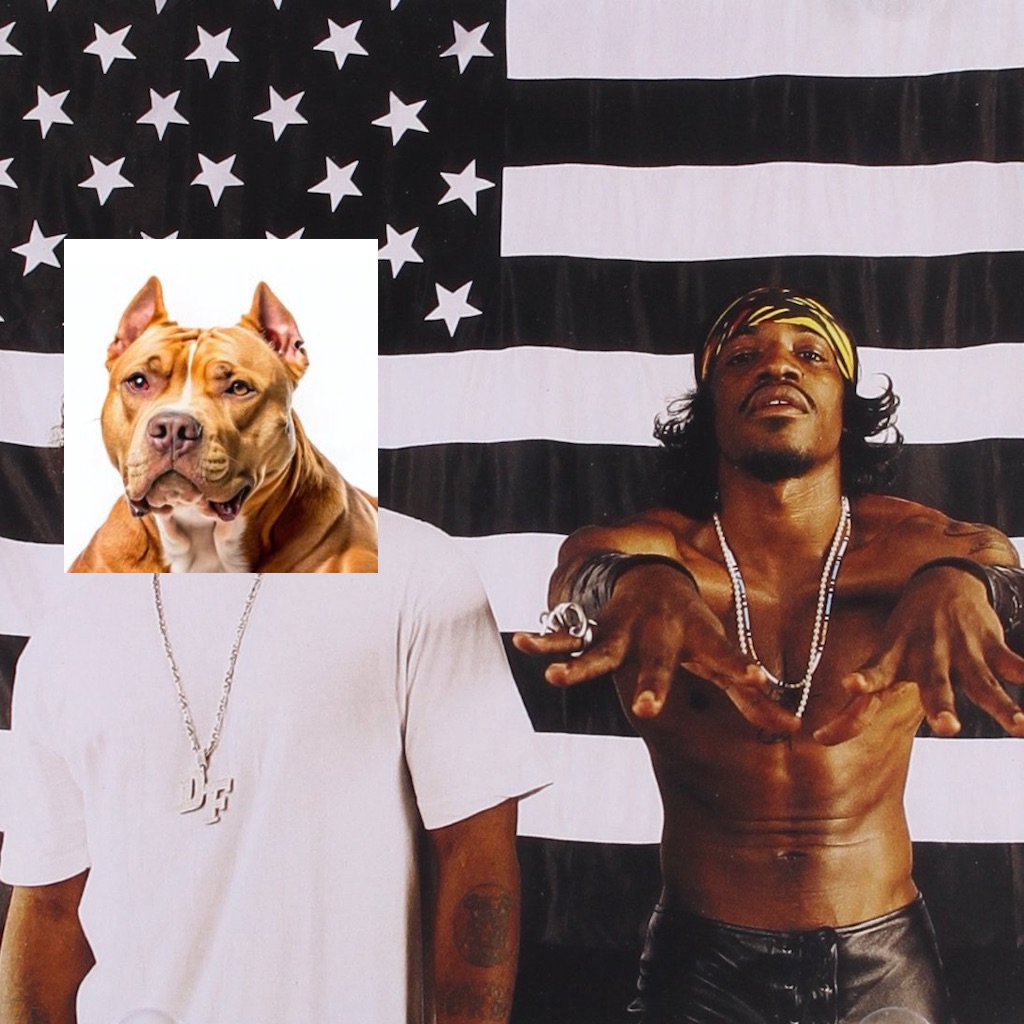} &
        \includegraphics[width=0.165\linewidth]{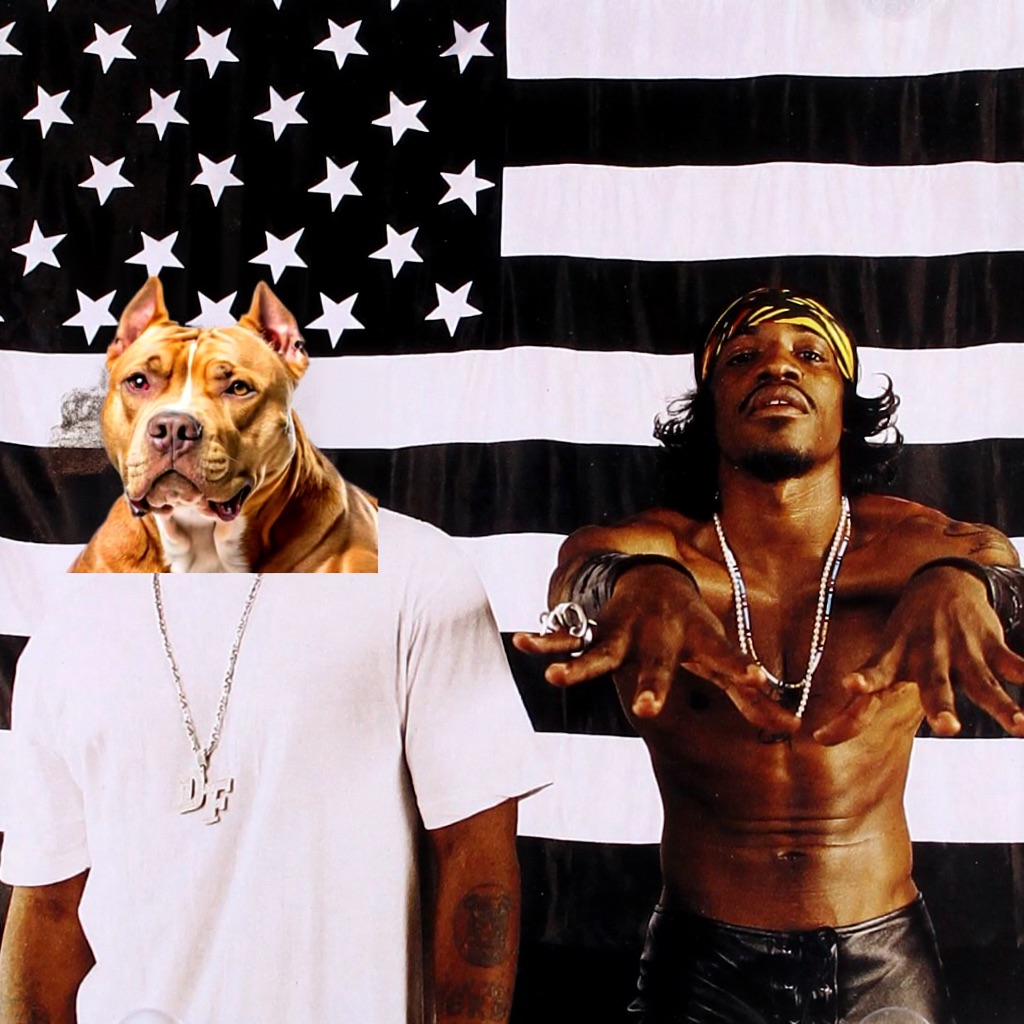} &
        \includegraphics[width=0.165\linewidth]{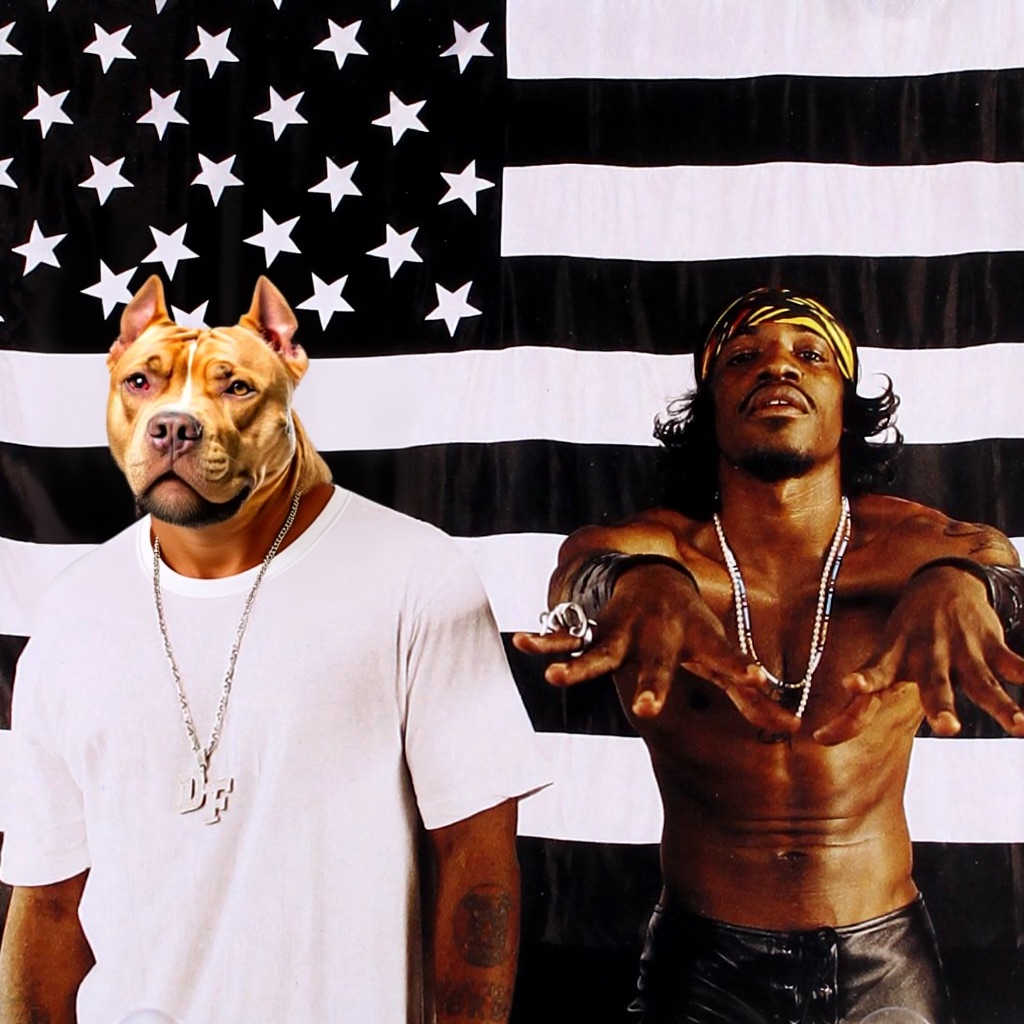} &
        &
        \includegraphics[width=0.165\linewidth]{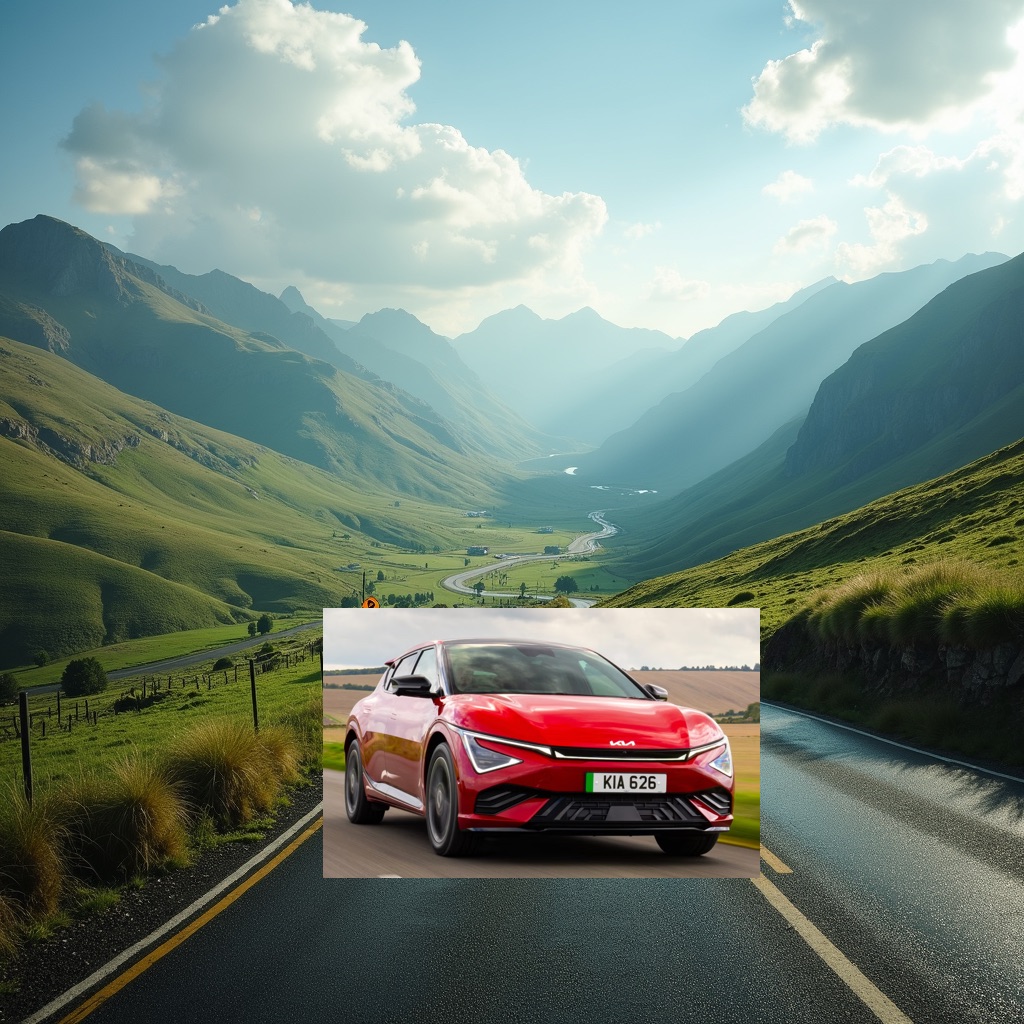} &
        \includegraphics[width=0.165\linewidth]{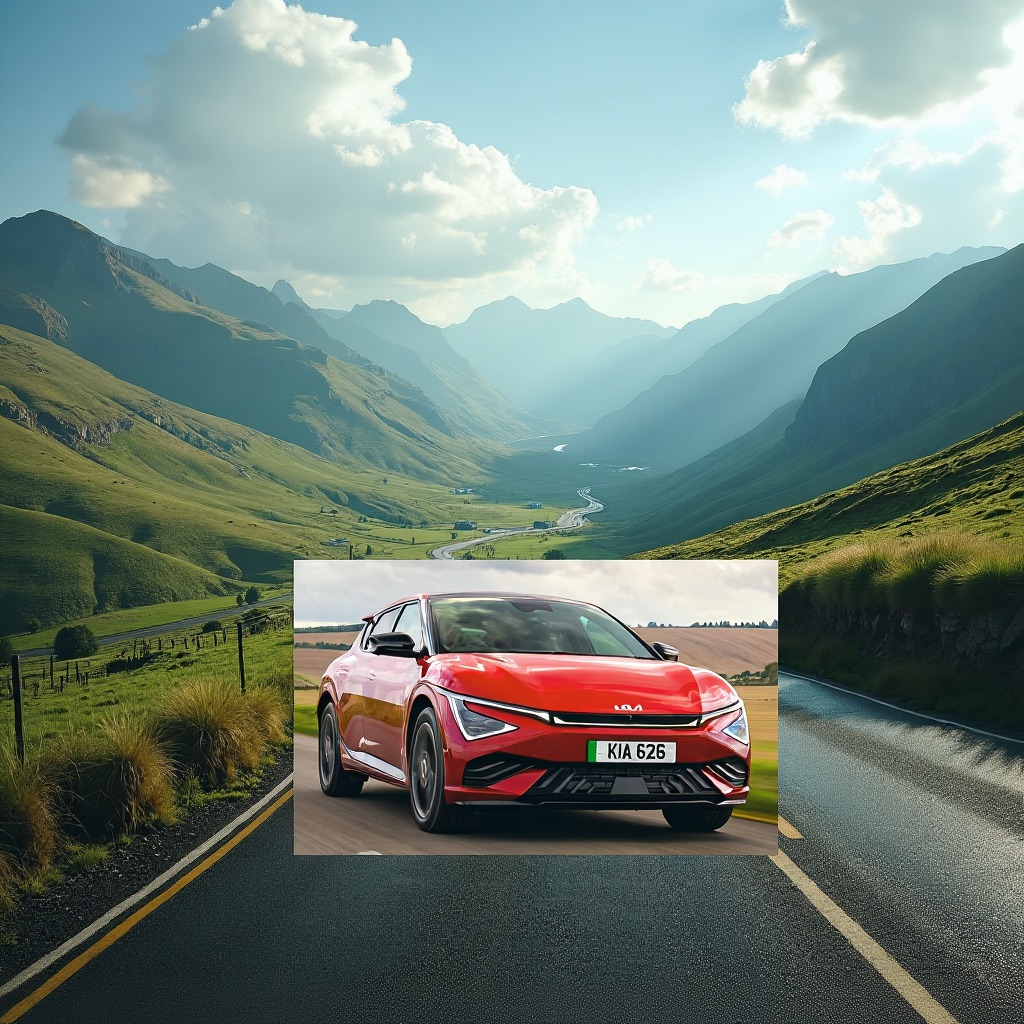} &
        \includegraphics[width=0.165\linewidth]{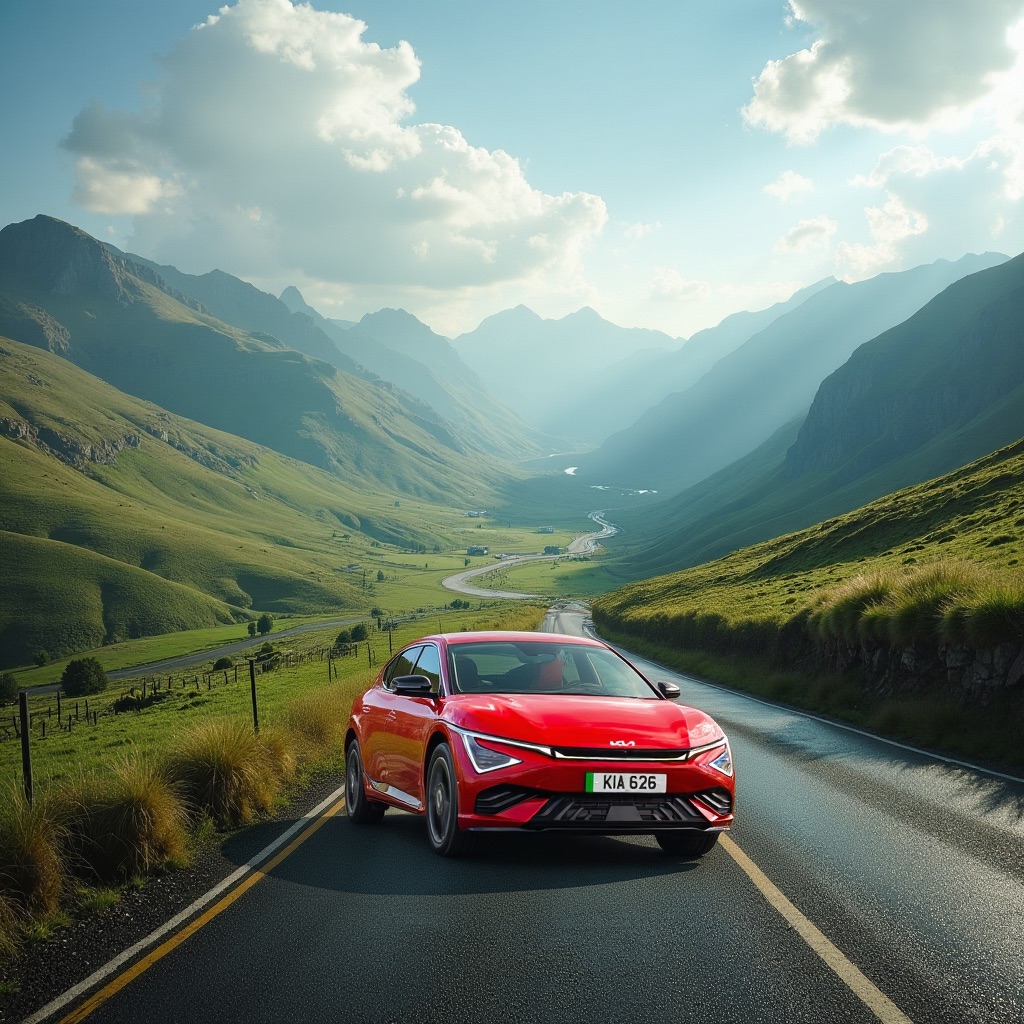} \\

    \end{tabular}
    \vspace{10pt}
    \caption{Additional LooseRoPE results, compared against our method's base model: FLUX Kontext.
    }
    \label{fig:supp_results}
\end{figure*}

\begin{figure*}[t] %
    \centering
    
    \setlength{\tabcolsep}{1pt}
    \begin{tabular}{ccccccc}
        Base & Edit \#1 & Output \#1 & Edit \#2 & Output \#2 & Edit \#3 & Output \#3 \\
        
        \includegraphics[width=0.135\linewidth]{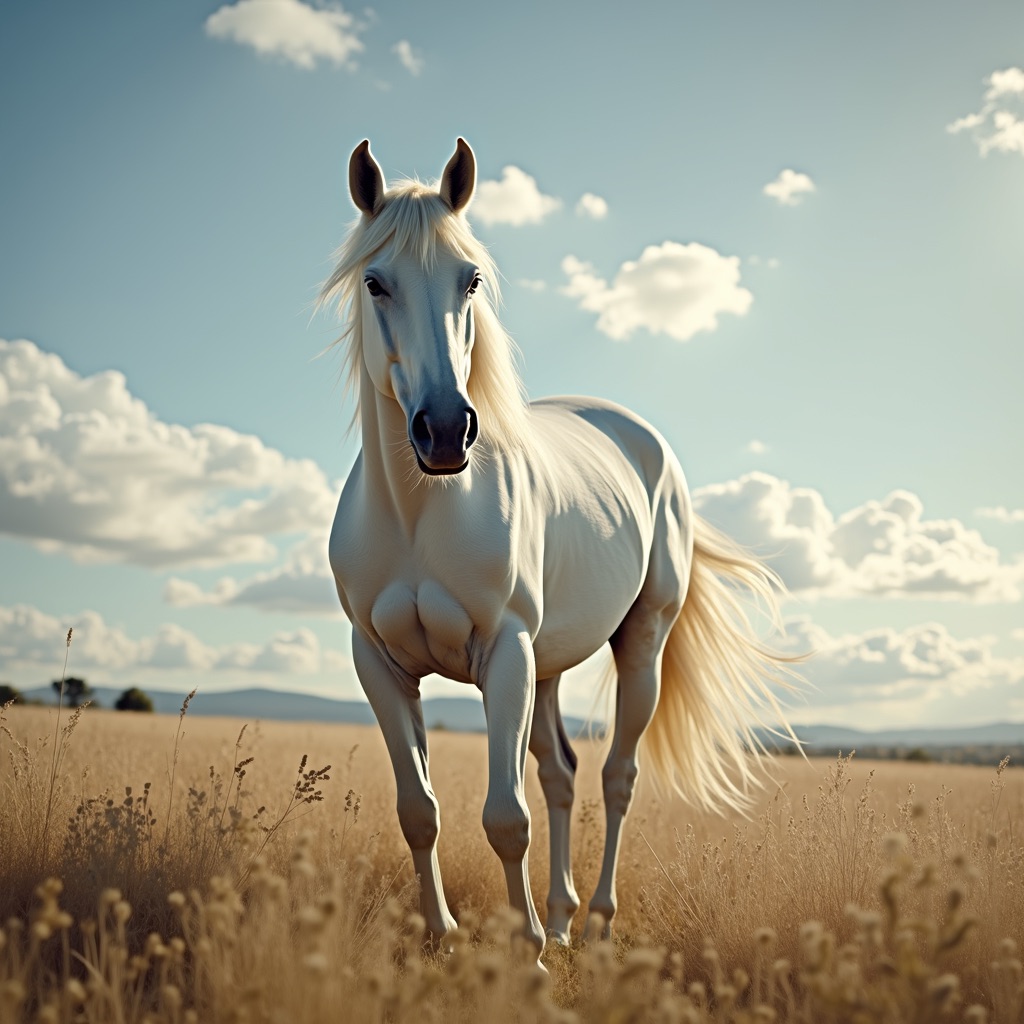} &
        \includegraphics[width=0.135\linewidth]{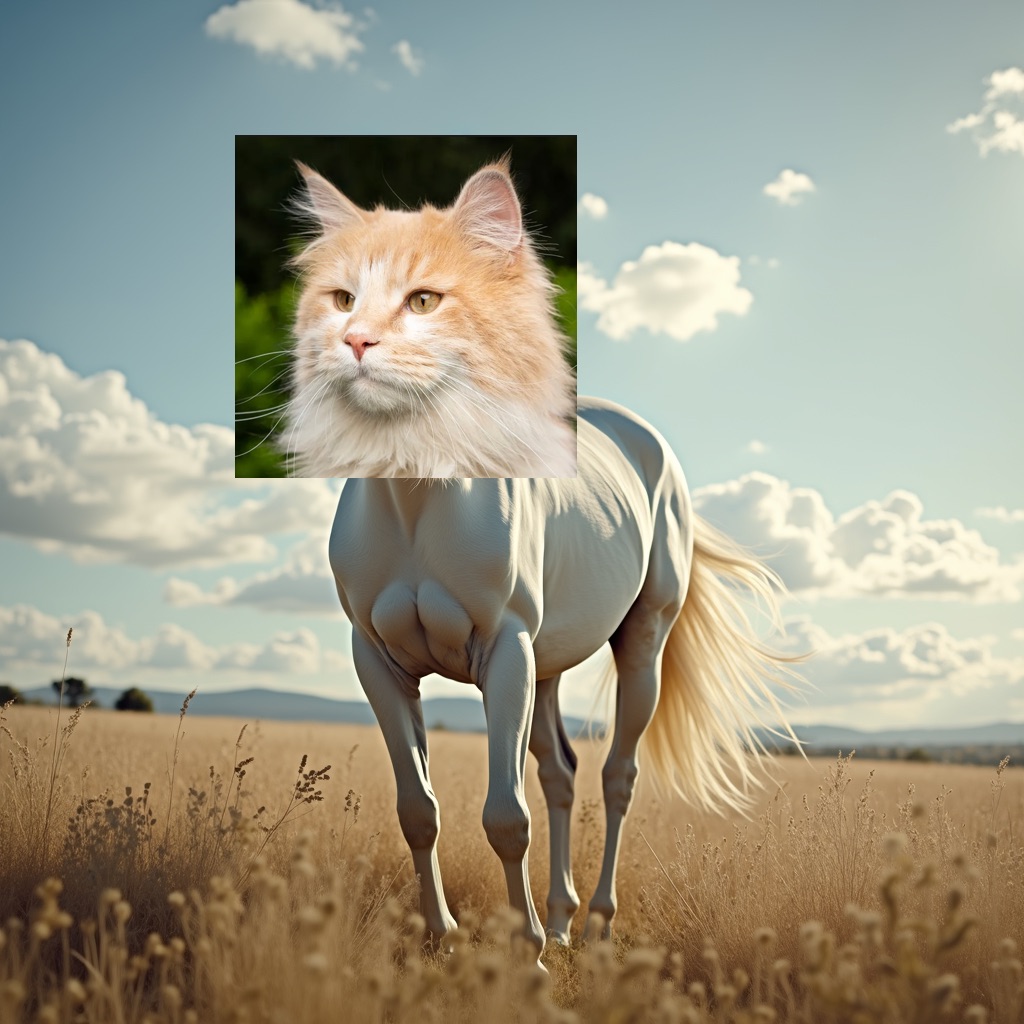} &
        \includegraphics[width=0.135\linewidth]{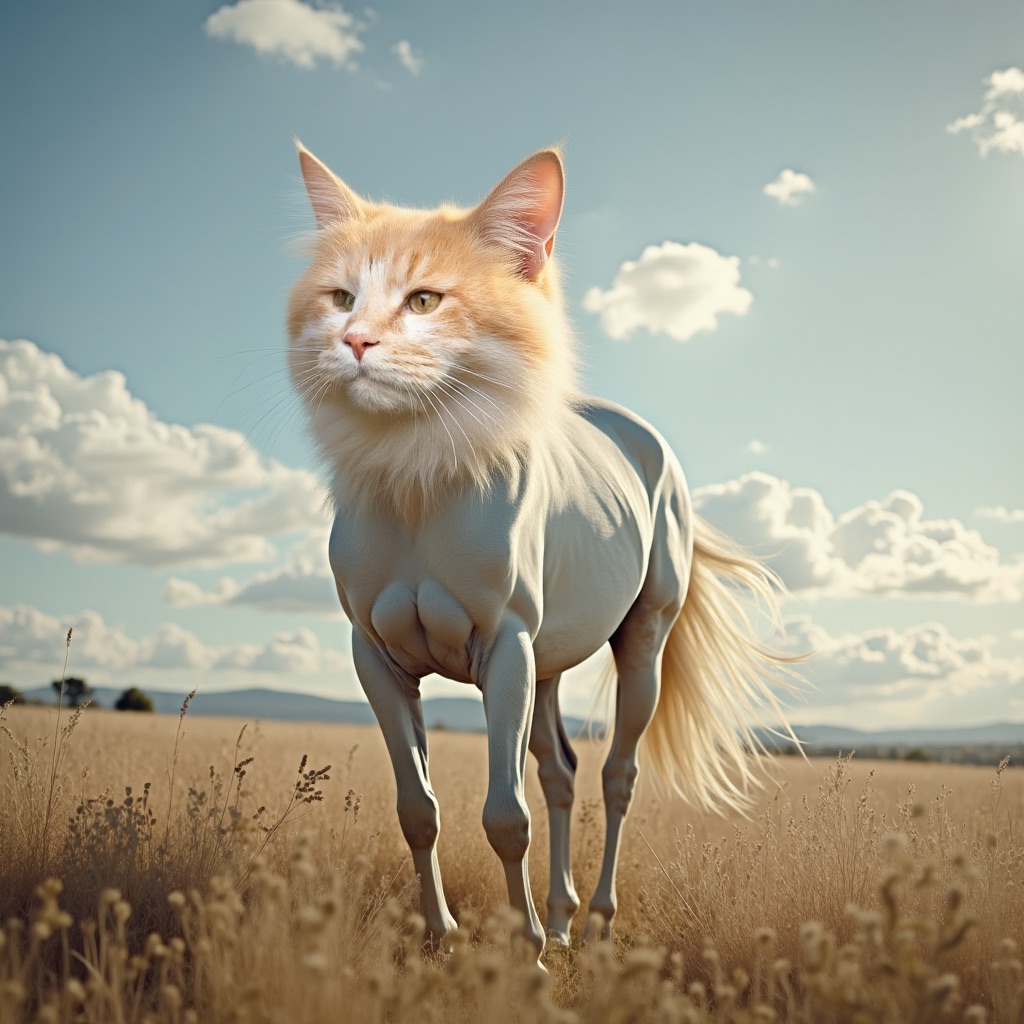} &
        \includegraphics[width=0.135\linewidth]{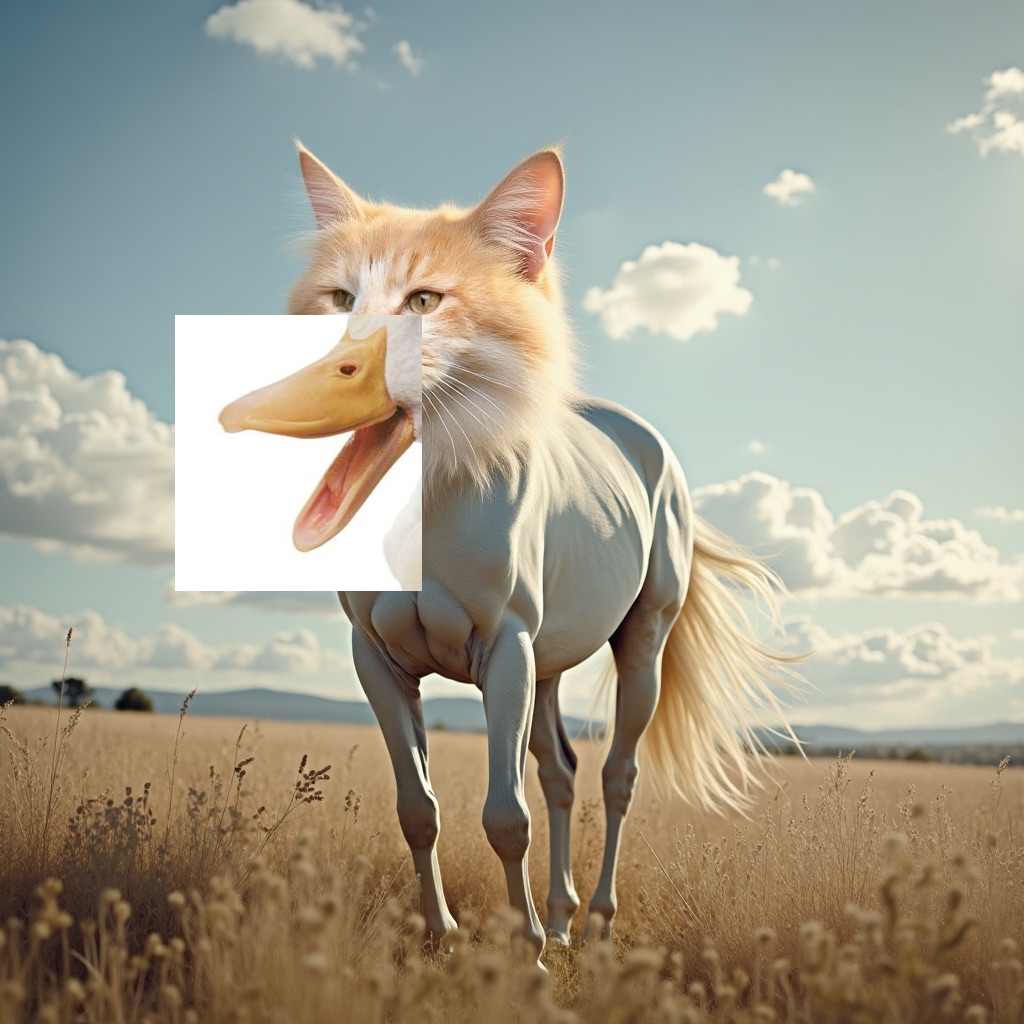} &
        \includegraphics[width=0.135\linewidth]{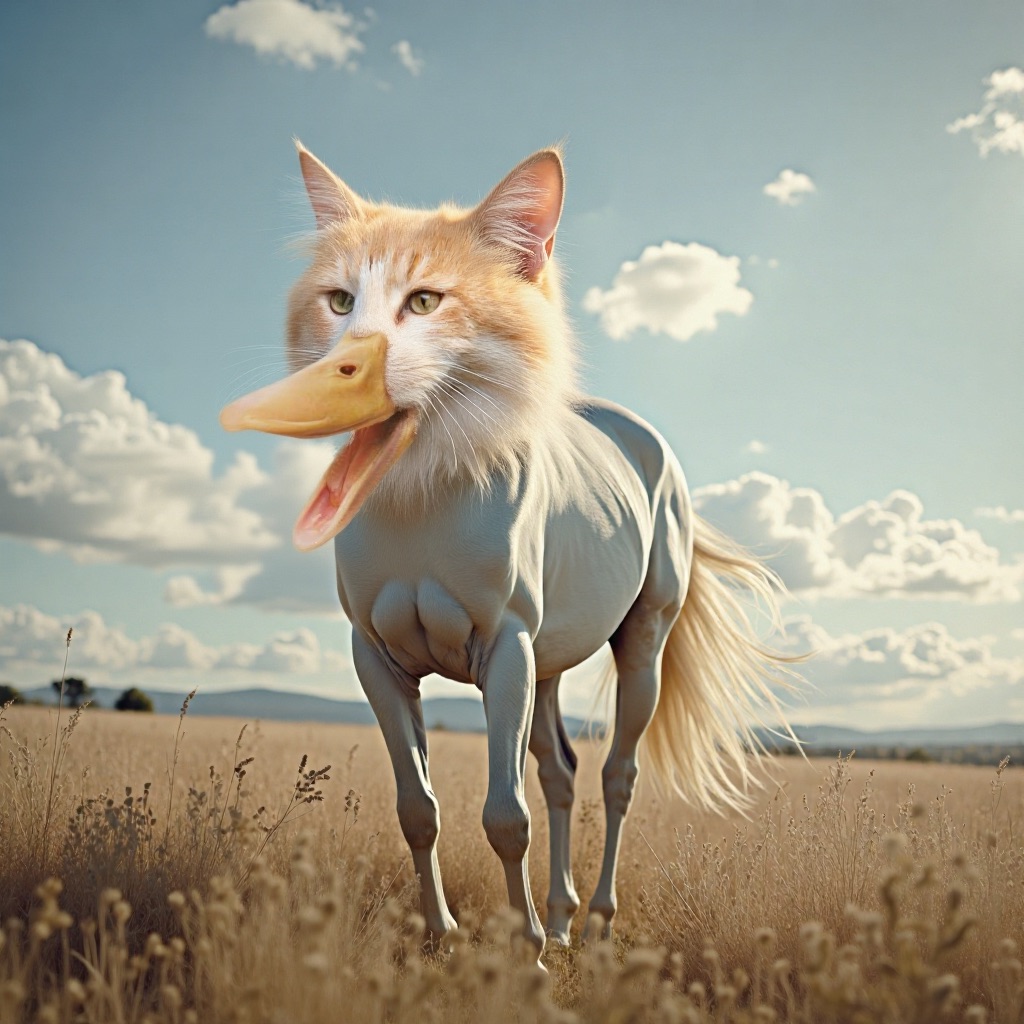} &
        \includegraphics[width=0.135\linewidth]{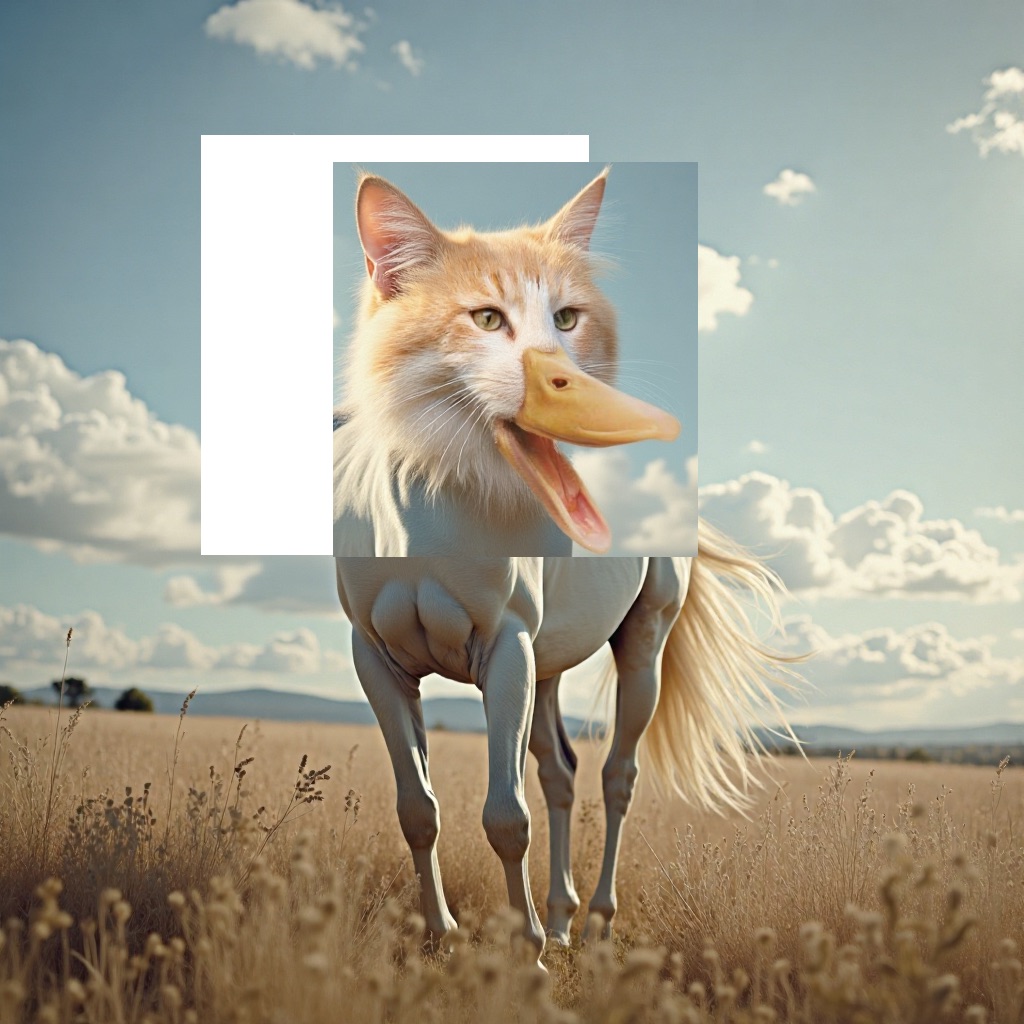} &
        \includegraphics[width=0.135\linewidth]{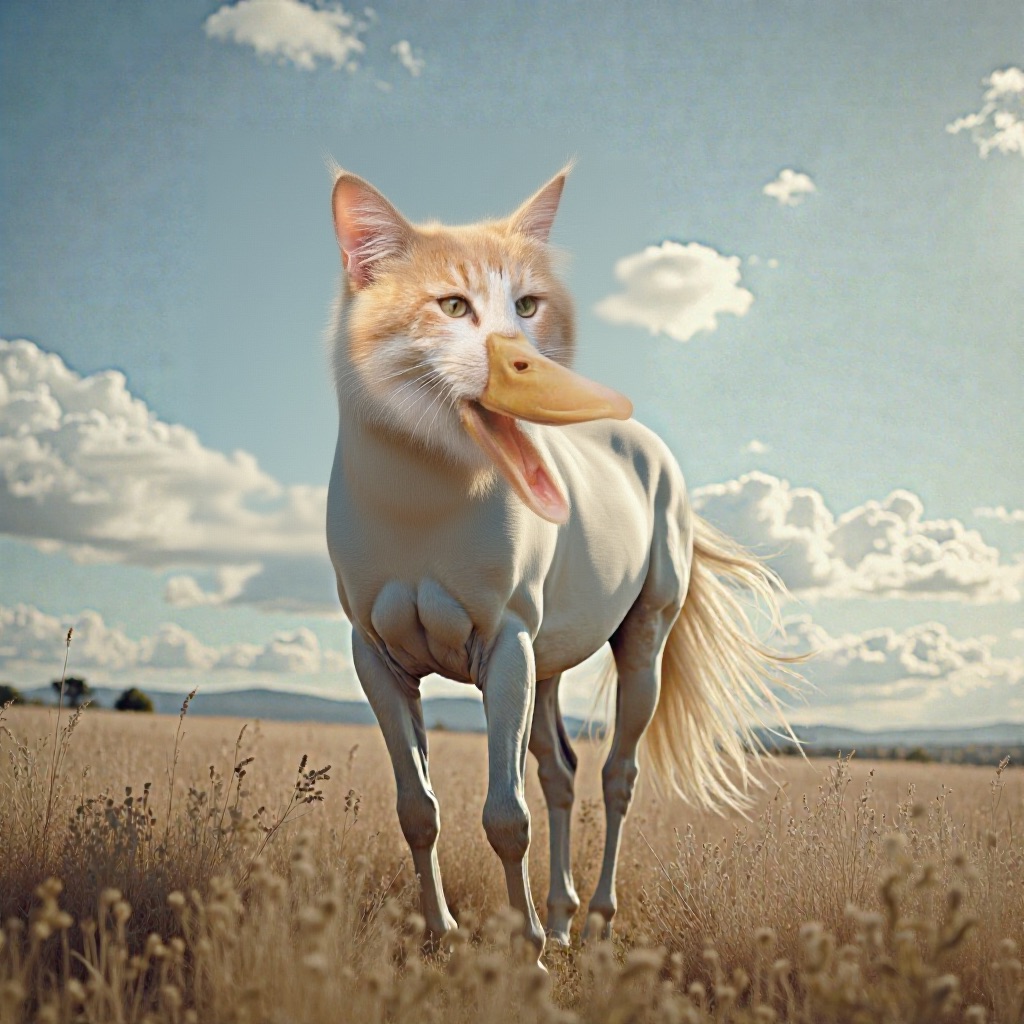} \\

        \includegraphics[width=0.135\linewidth]{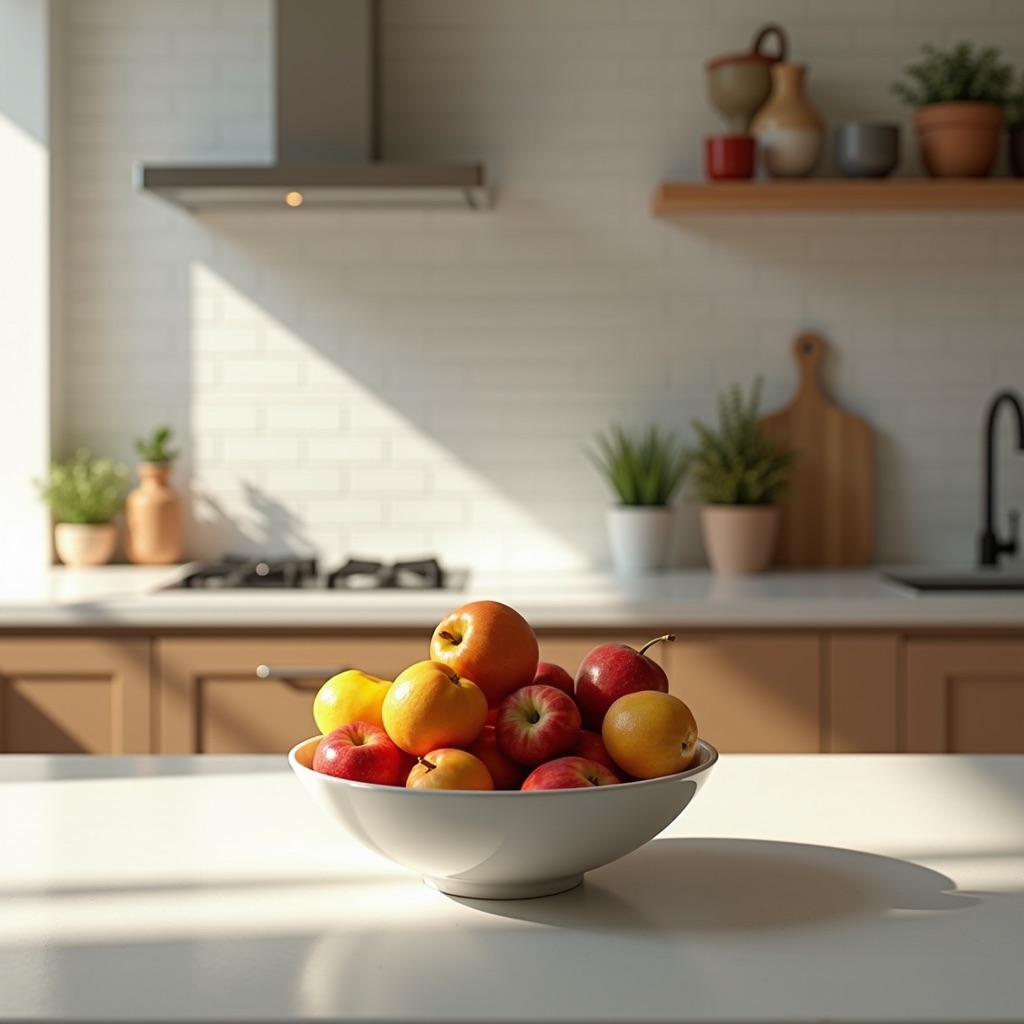} &
        \includegraphics[width=0.135\linewidth]{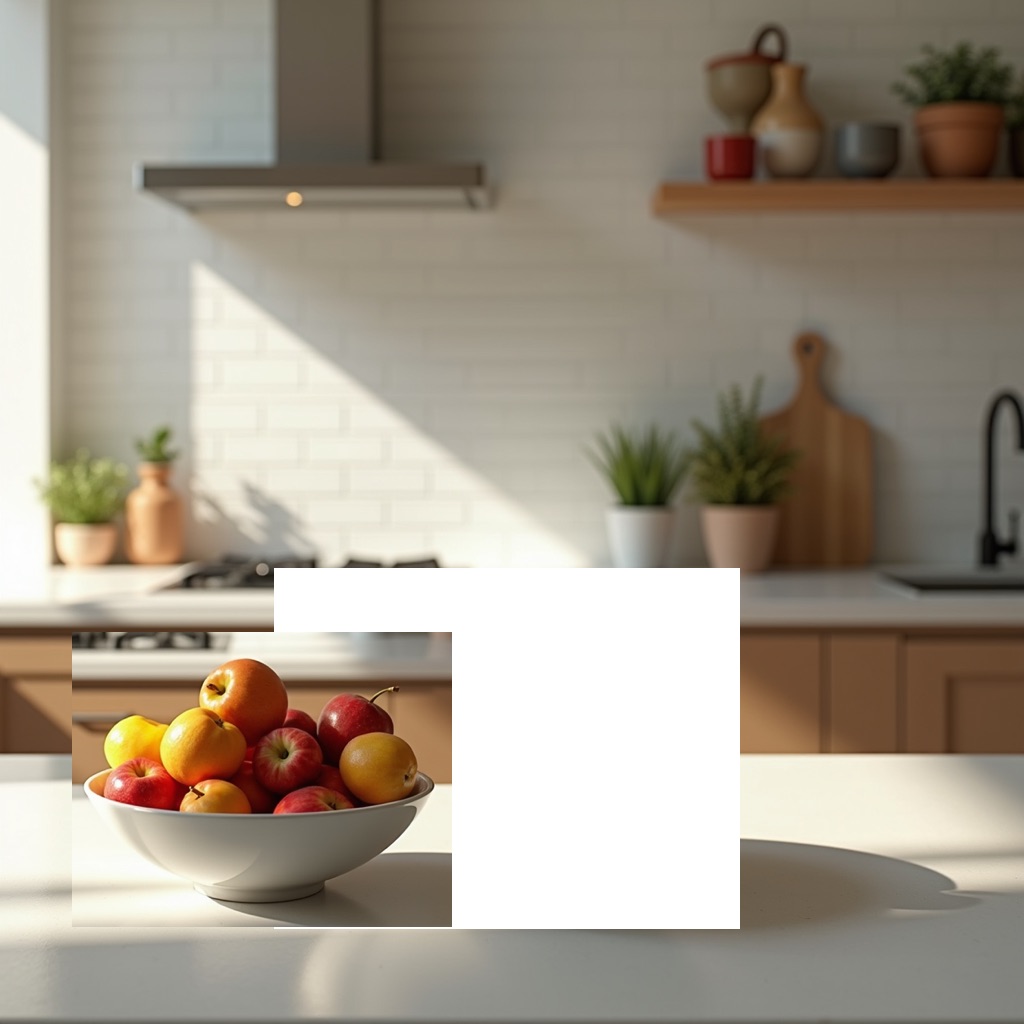} &
        \includegraphics[width=0.135\linewidth]{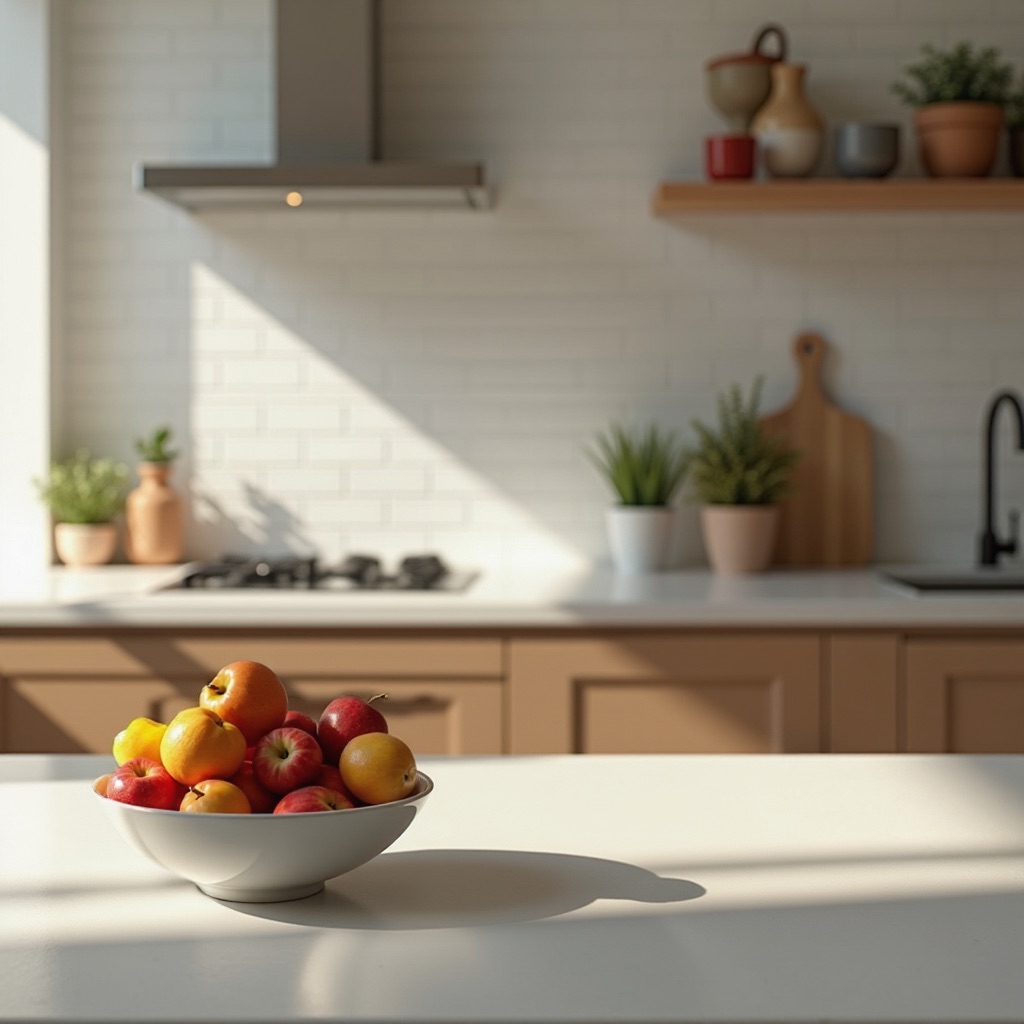} &
        \includegraphics[width=0.135\linewidth]{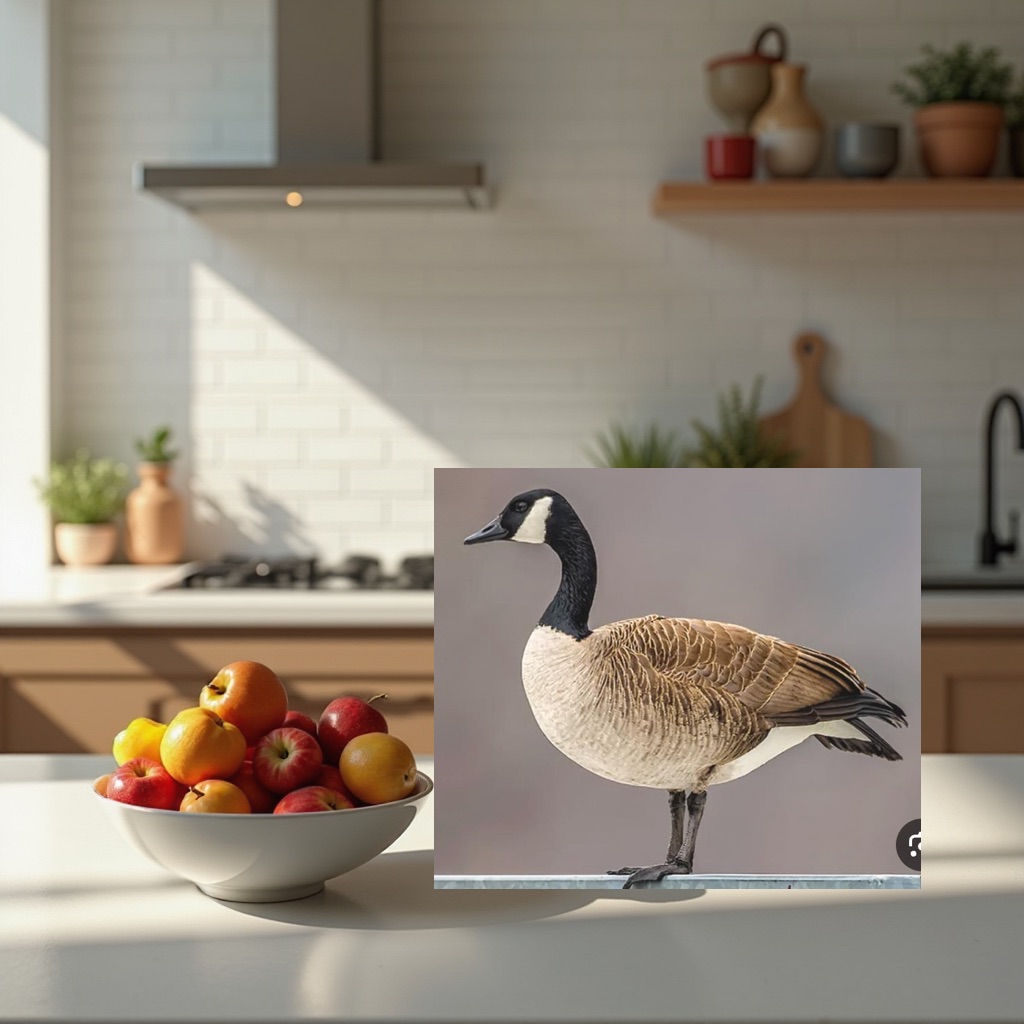} &
        \includegraphics[width=0.135\linewidth]{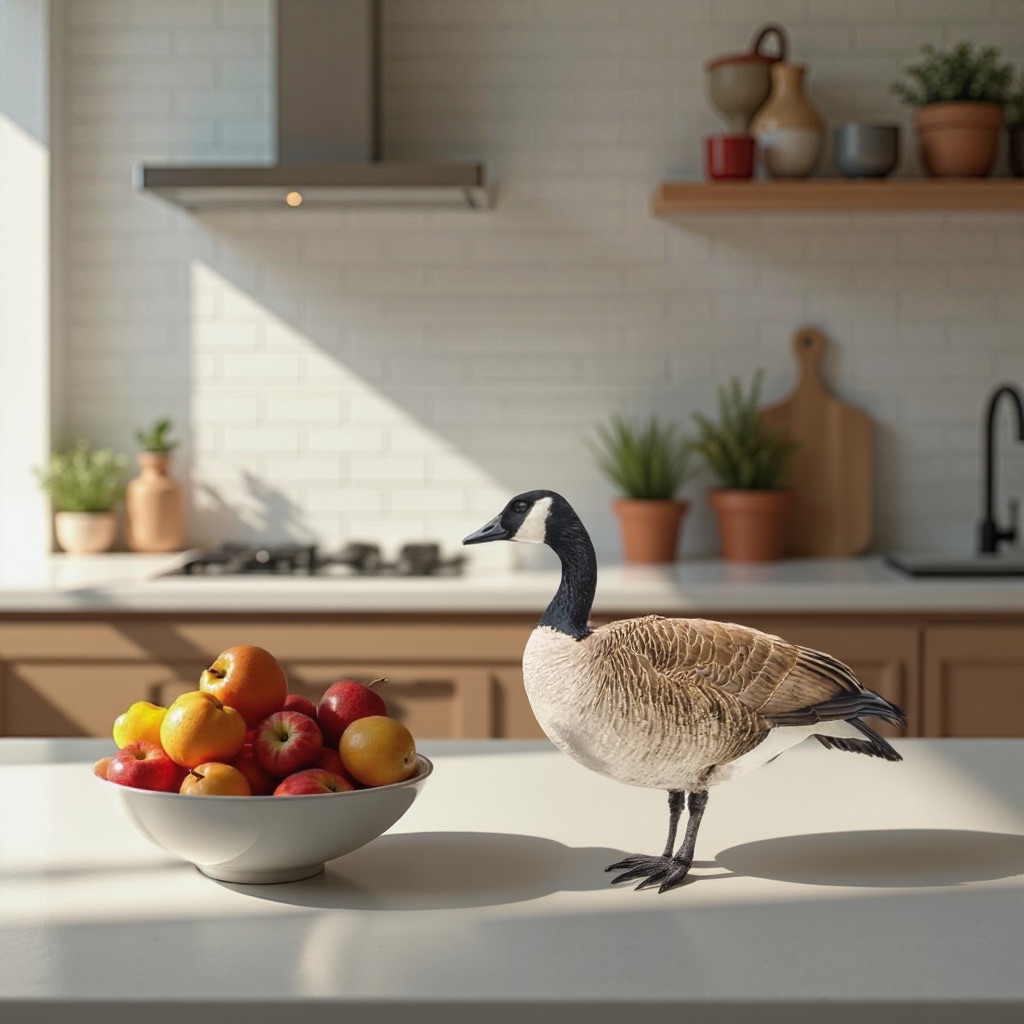} &
        \includegraphics[width=0.135\linewidth]{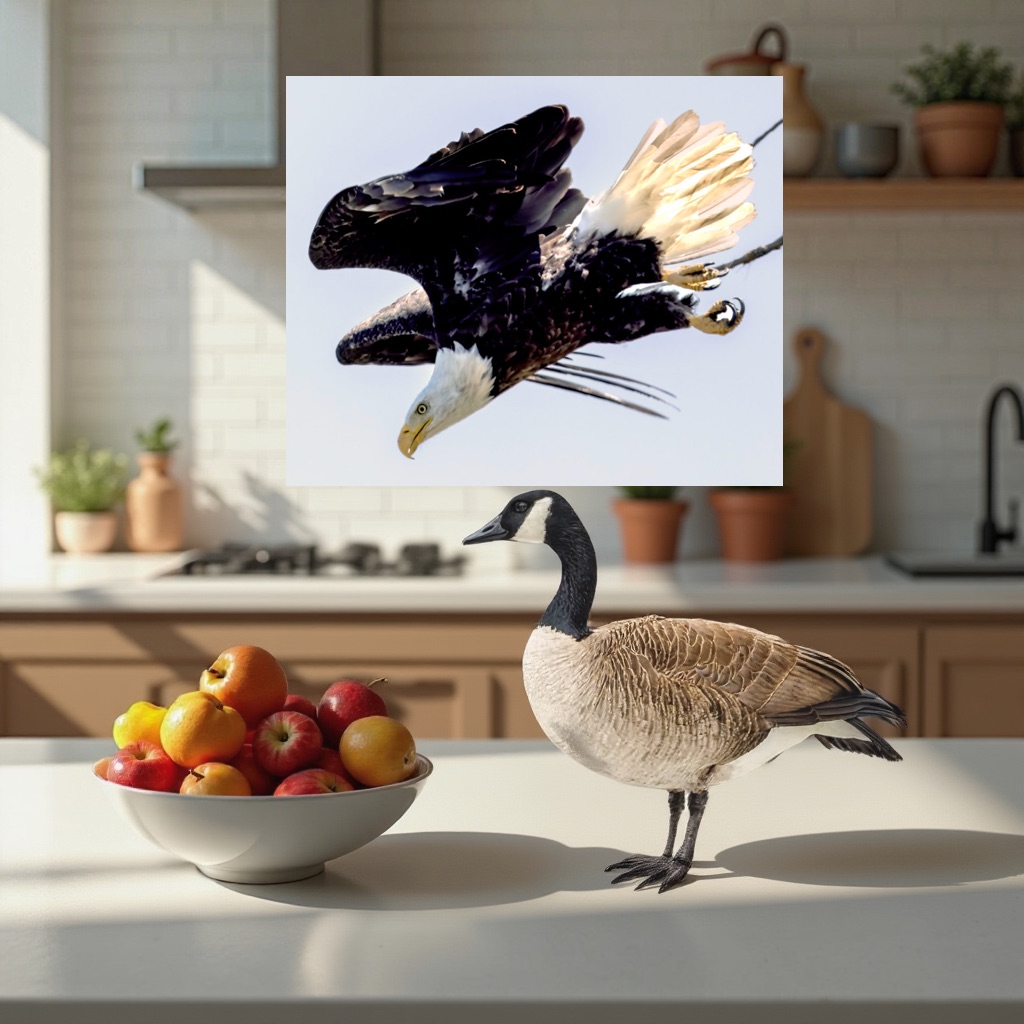} &
        \includegraphics[width=0.135\linewidth]{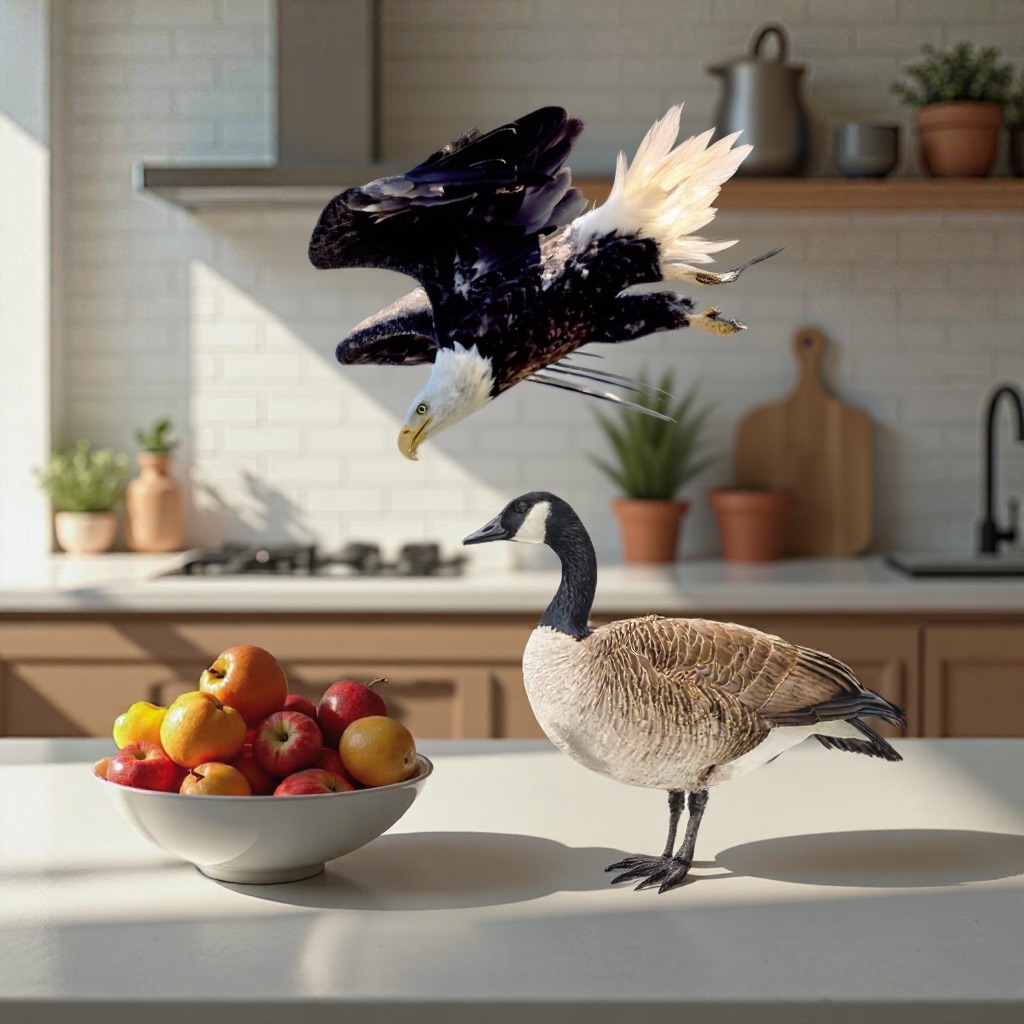} \\

        \includegraphics[width=0.135\linewidth, trim=20mm 40mm 20mm 0mm, clip]{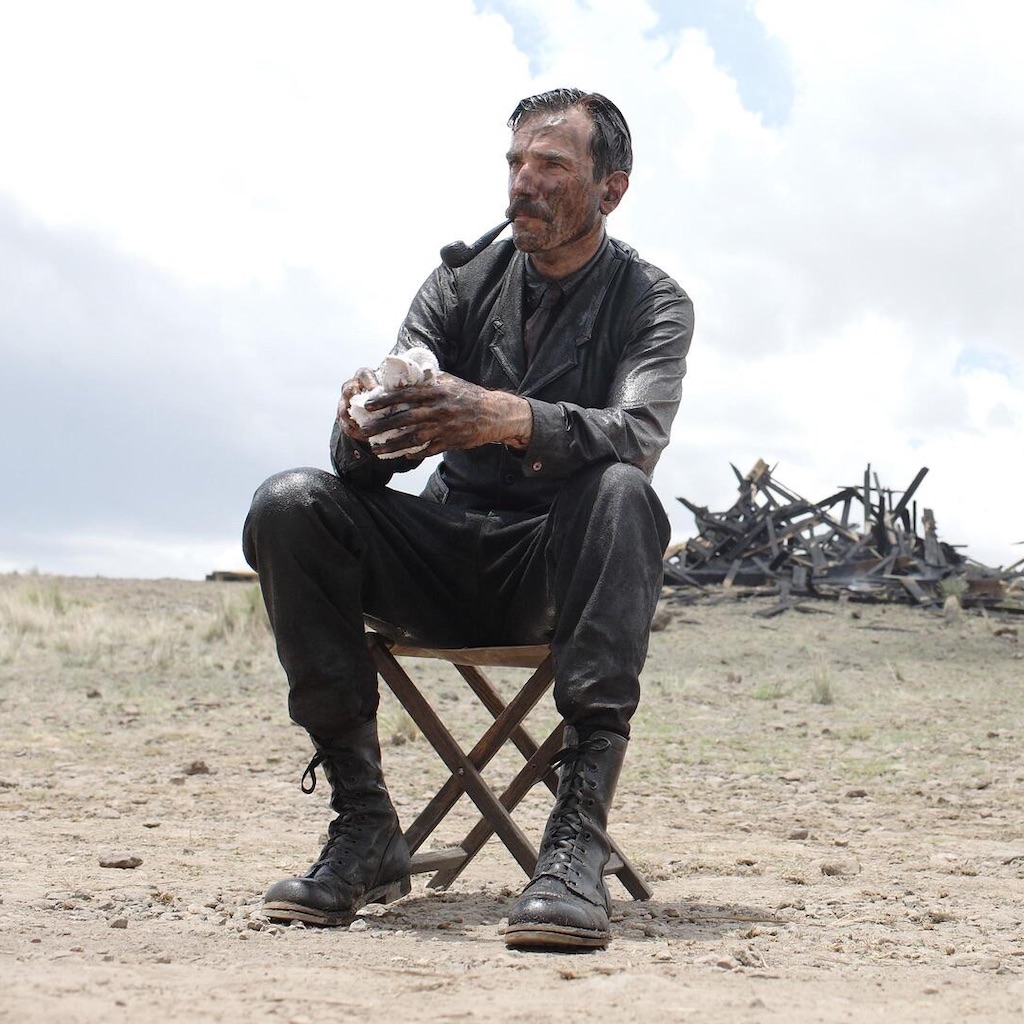} &
        \includegraphics[width=0.135\linewidth, trim=20mm 40mm 20mm 0mm, clip]{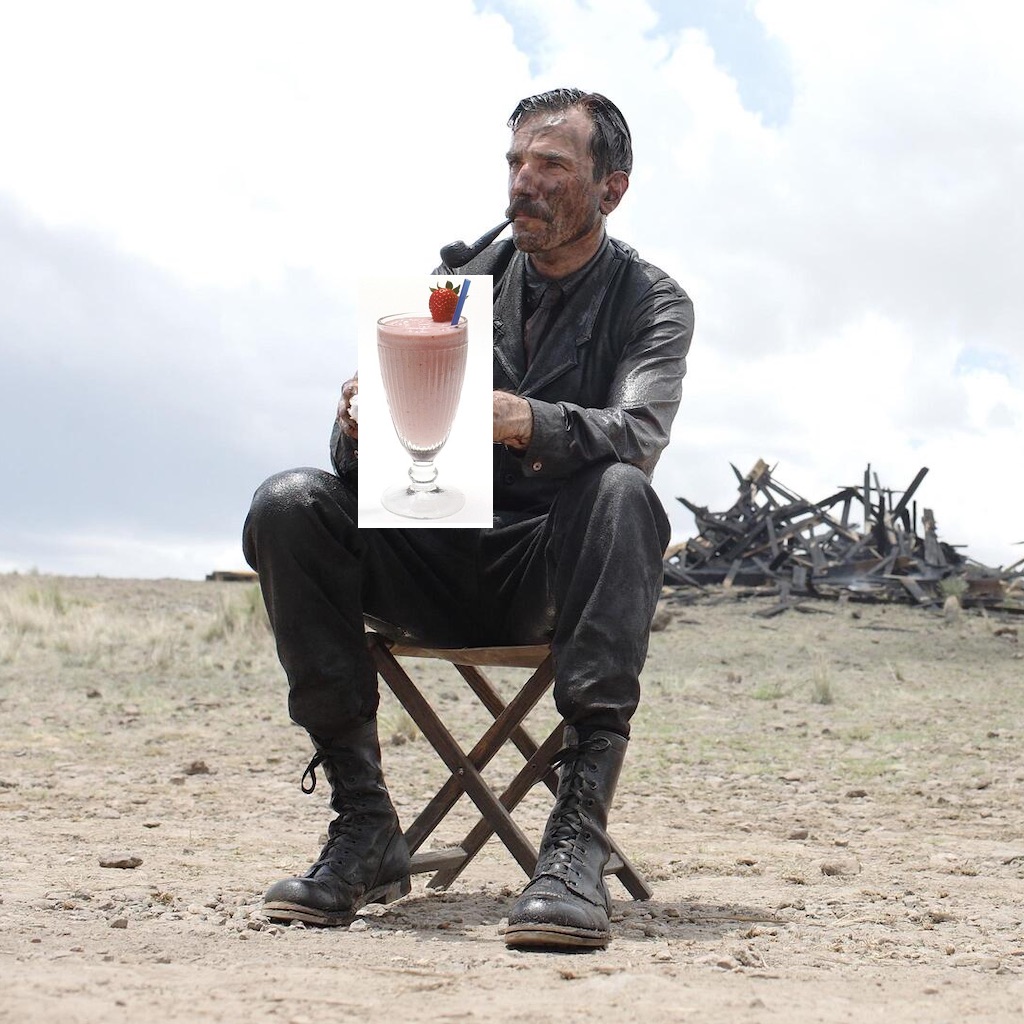} &
        \includegraphics[width=0.135\linewidth, trim=20mm 40mm 20mm 0mm, clip]{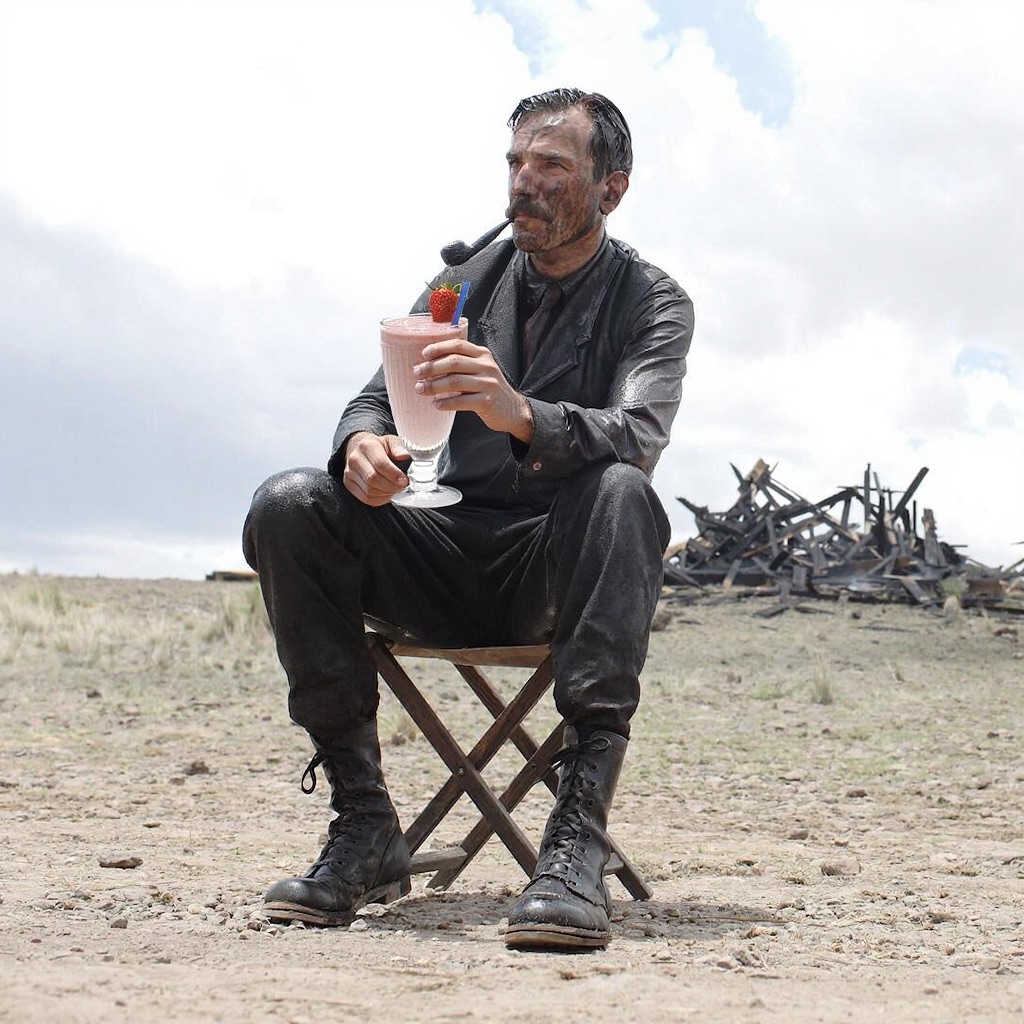} &
        \includegraphics[width=0.135\linewidth, trim=20mm 40mm 20mm 0mm, clip]{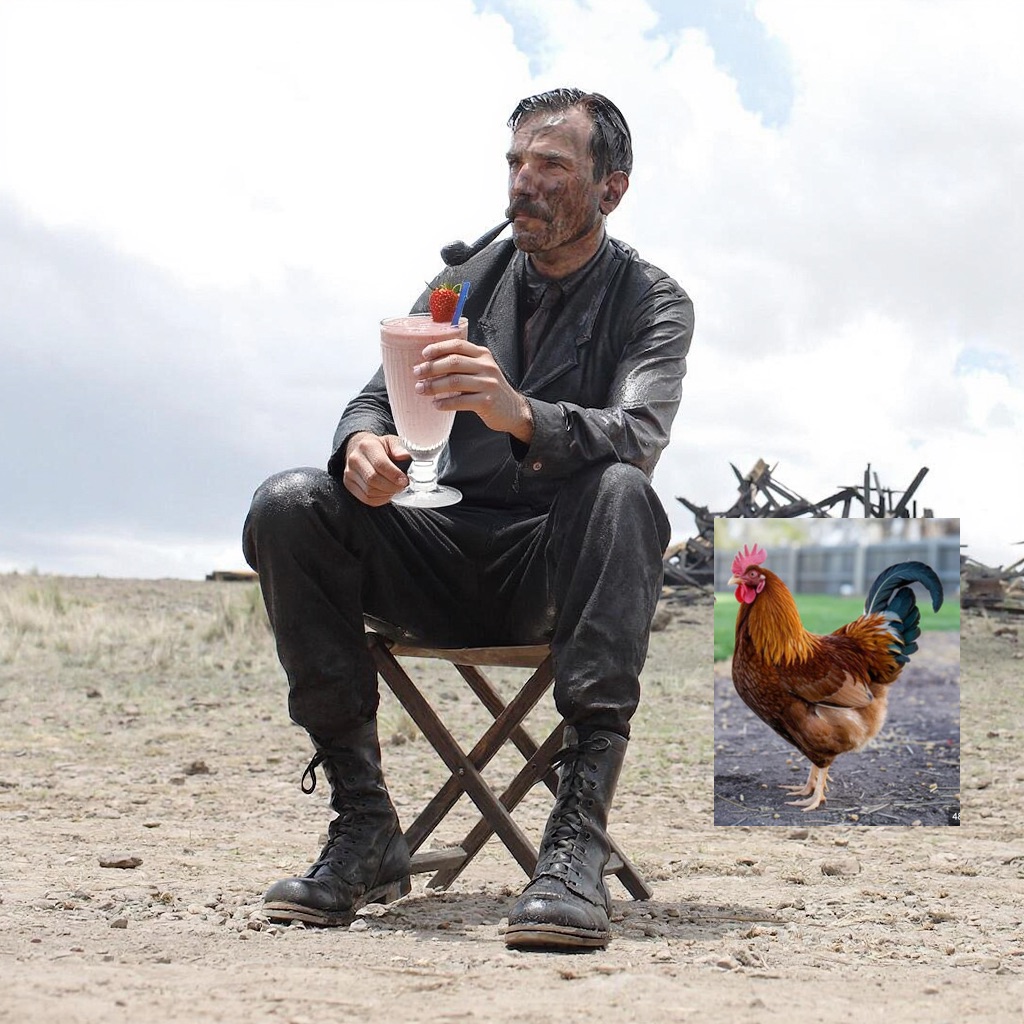} &
        \includegraphics[width=0.135\linewidth, trim=20mm 40mm 20mm 0mm, clip]{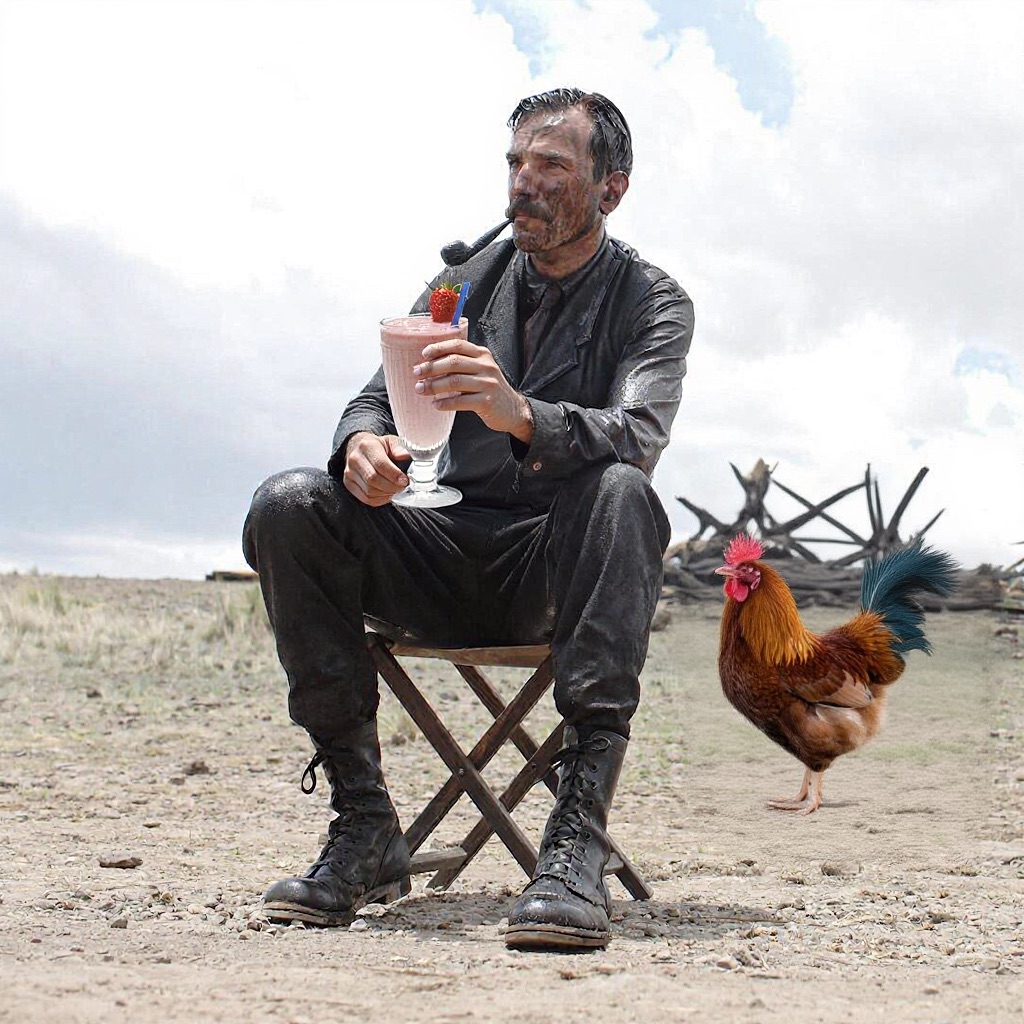} &
        \includegraphics[width=0.135\linewidth, trim=20mm 40mm 20mm 0mm, clip]{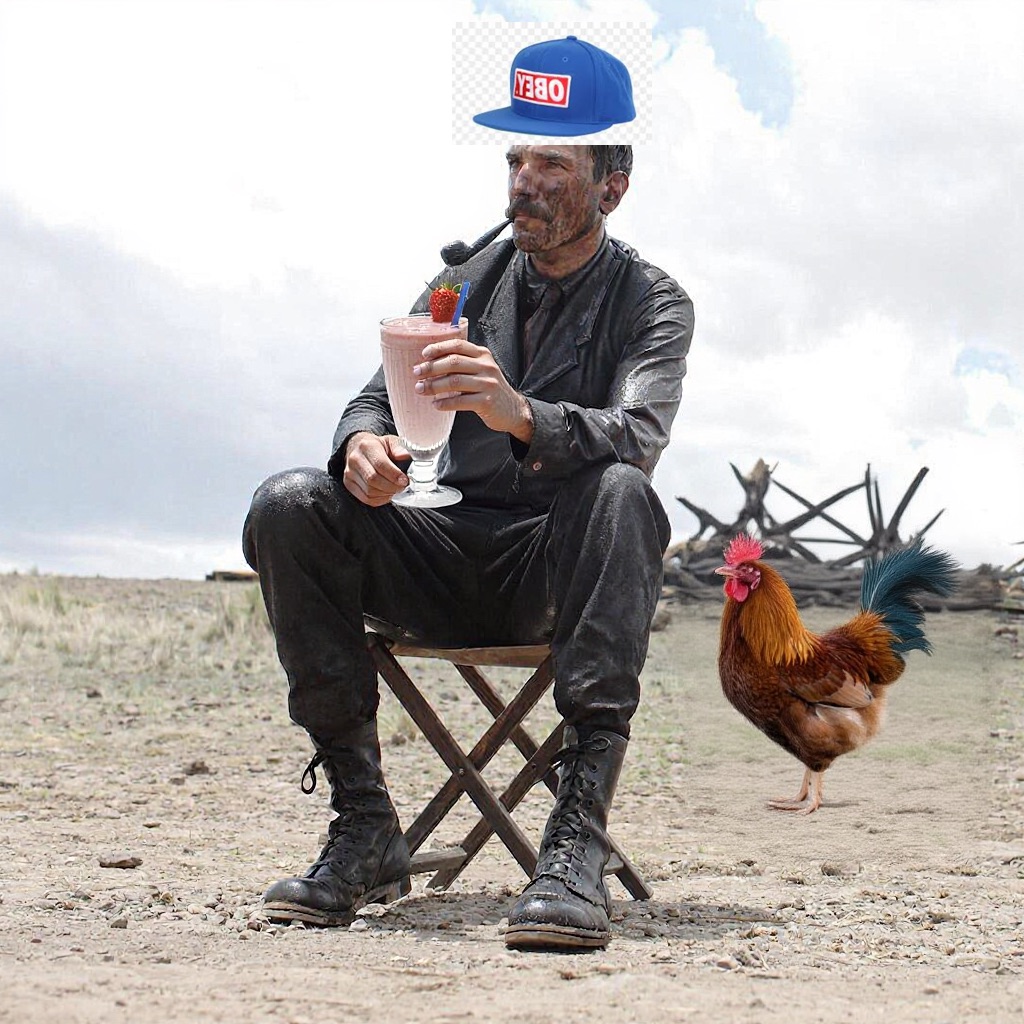} &
        \includegraphics[width=0.135\linewidth, trim=20mm 40mm 20mm 0mm, clip]{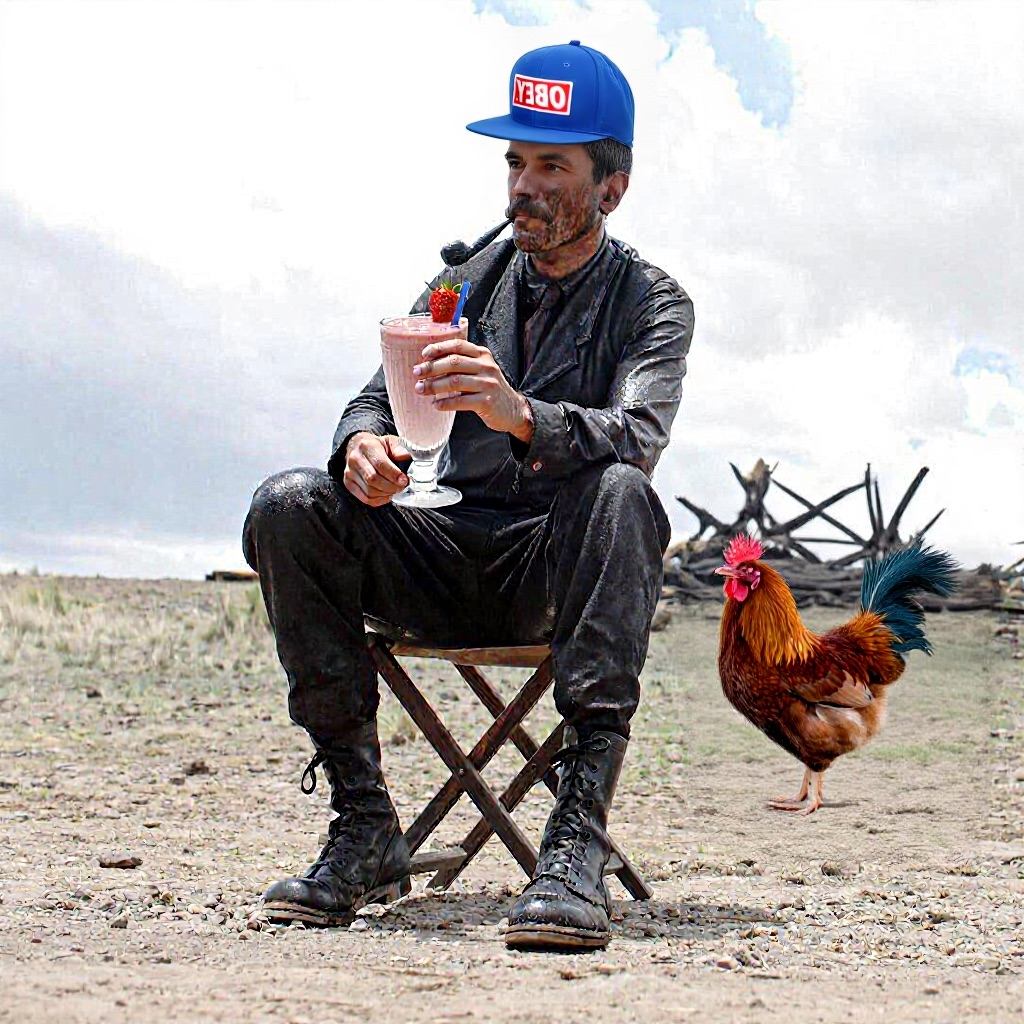} \\

        \includegraphics[width=0.135\linewidth, trim=40mm 80mm 40mm 0mm, clip]{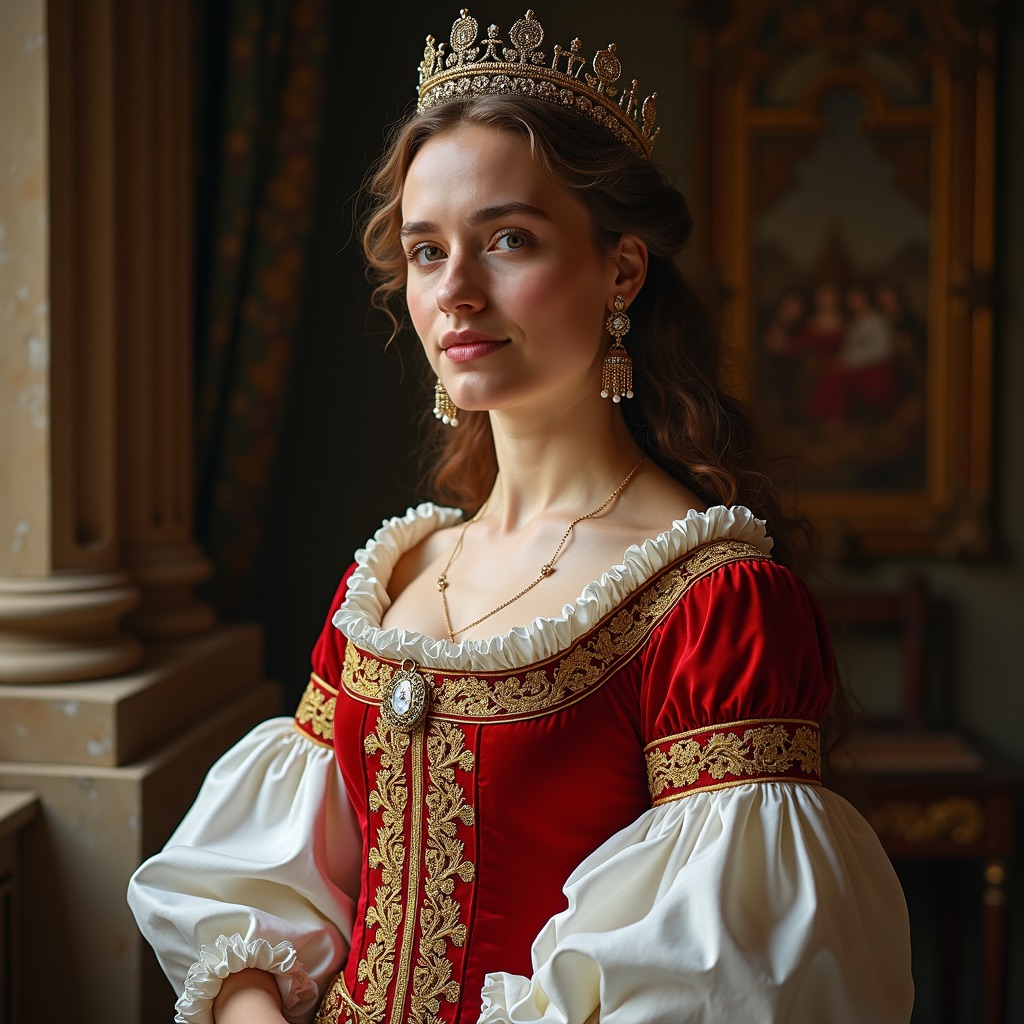} &
        \includegraphics[width=0.135\linewidth, trim=40mm 80mm 40mm 0mm, clip]{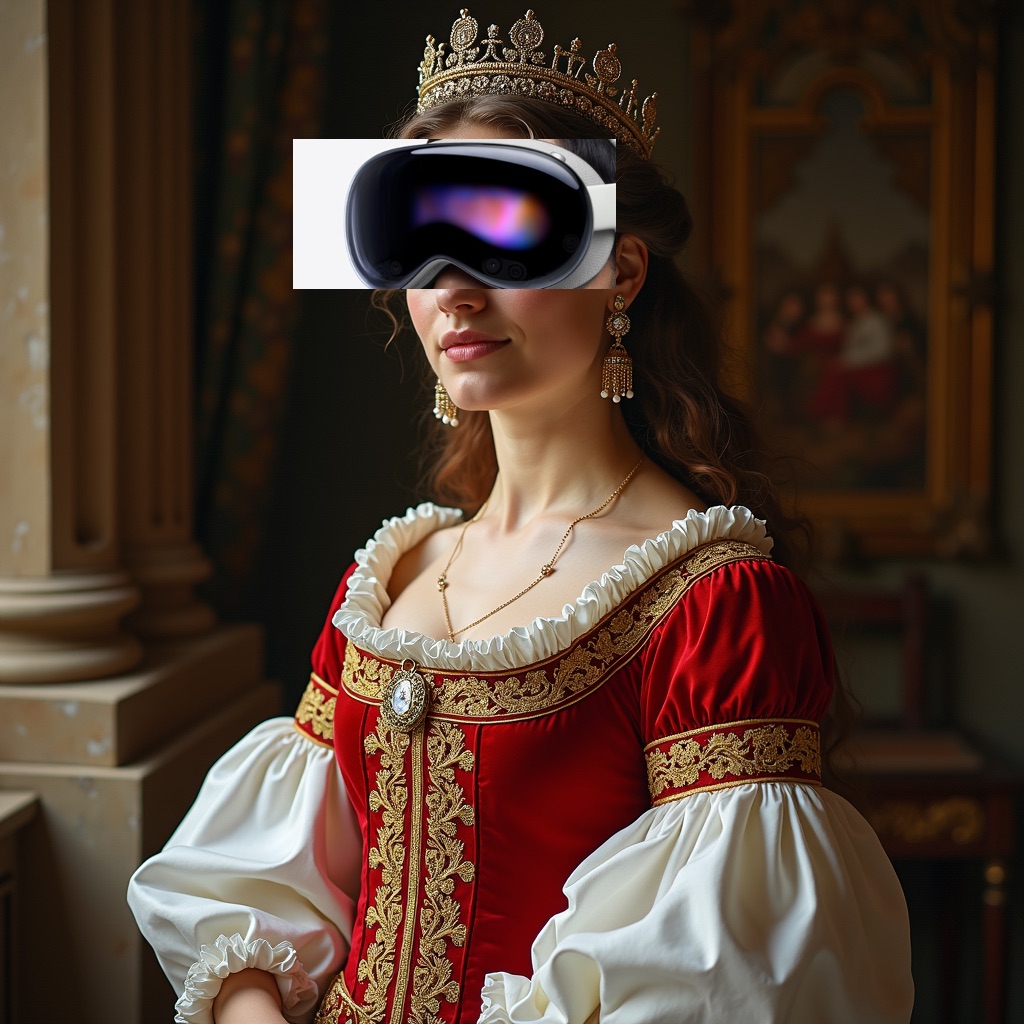} &
        \includegraphics[width=0.135\linewidth, trim=40mm 80mm 40mm 0mm, clip]{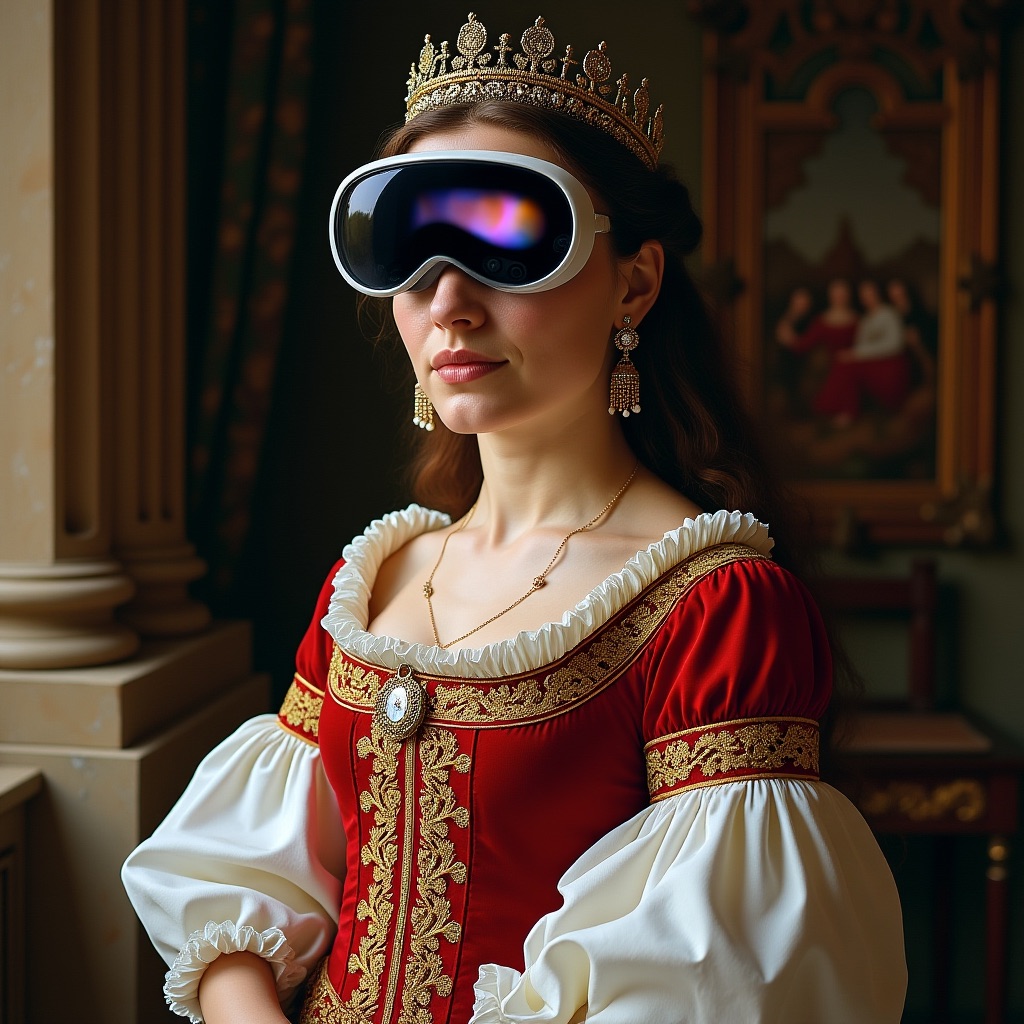} &
        \includegraphics[width=0.135\linewidth, trim=40mm 80mm 40mm 0mm, clip]{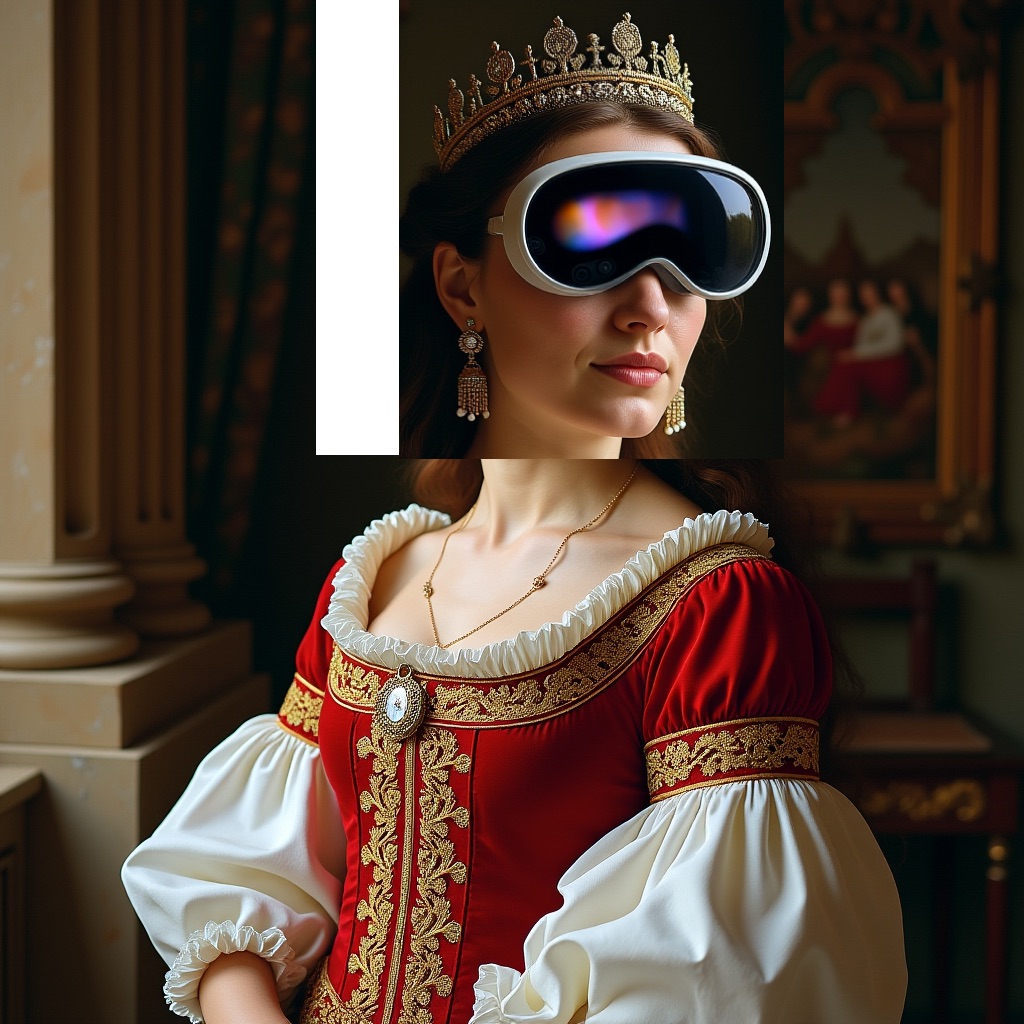} &
        \includegraphics[width=0.135\linewidth, trim=40mm 80mm 40mm 0mm, clip]{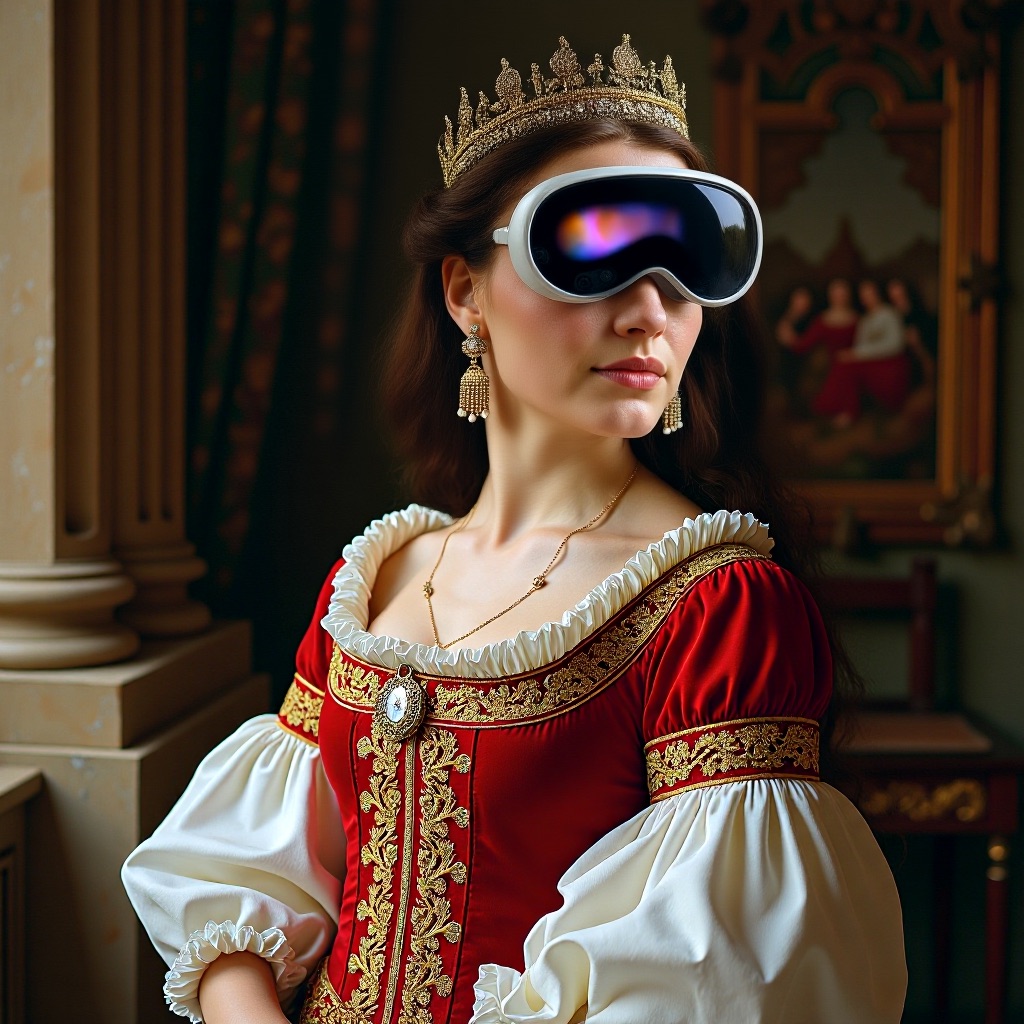} &
        \includegraphics[width=0.135\linewidth, trim=40mm 80mm 40mm 0mm, clip]{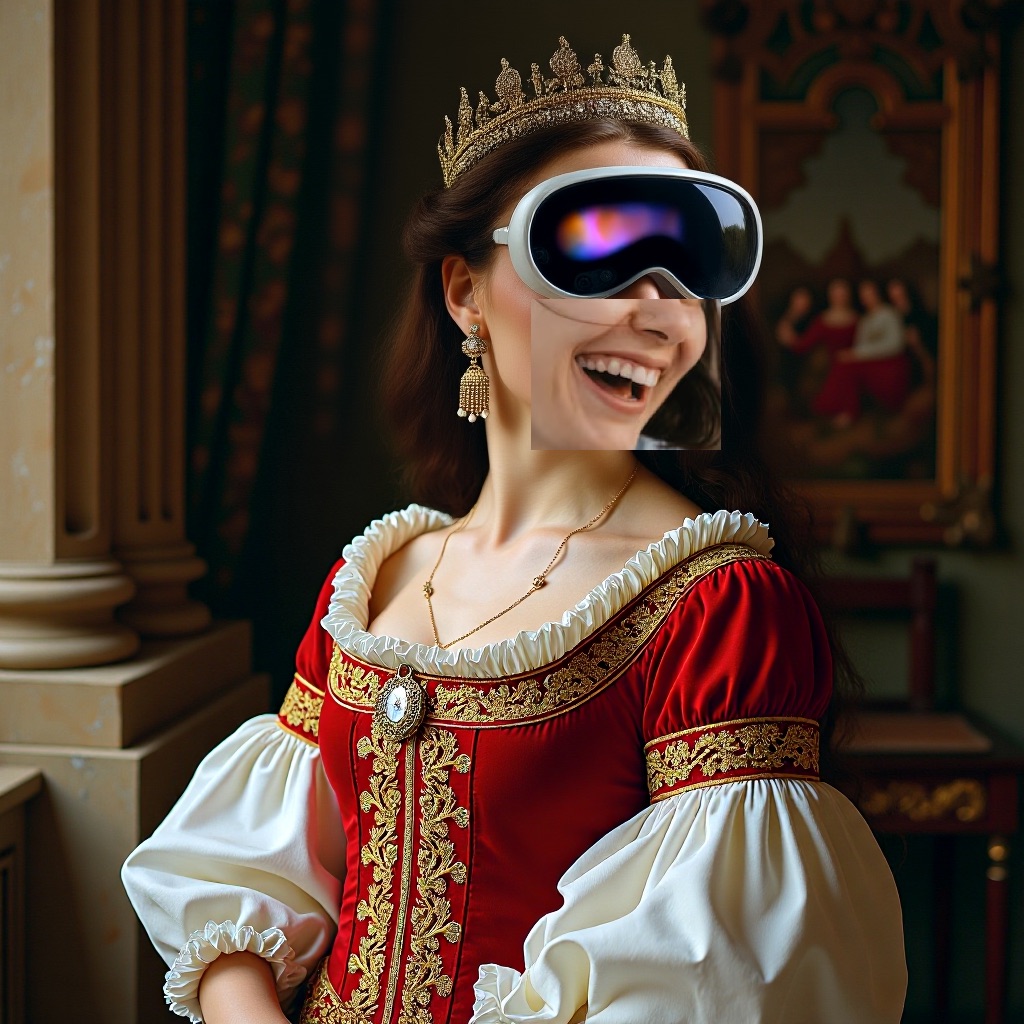} &
        \includegraphics[width=0.135\linewidth, trim=40mm 80mm 40mm 0mm, clip]{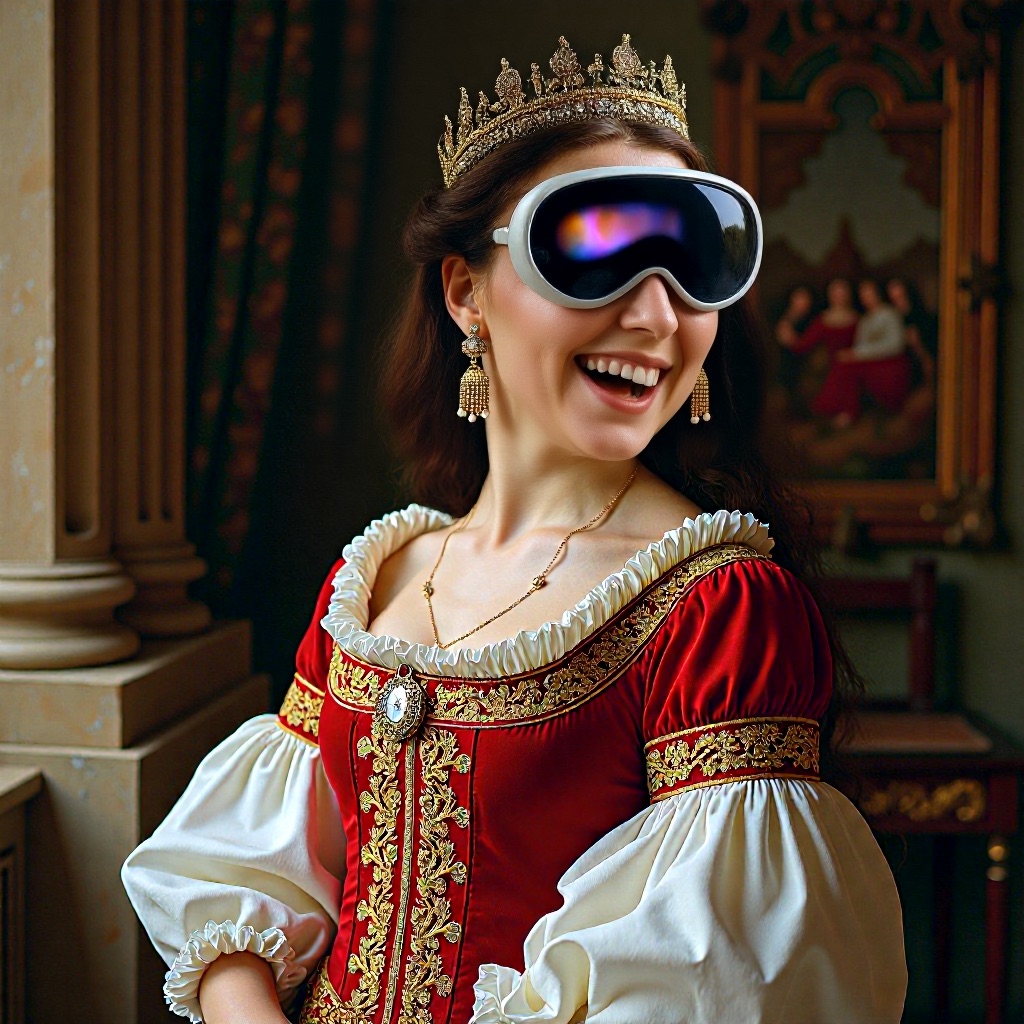} \\
    \end{tabular}
    \captionof{figure}{\textbf{Compound Editing}. We showcase our method's ability to make iterative compound edits.}
    \label{fig:supp_compound}

    \vspace{0.5em} %

    \begin{tabular}{cccccc}
        Input & Kontext & ObjectStitch & \new{SwapAnything} & SwapAnything-DB & Ours \\

        \includegraphics[width=0.155\linewidth]{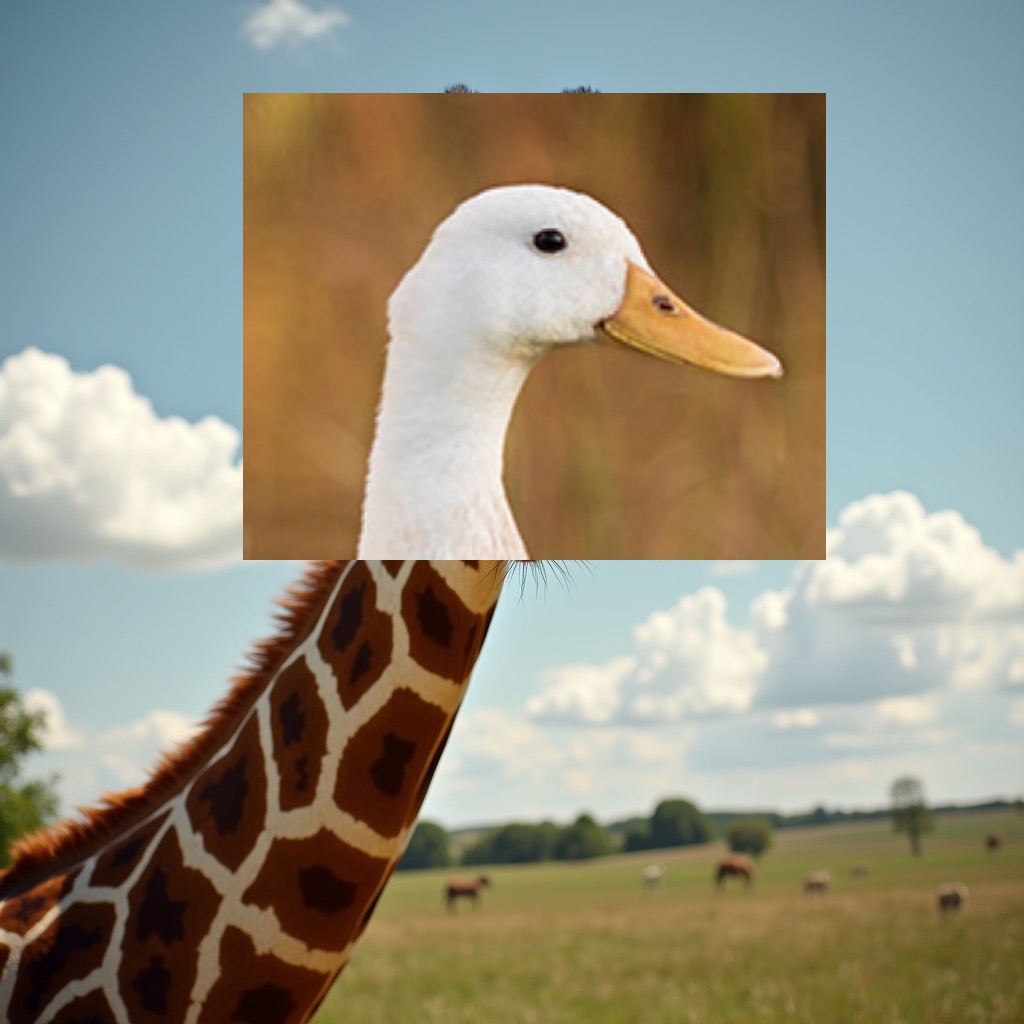} &
        \includegraphics[width=0.155\linewidth]{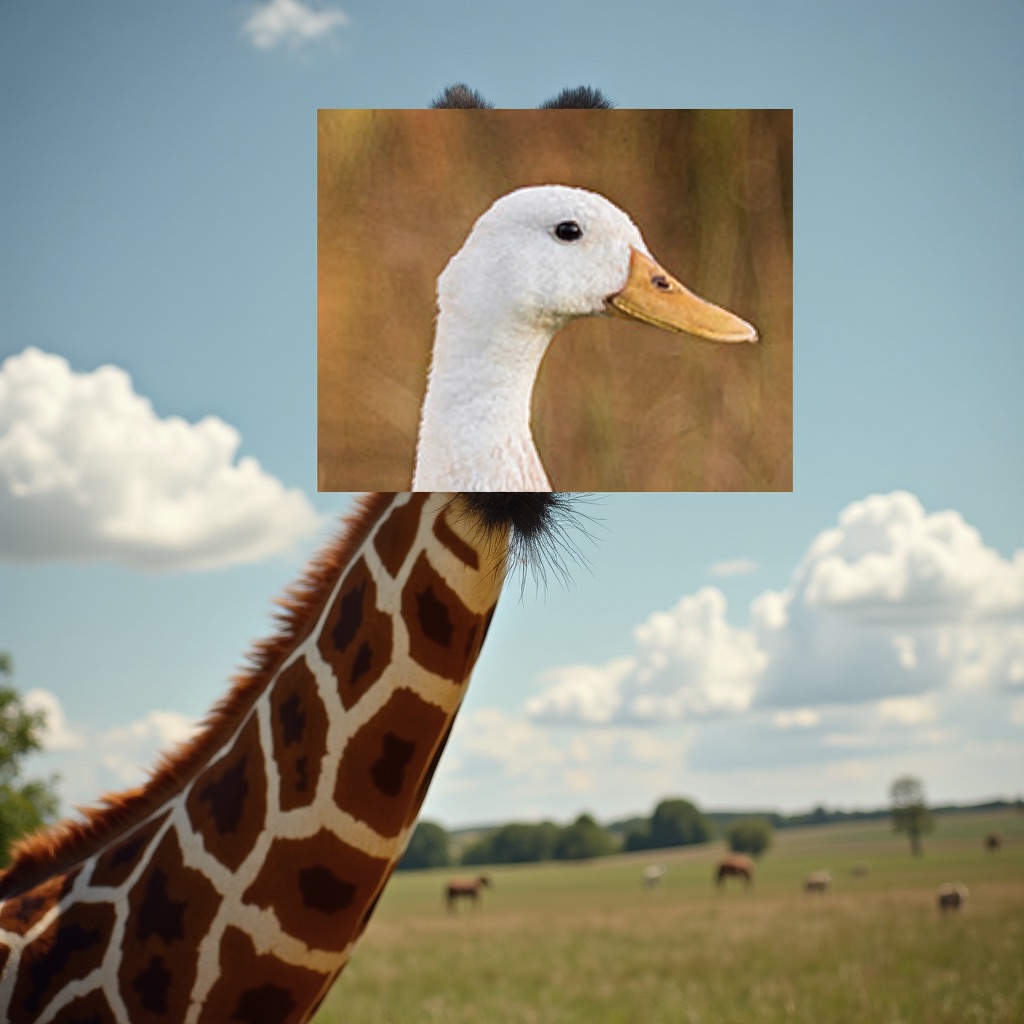} &
        \includegraphics[width=0.155\linewidth]{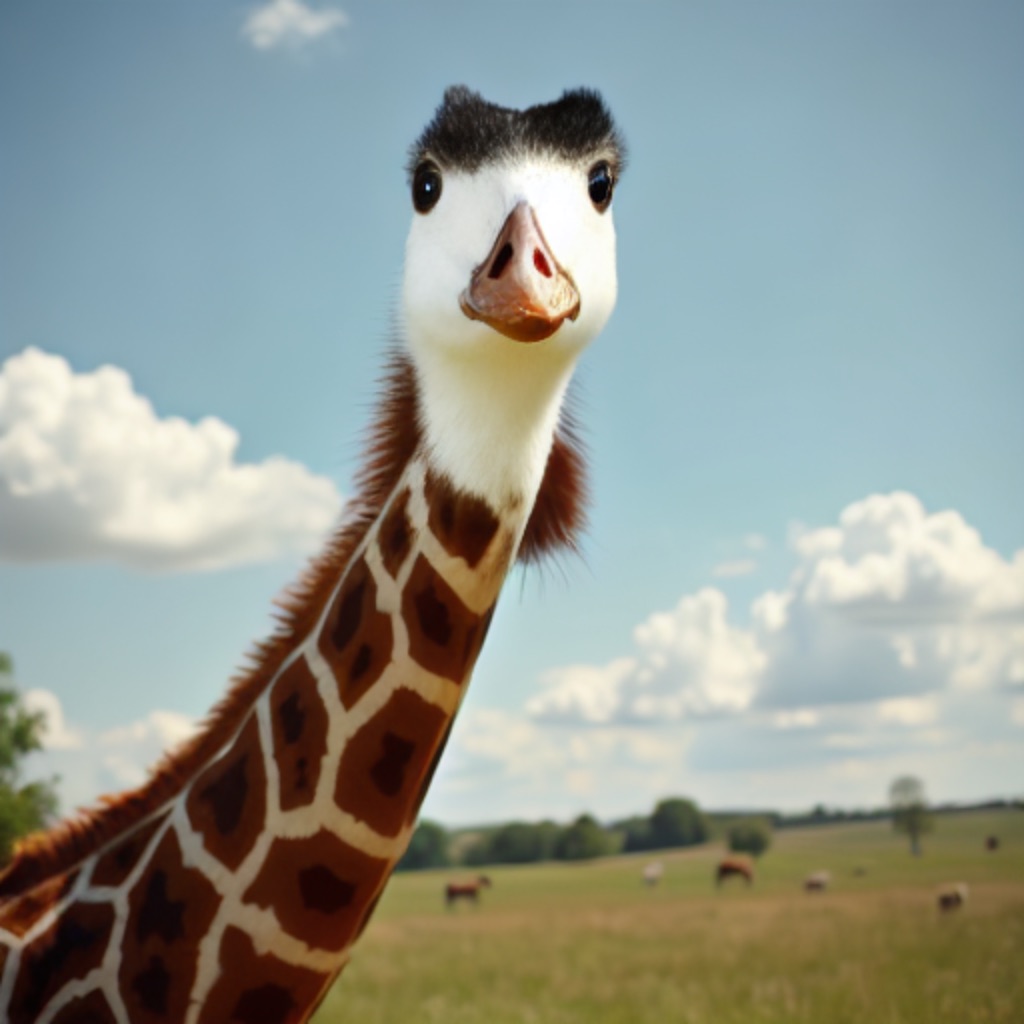} &
        \includegraphics[width=0.155\linewidth]{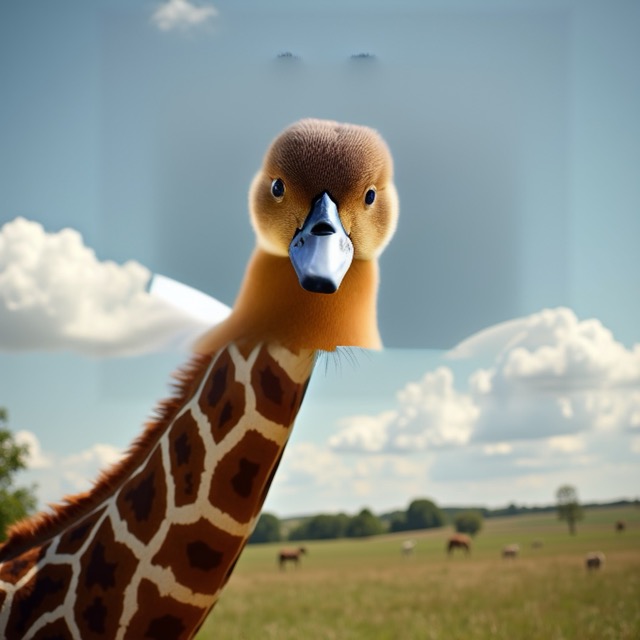} &
        \includegraphics[width=0.155\linewidth]{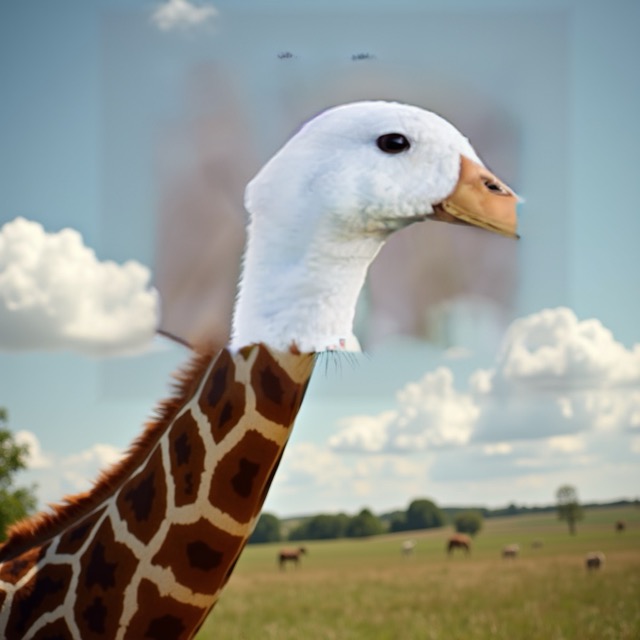} &
        \includegraphics[width=0.155\linewidth]{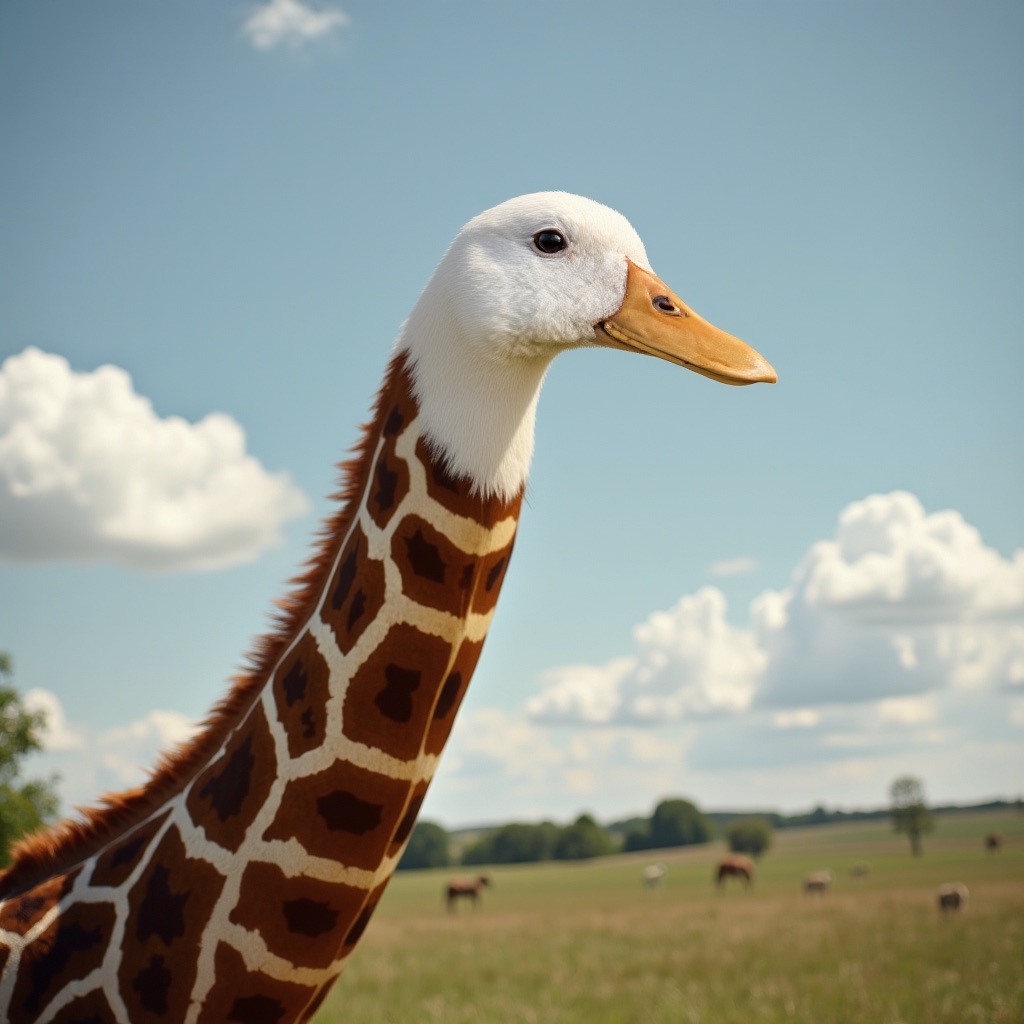} \\

        \includegraphics[width=0.155\linewidth]{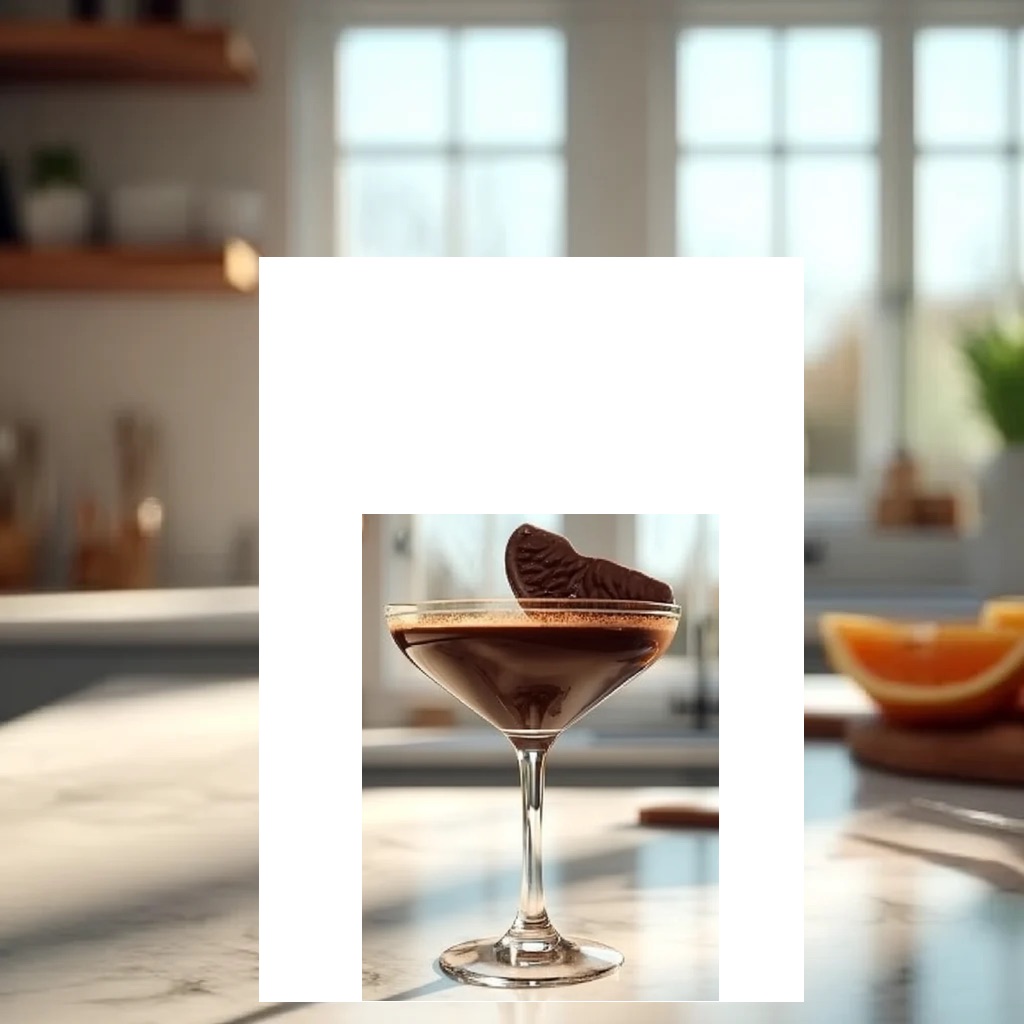} &
        \includegraphics[width=0.155\linewidth]{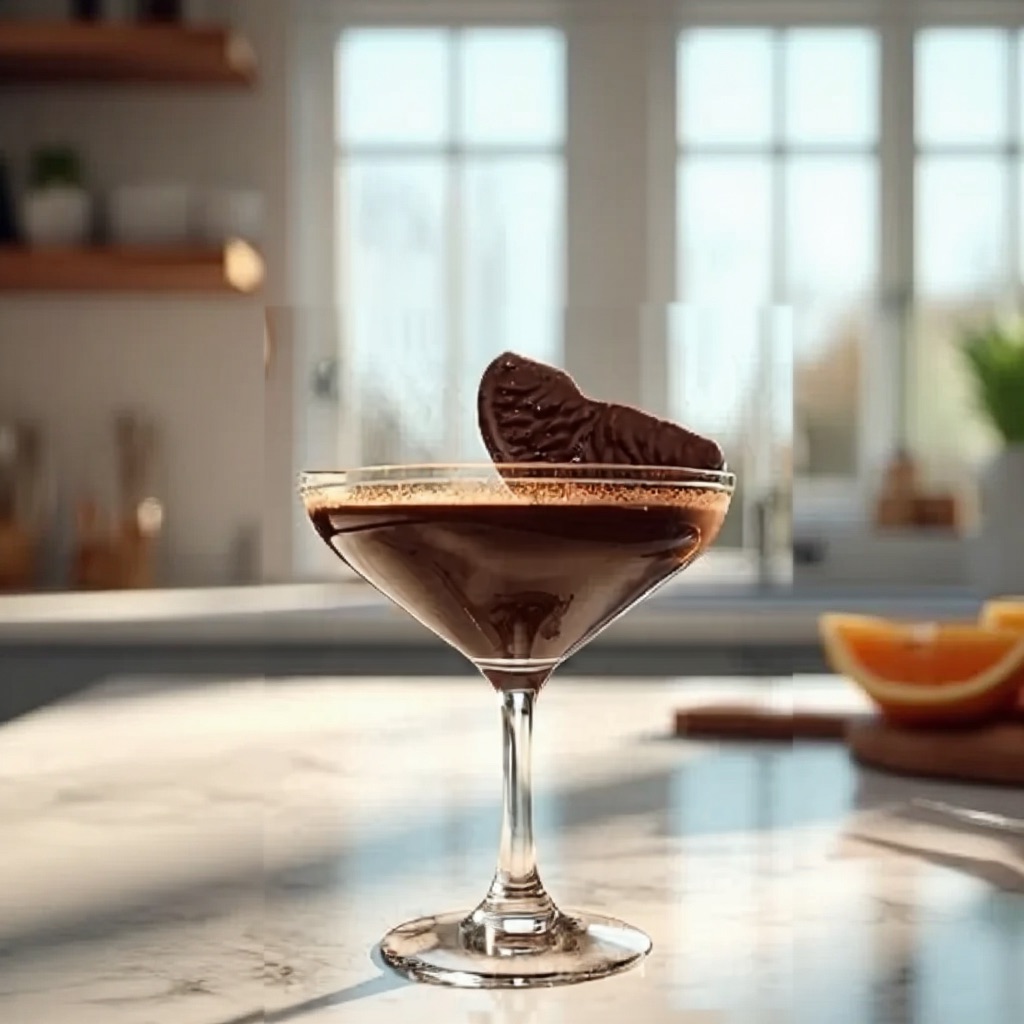} &
        \includegraphics[width=0.155\linewidth]{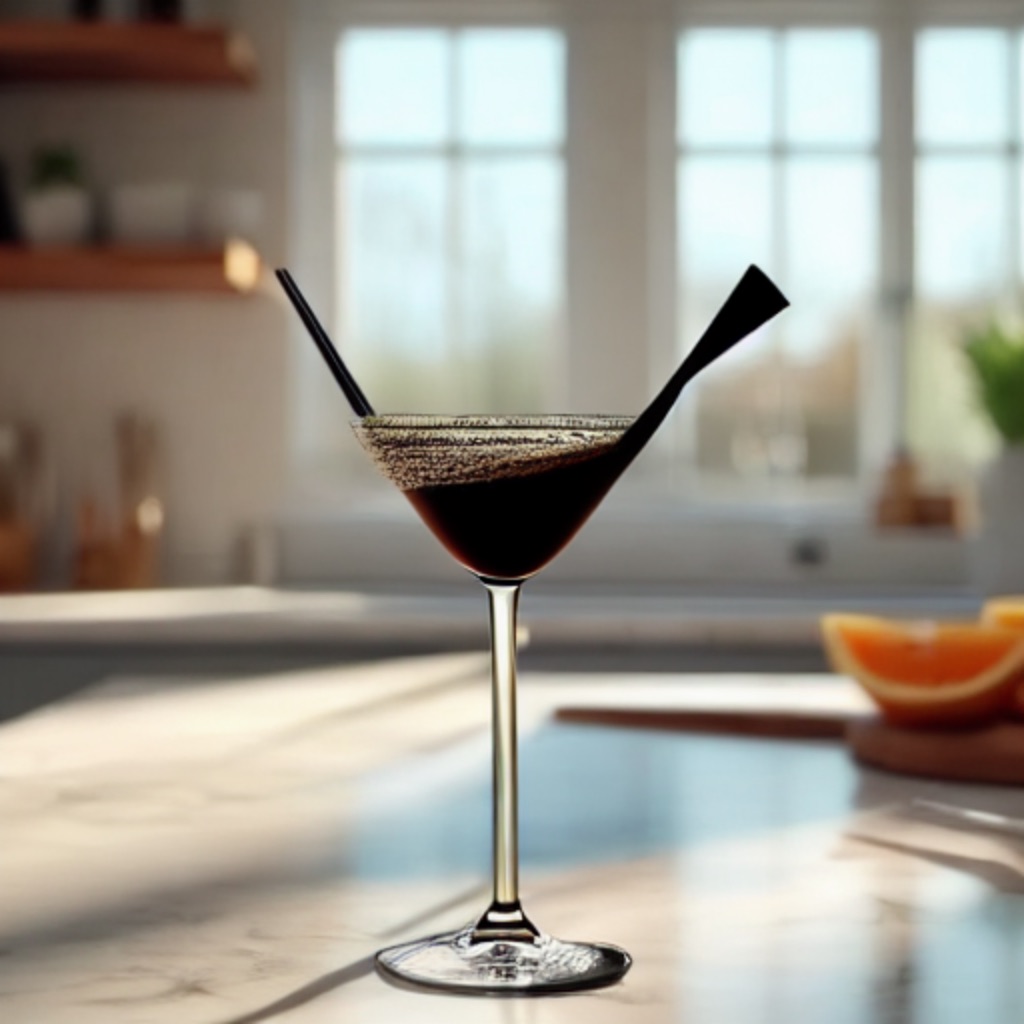} &
        \includegraphics[width=0.155\linewidth]{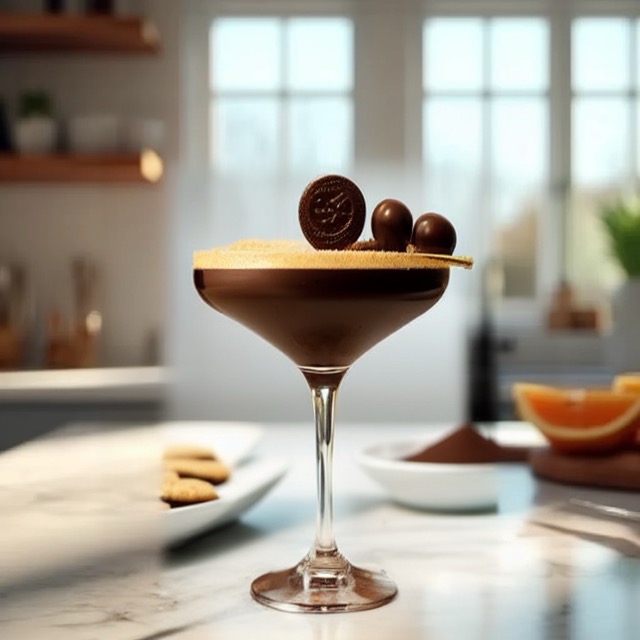} &
        \includegraphics[width=0.155\linewidth]{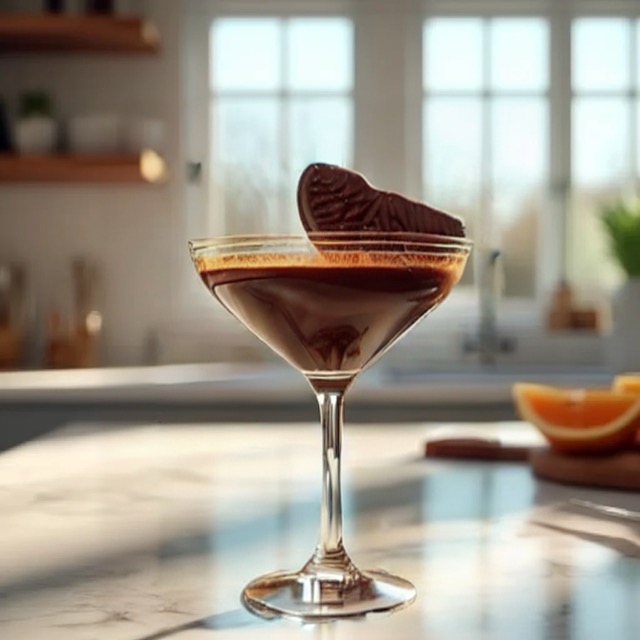} &
        \includegraphics[width=0.155\linewidth]{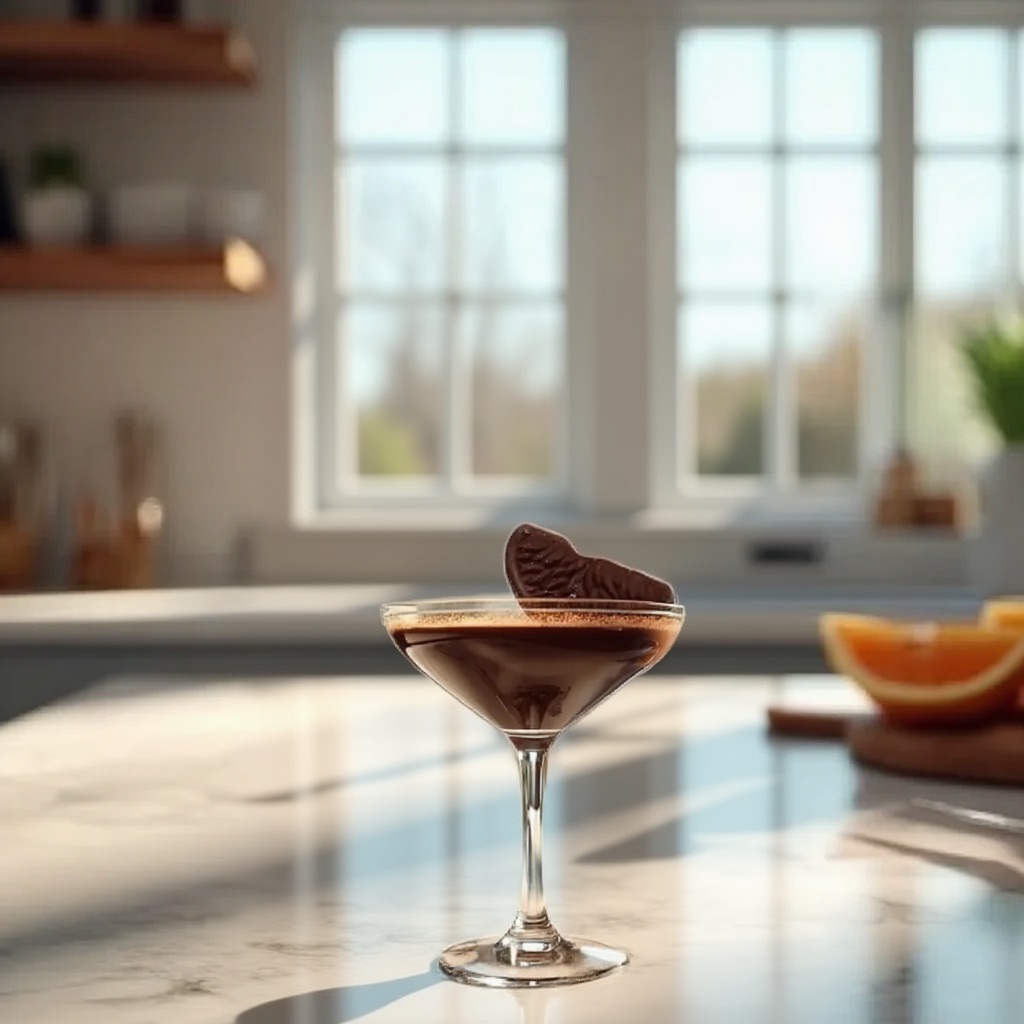} \\

        \includegraphics[width=0.155\linewidth]{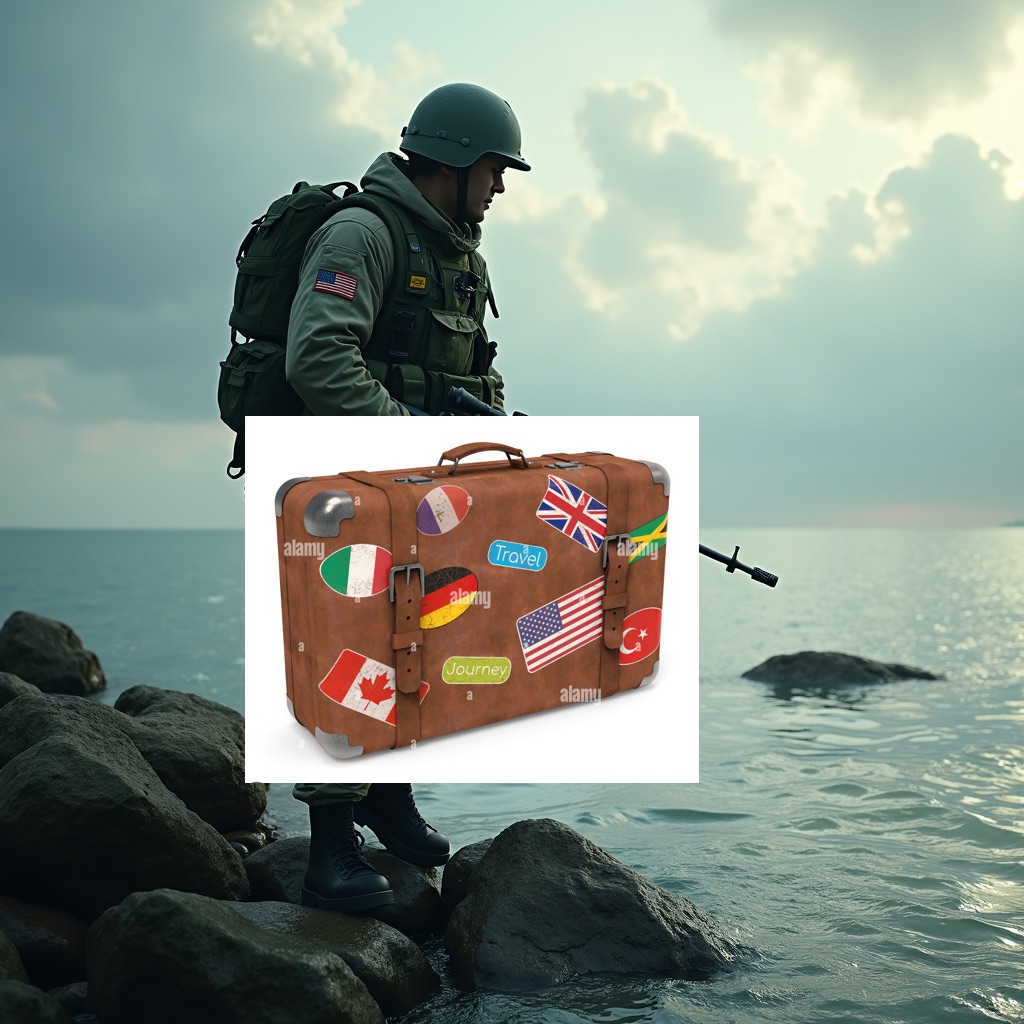} &
        \includegraphics[width=0.155\linewidth]{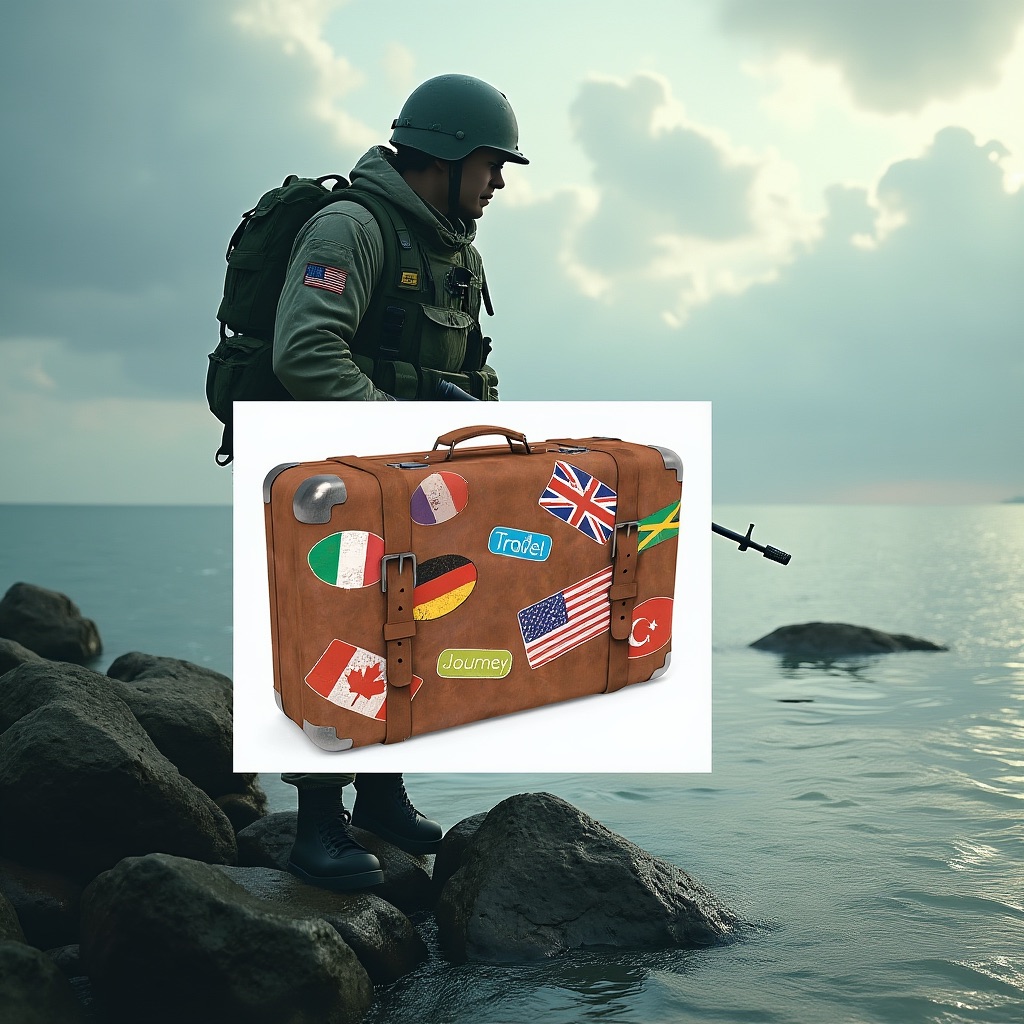} &
        \includegraphics[width=0.155\linewidth]{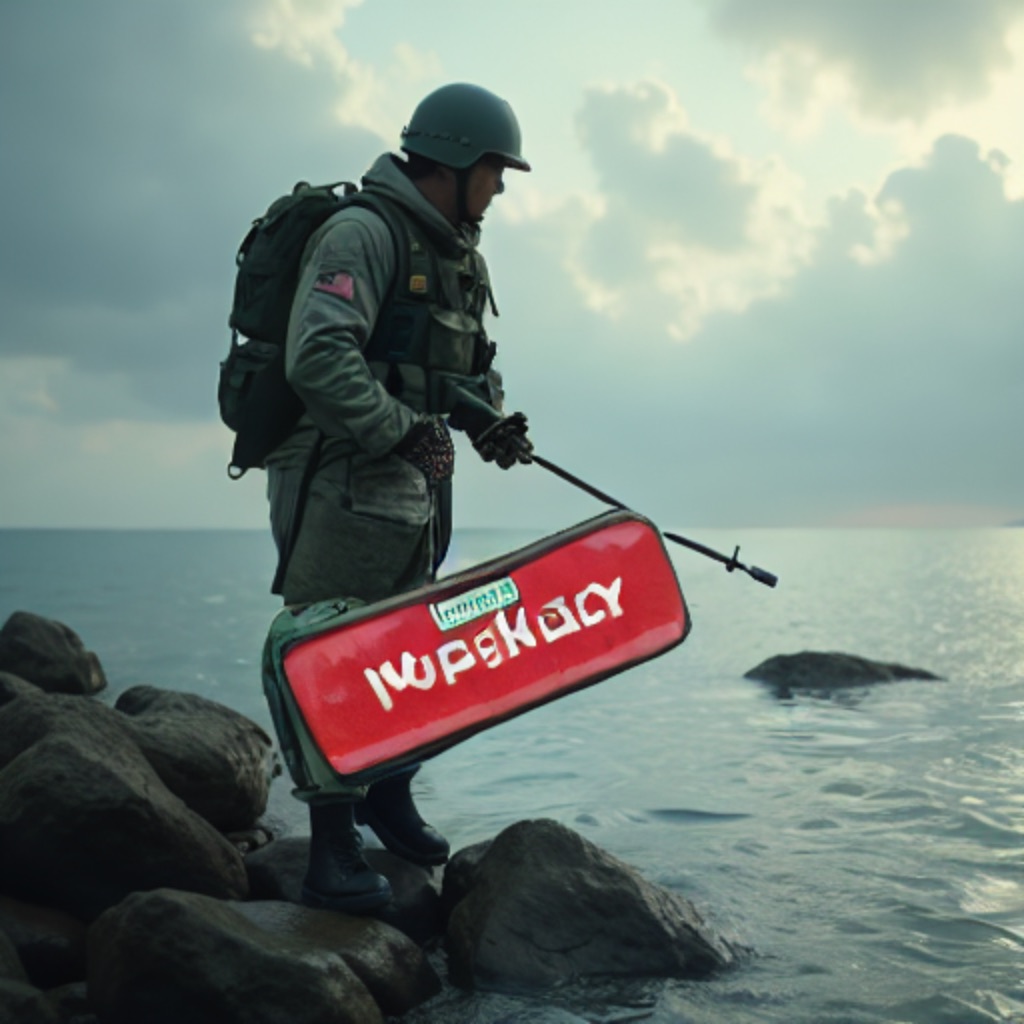} &
        \includegraphics[width=0.155\linewidth]{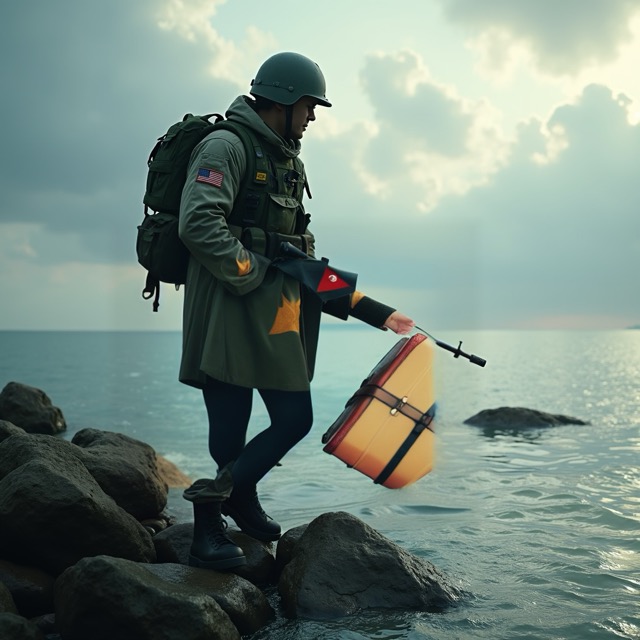} &
        \includegraphics[width=0.155\linewidth]{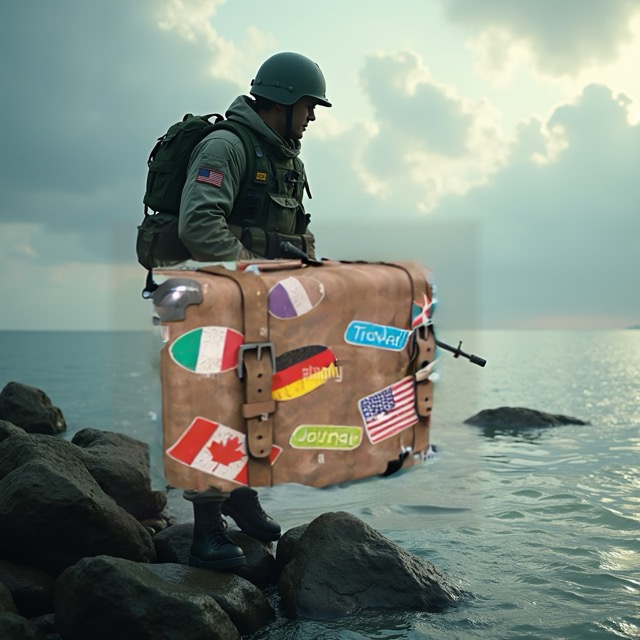} &
        \includegraphics[width=0.155\linewidth]{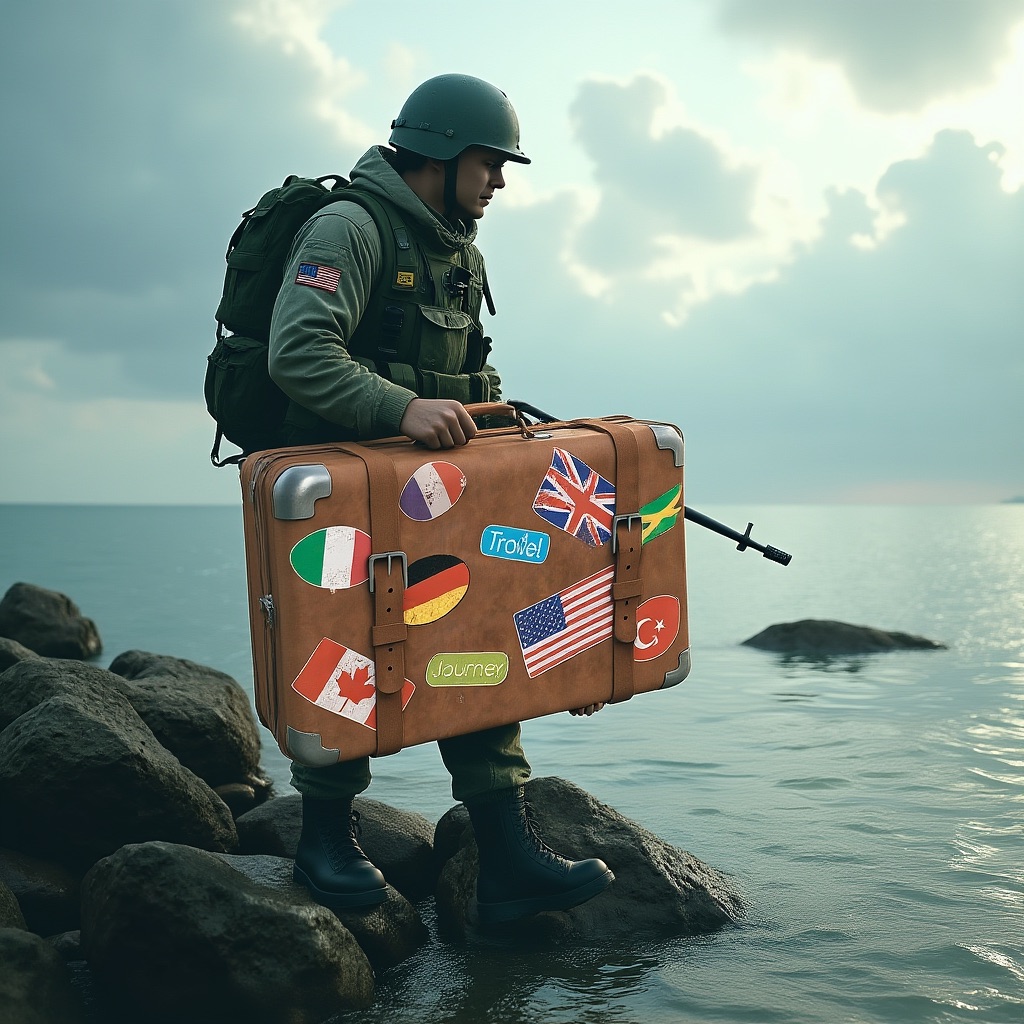} \\
    \end{tabular}
    \captionof{figure}{\textbf{Additional Comparisons}. We present comparisons against three additional baselines: ObjectStitch, SwapAnything and SwapAnything-DB. \new{With SwapAnything-DB being a SwapAnything variant which makes use of DreamBooth personlization}.  We also present FLUX Kontext results to emphasize our method's improvement over its base model.}
    \label{fig:supp_comparisons}

\end{figure*}

\clearpage
\twocolumn[
        \centering
        \Large
        \textbf{LooseRoPE: Content-aware Attention Manipulation \\ for Semantic Harmonization}\\
        \vspace{0.5em}Supplementary Material \\
        \vspace{1.0em}
    ] %

\renewcommand{\thesection}{\Alph{section}}
In this document, we present additional results and discussions (Section \ref{sec:supp_additional}), including limitations (Section \ref{sec:supp_limitations}), as well as providing implementation details for our method and experiments (Section \ref{sec:supp_details}).

\tableofcontents

\section{Additional Results and Discussions}
\label{sec:supp_additional}

\subsection{Additional Qualitative Results}
In Figure 9 in the main paper we present additional LooseRoPE outputs, compared against the outputs of our base model FLUX Kontext when given the same base prompt: \emph{``blend the cropped objects into the image in a convincing manner without changing the style of the image''}, and the input images presented in the ``Input'' columns. Additionally, we present several examples of compound edits—scenarios in which we iteratively alternate between crude editing and harmonization (Figure 10 in the main paper). These examples demonstrate the robustness and consistency of our method, which maintains high visual quality and coherent blending even across multiple successive editing steps.

\subsection{Additional Quantitative Evaluation}
\label{sec:supp_quant}
\begin{table}[htbp]
    \centering
    \caption{Quantitative comparison comparing LooseRoPE against competing methods. A subset of these results are also presented in Figure 7 of the main paper.
        }
    \resizebox{\columnwidth}{!}{%
        \begin{tabular}{lccc}
        
        \toprule
        \textbf{Method} & \textbf{CLIP-IQA (↑)} & \textbf{LPIPS (Full) (↓)} & \textbf{LPIPS (FG) (↓)} \\
        \midrule
        AnyDoor                     & 0.831 & 0.264 & 0.510 \\
        SwapAnything                & 0.854 & 0.161 & 0.609  \\
        SwapAnything - DB           & 0.846 & \textbf{0.120} & 0.528  \\
        TF-ICON                     & 0.885 & 0.403 & 0.619  \\
        Qwen-Image-Edit                  & 0.820 & 0.183 & 0.284  \\
        ObjectStitch                & 0.745 & 0.368 & 0.605  \\
        \new{MagicFixup}            & 0.791 & 0.281 & 0.379 \\
        FLUX Kontext     & 0.870 & 0.282 & 0.365 \\
        \textbf{LooseRoPE (Ours)} & \textbf{0.895} & 0.261 & \textbf{0.281} \\
        \bottomrule
        
        \end{tabular}
        }
        \label{tab:main_comparison}
\end{table}

\begin{table}[htbp]
    \centering
    \caption{Quantitative ablation study results. These results are also presented in Figure 7 of the main paper.
    }
    \resizebox{\columnwidth}{!}{%
    
        \begin{tabular}{lccc}
        \toprule
        \textbf{Model Variant} & \textbf{CLIP-IQA (↑)} & \textbf{LPIPS (Full) (↓)} & \textbf{LPIPS (FG) (↓)} \\
        \midrule
        w/o VLM         & 0.879 & 0.253 & \textbf{0.253} \\
        w/o RoPE      & 0.876 & \textbf{0.238} & 0.259 \\
        w/o Attention  & 0.889 & 0.305 & 0.423 \\
        \textbf{LooseRoPE (Ours)} & \textbf{0.895} & 0.261 & 0.281 \\
        \bottomrule
        \end{tabular}
    }
    
    \label{tab:ablation_study}
\end{table}

\noindent In this section, we provide the comprehensive metric tables supporting the analysis presented in the main paper (Section 4), offering an extensive comparison against a broader range of competing methods (Table \ref{tab:main_comparison}) and detailed ablation results (Table \ref{tab:ablation_study}). Beyond the baselines reported in the main text, Table \ref{tab:main_comparison} includes comparison results against Qwen-Image-Edit~\cite{wu2025qwen}, ObjectStitch \cite{song2023objectstitch}, \new{MagicFixup \cite{alzayer2025magic}} and Personalized SwapAnything with Dreambooth \cite{ruiz2023dreamboothfinetuningtexttoimage}. These results show that while SwapAnything-DB spends a considerable amount on learning the target concept (up to 20 minutes) it does not appear to improve its ability to harmonize. This is likely due to the fact that DreamBooth usually requires more than one image to effectively learn a concept. ObjectStitch appears to not be as well suited for our task as other competing methods, achieving the lowest CLIP-IQA scores out of all methods tested with relatively high LPIPS scores. \new{MagicFixup also appears less suitable for our task, achieving the second lowest CLIP-IQA scores and high LPIPS scores. This is somewhat unsurprising due to object insertion being scarce if at all present in MagicFixup's training data. } As for Qwen-Image-Edit, it seems more prone to neglect than the other image editing model we tested- FLUX-Kontext, acheiving lower LPIPS scores but a much lower CLIP-IQA score. This further justifies our choice of FLUX-Kontext as a base model. In addition to the quantitative results in Table \ref{tab:main_comparison} we also present qualitative results in Figure 11 in the main paper.

These extended results corroborate our primary findings, demonstrating the robustness of our method across diverse editing scenarios.

\begin{table}[htbp]
    \centering
    \caption{Comparing Gemini Flash 2.5 and QWEN3-VL as the VLM model used in the VLM based parameter steering mechanism component of our method. Due to usage limitations, this experiment was conducted on a subset of our benchmark.
    }
    \resizebox{\columnwidth}{!}{%
    
        \begin{tabular}{lccc}
        \toprule
        \textbf{VLM Backbone} & \textbf{CLIP-IQA (↑)} & \textbf{LPIPS (Full) (↓)} & \textbf{LPIPS (FG) (↓)} \\
        \midrule
        Gemini Flash 2.5     & \textbf{0.899} & \textbf{0.251} & 0.342 \\
        Qwen & 0.892 & 0.256 & \textbf{0.346} \\
        \bottomrule
        \end{tabular}
    }
    
    \label{tab:gemini_as_vlm}
\end{table}

Furthermore, we isolate the impact of the Vision Language Model (VLM) used in our ``VLM-Based Parameter Steering'' mechanism by comparing our default model \texttt{Qwen3-VL} with \texttt{Gemini Flash 2.5}. Due to usage limitations, this experiment was conducted on a 65-sample subset of our benchmark. The results, reported in Table~\ref{tab:gemini_as_vlm}, show that while Gemini Flash 2.5 slightly outperforms Qwen3-VL, the performance gap is marginal. This suggests that VLM reasoning capability is not a limiting factor in our framework, and that our method is largely robust to the choice of VLM backend.

We detail the exact implementation and settings for this and all other experiments conducted in this work in the next section (Section \ref{sec:supp_details}).

\begin{figure}
    \centering
    \includegraphics[width=0.98\linewidth]{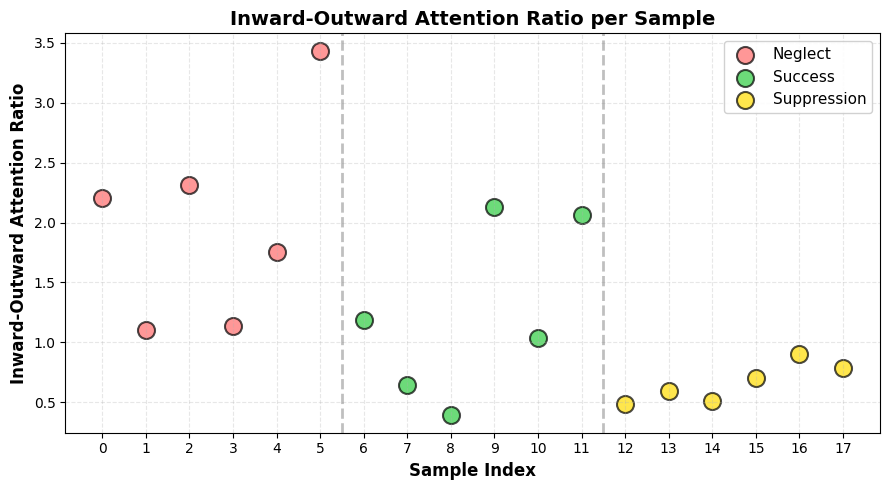}
    \caption{Inward–outward attention ratio (total attention from crop-region queries to keys inside the crop mask divided by the attention directed outside the mask) per FLUX Kontext result sample. We evaluate FLUX Kontext on our benchmark, recording the inward-outward attention ratio for each sample and categorizing the end result (either Neglect, Suppression or Success).}
    \label{fig:supp_attn_locality}
\end{figure}

\begin{figure*}
    \centering
    \setlength{\tabcolsep}{1pt}
    \begin{tabular}{c|cc|cc}
        & \multicolumn{2}{>{\centering\arraybackslash}p{0.38\linewidth}|}{\emph{Blend the cropped objects into the image in a convincing manner.}} 
        & \multicolumn{2}{>{\centering\arraybackslash}p{0.38\linewidth}}{\emph{Blend the colorful umbrella seamlessly into the beach scene, remove the umbrella's original cityscape background.}} \\[5pt]
        
        \includegraphics[width=0.19\linewidth]{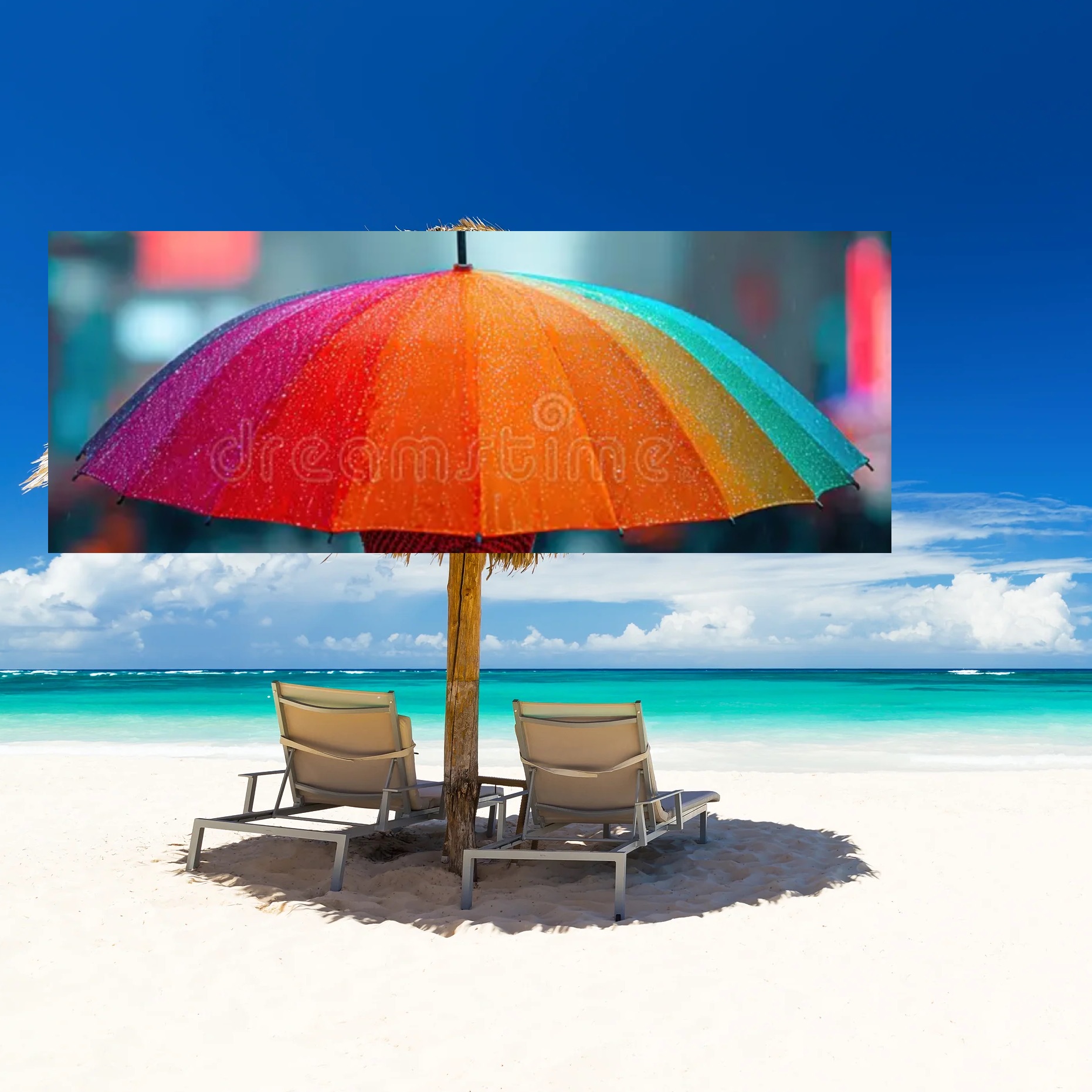} &
        \includegraphics[width=0.19\linewidth]{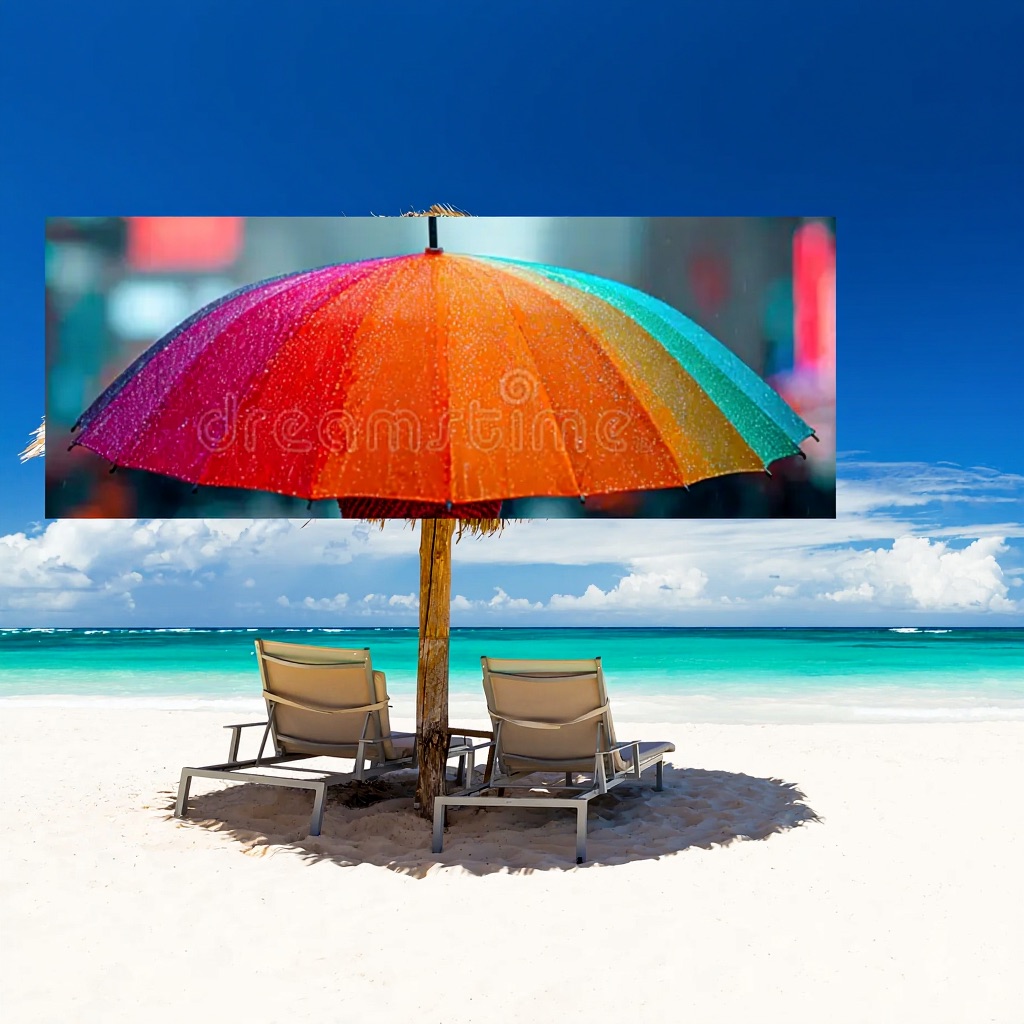} &
        \includegraphics[width=0.19\linewidth]{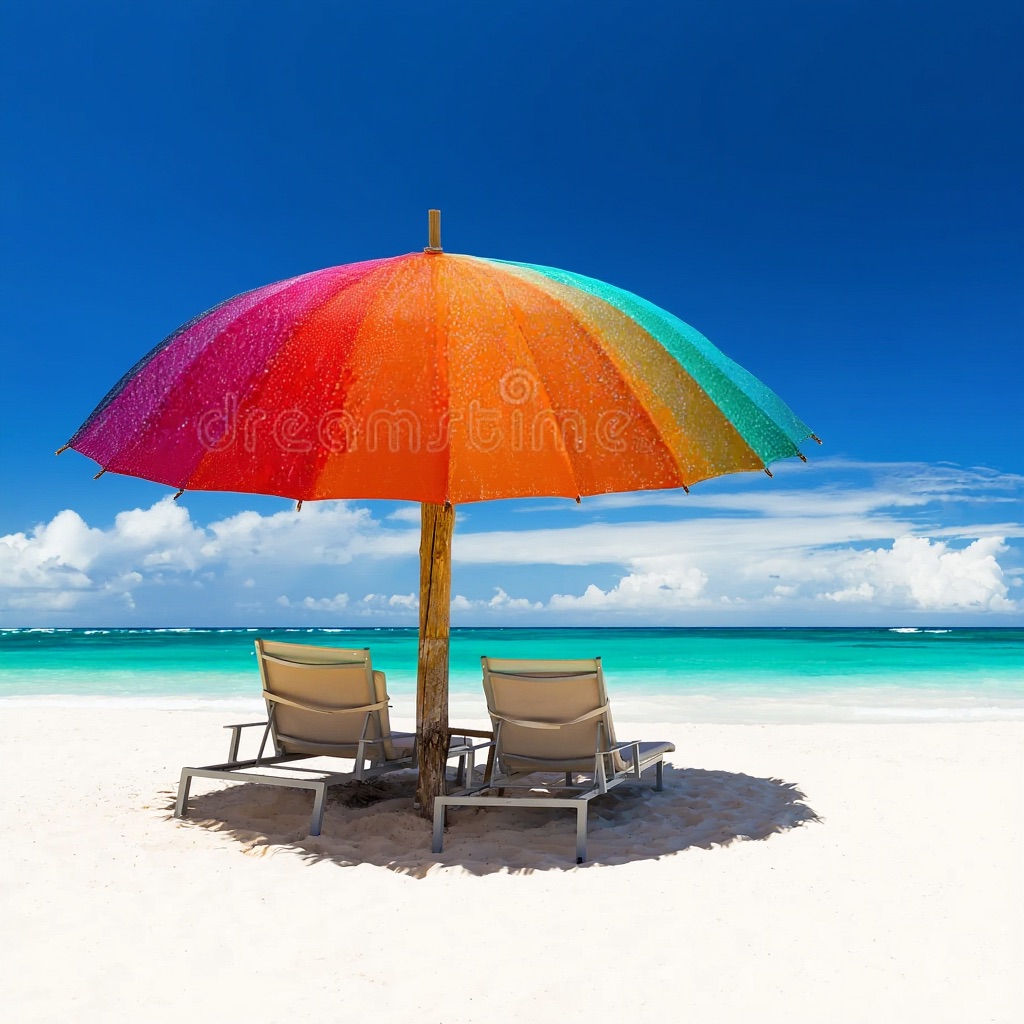} &
        \includegraphics[width=0.19\linewidth]{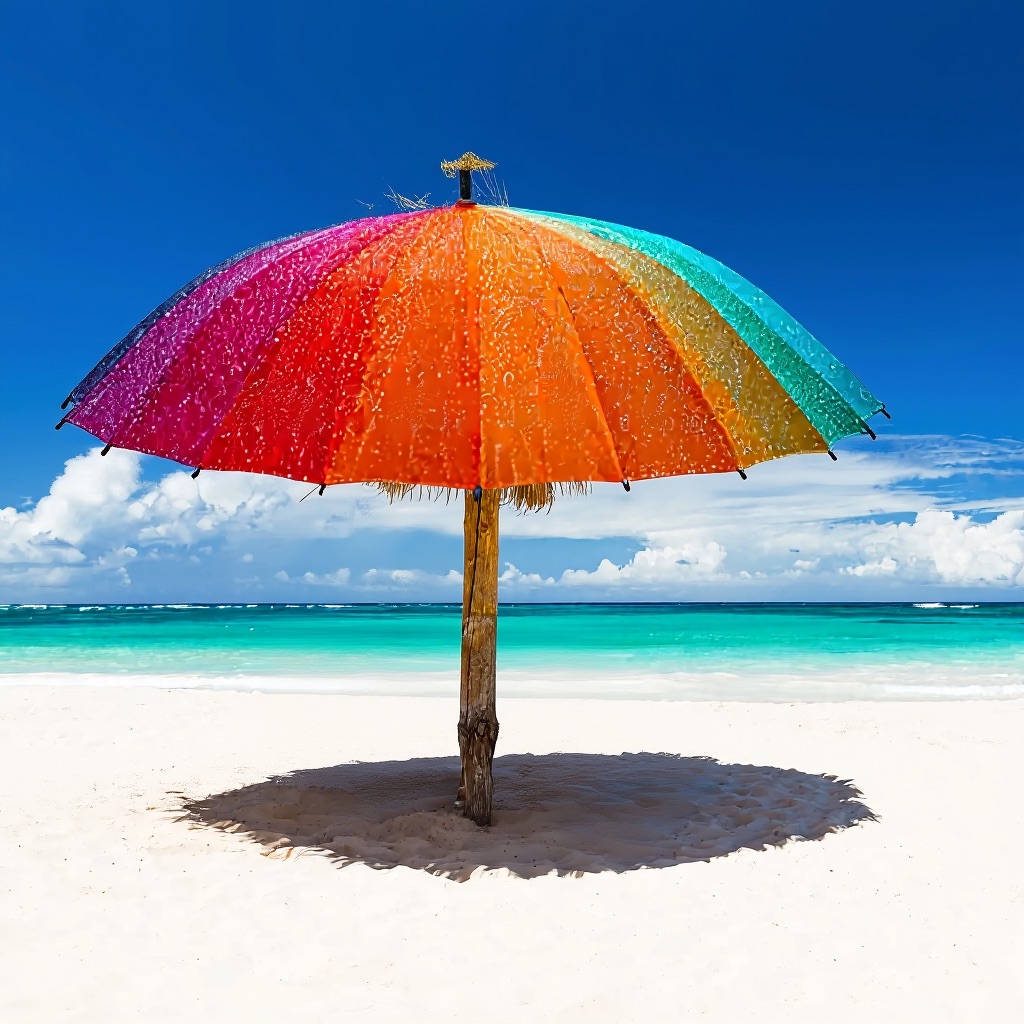} &
        \includegraphics[width=0.19\linewidth]{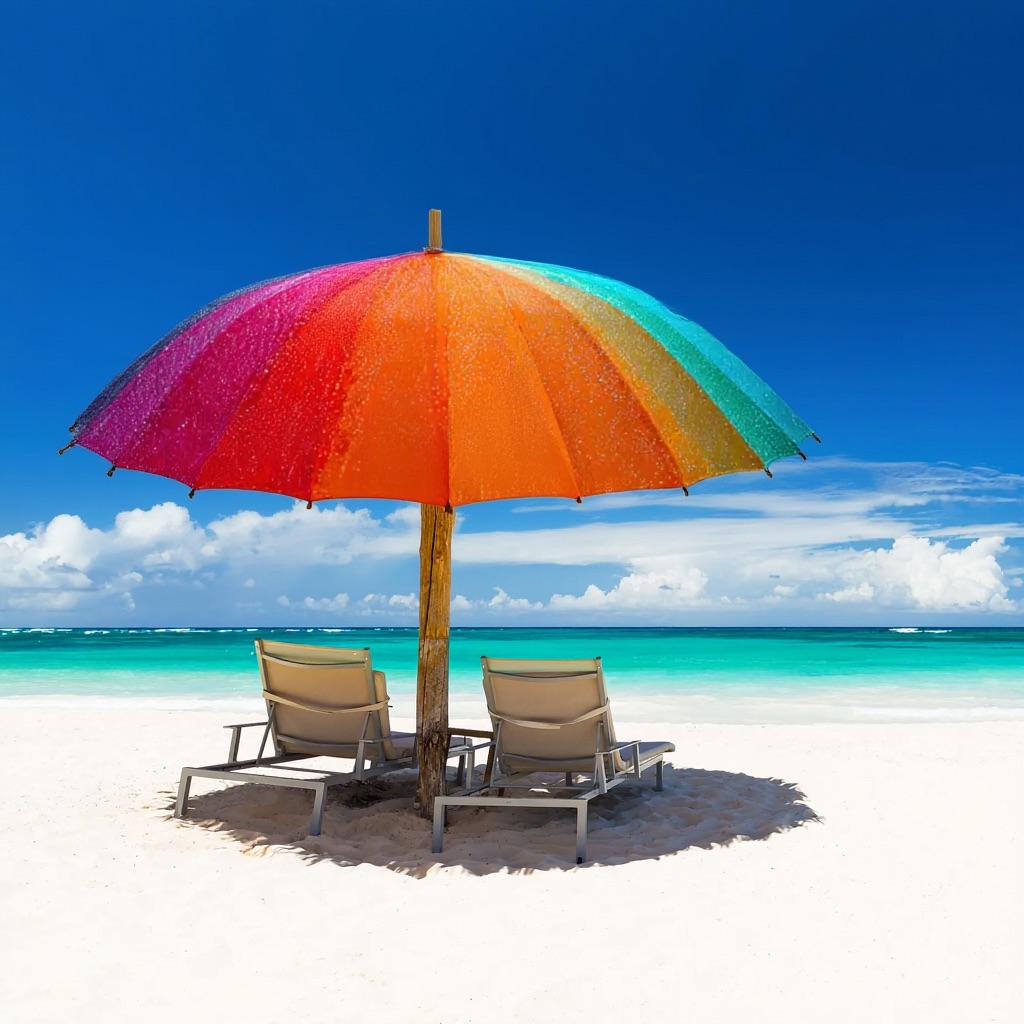}
        \\ \noalign{\vspace{10pt}}

        & \multicolumn{2}{>{\centering\arraybackslash}p{0.38\linewidth}|}{\emph{Blend the cropped objects into the image in a convincing manner.}} 
        & \multicolumn{2}{>{\centering\arraybackslash}p{0.38\linewidth}}{\emph{Blend the alpaca's head seamlessly into the dog's body, remove the alpaca's original landscape backdrop.}} \\%[5pt]
        
        \includegraphics[width=0.19\linewidth]{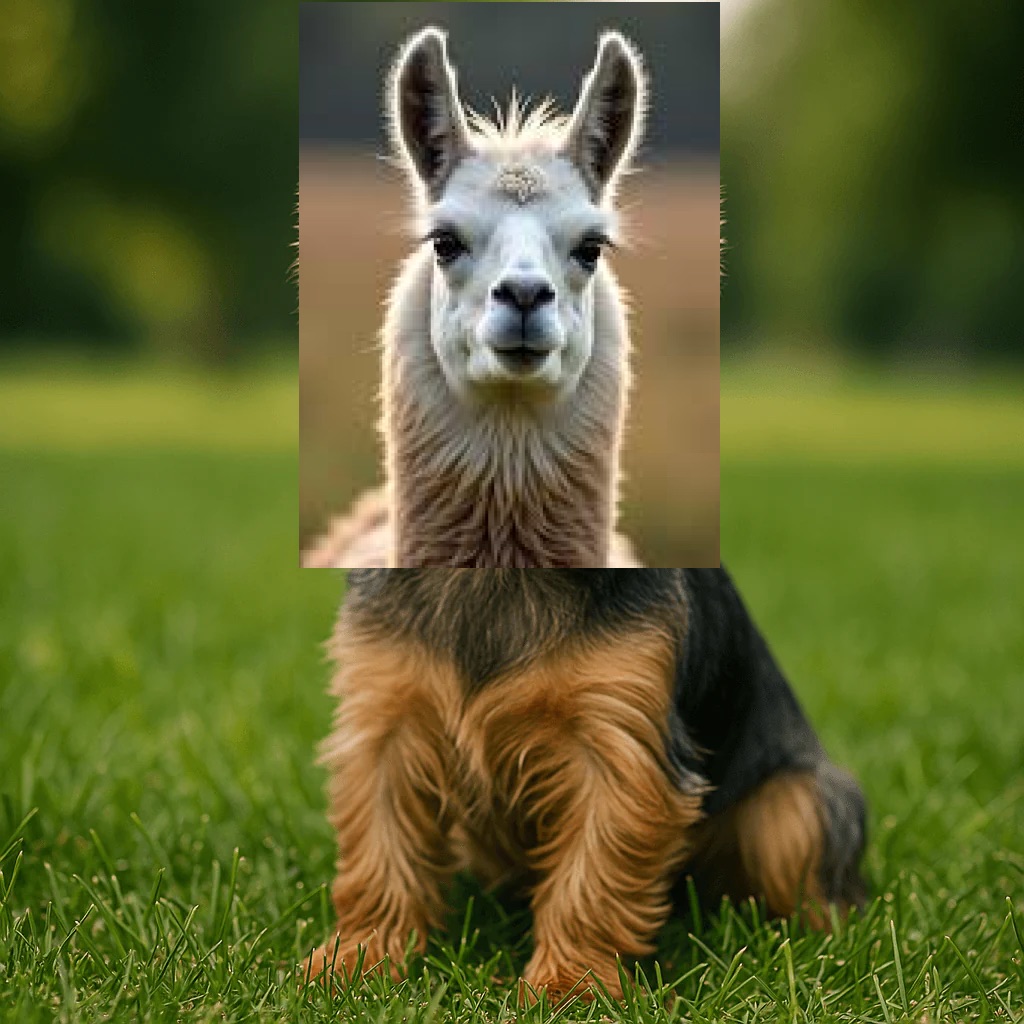} &
        \includegraphics[width=0.19\linewidth]{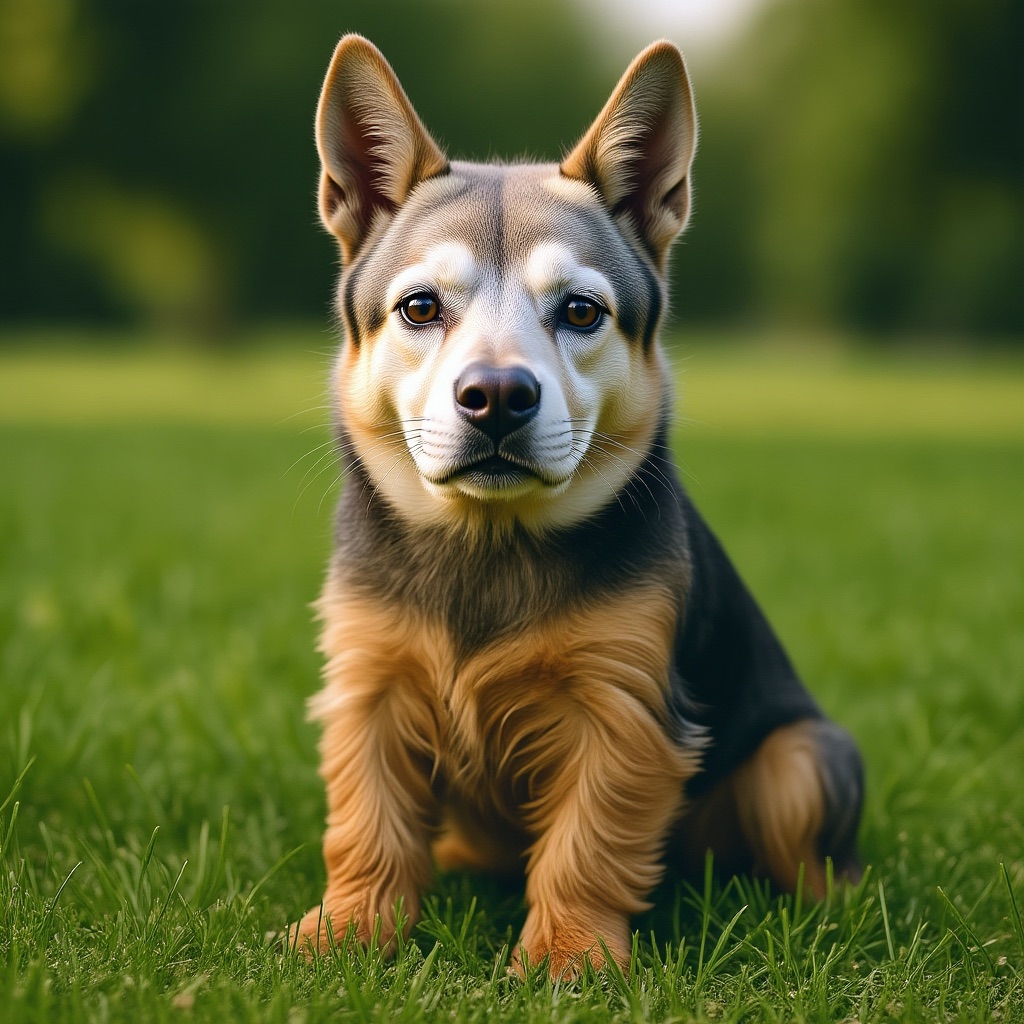} &
        \includegraphics[width=0.19\linewidth]{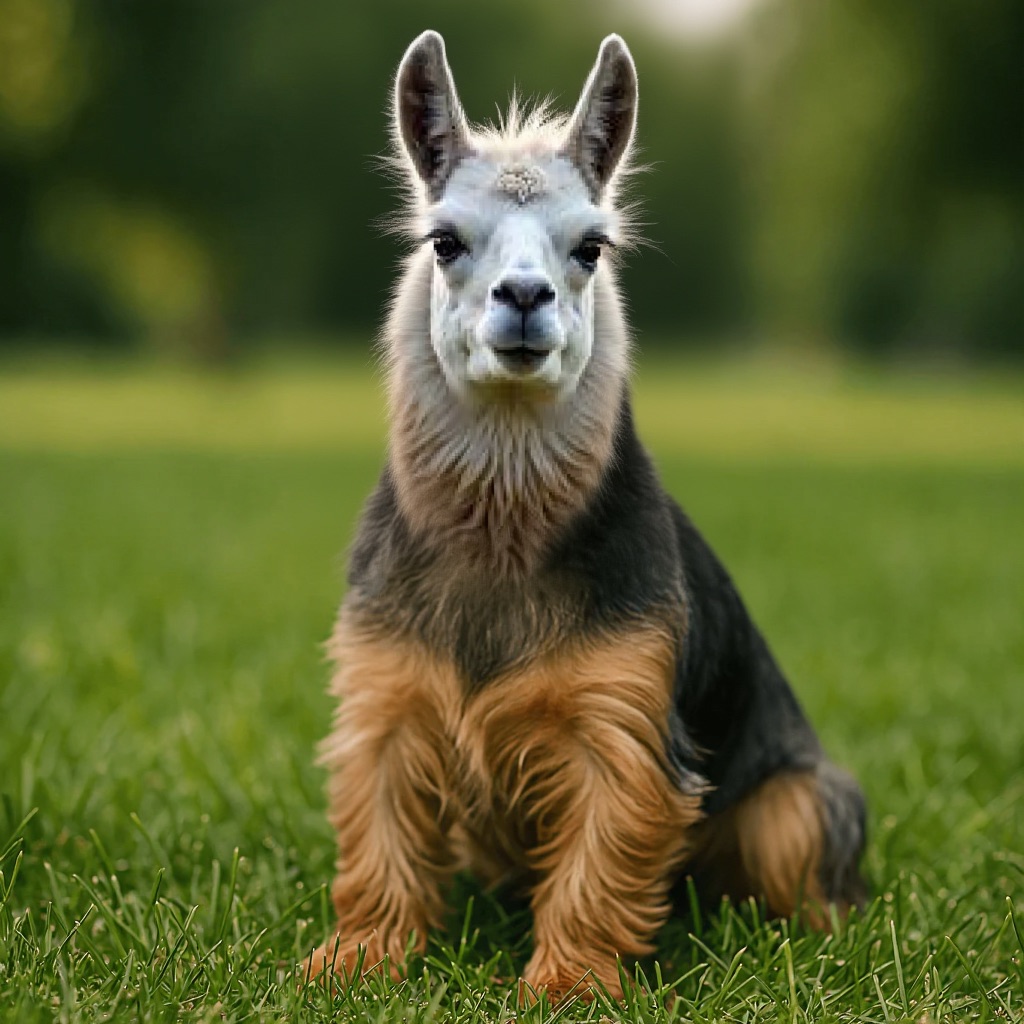} &
        \includegraphics[width=0.19\linewidth]{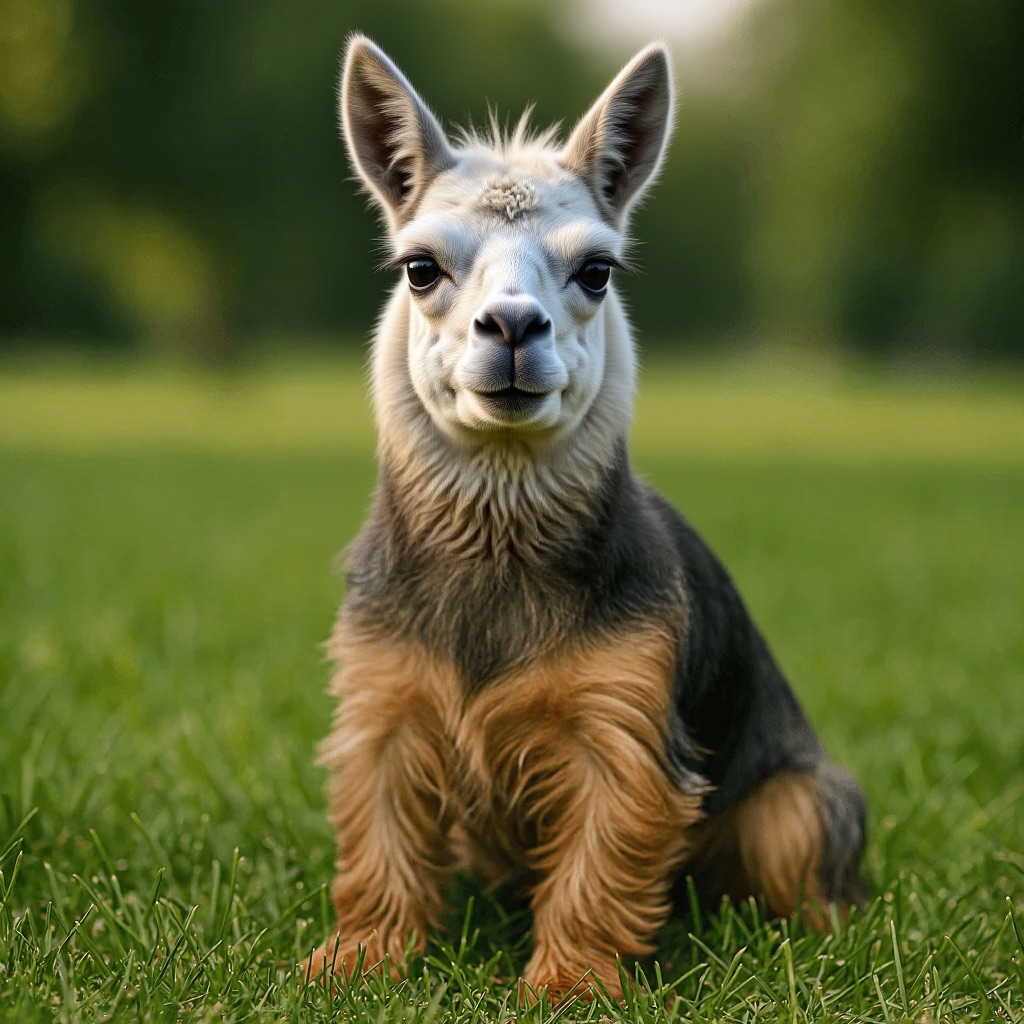} &
        \includegraphics[width=0.19\linewidth]{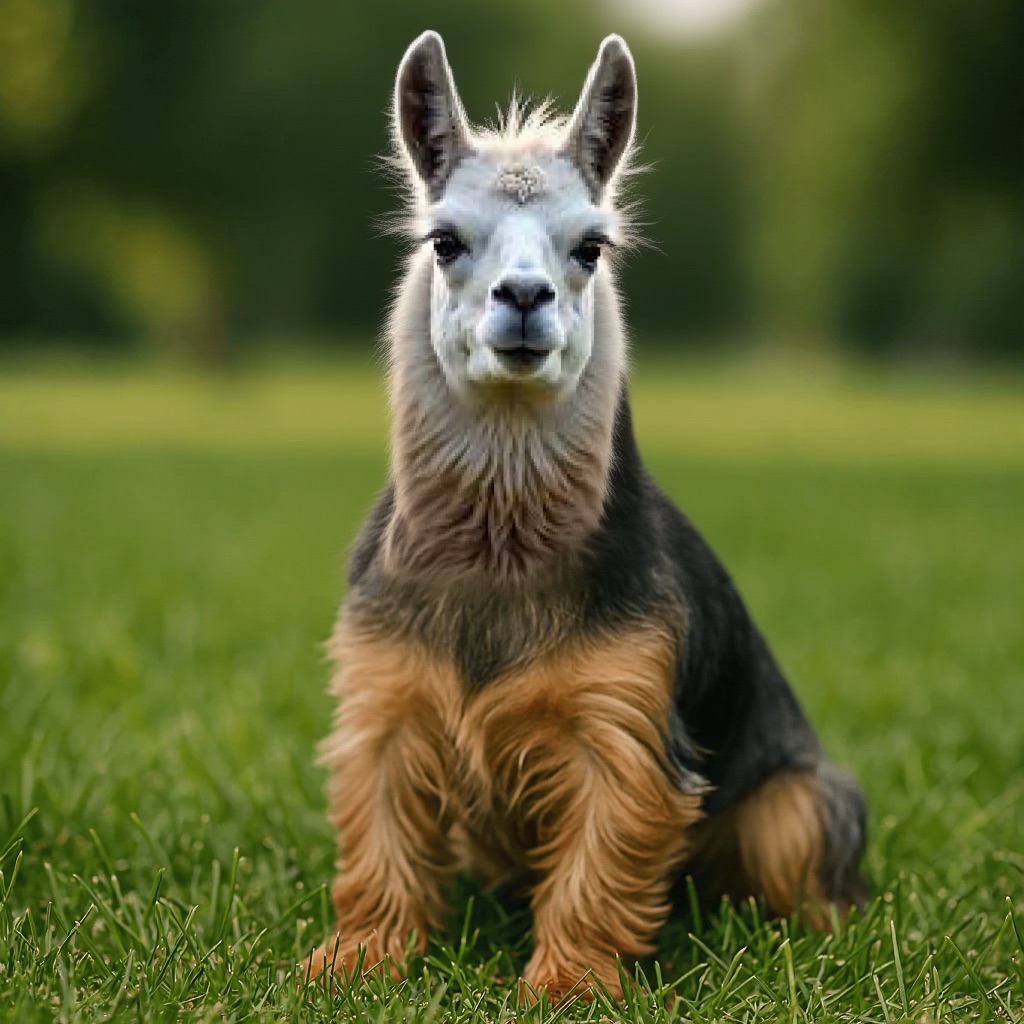}
        \\ \noalign{\vspace{10pt}}

        & \multicolumn{2}{>{\centering\arraybackslash}p{0.38\linewidth}|}{\emph{Blend the cropped objects into the image in a convincing manner.}} & 
        \multicolumn{2}{>{\centering\arraybackslash}p{0.38\linewidth}}{\emph{Blend the bulldozer seamlessly into the parking lot, remove the bulldozer's original white background.}} \\[5pt]
        
        \includegraphics[width=0.19\linewidth]{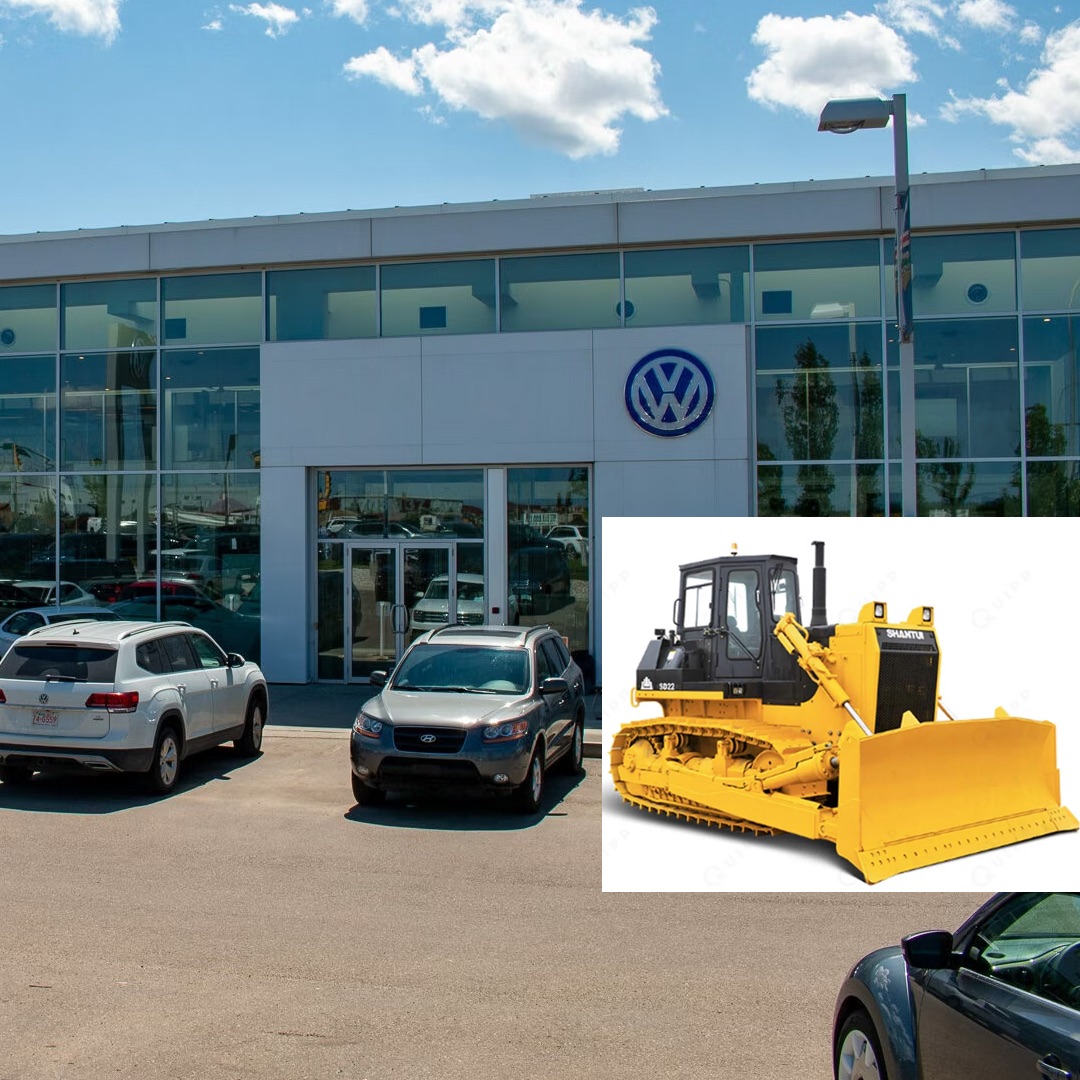} &
        \includegraphics[width=0.19\linewidth]{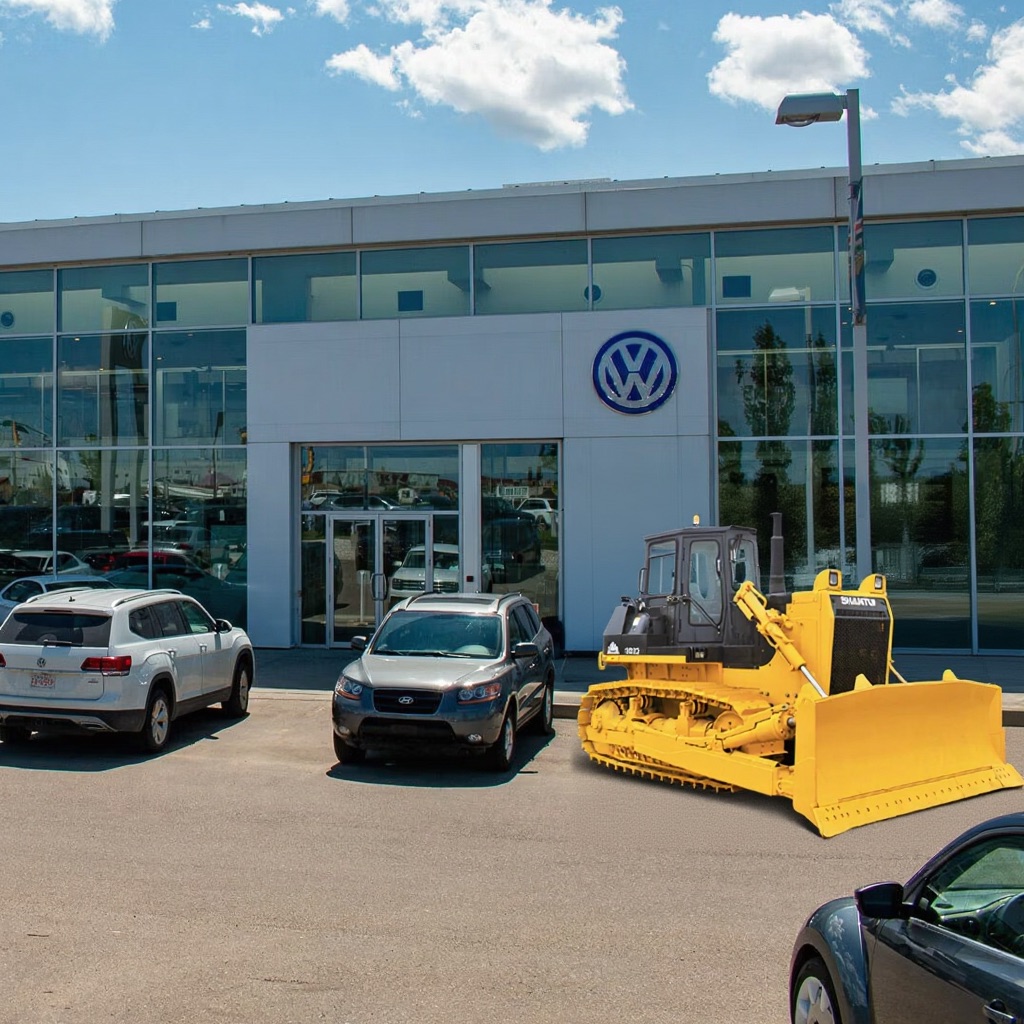} &
        \includegraphics[width=0.19\linewidth]{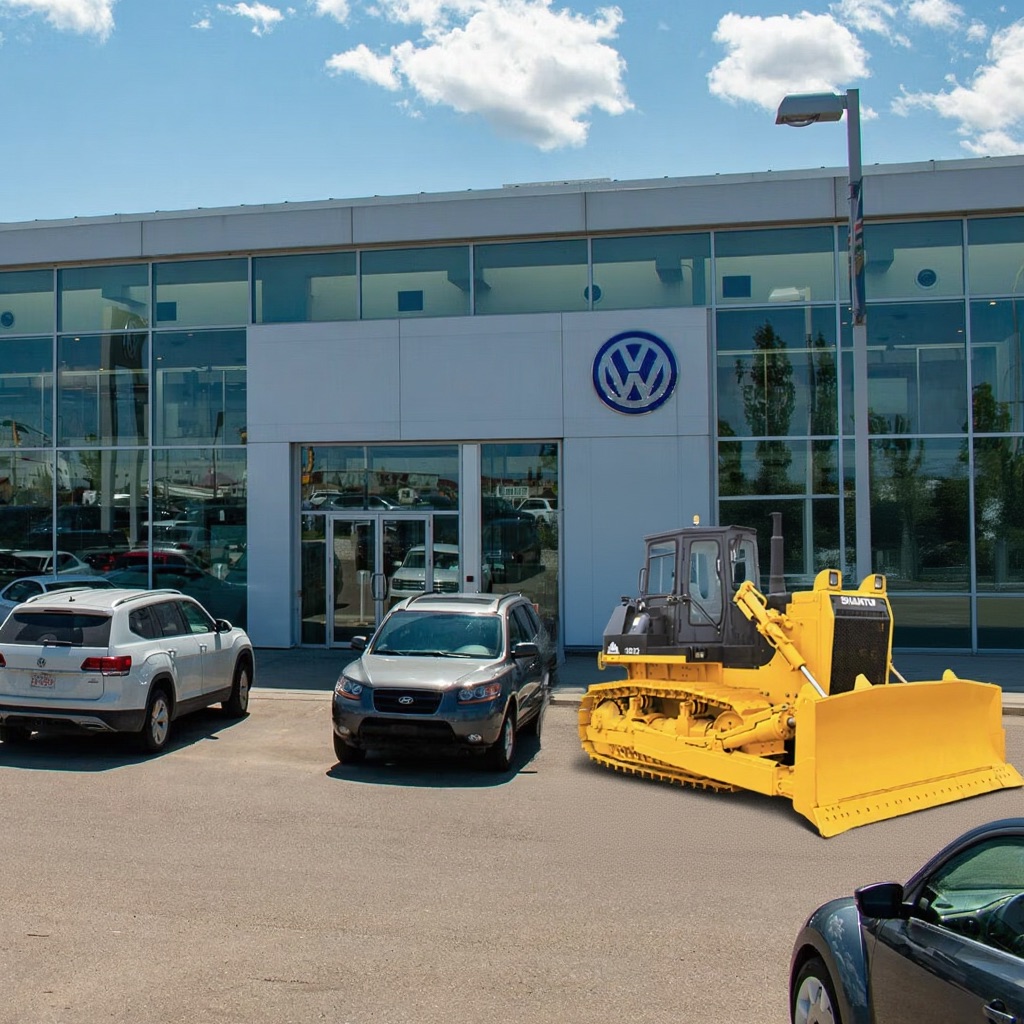} &
        \includegraphics[width=0.19\linewidth]{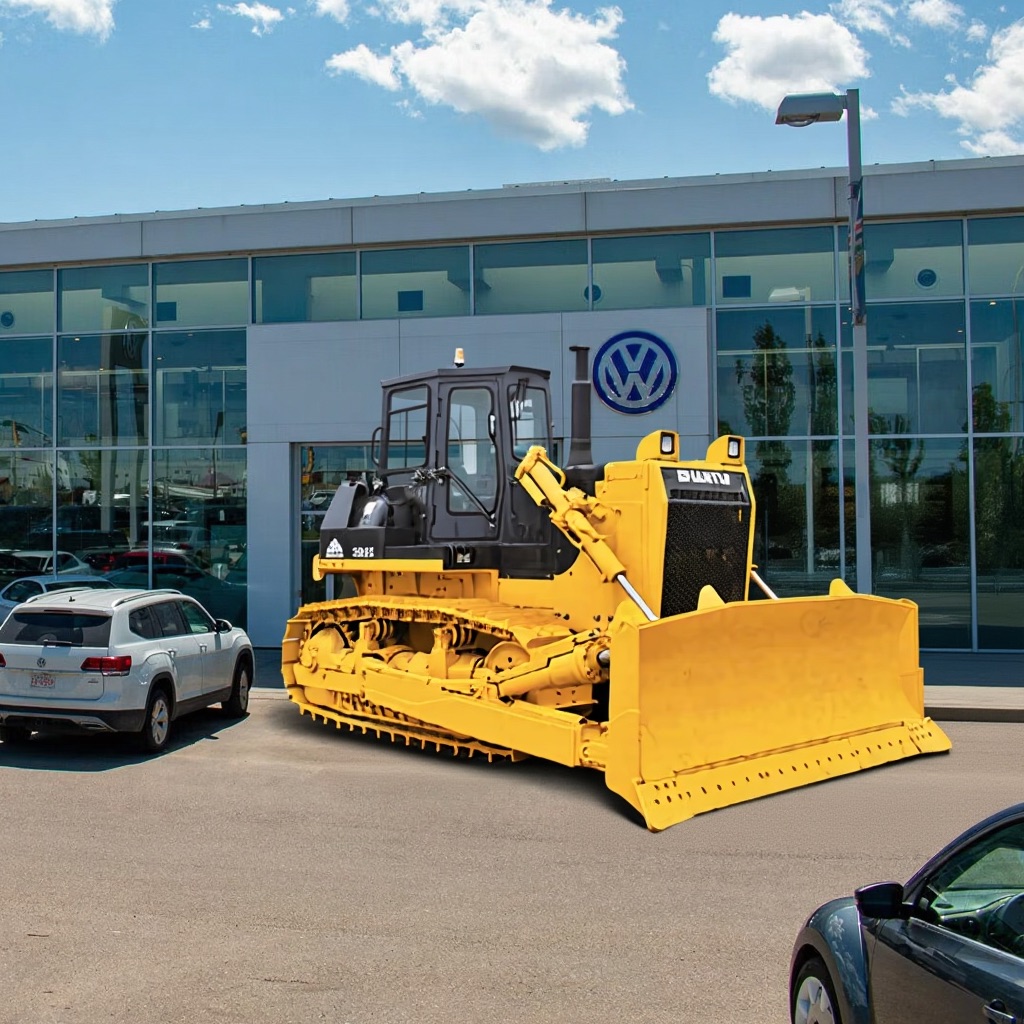} &
        \includegraphics[width=0.19\linewidth]{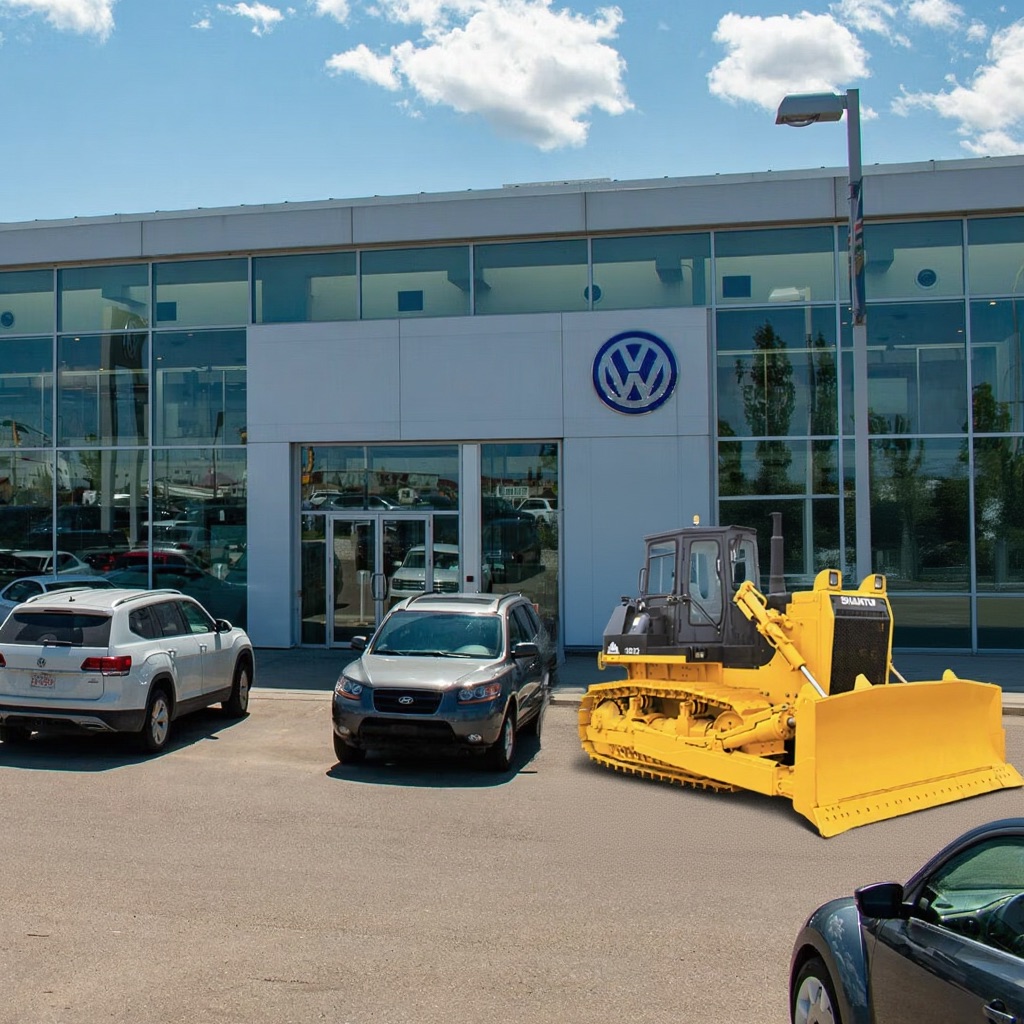}
        \\ \noalign{\vspace{6pt}}
        
        \small Input & \small Kontext & \small Ours & \small Kontext & \small Ours
    \end{tabular}
    \caption{\new{\textbf{Effect of tailored prompts.} We display editing results for the baseline (Kontext) and our method across three scenarios using a generic default prompt (second and third columns from the left) versus a descriptive tailored prompt (fourth and fifth columns from the left). While tailored prompts can reduce object "neglect" in the baseline, they often trigger identity loss (e.g., removed seating in the top row) or unfaithful object placement (bottom row). Conversely, our method demonstrates prompt-invariance, maintaining consistent identity preservation and seamless blending regardless of the prompt specificity.}}
    \label{fig:tailored_prompts}
\end{figure*}

\begin{figure}
    \centering
    \setlength{\tabcolsep}{0.5pt} 
    
    \begin{tabular}{cccc}
        & \textbf{Input} & \textbf{Kontext} & \textbf{Ours} 
        \vspace{-5pt}
        \\[6pt]

        \rotatebox[origin=c]{90}{\small Segmented} &
        \raisebox{-.5\height}{\includegraphics[width=0.31\linewidth]{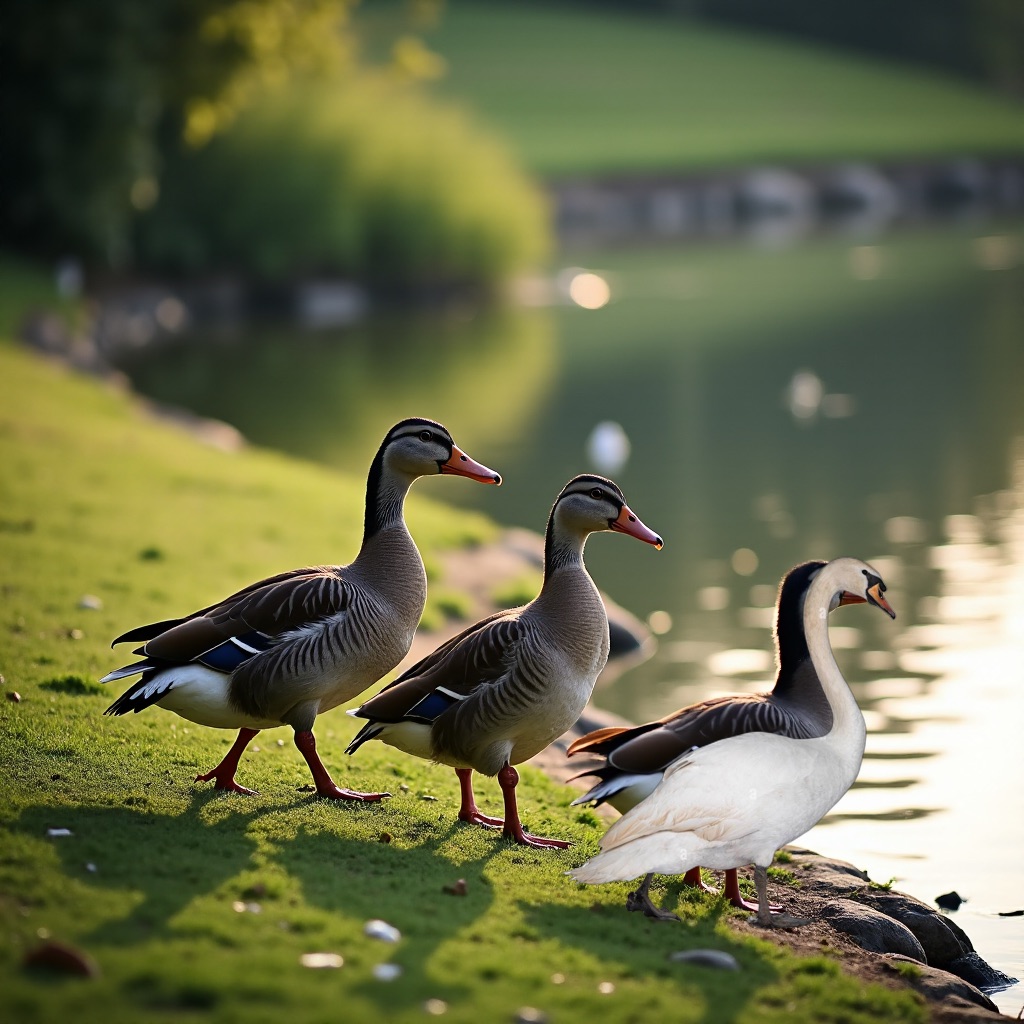}} &
        \raisebox{-.5\height}{\includegraphics[width=0.31\linewidth]{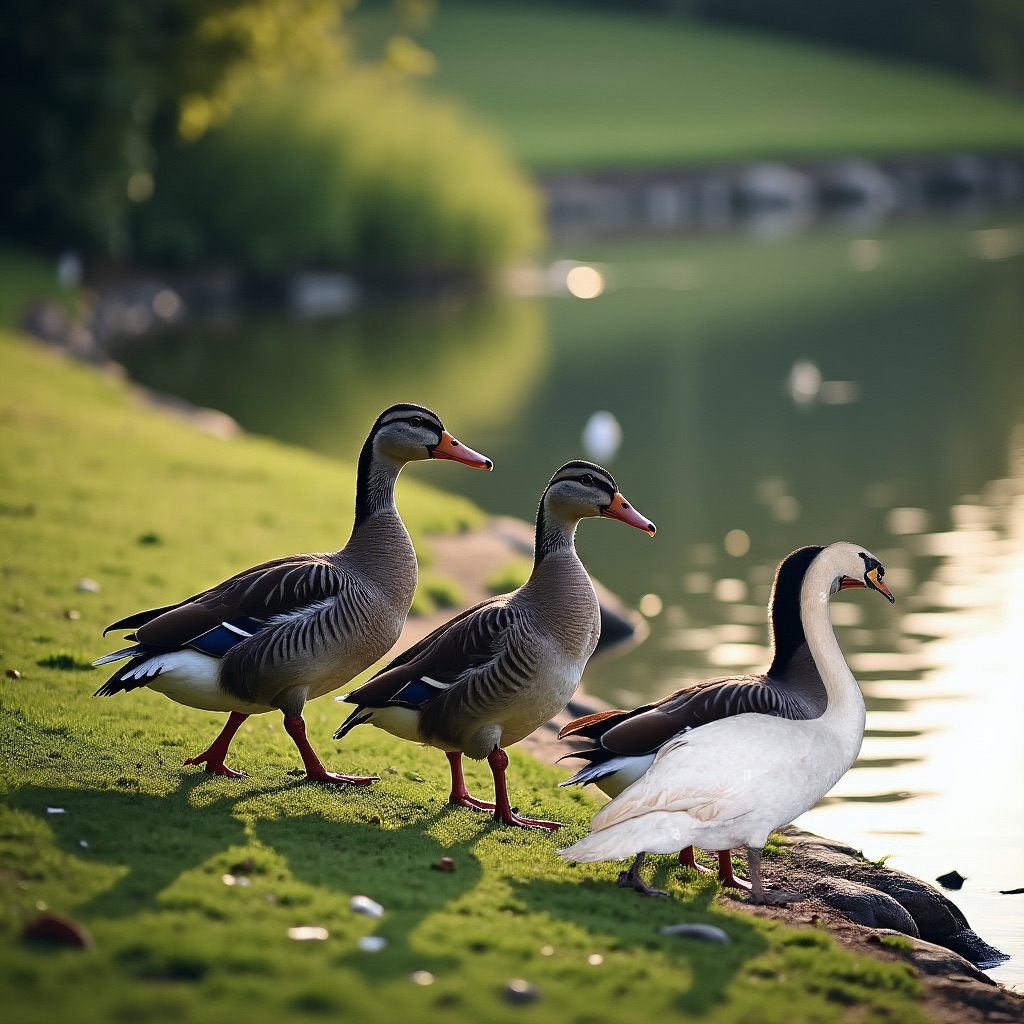}} &
        \raisebox{-.5\height}{\includegraphics[width=0.31\linewidth]{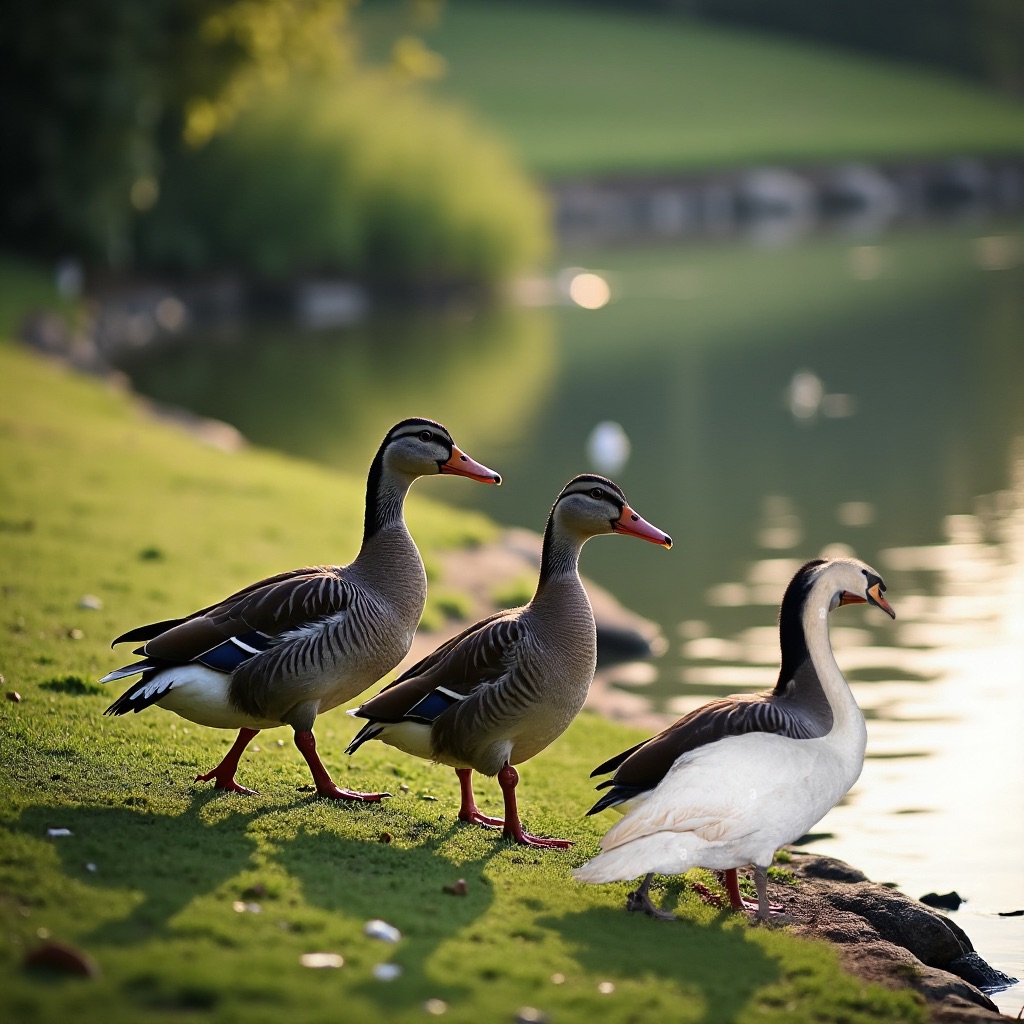}} 
        \\[2pt] %

        \rotatebox[origin=c]{90}{\small Bounding Box} &
        \raisebox{-.5\height}{\includegraphics[width=0.31\linewidth]{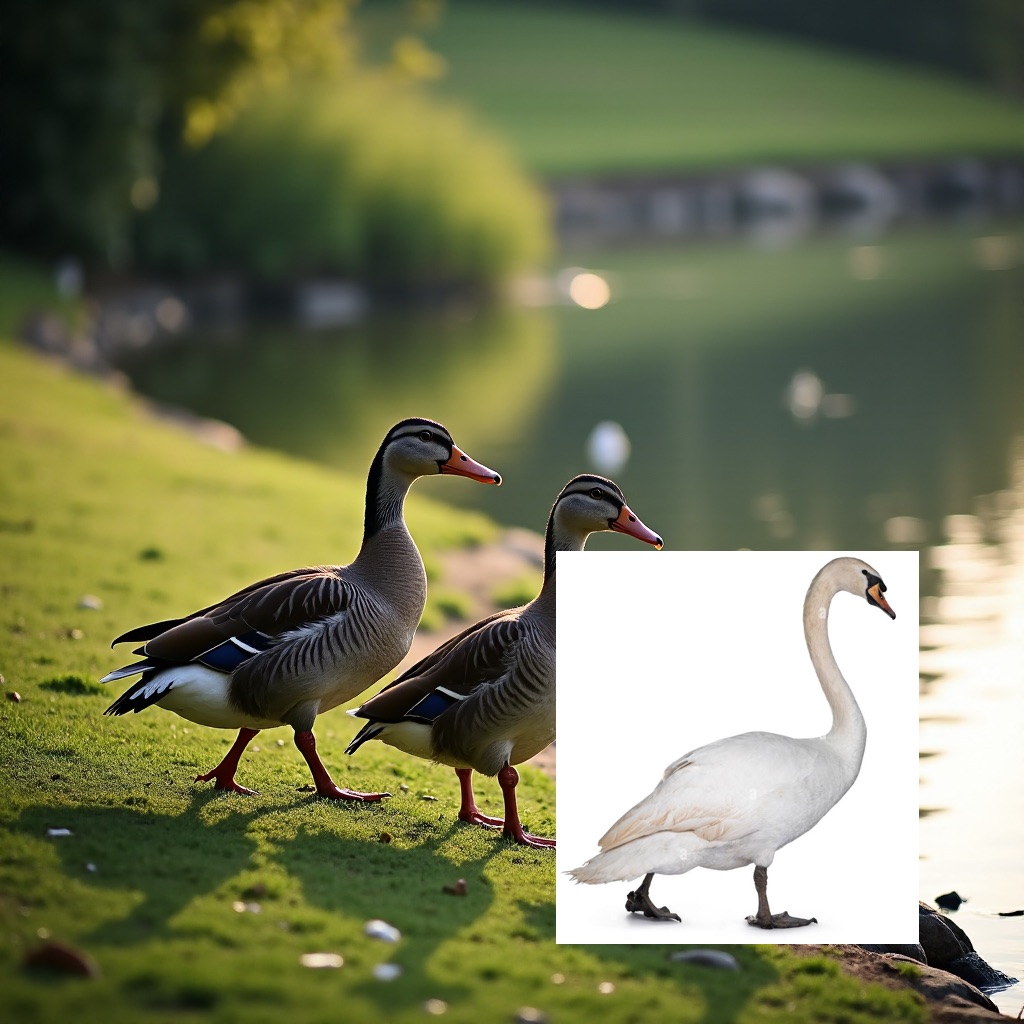}} &
        \raisebox{-.5\height}{\includegraphics[width=0.31\linewidth]{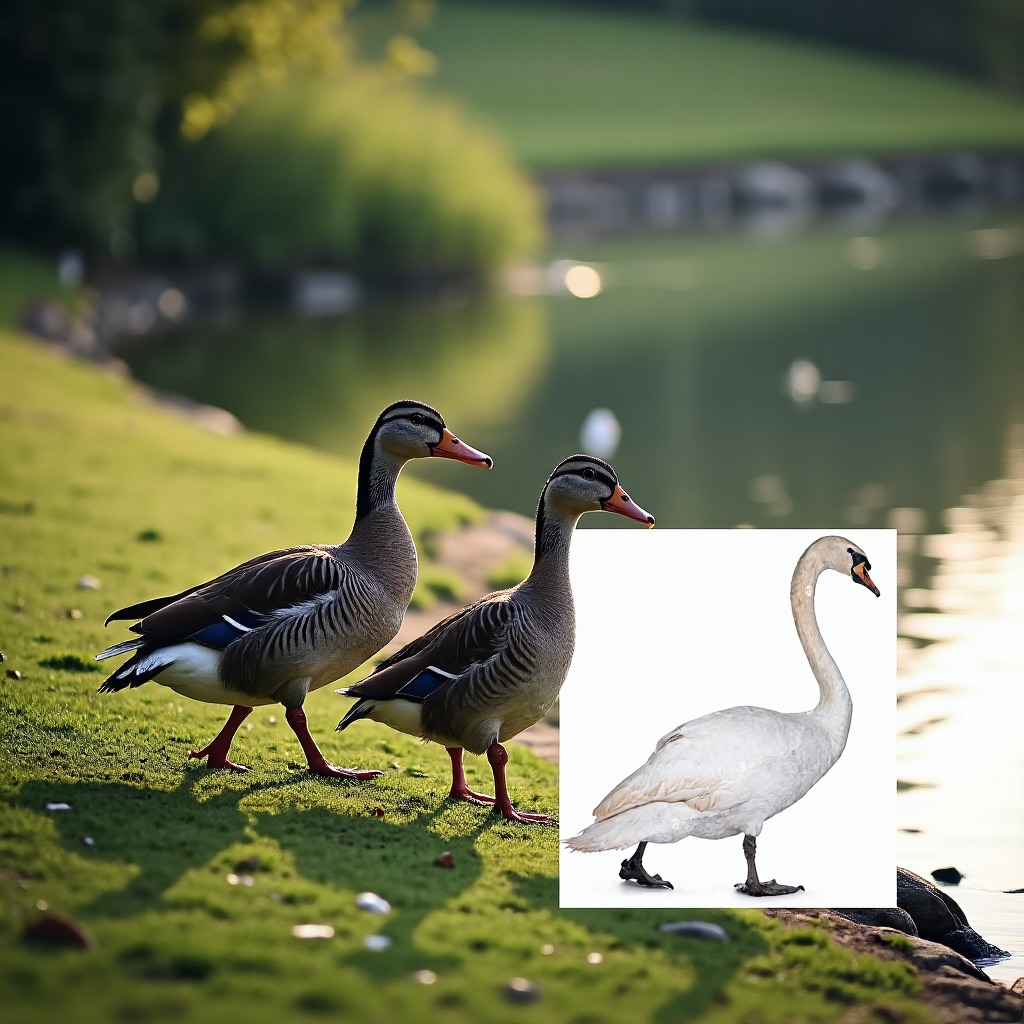}} &
        \raisebox{-.5\height}{\includegraphics[width=0.31\linewidth]{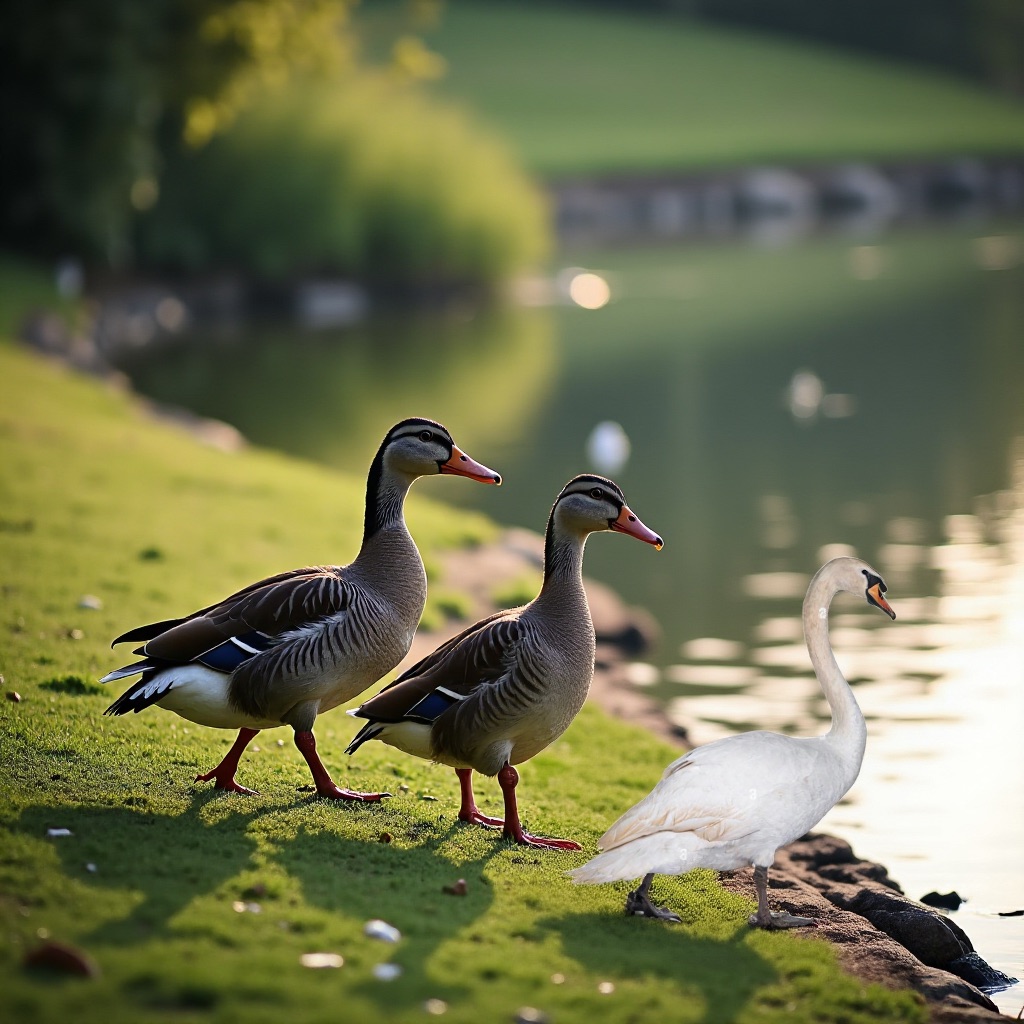}} 
        \\[15pt] %
        \\

        \rotatebox[origin=c]{90}{\small Segmented} &
        \raisebox{-.5\height}{\includegraphics[width=0.31\linewidth]{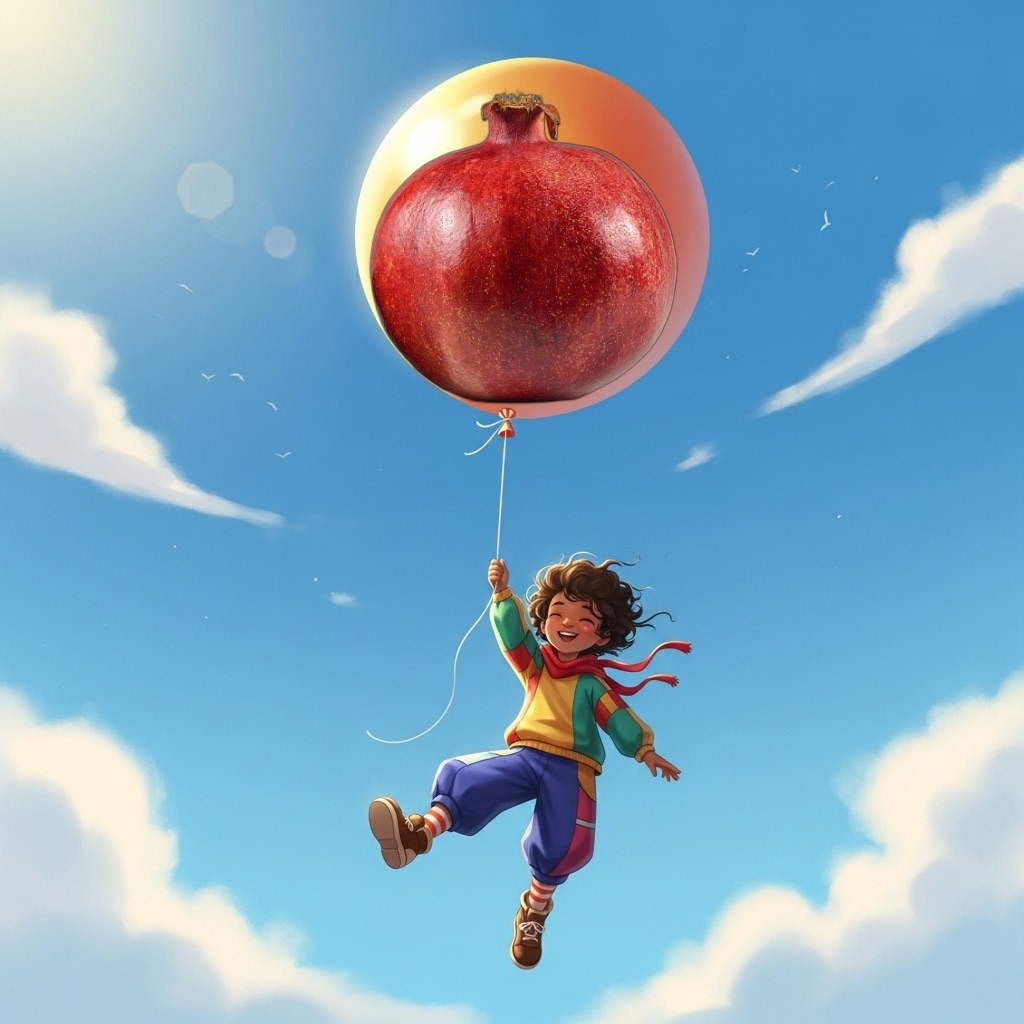}} &
        \raisebox{-.5\height}{\includegraphics[width=0.31\linewidth]{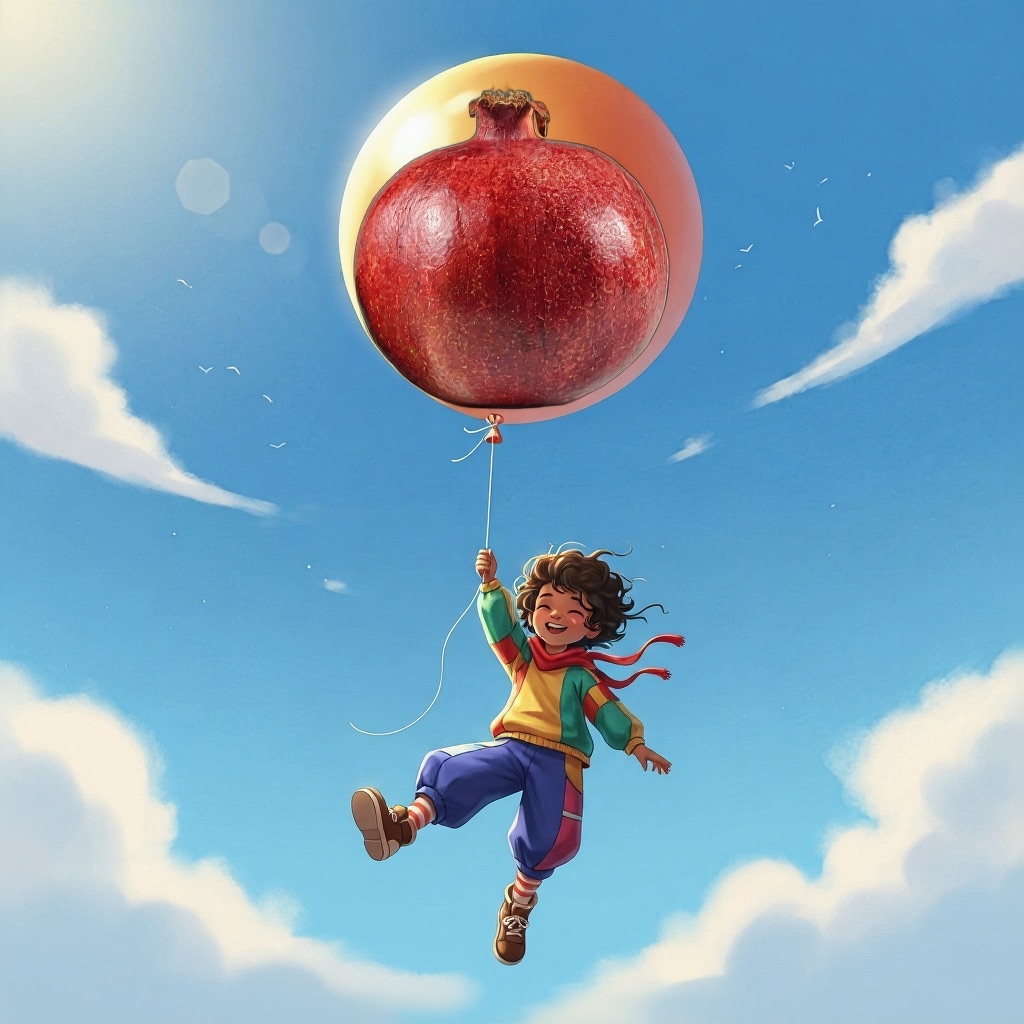}} &
        \raisebox{-.5\height}{\includegraphics[width=0.31\linewidth]{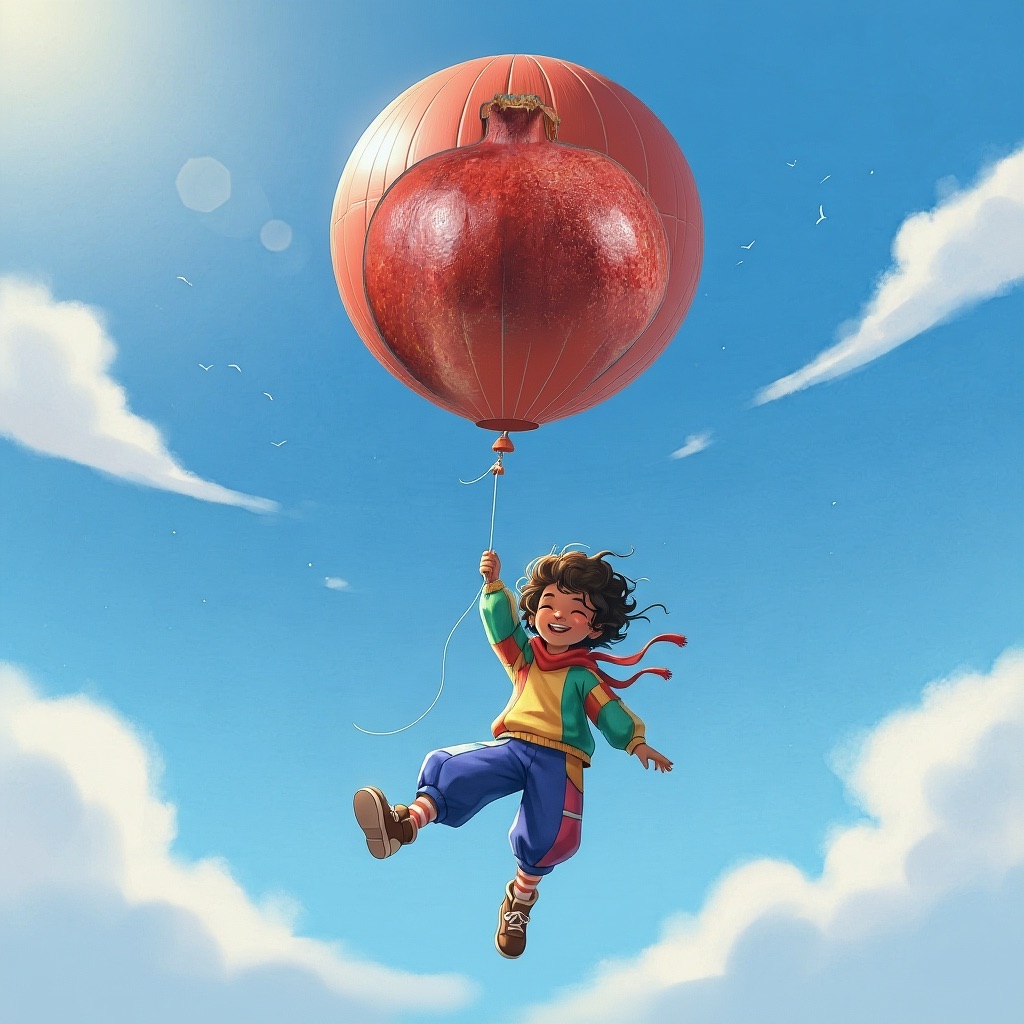}} 
        \\[2pt]

        \rotatebox[origin=c]{90}{\small Bounding Box} &
        \raisebox{-.5\height}{\includegraphics[width=0.31\linewidth]{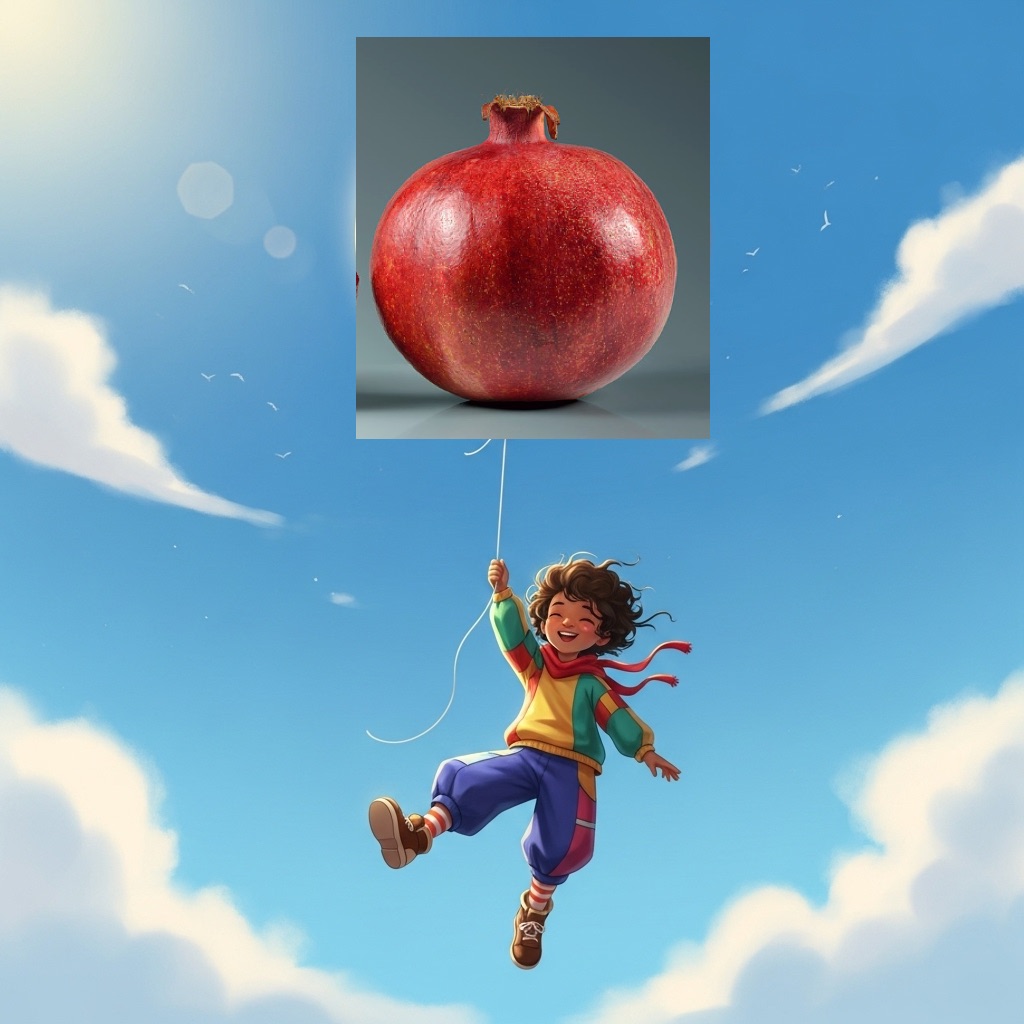}} &
        \raisebox{-.5\height}{\includegraphics[width=0.31\linewidth]{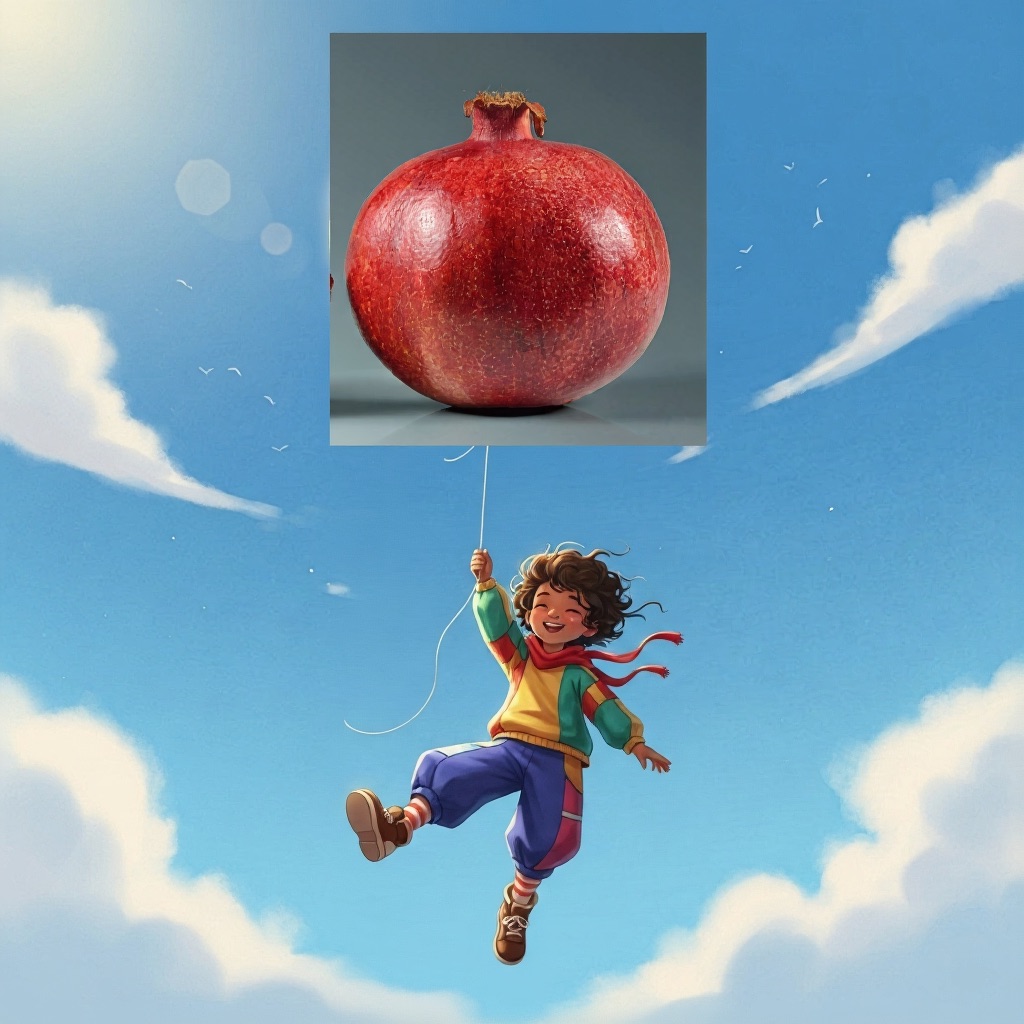}} &
        \raisebox{-.5\height}{\includegraphics[width=0.31\linewidth]{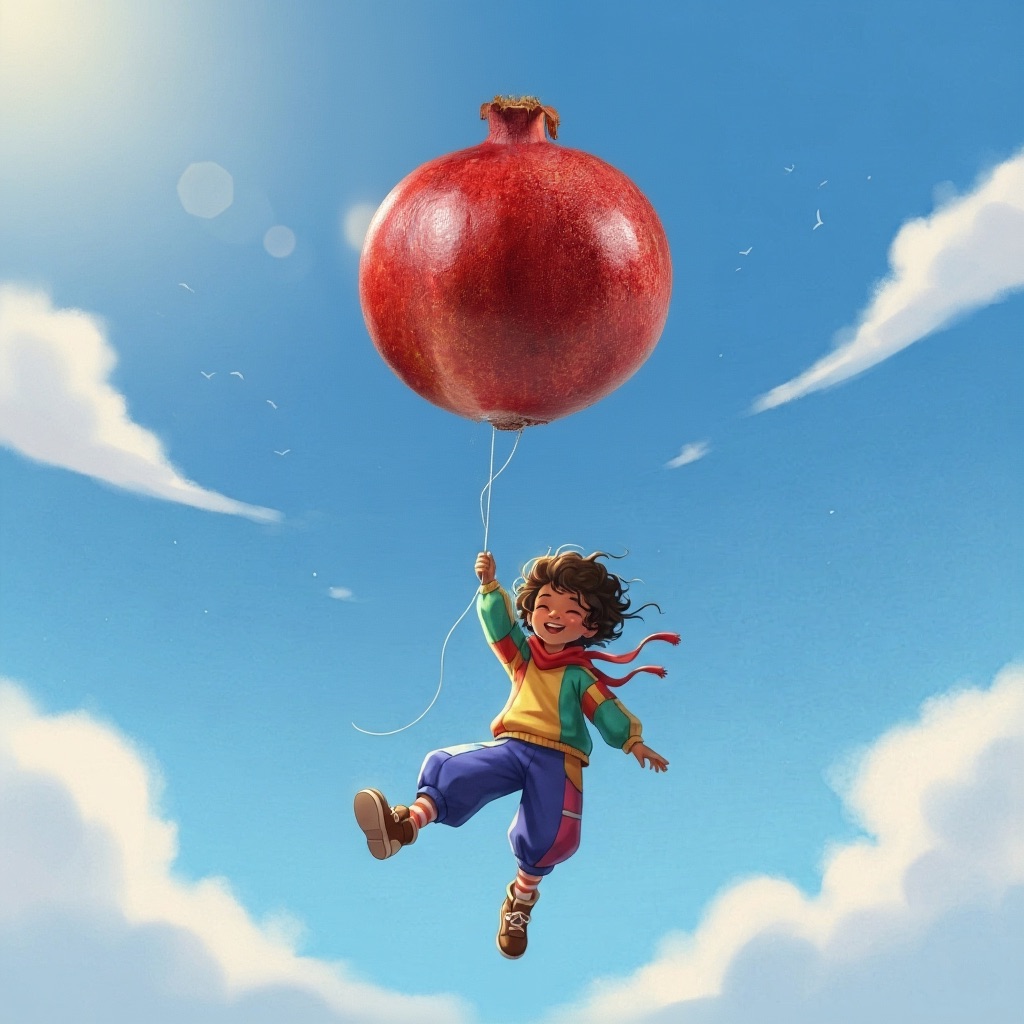}} 
        \\[15pt]
        \\
        
        \rotatebox[origin=c]{90}{\small Segmented} &
        \raisebox{-.5\height}{\includegraphics[width=0.31\linewidth]{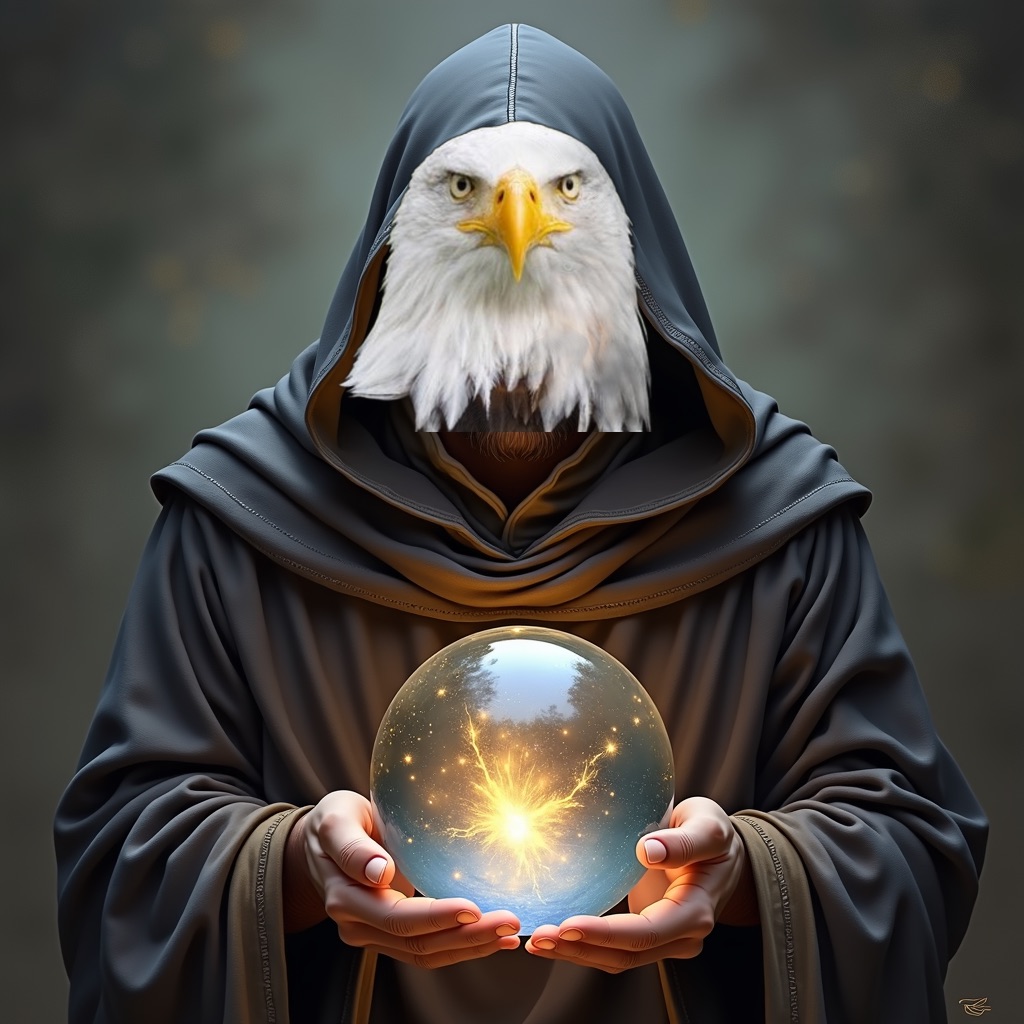}} &
        \raisebox{-.5\height}{\includegraphics[width=0.31\linewidth]{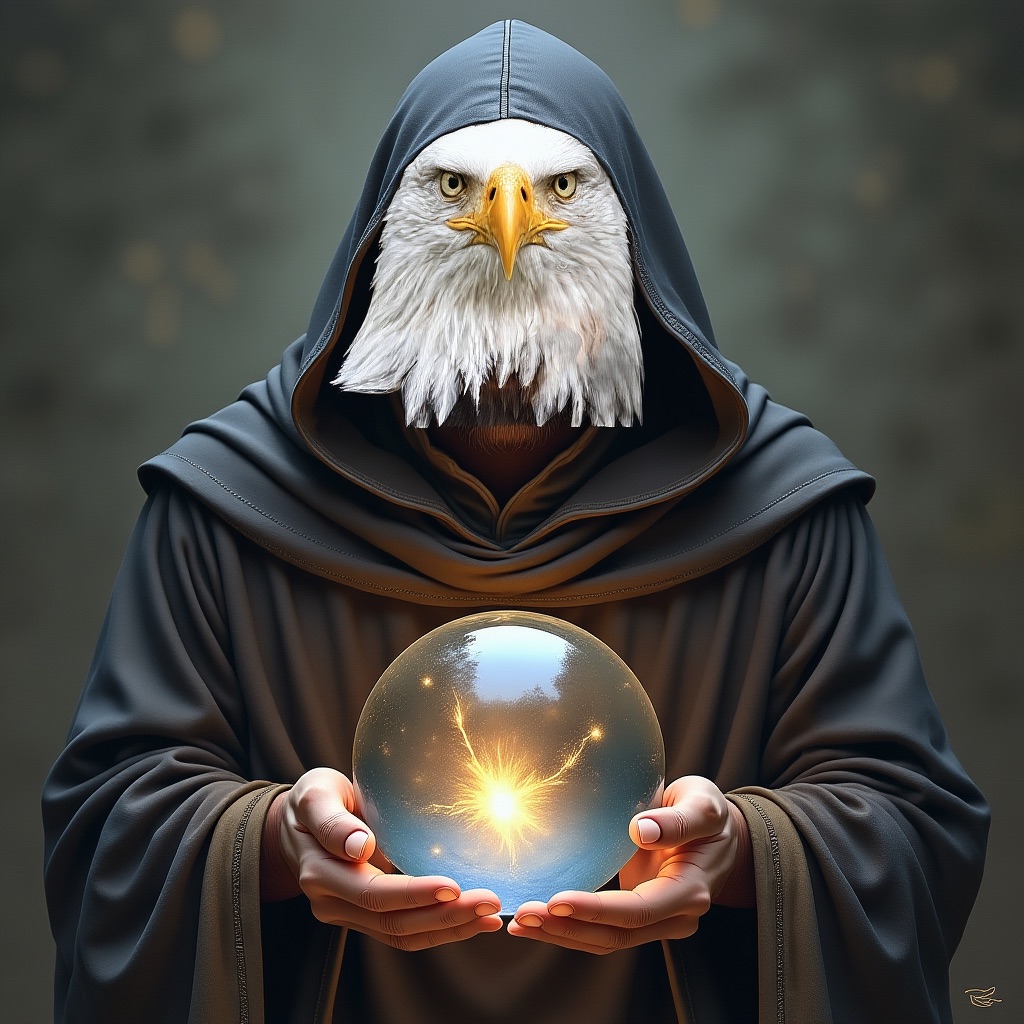}} &
        \raisebox{-.5\height}{\includegraphics[width=0.31\linewidth]{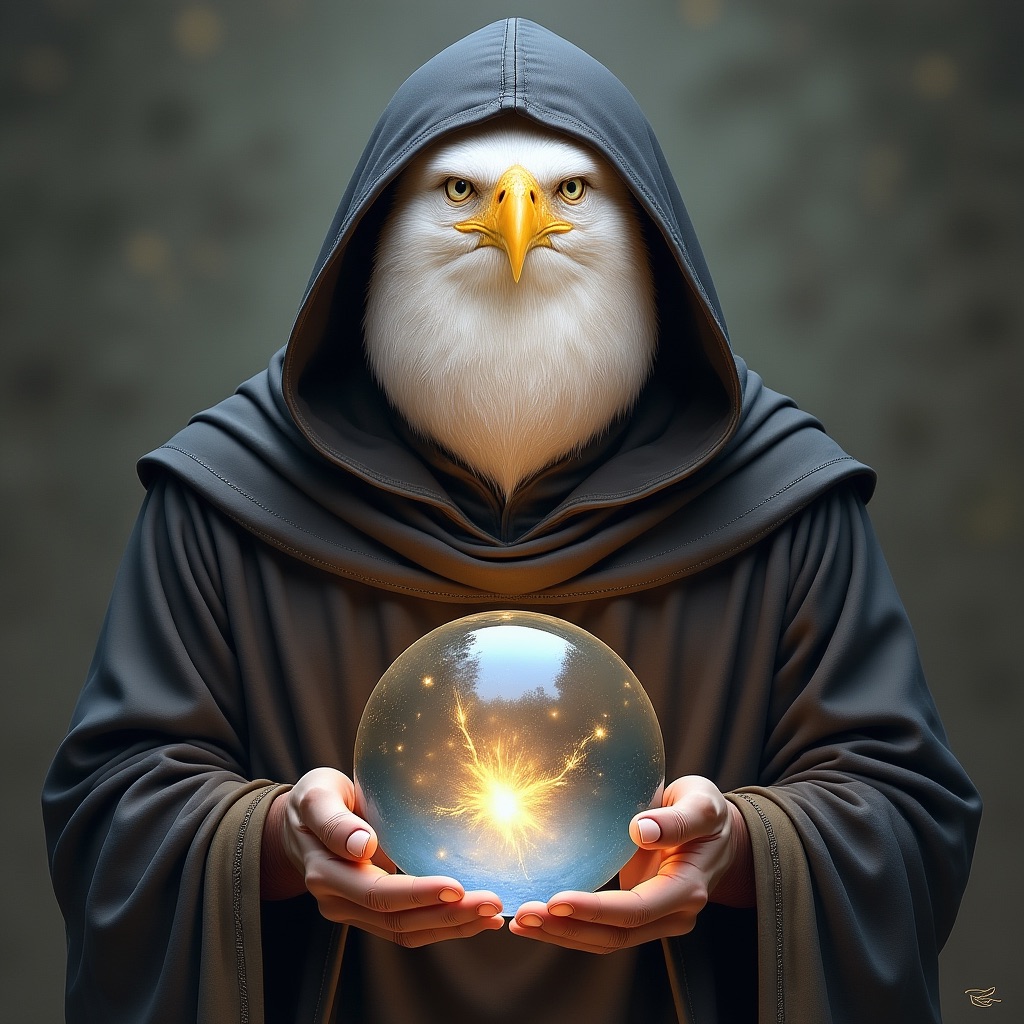}} 
        \\[2pt]

        \rotatebox[origin=c]{90}{\small Bounding Box} &
        \raisebox{-.5\height}{\includegraphics[width=0.31\linewidth]{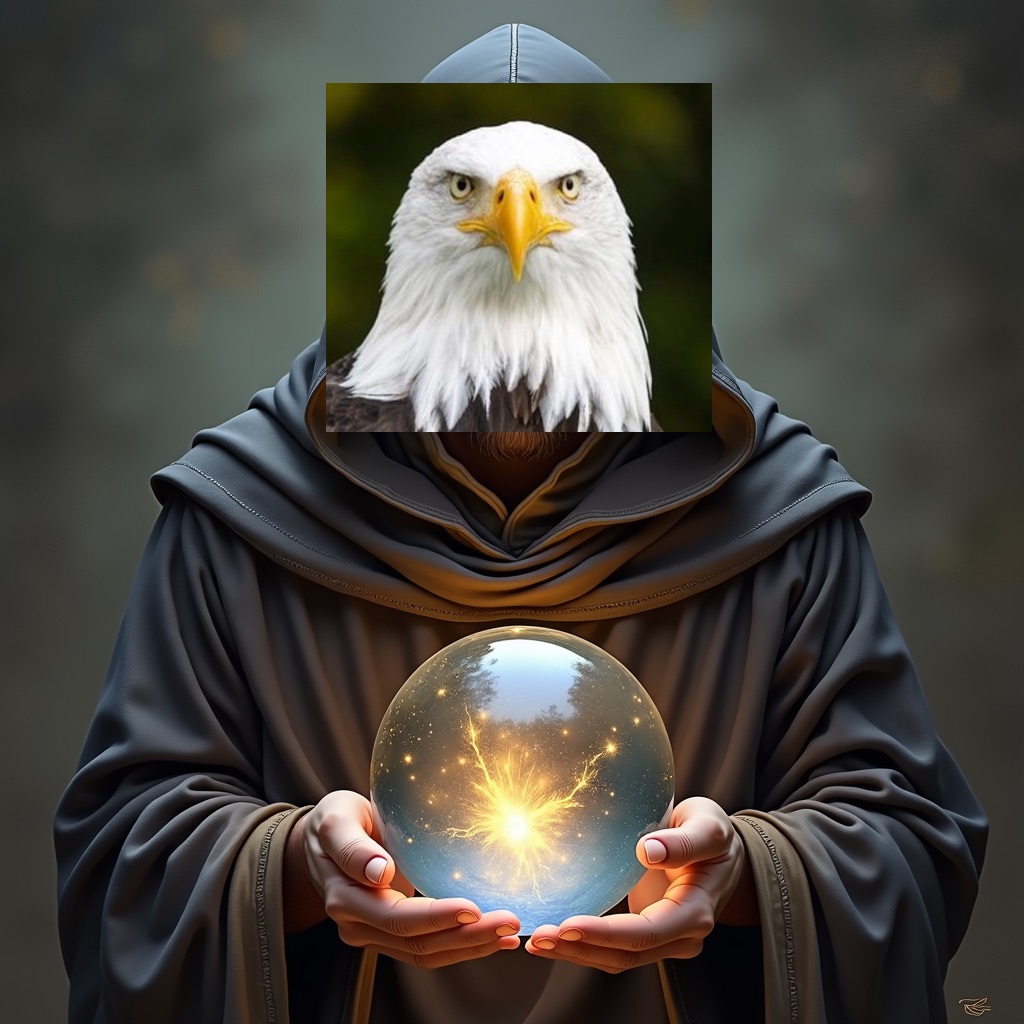}} &
        \raisebox{-.5\height}{\includegraphics[width=0.31\linewidth]{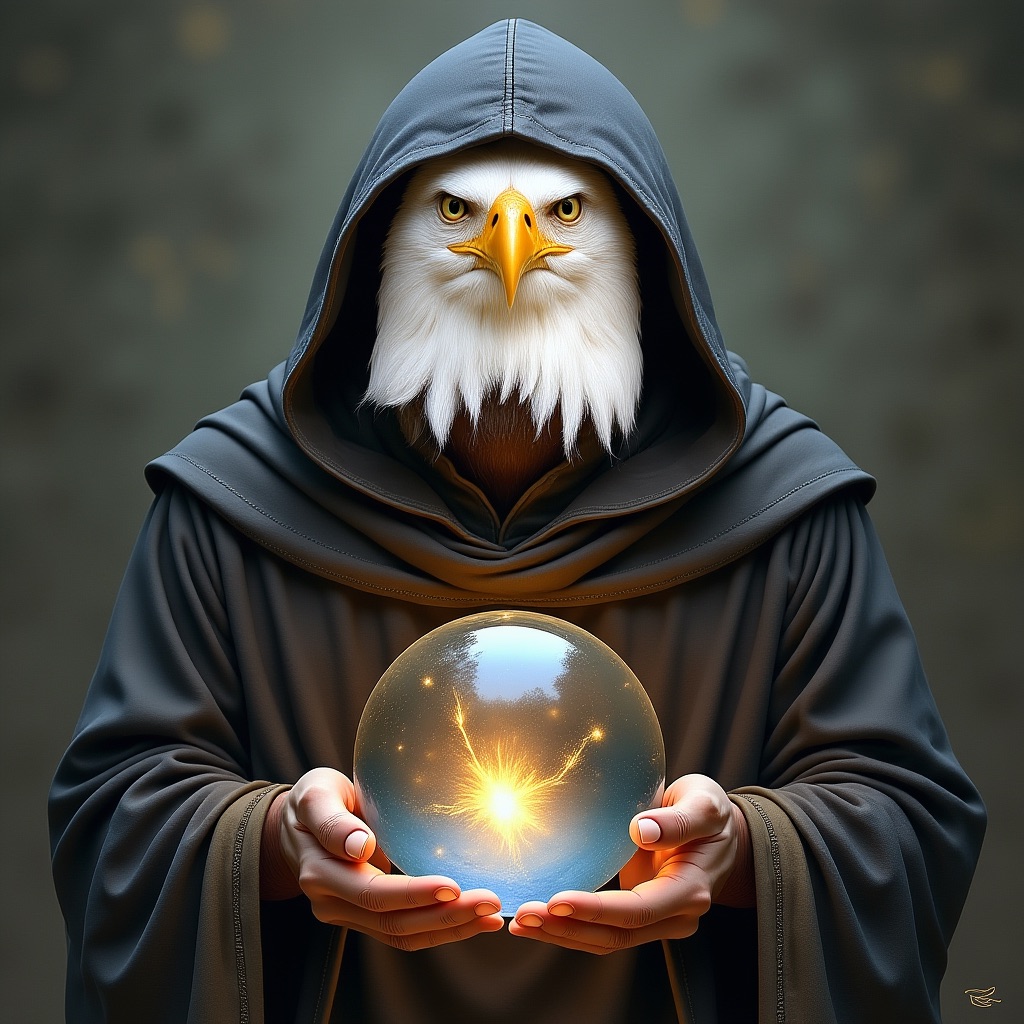}} &
        \raisebox{-.5\height}{\includegraphics[width=0.31\linewidth]{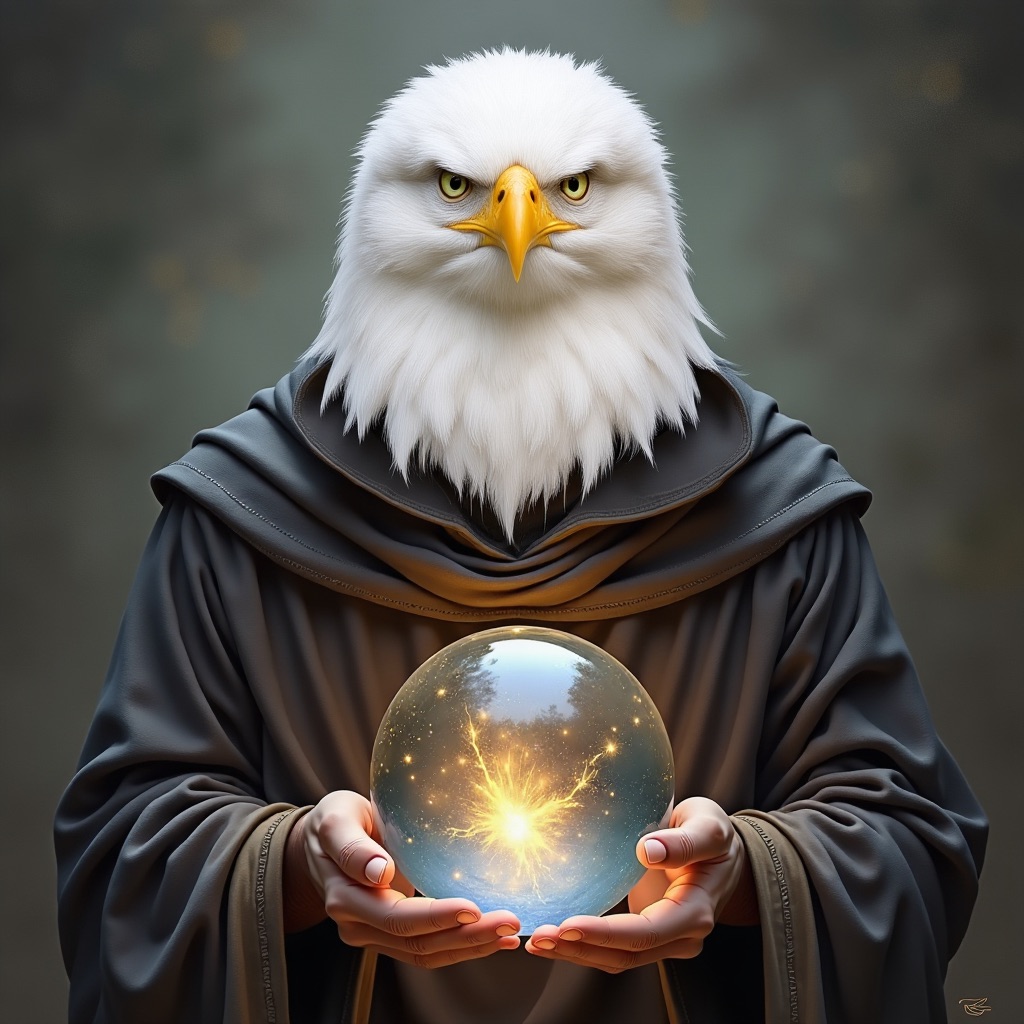}}
    \end{tabular}
    \vspace{-5pt}
    \caption{
    \new{
        \textbf{Bounding-box versus segmented input crops.}
We compare results for the base model (FLUX Kontext) and our method using two input types: the default bounding-box crop and a segmented crop obtained with the \texttt{rembg} model. In the top two examples (four top rows), the segmented input creates ambiguity about which object should be preserved. In the bottom example (bottom two rows), the segmented input makes FLUX Kontext more prone to neglect, as if the blending were already complete.
        }
    }
    \label{fig:rembg}
\end{figure}

\subsection{Attention Locality and Harmonization Outcomes}
To support the claim made in the main paper, that the attention maps of instruction-based editing models inherently govern whether a pasted region is copied from the input image or modified for harmonization, we analyze the attention behavior of FLUX Kontext across our benchmark. Following the notations defined in the main paper (Section 3.2), let the queries within the pasted region be denoted as $Q_{\text{out}}[M]$, the keys of the input image as $K_{\text{in}}$, and the resulting attention weights as
\[
W_{\text{in}} = \operatorname{softmax}\!\left(\frac{Q_{\text{out}}[M]K_{\text{in}}^\top}{\sqrt{d}}\right),
\]
where $d$ is the feature dimension. We define the \emph{inward–outward attention ratio} $R$ as
\[
R = \frac{\sum\limits_{q \in Q_{\text{out}}[M]} \sum\limits_{k \in M} W_{\text{in}}(q,k)}
{\sum\limits_{q \in Q_{\text{out}}[M]} \sum\limits_{k \notin M} W_{\text{in}}(q,k)},
\]
measuring the relative amount of attention directed inside versus outside the crop mask.

As shown in Figure~\ref{fig:supp_attn_locality}, this ratio correlates strongly with the blending outcome. Neglect cases exhibit high ratios, indicating predominantly inward attention that causes the model to over-copy the pasted region. Suppression cases yield low ratios, reflecting outward attention that overwhelms and overwrites the inserted content. Successful edits cluster around intermediate ratios, where attention is neither overly localized nor overly dispersed. Notably, there is no clear threshold separating these regimes, suggesting that effective harmonization requires fine-grained and content-aware modulation of attention rather than a simple global increase or decrease of this ratio. This analysis demonstrates that the locality pattern of attention itself is a strong indicator—and likely driver—of whether a region is effectively harmonized or simply copied.

\new{
\subsection{Impact of Prompt Engineering on Harmonization}
To further analyze the role of text guidance, we compare our method against the baseline using both a fixed generic prompt and detailed, ``tailored'' prompts (Fig. \ref{fig:tailored_prompts}). We observe that the baseline model is highly sensitive to prompt phrasing: while meticulous prompt engineering can occasionally mitigate neglect, it often introduces a significant trade-off. For instance, in the umbrella example (Fig. \ref{fig:tailored_prompts}, top row), the tailored prompt improves blending but triggers significant identity loss in the surrounding scene, such as the removal of the seats under the shade. Similarly, the tailored prompt for the bulldozer (bottom row) causes the model to over-rely on the text description, resulting in a placement that is unfaithful to the original scene.}

\new{
Crucially, Fig. \ref{fig:tailored_prompts} demonstrates that our method is significantly less sensitive to the specific choice of input prompt. Across all subjects, our model consistently preserves object identity and achieves seamless blending regardless of whether a generic or descriptive prompt is provided. This ``prompt-invariance'' suggests that our approach successfully prioritizes image-grounded features over linguistic cues. While describing the object can sometimes assist the baseline (e.g., the llama in the middle row), achieving a successful edit this way typically requires multiple manual iterations and specific domain knowledge. In contrast, our method provides a more robust and user-friendly solution by maintaining high-fidelity harmonization without the need for iterative prompt tuning.
}

\begin{figure}
    \centering
    \setlength{\tabcolsep}{0.5pt}
    \begin{tabular}{ccc}
        Base Image & Input & Ours  \\

        \includegraphics[width=0.325\linewidth]{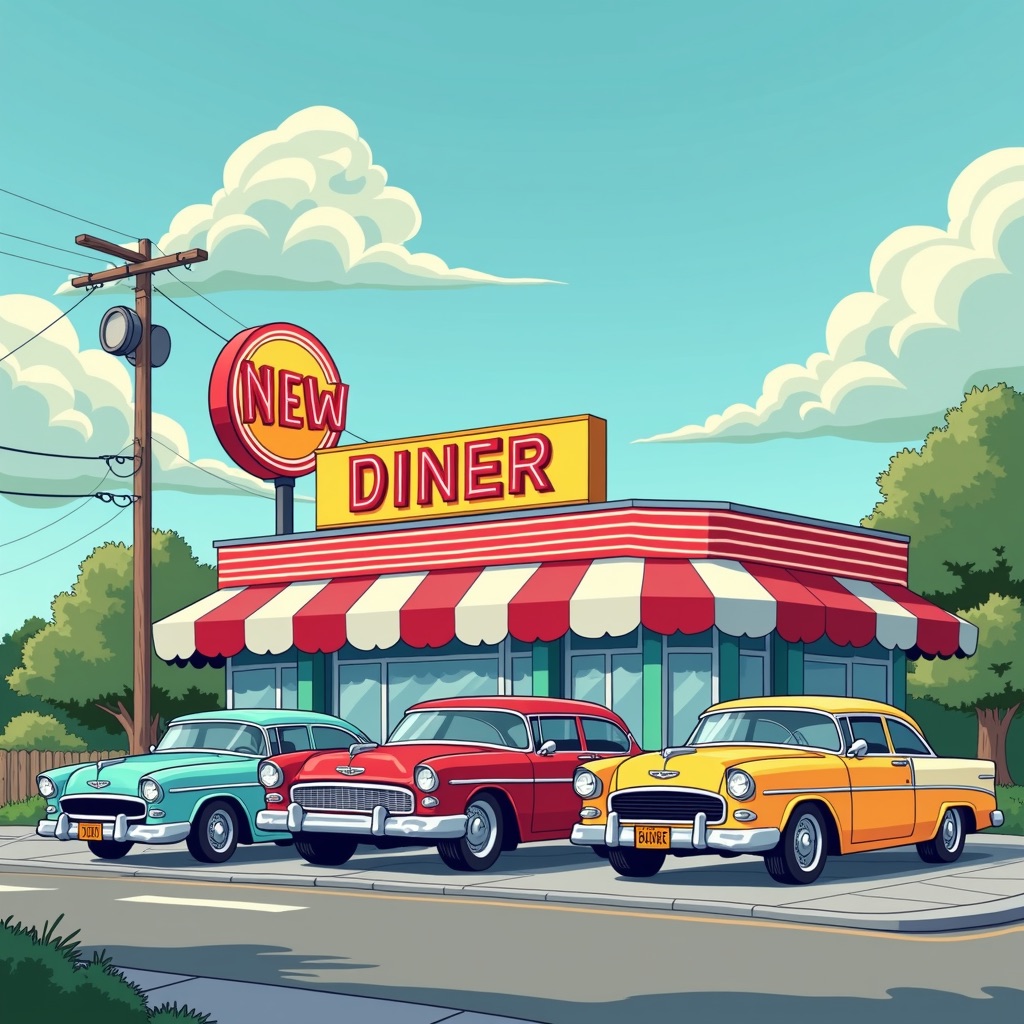} &
        \includegraphics[width=0.325\linewidth]{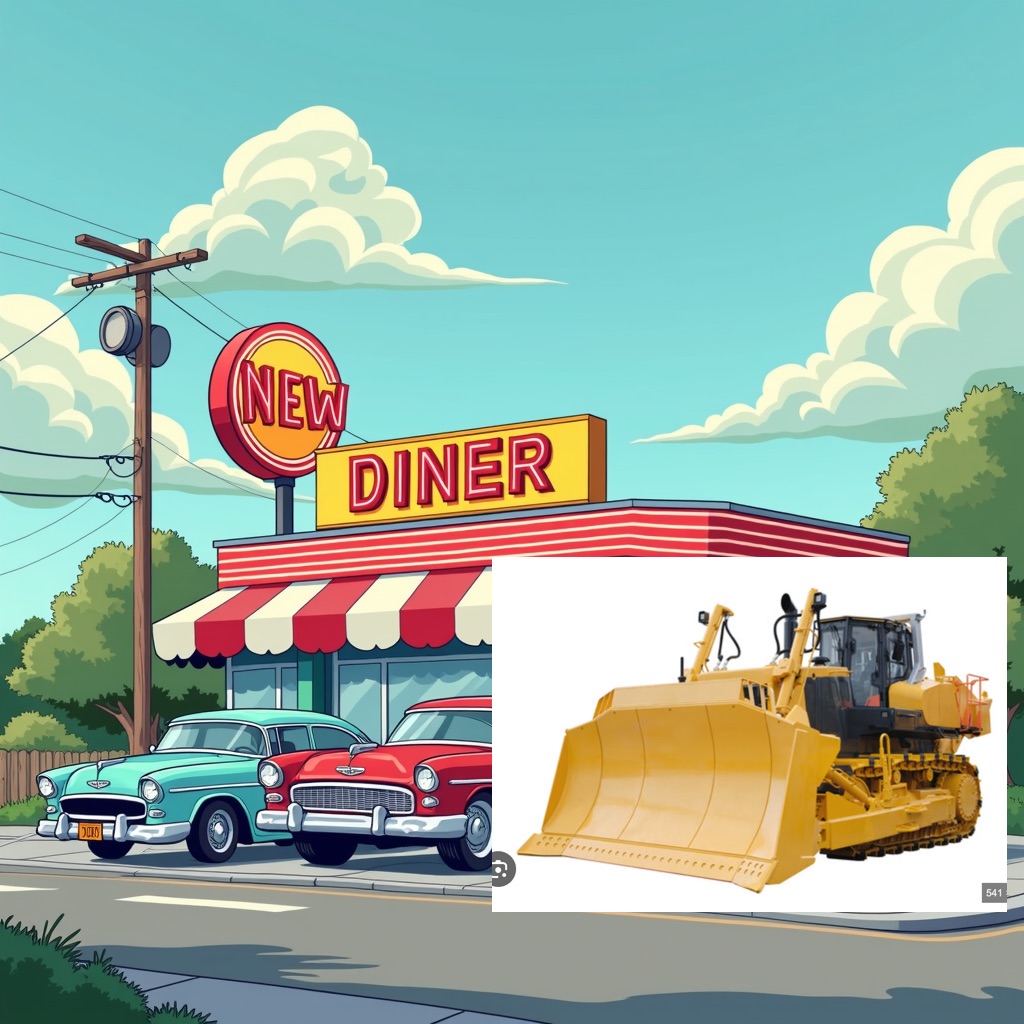} &
        \includegraphics[width=0.325\linewidth]{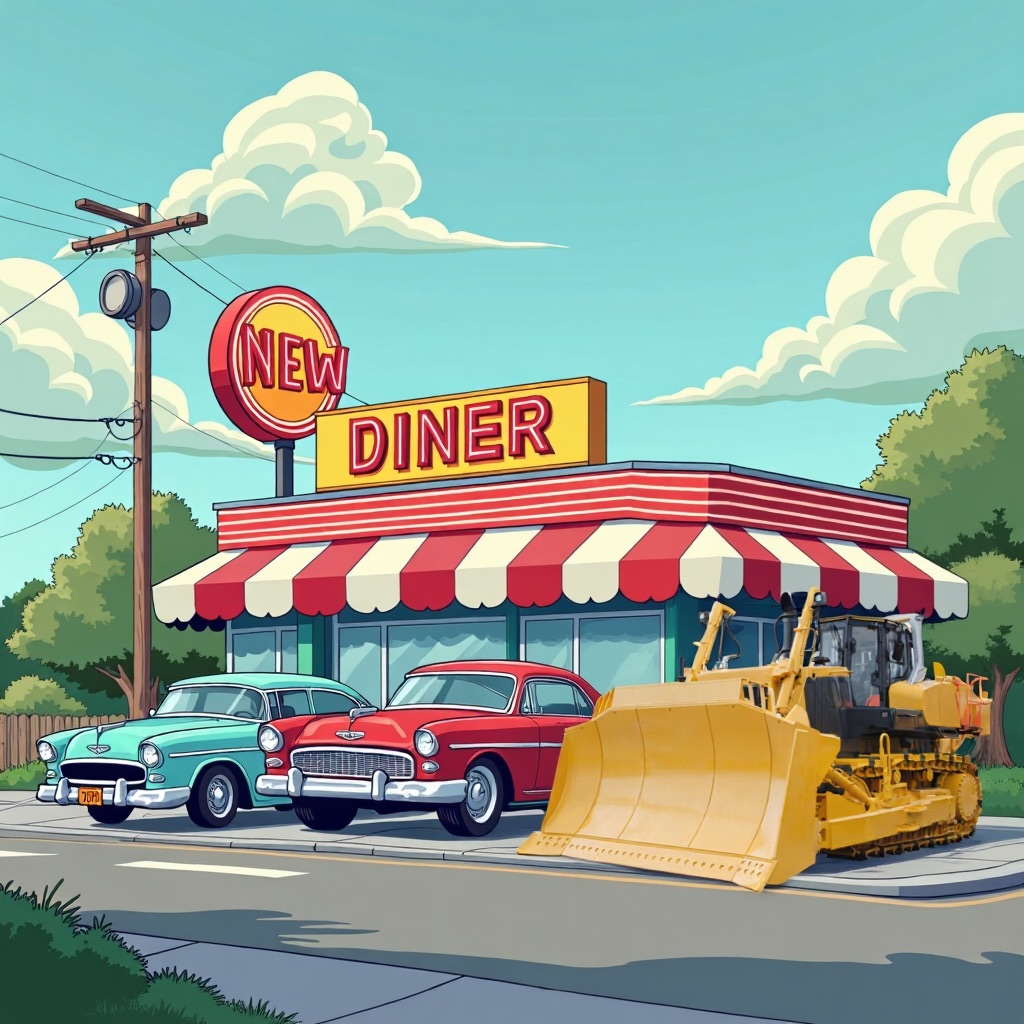}
        \\

        \includegraphics[width=0.325\linewidth]{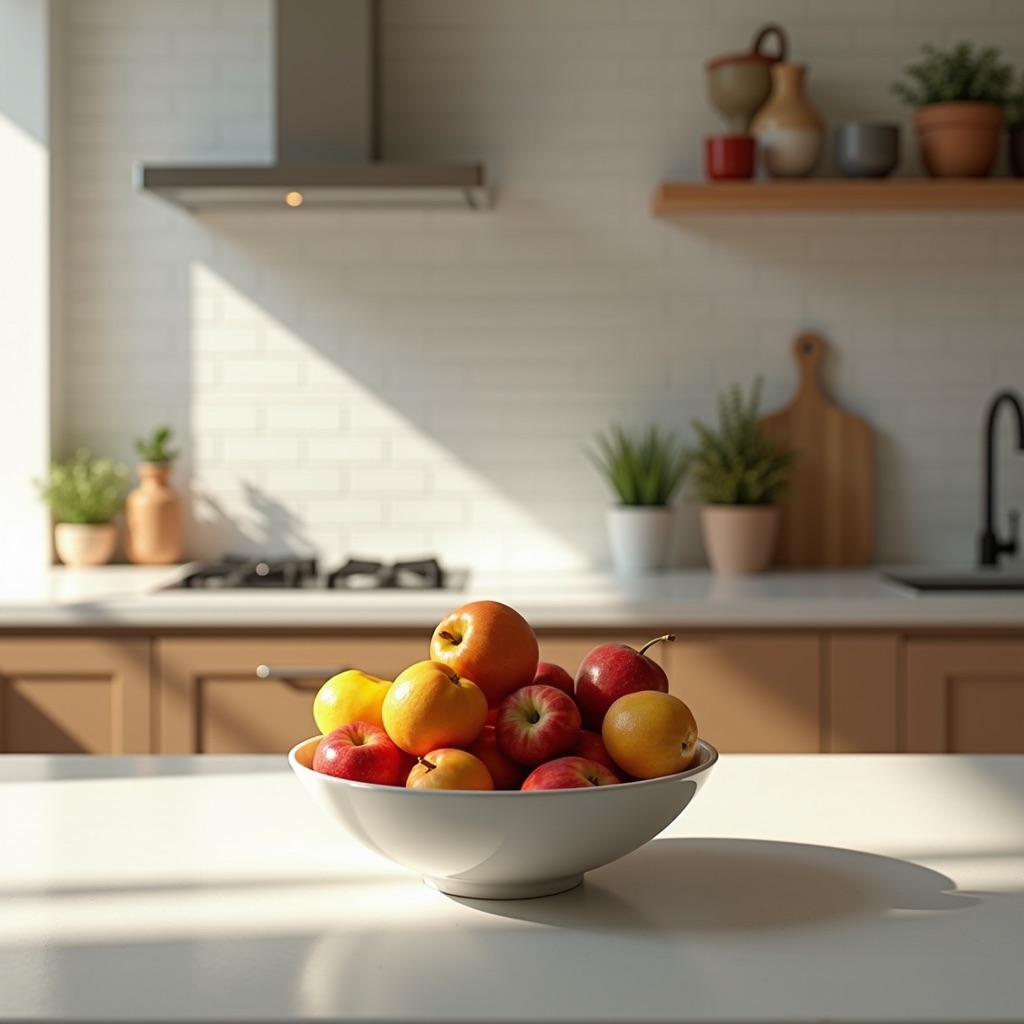} &
        \includegraphics[width=0.325\linewidth]{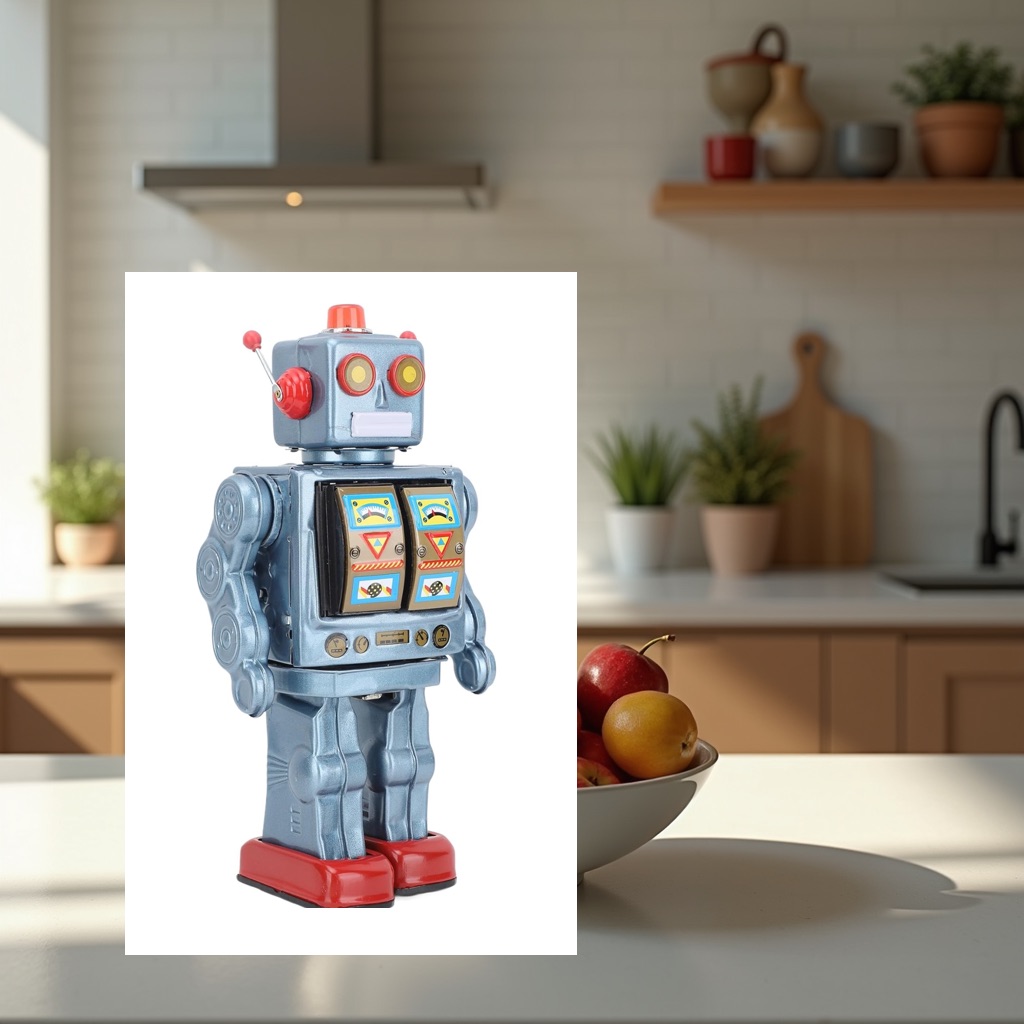} &
        \includegraphics[width=0.325\linewidth]{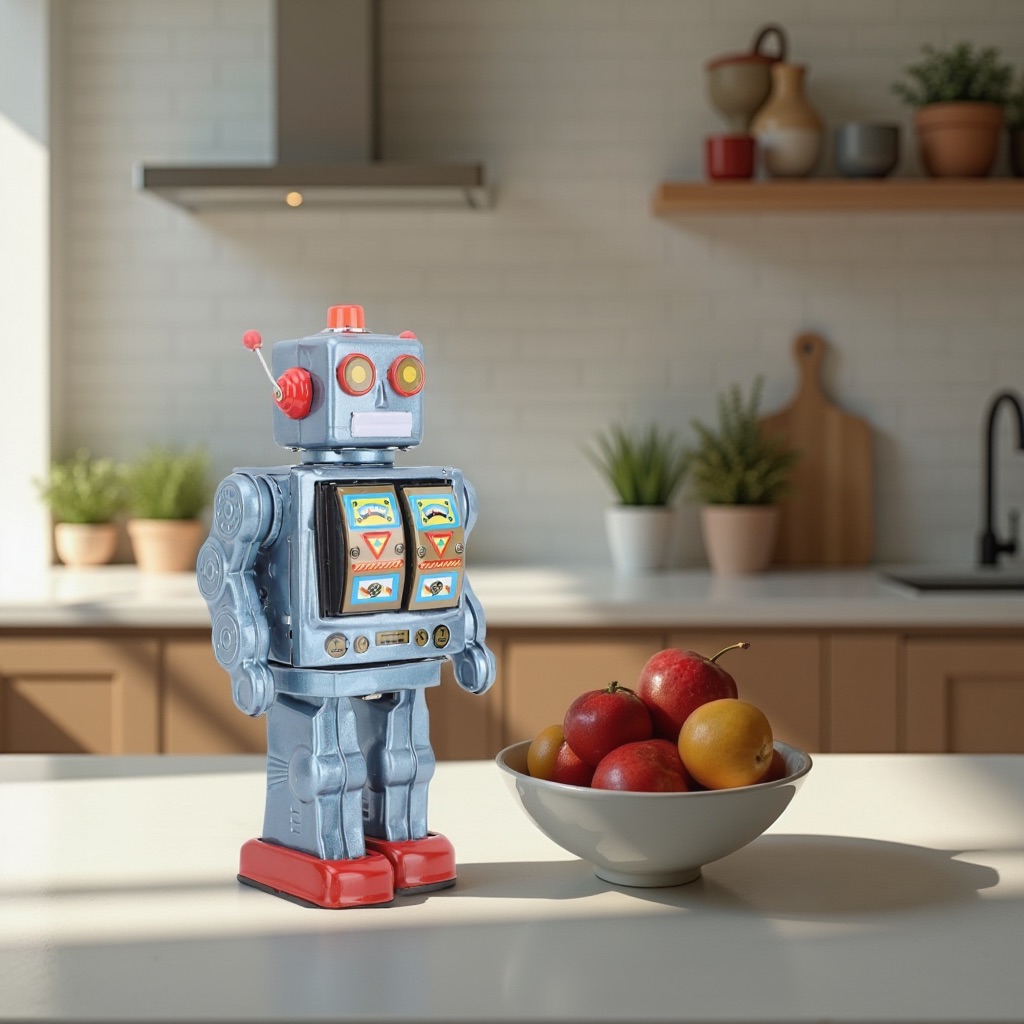}
        \\

        \includegraphics[width=0.325\linewidth]{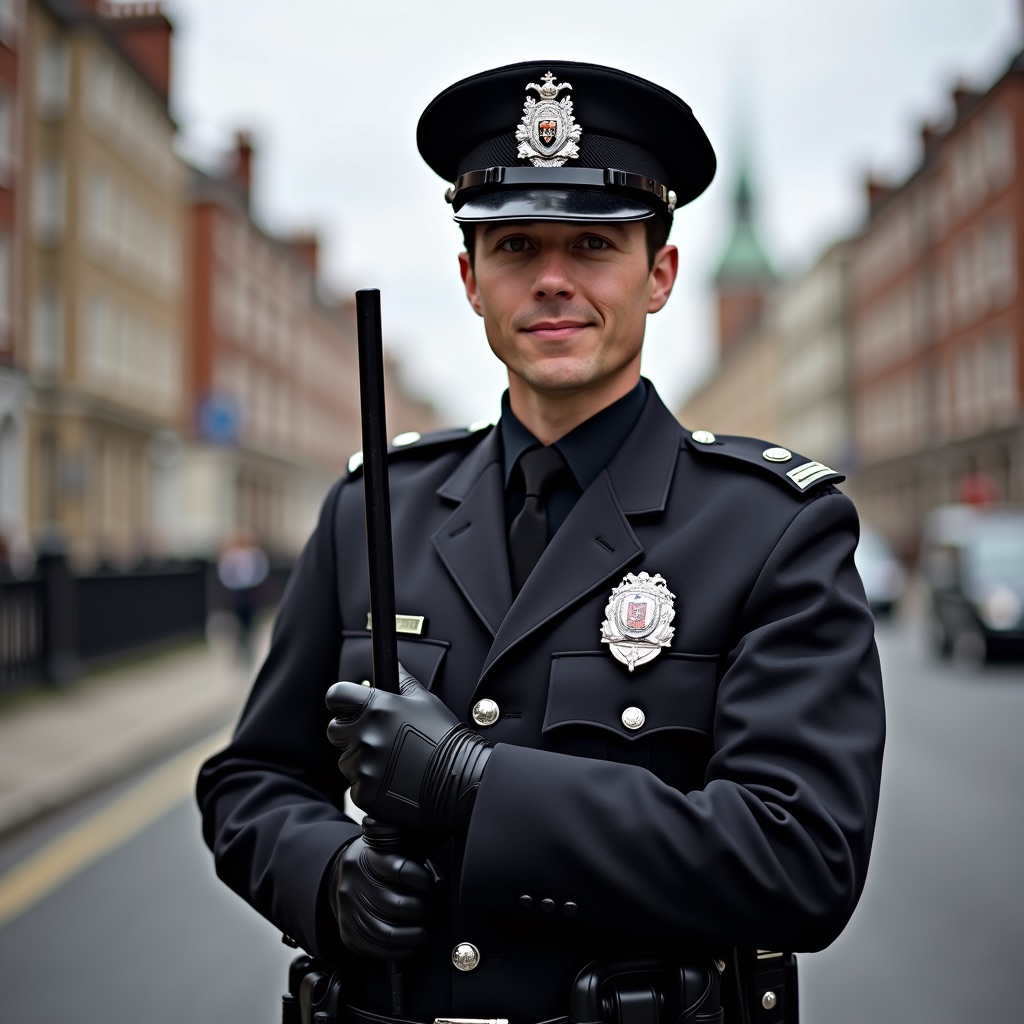} &
        \includegraphics[width=0.325\linewidth]{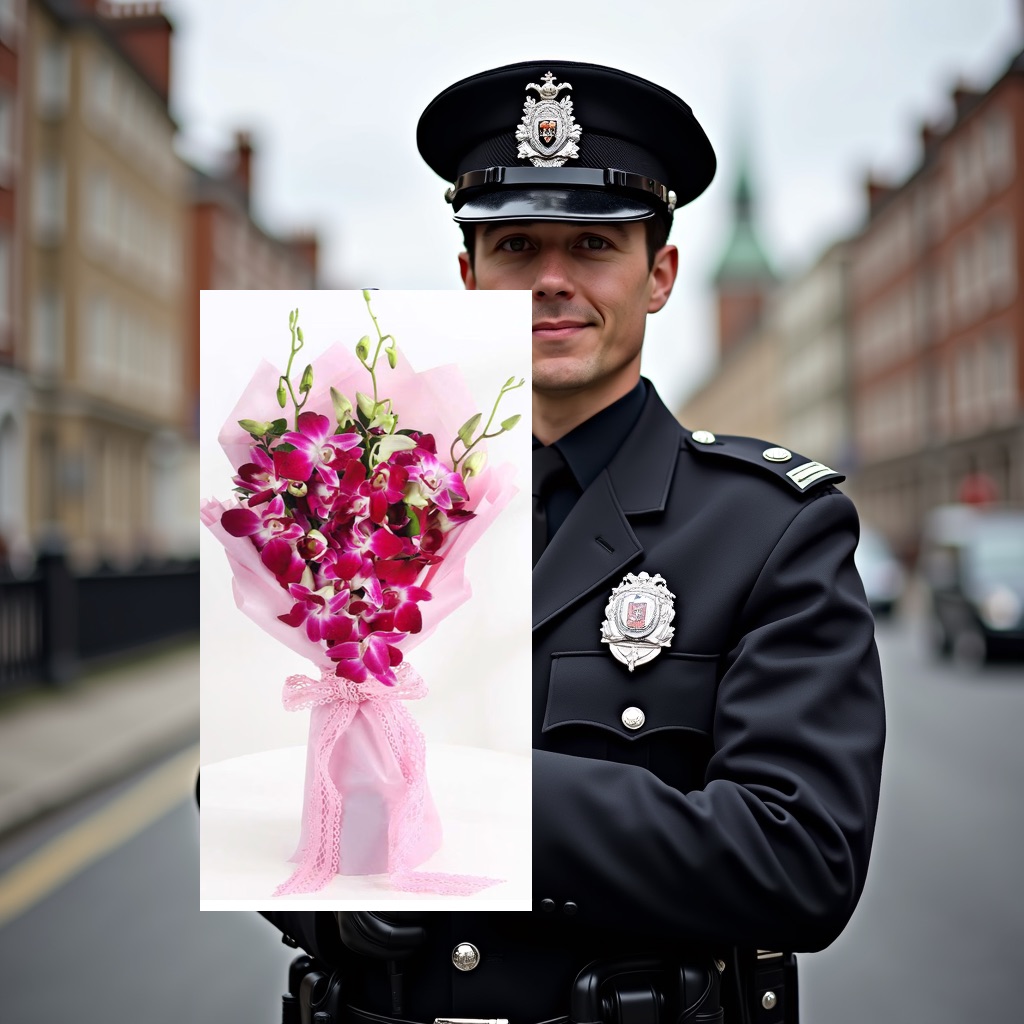} &
        \includegraphics[width=0.325\linewidth]{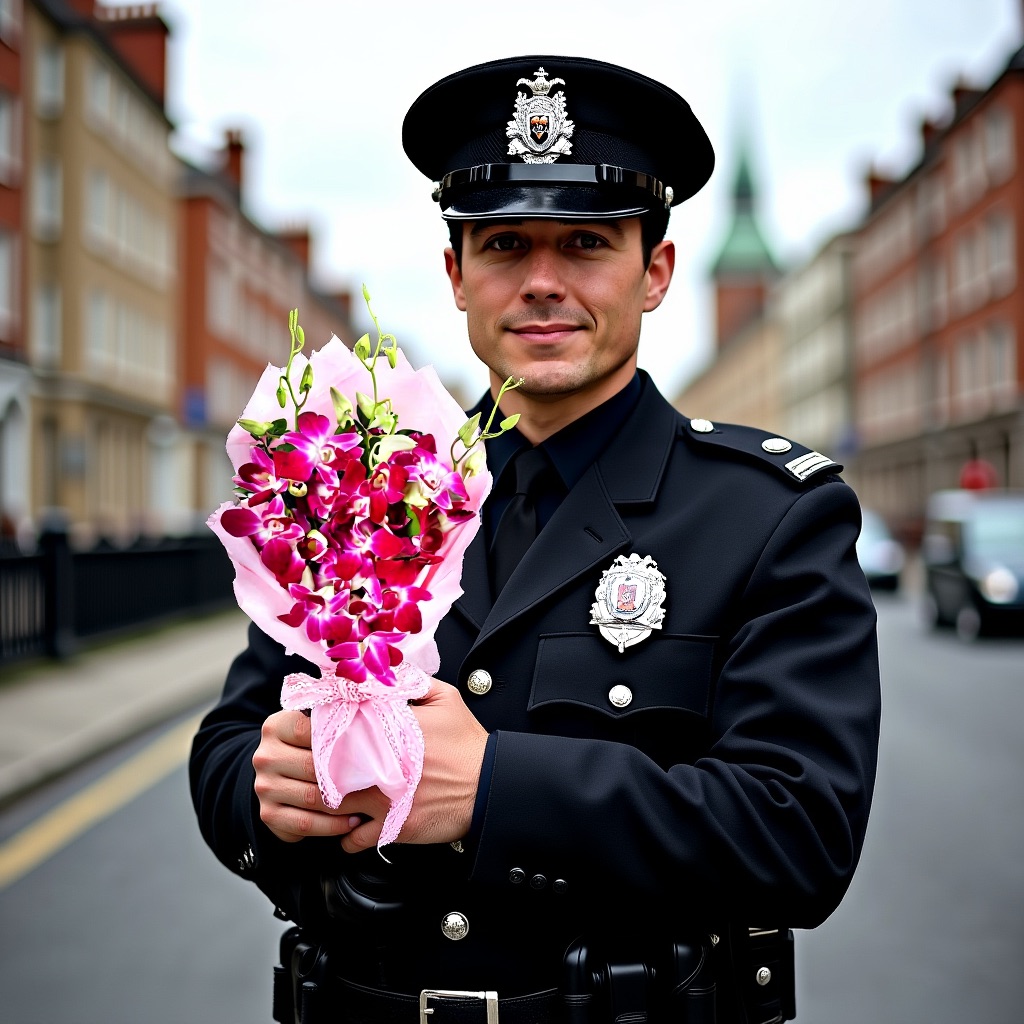}
        \\

        \includegraphics[width=0.325\linewidth]{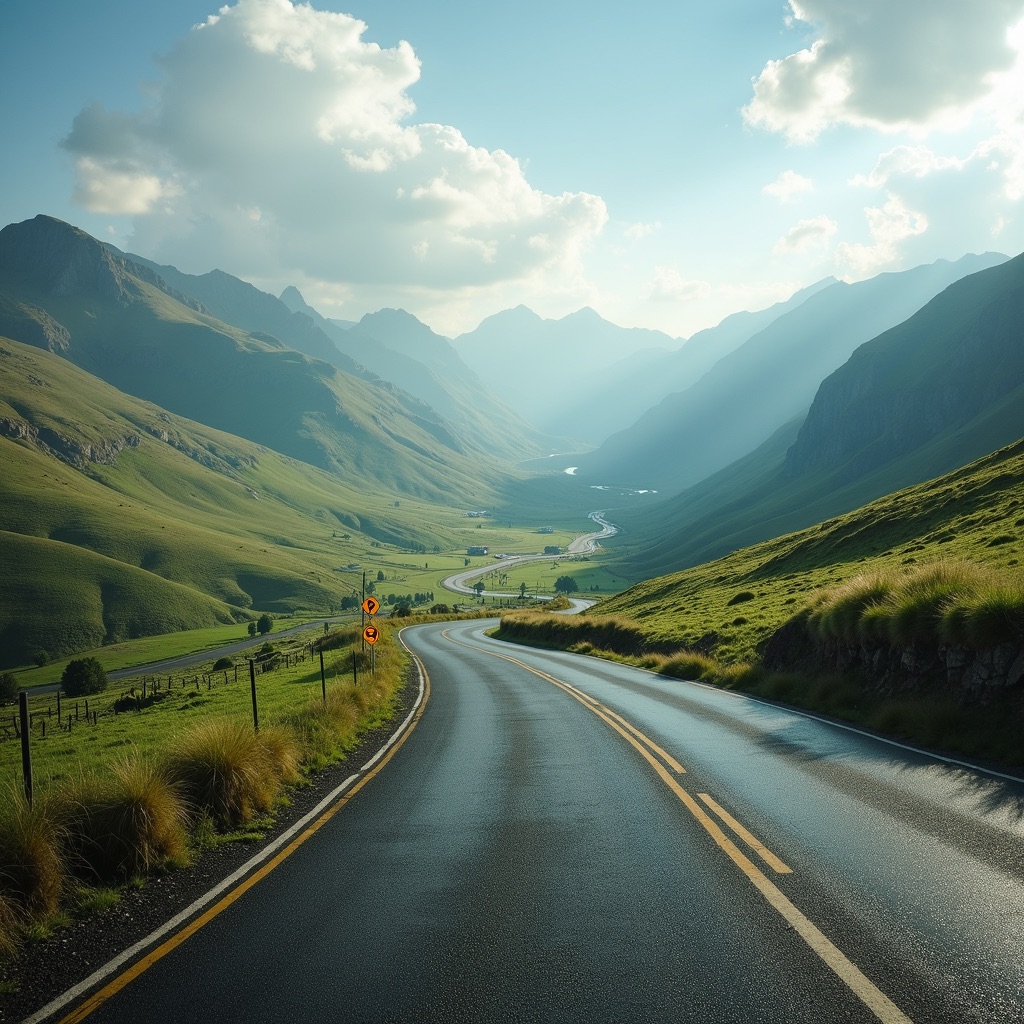} &
        \includegraphics[width=0.325\linewidth]{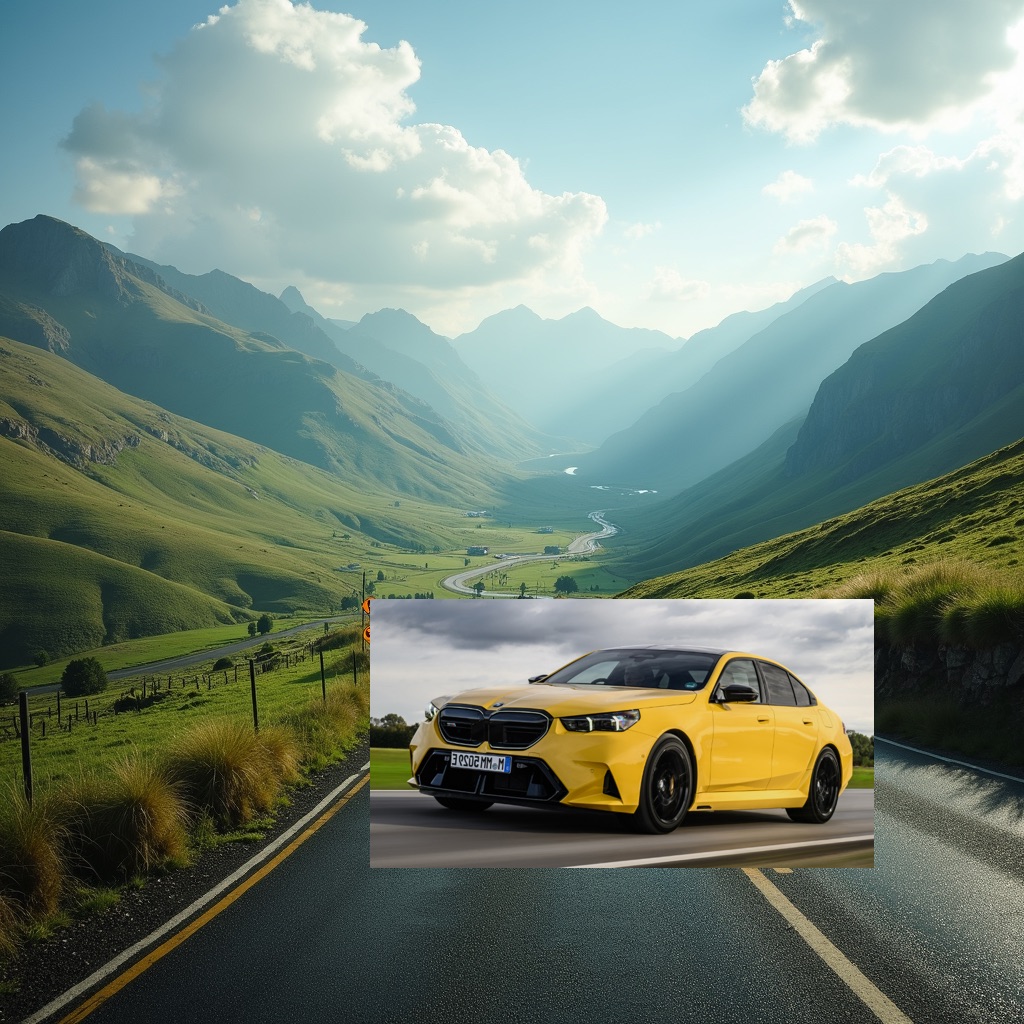} &
        \includegraphics[width=0.325\linewidth]{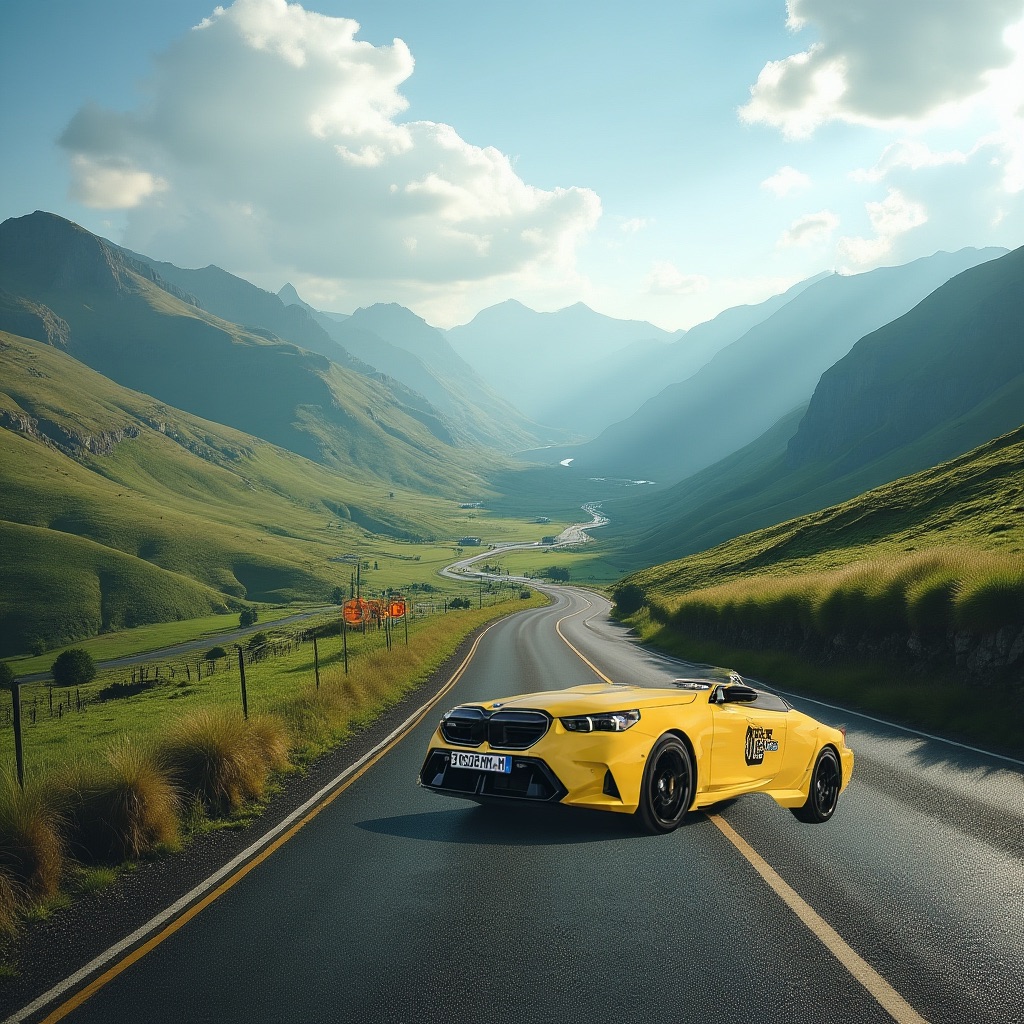}
        \\
    \end{tabular}
    \caption{
    \textbf{Limitations.} 
While our method achieves strong semantic blending and identity preservation, it exhibits limited stylization flexibility (top row), \new{limited ability to make global lighting changes on pasted objects (second row),} struggles with occlusions (third row), and has reduced capacity to accommodate large pose changes (bottom row). We also inherit characteristic artifacts from FLUX Kontext, such as slight enlargement and contrast shifts in preserved regions (middle row).
    }
    \label{fig:supp_limitations}
\end{figure}

\new{
\subsection{Bounding Box vs Segmented Mask Conditions}
To further examine our choice of bounding-box crops over segmented masks, we experimented with both input types by using the ``rembg'' model (available \href{https://github.com/danielgatis/rembg/blob/main/onnxruntime-installation-matrix.png}{here}) to segment the inserted objects or parts in our dataset, creating a segmented variant of the benchmark that we compare against the original bounding-box version. We present representative examples in Figure~\ref{fig:rembg}. In the first two examples, where the edit replaces an existing object, segmented inputs create an ambiguous preservation signal: because the pasted object does not fully override the original one, both our method and the base model struggle to determine what should be preserved. This is evident in the swan-over-duck and pomegranate-over-balloon examples, where segmented inputs lead to failures, while bounding-box inputs provide a clearer editing signal and enable more coherent harmonization. In principle, one could ask the user to additionally specify both the object to insert and the object to remove, but this would make the workflow more cumbersome and may introduce further errors when these annotations are inaccurate or incomplete. In the third example, both our method and Kontext yield reasonable results with the bounding box inputs. However, segmented inputs makes FLUX Kontext more prone to neglect, as if blending were already complete, while allowing our method to preserve slightly more of the original background. Overall, these comparisons highlight the trade-off between tighter background preservation and a clearer editing signal, supporting our use of bounding-box crops as a simple, intuative and robust default.}

\subsection{Limitations}
\label{sec:supp_limitations}
While our approach enables robust and intuitive text-free image editing, there are several limitations to consider. 
First, our strong emphasis on identity preservation in salient regions often results in limited stylization flexibility. As a consequence, the visual style of the pasted object may remain inconsistent with the target scene, as seen in the first row of Figure~\ref{fig:supp_limitations}, where the bulldozer blends spatially but retains a mismatched visual style relative to the illustration-like background. \new{Similarly, while our method can make lighting changes to the scene in order to blend in cropped objects, its ability to make drastic global lighting changes to inserted objects is limited by its identity preserving objective. This is can be seen in the ``robot'' example in the second row of the figure, in which our method generates an acceptable shadow for the inserted robot, however the lighting on the robot itself does not match the scene's lighting direction and intensity.}

Second, our method struggles with occlusions introduced by the pasted object. When important regions of the base image are covered, such as the police officer’s gloves in the middle example, the model cannot recover or reason about the hidden content, leading to diminished identity preservation in the final output. Leveraging information from the occluded base image region remains an important direction for future work.

Third, our method has limited ability to accommodate significant pose changes. Since pose is partially preserved as part of the object identity, mismatches between the object and scene geometry can lead to unnatural warping or visible artifacts. This is evident in the bottom example, where the car is forced into a perspective that does not align naturally with the road geometry.

Finally, as our approach builds on FLUX Kontext, we inherit some of its characteristic limitations. These include slight enlargement of preserved regions and increased contrast (see the middle example in Figure \ref{fig:supp_limitations}), which can introduce subtle distortions even when identity retention is desired.

\begin{table}[t]
\caption{\new{\textbf{Runtime Comparison}. Average runtime per sample for baseline methods tested in this work. All runtimes were measured on a single NVIDIA A100 80GB GPU.}}
\centering
\small
\setlength{\tabcolsep}{8pt}
\begin{tabular}{lc}
\toprule
\textbf{Method} & \textbf{Average Runtime} \\
\midrule
TFICON & 21s \\
SwapAnything & 1m 50s \\
SwapAnything-DB &  18m 12s \\
AnyDoor & 39s, \\
ObjectStitch & 15s \\
MagicFixup & \textbf{6s} \\
Qwen-Image-Edit & 2m 3s \\
FLUX-Kontext & 57s \\
\midrule
\textbf{LooseRoPE (Ours)} & 2m 48s \\
\bottomrule
\end{tabular}
\label{tab:runtime}
\end{table}

\new{
\subsection{Method Runtimes}
We report the average runtime per sample for all baseline methods in Table \ref{tab:runtime}. SwapAnything-DB is the most computationally intensive method (18m 12s) as it requires a DreamBooth-based concept inversion to preserve the identity of the inserted objects. Conversely, training-based methods such as MagicFixup and ObjectStitch are significantly faster (6s and 15s, respectively); however, this speed comes at the expense of generalization outside the training distribution, as shown by their lower quantitative performance in Table \ref{tab:main_comparison}. Our method, LooseRoPE, has an average runtime of 2m 48s, which includes the VLM steering phase. This steering process takes approximately 18s per iteration with a maximum limit of four attempts. Notably, 56\% of cases succeed on the first attempt, while only 14\% reach the maximum iteration limit without success.
}

\section{Implementation Details}
\label{sec:supp_details}
\subsection{LooseRoPE}
\paragraph{Base Model.}{We base our method on the \texttt{black-forest-labs/ FLUX.1-Kontext-dev} image editing diffusion model, specifically using the distribution available on HuggingFace at \href{https://huggingface.co/black-forest-labs/FLUX.1-Kontext-dev}{this URL}. For all experiments and results presented in this paper we use a crudely edited image and the base prompt: \emph{``blend the cropped objects into the image in a convincing manner without changing the style of the image''} as input. We use the default guidance scale of $2.5$ and no negative prompts for the default $28$ reverse diffusion steps.
}

\paragraph{Saliency Estimation.}{
 As discussed in Section 3.2 of the main paper, at this stage we evaluate the saliency distribution map of the crop area by extracting feature activations from a pre-trained instance detection network. Specifically, we set all pixels outside of the crop mask $M$ in the input image to $[0,0,0]$ and extract features from the first and second layers of the \texttt{COCO}-\texttt{InstanceSegmentation/mask\_ rcnn\_R\_50\_FPN\_3x} model (available in the \href{https://github.com/facebookresearch/detectron2}{Detectron2 distribution}) when passing the masked image through it. The features are rescaled to fit the latent image resolution of $64\times64$ and averaged with eachother, after which we pass the resulting map through a $2D$ Gaussian filter with kernel size of $5\times 5$ and $\sigma_x = \sigma_y = 1.1$ to obtain the saliency map.
}

\begin{figure}
    \centering
    \includegraphics[width=0.95\linewidth]{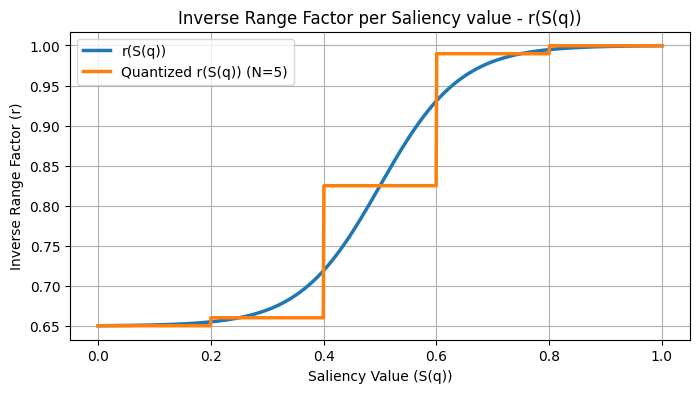}
    \caption{Inverse Range factor $r$ as a function of a query's saliency value $S(q)$. In practice, we quantize saliency values to $N=5$ different values, resultin in the step function shown in \textcolor{orange}{orange}.}
    \label{fig:supp_rsq}
\end{figure}

\begin{figure}
    \centering
    \includegraphics[width=0.95\linewidth]{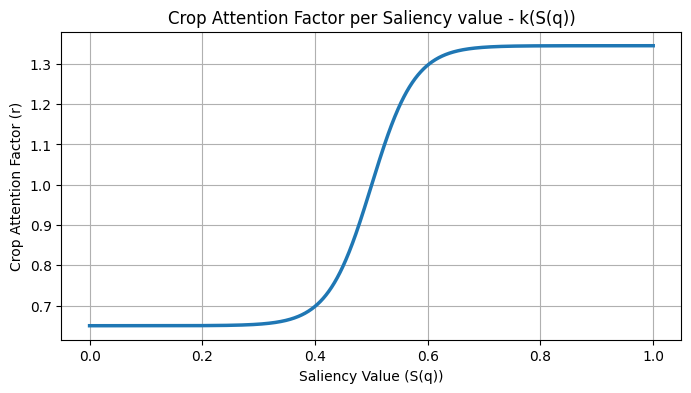}
    \caption{Attention Scale Factor $k$ as a function of a query's saliency value $S(q)$.}
    \label{fig:supp_ksq}
\end{figure}

\paragraph{Content-Aware Attention Manipulation.}{
Given the saliency estimation map $S$, in this stage we modify the attention distribution and RoPE parameters for queries within the crop mask $M$ according to their saliency values. This mechanism is summarized in Algorithm 1 in the main paper. In this algorithm the inverse range value and attention scale factor assigned to each query in $M$ are defined by the $r(S(q))$ and $k(S(q))$ functions accordingly. These functions can be parameterized as:

\begin{equation}
    \left(\frac{\tanh(S(q)*G)}{2} + \frac{1}{2}\right)*(v_{max}-v_{min}) + v_{min}
\end{equation}

\noindent with $v_{max}$ and $v_{min}$ being the maximal and minimal values the function can reach and $G$ being a constant \emph{steepness factor}. For $r(S(q))$ we use $v_{max} = r_{high} = 1.0$, $v_{min} = r_{low}= 0.65$ and $G = 3.5$. For $k(S(q))$ we use $v_{max} = k_{high} = 1.34$, $v_{min} = k_{low} =0.65$ and $G = 6.5$. In our algorithm, each different value $k(S(q))$ requires rotating the query $q$ and $K_{in}$ accordingly. As such, this process can become very computationally inefficient. To overcome this, before passing $S(q)$ through $r(S(q))$ and $k(S(q))$ we quantize it to $N=5$ possible values evenly split between $0$ and $1$, resulting in $r(S(q))$ and $k(S(q))$ functioning as step functions. We plot $r(S(q)$ before and after quantization in Figure \ref{fig:supp_rsq} and $k(S(q))$ in Figure \ref{fig:supp_ksq}.

Our algorithm operates on each of FLUX Kontext's $58$ attention layers over the first $22$ of $28$ diffusion timesteps. Over time, we gradually relax inverse range and attention scaling factors towards their equivalent value in the default FLUX Kontext model - $1.0$. Specifically, at timestep $10$ we relax $r_{low}$ to $0.9$, $k_{low}$ to $0.76$ and $k_{high}$ to $1.24$ and at timestep $18$ we relax $r_{low}$ to $1.0$, $k_{low}$ to $0.84$ and $k_{high}$ to $1.17$.
}
\new{
\paragraph{LooseRoPE on Qwen-Image-Edit}
In Figure 6 of the main manuscript we showcase results of our method when implemented on the Qwen-Image-Edit base model. To obtain these results we used the \texttt{Qwen/Qwen-Image-Edit} model available in its \href{https://huggingface.co/Qwen/Qwen-Image-Edit}{HuggingFace diffusers distribution}. Using its default $5.0$ guidance scale and $50$ diffusion steps. While Qwen-Image-Edit and FLUX Kontext share the same architecture and general functionality, they were trained differently and use different inference processes, as such we changed the following LooseRoPE hyper-paramters to optimize performance on this model: $r_{low} = 0.75$, $k_{high} = 1.36$, $k_{low} = 0.75$, $\lambda = 1.1$. %
}
\paragraph{VLM Based Parameter Steering}{
    
We employ a Vision-Language Model (VLM) to dynamically assess harmonization quality during inference and adjust attention modification parameters accordingly. The VLM evaluates the $x_0$ prediction at a specific timestep during the denoising process and classifies the harmonization quality of the prediction into one of three categories: \textit{Success}, \textit{Neglect}, or \textit{Suppression} (See section 3.2 of the main paper).

The model used for this task is \texttt{Qwen3-VL-4B-Instruct} (available on \href{https://huggingface.co/Qwen/Qwen3-VL-4B-Instruct}{HuggingFace}), a 4-billion parameter vision-language model. At runtime, we provide the model with the $x_0$ prediction at the current timestep, the input image and 6 in-context examples (2 \textit{Success}, 2 \textit{Neglect}, and 2 \textit{Suppression}), resulting in 14 total images in the VLM input (each example includes an input image and an $x_0$ prediction).
We instruct the VLM first with the definitions for each possible case and then with guiding questions for correctly identifying the setting, alongside common cases it might encounter. We include the full instruction prompt given to the VLM alongside this manuscript (see \texttt{vlm-instruction.txt}). %

VLM evaluation is triggered at a configurable timestep during the denoising process; in our experiments, this was set to $t_s=2$. 
This provides sufficient signal about the harmonization progress while requiring minimal backtracking in the case of a failed outcome. The VLM performs a single inference per timestep evaluation (one try), generating up to $\text{2048}$ new tokens which are subsequently parsed to extract the verdict. In addition to the final verdict, the VLM was also instructed to provide its reasoning behind it. While this reasoning is not used in any way by our method, we found it useful for development and debugging purposes. Given the VLM's verdict, unless it determined a successful outcome, we either scale up or scale down the saliency map $S$. Specifically, we define $S$ as: 
\begin{equation}
    S = \max\{\min\{\lambda\cdot S_{original},1\},0\}
\end{equation}
\noindent with $S_{original}\in[0,1]$ being the saliency extracted in the ``Saliency Estimation'' stage (detailed in the previous paragraph) and $\lambda$ being a scaling factor set to $0.83$ by default. If the VLM determines \textit{Neglect} $\lambda$ is decreased by a constant of $0.045$, thereby down-scaling the saliency and as a result further encouraging blending. If \textit{Suppression} is determined $\lambda$ is increased by $0.05$, up-scaling the saliency and encouraging preservation as a result.
We limit the VLM steering attempts to a maximum of 4 tries. 

}

\subsection{Experiments}
\subsubsection{Baselines}
In this section we communicate the technical details regarding each of the methods used in comparison to ours throughout our work. 

\paragraph{AnyDoor.}
We evaluate AnyDoor using the official pre-trained model available on AnyDoor's official \href{https://github.com/ali-vilab/AnyDoor}{GitHub repository}. We provide AnyDoor with an image of the inserted object (or sub-object) by using the crop mask $M$ to crop it from the input image. We then run AnyDoor with its default diffusion parameters (DDIM sampling for $50$ steps and a $5.0$ guidance scale).

\paragraph{SwapAnything.} We evaluate our method against SwapAnything in two distinct configurations: non-personalized and personalized. %
For the non-personalized variant (SwapAnything's default configuration), we employ the \texttt{Qwen2.5-VL-3B-Instruct} model (using its  \href{https://huggingface.co/Qwen/Qwen2.5-VL-3B-Instruct}{HuggingFace distribution}) to identify the subject within the crop area. Subsequently, we construct a prompt based on the identified subject, adhering to the recommendations for general object insertion provided in SwapAnything's official  \href{https://github.com/eric-ai-lab/swap-anything}{GitHub repository}. 

For the personalized variant, we train a separate DreamBooth \cite{ruiz2023dreamboothfinetuningtexttoimage} personalized model (using DreamBooth's \href{https://huggingface.co/docs/diffusers/en/training/dreambooth}{HuggingFace distribution} with default settings) for each sample, utilizing the single cropped source object as the sole training instance. The class name required for training (e.g. ``dog'', ``chair'') is derived automatically from the VLM-based subject identification step outlined above.
The resulting model checkpoint and its unique identifier token are then employed during inference. 

Other than using different inputs (either a textual description or a personalized model) both modes use the default settings provided in SwapAnything's repository.

\paragraph{FLUX Kontext.}
We run the \texttt{black-forest-labs/FLUX.1-Kontext -dev} model (same as our base model) in its \href{https://huggingface.co/black-forest-labs/FLUX.1-Kontext-dev}{HuggingFace diffusers distribution} with the crudely edited image as input and the base prompt - \emph{``blend the cropped objects into the image in a convincing manner without changing the style of the image''}. We use the default diffusion settings ($2.5$ guidance scale, $28$ steps and no negative prompts).

\paragraph{Nano Banana.}
Results for Nano Banana were acquired from the \href{https://gemini.google.com/app}{Gemini interface}. Each image was generated in a new chat in which we provided the crudely edited input image and instructed the model with the prompt with the same prompt we use in our method: \emph{``blend the cropped objects into the image in a convincing manner without changing the style of the image''}.

\paragraph{TF-ICON.} We run the pipeline provided in TF-ICON's official \href{https://github.com/Shilin-LU/TF-ICON}{GitHub repository}. We provide the model with an image of the foreground object or sub-object using the crop mask $M$ and an estimated foreground mask (within the crop region) extracted with the \texttt{COCO}-\texttt{InstanceSegmentation/mask\_rcnn\_R\_50\_FPN\_3x} model. If the model did not detect a foreground object in the crop area, we assumed the entire crop region is a foreground object. We run the model in its ``cross-domain'' setting as we empirically found it to perform better in our setting.

\paragraph{Qwen-Image.} We run the \texttt{Qwen/Qwen-Image-Edit} model available in its \href{https://huggingface.co/Qwen/Qwen-Image-Edit}{HuggingFace diffusers distribution}. Similarly to how we run FLUX Kontext, with provide the crudely edited image as input and uset the base prompt - \emph{``blend the cropped objects into the image in a convincing manner without changing the style of the image''}. We use the default $5.0$ guidance scale, $50$ inference steps and no negative prompts.

\paragraph{ObjectStitch.} We run the pipeline provided in ObjectStitch's unofficial \href{https://github.com/bcmi/ObjectStitch-Image-Composition}{GitHub repository}. Similarly to TF-ICON, we provide the model with an image of the foreground object or sub-object using the crop mask $M$ and an estimated foreground mask (within the crop region) extracted with the \texttt{COCO}-\texttt{InstanceSegmentation/mask\_rcnn\_R\_ 50\_FPN\_3x} model, passing it the entire crop region as foreground if the model fails to detect an object. We use the default settings defined in the repository.
\new{
\paragraph{MagicFixup.} We run the default pipeline available in MagicFixup's official \href{https://github.com/adobe-research/MagicFixup}{GitHub repository.} using the original (un-edited) images as the input reference image and the crudely edited images as the edited inputs. For the masks we use the alpha channel of the crudely edited images, which indicates where the original edited object was present prior to being translated to another region of the image.
}

\subsubsection{Metrics}
We now detail the technical details and settings used when calculating the quantitative metrics utilized in our work.

\paragraph{CLIP-IQA.}
We utilize CLIP-IQA using the default settings provided in the implementation available on CLIP-IQA's official \href{https://github.com/IceClear/CLIP-IQA}{GitHub repository}.

\paragraph{LPIPS.}
We compute Perceptual Image Patch Similarity (LPIPS) scores
using the official pytorch implementation (available on \href{https://github.com/richzhang/PerceptualSimilarity}{GitHub}) using the traditional VGG \cite{simonyan2014very} features. We calculate the both the perceptual similarity of the entire output image to the entire input image (denoted as ``LPIPS (Full)'' and the similarity of the estimated foreground object (or sub-object) in the input image to the matching area in the output image (denoted as ``LPIPS (FG)''. For the latter, we use an estimated foreground mask (within the crop region) extracted with the \texttt{COCO}-\texttt{InstanceSegmentation/mask\_rcnn\_R\_50\_FPN\_3x} model, and assume the foreground is the entire crop region if the model fails to detect an object.

\begin{figure}
    \centering
    \includegraphics[width=0.98\linewidth]{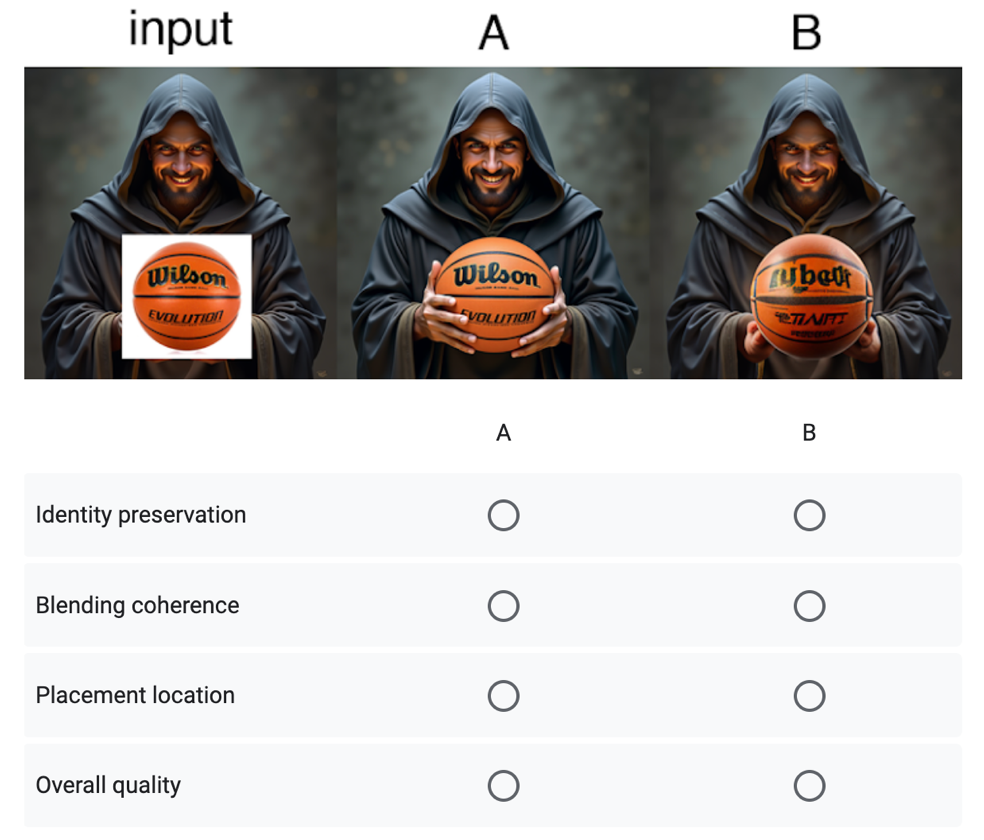}
    \caption{A sample comparison shown to users as part of our user study. Each user was shown $20$ such comparisons.}
    \label{fig:supp_userstudy}
\end{figure}

\paragraph{User Preference.}
To evaluate the perceptual quality of our edits, we conducted a user study utilizing three distinct forms, each containing 20 comparisons. We compared our method against four baselines: Kontext, TF-ICON, SwapAnything, and Anydoor, allocating 5 examples per method within each form. Participants rated the images based on identity preservation, blending coherence, placement location accuracy, and overall quality. In total, the study encompassed 60 unique comparisons that were sampled at random from the dataset ($3 \text{ forms} \times 20 \text{ comparisons}$). The exact instruction given to the users in the start of each survey is as follows: 
\begin{verbatim}
In this study, you will compare two dif-
ferent editing methods (labeled A and B).
Both methods aim to apply the edit given
on the left (input image), so that the 
crop will be inserted into the image in 
a convincing manner.
Some of the questions involve a transla-
tion task, in which we cut a region and 
move it to another location in the image.

The layout of every image provided is
input image, method a, method b

We ask you to judge which method does 
better, by answering four questions:

    1.Identity preservation - Which edit 
    better preserves the identity of the 
    pasted subject?

    2.Blending coherence - Which edit 
    executes the blend in a more convin-
    cing and coherent manner, without 
    artifacts?

    3.Placement location - Is the new 
    subject in the image located and or-
    iented correctly?

    4.Overall quality - Which edit do 
    you prefer overall?
\end{verbatim}

\noindent A sample comparison shown to users is presented in Figure \ref{fig:supp_userstudy}. Overall, the user study was answered by $27$ users, resulting in a total of $540$ responses per category.

\subsection{Benchmark}
Our benchmark consists of 150 examples in total, spanning a wide variety of settings, styles and compositions, each defined by a base image and a crudely edited version of it. 60\% of base images were synthesized and 40\% taken from the web. As for the crops pasted on the base images- 13\% originated from the base image itself, with the rest inserted from off-the-web images.

\end{document}